 \documentclass[final,5p,times,twocolumn,authoryear]{elsarticle}

\usepackage{amssymb}
\usepackage{lipsum}

\usepackage{siunitx}

\usepackage{aas_macros}
\usepackage{newtxtext,newtxmath}

\usepackage[T1]{fontenc}

\usepackage{graphicx}	
\usepackage{amsmath,mathtools}	
\usepackage{ragged2e}
\usepackage[table,xcdraw]{xcolor}
\usepackage{cprotect}
\usepackage{subfig}
\usepackage{float}
\usepackage{multirow}
\usepackage{placeins}
\usepackage{threeparttable}
\usepackage{xcolor}
\usepackage{bm}			
\usepackage{ulem}
\usepackage{xcolor}

\usepackage{url}
\usepackage{hyperref}

\usepackage{soul}
\def\red{\color{black}}

\journal{High Energy Astrophysics}

\begin{document}

\begin{frontmatter}



\title{{\textit{AstroSat} Observations of Stellar Mass Black Hole X-ray Binaries: Towards a Unified Accretion Disc Configuration}}

\author{Anuj Nandi$^{a}$}\ead{anuj@ursc.gov.in}
\author{Santabrata Das$^{b}$\corref{cor1}}\ead{sbdas@iitg.ac.in}
\cortext[cor1]{Corresponding author}
\author{Seshadri Majumder$^{c}$}
\author{Prajjwal Majumder$^{d}$}
\author{Sreetama Das Choudhury$^{b}$}
\author{Bhuvana G. R.$^{e}$}
\author{Ramiz Aktar$^{f}$}
\author{Samir Mandal$^{g}$}
\author{Broja G. Dutta$^{d}$}
\author{Sreehari H.$^{h}$}
\author{Geethu Prabhakar$^{i}$}
\author{Aneesha Unnikrishnan$^{j}$}
\author{Ravishankar B. T.$^{a}$}
\author{Swapnil Singh$^{a}$}
\author{Ankur Kushwaha$^{a}$}
\author{Athulya M. P.$^{k}$}
\author{Blessy E. Baby$^{l}$}
\author{Naﬁsa Aftab$^{m}$}
\author{Vivek Kumar Agrawal$^{a}$}
\author{M. C. Ramadevi$^{n}$}
\author{Tilak Katoch$^{o}$}
\author{D. Radhika$^{p}$}
\author{Nirmal Kumar Iyer$^{q}$}
\author{A. R. Rao$^{o}$}
\author{Kulinder Pal Singh$^{o,r}$}
\author{S. N. Tandon$^{s}$}
\author{J. S. Yadav$^{o,t}$}
\author{S. Seetha$^{m}$}

\affiliation[]{addressline={Space Astronomy Group, ISITE Campus, U. R. Rao Satellite Centre, Outer Ring Road, Marathahalli, Bangalore, 560037}, country={India}}
\affiliation[]{addressline={Department of Physics, Indian Institute of Technology Guwahati, Guwahati, 781039}, country={India}}
\affiliation[]{addressline={Department of Physics, Kyoto University, Kitashirakawa-Oiwake-cho, Sakyo-ku, Kyoto, 606-8502}, country={Japan}}
\affiliation[]{addressline={Department of Physics, Rishi Bankim Chandra College, Naihati, Kolkata, 743165}, country={India }}
\affiliation[]{addressline={Institute for Advanced Study, Gifu University, 1-1 Yanagido, Gifu, 501-1193}, country={Japan }}
\affiliation[]{addressline={Institute of Astronomy and Astrophysics, Academia Sinica, Taipei 10617}, country={Taiwan, R.O.C.}}
\affiliation[]{addressline={Department of Earth \& Space sciences, Indian Institute of Space science and Technology, Thiruvananthapuram, 695547}, country={India }}
\affiliation[]{addressline={Department of Electronics and Communication Engineering, Chinmaya Vishwa Vidyapeeth, Ernakulam, 686667}, country={India}}
\affiliation[]{addressline={Independent Researcher, Kottarakkara, Kollam, 691531}, country={India}}
\affiliation[]{addressline={Litmus7 Systems Consulting Private Limited, Phase III, Technopark, Kazhakkoottam, Trivandrum, 695582}, country={India}}
\affiliation[]{addressline={Centre for Astrophysics Research, University of Hertfordshire, College Lane, Hatfield, AL10 9AB}, country={UK }}
\affiliation[]{addressline={Compunnel Software Group Inc., 501/502, Dr Alvaro Costa Rd, Patto Centre, Panaji, 403001}, country={India }}
\affiliation[]{addressline={Raman Research Institute, C V Raman Avenue, Sadashivanagar, Bangalore, 560080}, country={India }}
\affiliation[]{addressline={Main Campus, U. R. Rao Satellite Centre, Old Airport Road, Bangalore, 560017}, country={India}}
\affiliation[]{addressline={Department of Astronomy and Astrophysics, Tata Institute of Fundamental Research, 1 Homi Bhabha Road, Mumbai, 400005}, country={India}}
\affiliation[]{addressline={Independent Researcher, Kaggadasapura, Bangaluru, 560093}, country={India }}
\affiliation[]{addressline={Excillum AB, Stockholm, 16440}, country={Sweden}}
\affiliation[]{addressline={Indian Institute of Science Education and Research Mohali, Knowledge City, Sector 81, Manauli P.O., SAS Nagar, 140306}, country={India }}
\affiliation[]{addressline={Inter-University Center for Astronomy and Astrophysics, Pune, 411007}, country={India }}
\affiliation[]{addressline={Space, Planetary \& Astronomical Sciences \& Engineering, IIT Kanpur, Kanpur, 208016}, country={India }}


\begin{abstract}

We present a comprehensive review of spectro-temporal properties of $34$ black hole X-ray binaries (BH-XRBs), classified as persistent-cum-variable, persistent, and transient, observed with \textit{AstroSat} during its first decade of operations. The study is based on more than $12$ Ms of observations and leverages mission's unique capabilities of multi-wavelength wide-band spectral coverage (in UV/optical and X-rays) and high temporal resolution ($\sim10~\mu$s). Through broadband spectral and timing analyses, we investigate accretion–ejection dynamics in strong gravity regime of BH-XRBs. Low-frequency Quasi-periodic Oscillations (LFQPOs; $0.1-8.0$ Hz) are detected in several sources and are traced up to energies exceeding $40$ keV in MAXI J1535$-$571 and H 1743$-$322, while Swift J1727.8$-$1613 exhibits the first detection of LFQPOs up to $100$ keV, indicating the dominance of Comptonized emission over thermal disc radiation. In GRS 1915$+$105, we report the discovery of a new variability class ($\tau$), increasing the known classes to fifteen, and detect high-frequency QPOs (HFQPOs; $\sim70$ Hz) in $\gamma$, $\kappa$, $\delta$, $\omega$, and $\tau$ classes. These findings suggest the presence of a `compact' oscillating corona located within $\lesssim 10~r_g$ of the BH. Among persistent sources, we report the first detection of a type-A LFQPO ($\sim2.7$ Hz) in 4U 1957$+$11 and type-B LFQPOs ($4-8$ Hz) in 4U 1630$-$472. Broadband spectra are well described by a combination of thermal disc and non-thermal Comptonization components, enabling detailed investigations of correlations among QPO frequency ($\nu_{\rm QPO}$), time-lag ($t_{\rm d}$), rms variability (rms$_{\rm QPO}\%$), photon index ($\Gamma_{\rm th}$), disc-to-Comptonized flux ratio ($F_{\rm ratio}$), and bolometric luminosity ($L_{\rm bol}$). These relations reveal a strong coupling between accretion disc, Componizing corona, and jet base, and delineate `canonical' spectral states of BH-XRBs, such as LHS, HIMS, SIMS, and HSS based on $\Gamma_{\rm th}$, $F_{\rm ratio}$ and $L_{\rm bol}$. We observe that type-C QPOs exhibit soft-lags dominated by Comptonized emission with occasional lag reversal at high coronal fraction ($F_{\rm Comp}/F_{\rm bol}>0.8$, $F_{\rm Comp}$ being Comptonized flux and $F_{\rm bol}$ is bolometric flux) reinforcing the role of corona in driving variability, whereas type-B QPOs generally exhibit hard-lags at lower coronal fraction $F_{\rm Comp}/F_{\rm bol}<0.5$, indicating to a possible connection with disc-jet coupling. Simultaneous UV and X-ray observations of LMC X$-$3 and MAXI J1820$+$070 further allow to study multi-wavelength SED from the outer accretion disc to the innermost X-ray emitting regions. Collectively, these findings support a unified accretion framework capable of explaining the diverse spectral and timing behaviour of BH-XRBs, further corroborated by accretion disc simulation and radiation driven spectral modelling. Finally, we highlight the limitations and challenges of the present analyses along with the potential aspects of new science with \textit{AstroSat} and future missions to probe accretion physics in extreme gravity regime.

\end{abstract}



\begin{keyword}

accretion, accretion disks -- black hole physics -- radiation mechanisms: general -- stars: black holes -- X-rays: binaries



\end{keyword}

\end{frontmatter}




\section{Introduction}\label{sec1:intro}

Black holes (BHs) serve as natural laboratories for investigating accretion physics and testing general relativity in the strong field regime. The extreme gravitational environments around stellar mass black holes provide a unique avenue for probing relativistic effects \cite[]{Shapiro-Teukolsky1983} through their rich timing and spectral behaviour, thereby offering valuable insights into the coupling between accretion dynamics, radiation processes, and jet formation \cite[and references therein]{White-etal1988, Mirabel-Rodrigue1994, Tanaka-etal1995, Chakrabarti-etal1996, Mirabel-etal1999,Nandi-etal2001b,Vadawale-etal2001a,Gallo-etal2003, Corbel-etal2003, Vadawale-etal2003,Fender-etal2004, Remillard-McClintock2006, Done-etal2007, Fender-etal2009,Belloni-etal2010, Uttley-etal2014, Das-etal2014,Radhika-etal2016a, Nandi-etal2018,Nandi-etal2024,Ingram-etal2019, Das-etal2021,Das-etal-2022,Kalemci-etal2022,Choudhury-etal2025}. Intense curvature of spacetime in the vicinity of BH allows the exploration of diverse relativistic phenomena, including gravitational lensing and frame dragging \cite[]{Lense-Thirring1918,Einstein1936,Bozza2007,Kraniotis2014,Narzilloev2023}. Beyond their role as key objects in the study of gravity, BHs are also crucial to understand the stellar cluster dynamics \cite[]{Miller-Colbert2004}, gravitational wave generation \cite[]{Einstein-Rosen1937,Thorne1995,Oleary-etal2009,Lai-etal2018}, and the co-evolution of galaxies \cite[]{Dimatteo-etal2005}.

The study of gravitational waves due to black hole mergers has opened a new era of gravitational wave astronomy, enabling tests of general relativity in regimes previously inaccessible \cite[]{Centrella-etal2010}. While such detections confirm the existence of black holes, our interest lies in exploring their electromagnetic counterparts, which provide complementary insights into the accretion physics and radiative processes governing these extreme systems. It is worth mentioning that the electromagnetic emissions have continued to offer a powerful window into the effects of strong gravity for more than six decades. The discovery of the first non-solar X-ray source, Scorpius X-1 \cite[]{Giacconi-etal1962}, marked the beginning of X-ray astronomy and demonstrated the potential of high energy observations to probe the physics of accretion onto compact objects.

Interestingly, black hole X-ray binaries (BH-XRBs) can be studied when the black hole resides in a binary system with a stellar companion. The intense gravitational pull of the black hole draws matter from the companion star forming an accretion disc around it. As the infalling material spirals inward through viscous transport of angular momentum, it is heated to extreme temperatures releasing copious X-rays in the process \cite[]{Frank-etal2002,Longair2011,Kato-etal2020}. The first confirmed stellar-mass black hole, Cygnus X-1 (hereafter Cyg X$-$1), was discovered with the \textit{Uhuru} satellite \cite[]{Oda-etal1971,Tananbaum-etal1972}, and long-term monitoring of BH-XRBs has been enabled by all-sky X-ray monitors onboard space observatories such as \textit{Ginga} ($1987-1991$), \textit{RXTE} ($1996-2012$), \textit{INTEGRAL} ($2002-2025$), \textit{Neil Gehrels Swift} ($2004-2026$), and \textit{MAXI} (since 2009). These BH-XRBs are broadly classified into high mass X-ray binaries (HMXBs) and low mass X-ray binaries (LMXBs) according to the mass of the companion star. The rate of mass transfer from the companion to the black hole predominantly governs the observable characteristics of the system and further distinguishes them into persistent and transient sources \cite[]{Tanaka-Shibazaki1996,Chen-etal1997b}. Persistent systems are generally associated with sustained, high accretion rates ($\dot{M} \sim 10^{-8}~{\rm M}_{\odot}$ yr$^{-1}$), whereas transient sources accrete episodically with accretion rate $\dot{M} \leq 10^{-9}~{\rm M}_{\odot}$ yr$^{-1}$, spending most of their time in a quiescent phase \cite[]{Tanaka-Shibazaki1996}. During quiescence, as the disc temperature gradually rises, a thermal-viscous instability \cite[]{Lasota2001} may be triggered within the disc, causing a rapid inward transport of matter towards the black hole and producing a luminous X-ray outburst \cite[]{Shakura-Sunyaev1976, Hamuery-etal1990, Lasota-etal1996}. 

In addition to the classical transient systems, there exist a few sources that behave neither as strictly persistent nor purely transient. For example, GRS 1915$+$105 remained persistently active from 1996 until it entered an obscured phase in 2019 \cite[]{Athulya-etal2023}, while Cyg X–1 has exhibited persistent emission since its discovery. However, both sources display significant long term variability in their lightcurves, and hence, they are classified as persistent-cum-variable systems (see also \S\ref{sec3:BH_overview}). The luminosity of such sources typically varies between $0.4$ and $3.0$ Crab in the $\sim 2-20$ keV energy band \cite[]{Jiang2024,Majumder-etal2025a}. In contrast, typical persistent sources exhibit luminosities ranging from $\sim 20$ mCrab to $\sim 1$ Crab \cite[]{Gou-etal2009,Boyd-Smale2004,Bhuvana-etal2023,Bhuvana-Nandi2025}. For transient sources, the luminosity usually ranges from $\sim 30$ mCrab during the quiescent phase to as high as $\sim 50$ Crab during the outburst phase \cite[][see also Table~\ref{tab:Table-01}]{McClintock-Remillard2006,Segreto-etal2015}.

During an outburst, a BH-XRB evolves sequentially through the canonical spectral states, represented in the Hardness–Intensity Diagram (HID). The HID characterizes the relationship between the source intensity and spectral hardness, defined as the ratio of fluxes in the high- and low-energy X-ray bands. Typically, the HID traces a `q'-shaped trajectory in an anti-clockwise direction  \cite[][and references therein]{Chakrabarti1998,Homan-etal2001, Maccaroni-Coppi2003, Casella-etal2005,Belloni-etal2005, Nandi-etal2012, Sreehari-etal2018,Sreehari2021}. Temporal variability on timescales ranging from milliseconds to a few hundred seconds is commonly observed along the right and upper branches of the `q'-track and often manifests as Quasi-periodic Oscillations (QPOs) \cite[]{VanderKlis1989a} in the power density spectrum (PDS) \cite[]{Leahy-etal1983}, providing crucial insights into the dynamics of the accretion flow. Depending on the QPO properties and PDS morphology, QPOs are generally classified as type-C, type-B and type-A \cite[][see also Table \ref{tab:Table-06} for details]{Wijnands-etal1999,Casella-etal2004,Casella-etal2005}. The initial phase of the outburst corresponds to the Low/Hard State (LHS), where the X-ray spectrum is dominated by a Comptonised component with low bolometric flux. In this state, the PDS exhibits a strong fractional root mean squared (rms) variability $> 10\%$ \cite[]{Remillard-McClintock2006} in the presence of type-C QPOs with centroid frequencies typically span $\sim 0.1 - 30.0$ Hz. As the outburst evolves, the source undergoes transitions through the Hard Intermediate State (HIMS) and the Soft Intermediate State (SIMS), during which the spectra gradually soften owing to an increasing contribution from the thermal emission of the accretion disc. Note that we refer Intermediate State (IMS) whenever HIMS and SIMS remain indistinguishable. Type-C QPOs are predominantly observed during the LHS and HIMS, while type-B and type-A QPOs are more commonly detected in the SIMS. Eventually, the source reaches the High/Soft State (HSS), characterized by a spectrum dominated by multi-temperature blackbody emission from an optically thick and geometrically thin accretion disc along with PDS appearing featureless and exhibiting minimal variability. It is noteworthy that some sources do not follow canonical outburst profile of BH-XRBs, which are often named as `failed' outburst \cite[and references therein]{Sturner-Shrader2005,Ramadevi-Seetha2007,Capitanio-etal2009b,Zhang-etal2019,Baby-etal2020,Alabarta-etal-2021,Wang-etal2022b,Debnath-etal2023}.

In addition to these spectral and temporal variations, BH-XRBs frequently exhibit jet activities that are prominent in the radio band. Quasi-steady and persistent jets are often observed during LHS and HIMS, while transient relativistic ejections are observed during SIMS \cite[]{Brocksopp-etal2002, Fender-etal2004, Fender-etal2009, Radhika-Nandi2014, Nandi-etal2018, Homan-etal2020}. The radio luminosity ($L_{R}$) of BH-XRBs is often found to correlate with their X-ray luminosity ($L_{X}$) through a power-law relation \cite[]{Fender2001, Corbel-etal2003, Gallo-etal2003}, highlighting strong disc-jet coupling established via multi-wavelength studies. This correlation suggests a link between jet power and accretion rate \cite[]{Gallo-etal2003, Corbel-etal2003}. \citet{Fender-etal2004} proposed a unified model where steady compact jets are present in the hard state, quenched in the soft state, and powerful transient ejections occur during state transitions \cite[see also][]{Choudhury-etal2025}. \citet{Radhika-etal2016a} observed a `local' softening in spectro-temporal properties just before peak radio flux of transient jets in several BH-XRBs. Furthermore, \citet{Nandi-etal2018} estimated that $8-16\%$ of the mass outflow contributes to the kinetic jet power in XTE J1859$+$226. The observed coupling between jet activity and spectral state transitions provides strong evidence for a fundamental connection between accretion dynamics and jet ejection processes \cite[]{Feroci-etal1999,Chakrabarti-1999,Das-etal-2001,Nandi-etal2001b,Vadawale-etal2001a,Chattopadhyay-etal2004,Chattopadhyay-etal-2005,Chattopadhyay-Das07,Das-Chattopadhyay08,Aktar-etal2015,Aktar-etal18,Jana-Das2024,Jana-Das2026}. Despite these insights, the physical origin of jets remains an open question. 

Wide-band energy spectral modelling of various BH–XRBs reveal the presence of a multi-temperature accretion disc \cite[]{Shakura-Sunyaev1973}, which primarily contributes to the thermal component of the observed emission. In addition, the high energy non-thermal emission is possibly originated from a ‘hot’ corona located in the vicinity of the black hole, where seed blackbody photons from the disc undergo inverse-Compton scattering by highly energetic electrons \cite[]{Sunyaev-Titarchuk1980,Chakrabarti-Titarchuk1995,Zdziarski-etal1996,Mandal-Chakrabarti2005}. In the hard state, the optically thick and geometrically thin disc is truncated at relatively larger radii, resulting in a spectrum dominated by high energy non-thermal photons. As the disc gradually extends inward toward the black hole, the thermal component becomes increasingly dominant, signifying the transition of the source into the softer state \cite[]{Chakrabarti-Titarchuk1995,Chakrabarti-etal2009,Dutta-Chakrabarti2010,Nandi-etal2012,Iyer-etal2015,Baby-etal2020}. During HSS, the inner edge of the accretion disc is typically observed to lie close to the innermost stable circular orbit (ISCO) of the black hole. Often, the X-ray spectra exhibit a prominent Fe K$\alpha$ emission line at $\sim 6.4$ keV, accompanied by a characteristic ‘Compton hump’ feature in the $\sim 10-40$ keV range. These features arise due to the reflection of high energy X-rays from the corona irradiating the accretion disc \cite[][and references therein]{Ross-Fabian2007,Garcia-etal2010, Miller-etal2013,Majumder-etal2025a}. Moreover, the Fe K$\alpha$ emission line often appears broadened and skewed owing to the strong relativistic effects induced by the black hole’s spin \cite[]{Risaliti-etal2013,Reynolds2014}. Consequently, the study of the iron line emission characteristics serves as a crucial diagnostic tool for constraining the spin parameter of the black hole.

Observations of BH-XRBs reveal complex X-ray timing features, with rapid variability in the lightcurve. Depending on the frequency range, QPOs are categorized as (i) Low-frequency QPO (LFQPO) with centroid frequency $< 40$ Hz and (ii) High-frequency QPO (HFQPO) with centroid frequency $> 40$ Hz \cite[]{Remillard-McClintock2006}. The LFQPO frequencies are found to correlate strongly with the spectral states of BH-XRBs, typically increasing as the source evolves from LHS to HIMS and eventually saturates at SIMS \cite[]{Casella-etal2004,Titarchuk-Shaposhnikov2005,Debnath-etal2008,Motta-etal2009, Radhika-Nandi2014,Iyer-etal2015, Sreehari-etal2018}. In addition to frequency evolution, time- or phase-lag analysis serves as an important diagnostic tool, offering insights into the physical mechanisms responsible for QPO generation. The observed time-lags at QPO frequencies can arise from various processes, including Comptonization, reflection, gravitational light bending, and the geometry of the accretion flow \cite[and references therein]{Dutta-etal2016}. For LFQPOs, high inclination systems generally exhibit soft-lags at higher QPO frequencies ($>2$ Hz), while low inclination sources tend to show hard-lags \cite[and references therein]{Dutta-etal2016, Eijnden-etal2017,Choudhury-etal2025}. Interestingly, the high inclination source GRS 1915$+$105 displays a transition in lag behavior, showing soft-lags above $\sim 2.2$ Hz and hard-lags below that frequency \cite[]{Reig-etal2000,Zhang-etal2020b}. In contrast, HFQPOs are generally observed in softer spectral states \cite[]{Majumder-etal2022, Majumder-etal2025a} and do not show any clear frequency dependent lag correlation. 

\citet{Mendez-etal2013} investigated the phase-lag properties of HFQPOs in several BH–XRBs using \textit{RXTE} observations and found that the phase-lag/time-lag varies with energy up to $\sim 30$ keV. This energy dependent variation in time-lag is generally attributed to the processes of Compton up- and down-scattering. However, the extent of this variation at higher energies is still uncertain as timing analyses beyond $\sim 20$ keV were not feasible with \textit{RXTE} due to its limited effective area in the hard X-rays. Investigating the timing behavior at these higher energies is therefore crucial, as it can provide valuable insights into the physical mechanisms governing accretion and high energy emission near black holes.

Prior to 2015, most space science missions offered limited energy coverage, often requiring coordinated multi-mission observations for broad-band studies. For example, \textit{XMM-Newton} (1999) covers optical$-$UV ($170-650$ nm) and X-rays ($0.1-15.0$ keV), making it ideal for imaging and spectroscopy in the soft X-ray band, but its low timing resolution and modest effective area limit its timing capabilities. On the other hand, \textit{NuSTAR} (2012) provides high resolution spectral data in the $3-70$ keV range but suffers from a $2.5$ ms dead-time \cite[]{Bachetti-etal2015} and reduced effective area, hindering fast timing studies. In contrast, \textit{Insight-HXMT} (2017) offers significantly broad spectral coverage ($1-250$ keV) with a large effective area and $\sim 25~\mu$s \cite[]{Zhang-etal2020a} time resolution in the $60-200$ keV range, makes it optimal for timing analysis in the hard X-ray band. However, its limited sensitivity at lower energies restricts timing studies in energy range $3-60$ keV. Missions like \textit{INTEGRAL} (2002) and \textit{Neil Gehrels Swift} ($2004-2026$) aimed for multi-wavelength coverage from optical/UV to $\gamma$-rays with $\sim 100~\mu$s time resolution, but their moderate effective areas constrained detailed timing studies.

\textit{AstroSat}, in this context, has effectively bridged the gap by providing a multi-wavelength observational platform with superior time resolution ($10~\mu$s) and a higher effective area ($\sim 6000~{\rm cm}^2$), which are essential for investigating the wide-band spectral and timing characteristics of astrophysical sources. \textit{AstroSat} offers simultaneous coverage from the optical to the $\gamma$-rays, enabling the investigation of correlations across different energy bands of the electromagnetic spectrum. Furthermore, its wide X-ray spectral range of $0.3-150.0$ keV facilitates detailed studies of all spectral states of BH–XRBs. However, the \textit{AstroSat}’s moderate spectral resolution imposes certain limitations in precisely characterizing the reflection components. In this work, we present an extensive and systematic investigation of \textit{AstroSat} observations of $34$ BH–XRBs over the past decade. Through detailed spectral and temporal modelling, we unveil key observational signatures and underlying physical processes governing accretion and emission in these BH systems. This comprehensive analysis encapsulates a decade of \textit{AstroSat}’s contributions, offering new insights into the complex interplay between accretion dynamics, relativistic effects, and high energy radiation in the vicinity of stellar-mass black holes.

The manuscript is organized as follows. In \S \ref{sec2:astrosat}, we provide a brief overview of the suite of scientific instruments onboard \textit{AstroSat}, along with the sky distribution of black hole (BH) sources observed by \textit{AstroSat}. In \S \ref{sec3:BH_overview}, we present a concise description of the $34$ BH-XRBs observed by \textit{AstroSat}, with source characteristics obtained from previous observational studies. The observational details of the selected BH-XRBs are given in \S \ref{sec4:obs}. A comprehensive account of the data reduction procedures is presented in \S \ref{sec5:methods}. The detailed methodologies for timing and spectral analyses are described in \S \ref{sec6:analysis_modelling}. The results and findings are presented in \S \ref{sec7:results}, followed by an investigation of the spectro-temporal correlations in \S \ref{sec8:correlation}. The observational results are further supported by simulations and spectral modelling in \S \ref{sec9:accretion_model}. In \S\ref{sec10:discussion}, we discuss and summarise the key results in the context of accretion geometry. Finally, instrument specific limitations, operational challenges, and future prospects are outlined in \S\ref{sec11:limitations}.

\cprotect\section{\textit{AstroSat} at A Glance}\label{sec2:astrosat}

\begin{figure}
    \begin{center}
    \includegraphics[width=\columnwidth]{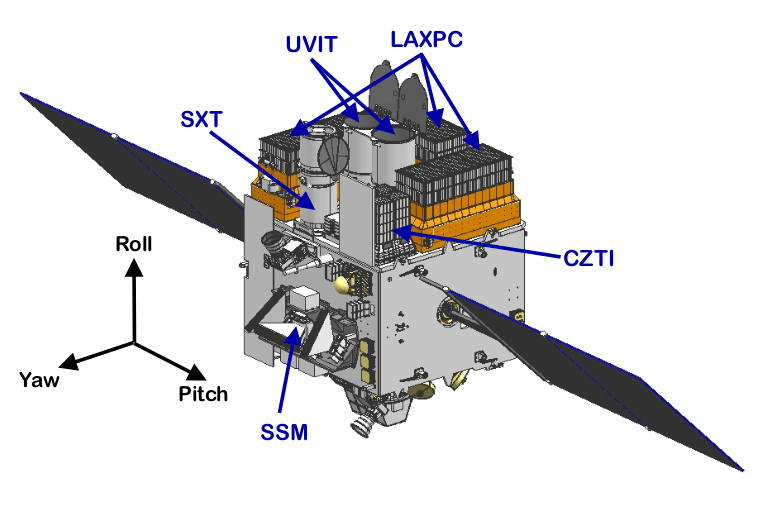}
    \end{center}
    \caption{Three dimensional (3D) schematic of the \textit{AstroSat} spacecraft in the solar panel deployed configuration showing all its primary scientific instruments. The four co-aligned instruments, namely UVIT, SXT, LAXPC, and CZTI are point towards the $+$Roll axis of the spacecraft. See text for details.
    }
    \label{fig:fig1}
\end{figure}

\textit{AstroSat} \cite[]{Singh-etal2002,Agrawal2006,Singh-etal2014,Agrawal2017,Seetha-Megala2017}, launched in 28 September 2015, is the first dedicated Indian astronomy mission aimed at multi-wavelength studies in the optical ($350-550$ nm), near- ($200-300$ nm) and far-ultraviolet ($130-180$ nm) and a broad X-ray spectral band covering $0.3-150.0$ keV. This unique capability of \textit{AstroSat} allows the astronomical community to study a wide range of astrophysical phenomena starting from star formations to black holes with unprecedented detail. The spacecraft is equipped with five payloads, each specially designed for targeted observations. These include the Ultraviolet Imaging Telescope (UVIT), Soft X-ray Telescope (SXT), Large Area X-ray Proportional Counter (LAXPC), Cadmium Zinc Telluride Imager (CZTI), and Scanning Sky Monitor (SSM). The 3D model of \textit{AstroSat} spacecraft with the solar panels in deployed configuration is shown in Fig. \ref{fig:fig1}.

The four instruments (UVIT, SXT, LAXPC and CZTI) are co-aligned and point towards the $+$Roll axis of the spacecraft (see Fig. \ref{fig:fig1}). In comparison, the SSM, an assembly of three X-ray cameras, is mounted on a rotating platform with the rotation axis aligned towards $+$Yaw axis of the spacecraft. While the four pointed instruments observe a common source, the SSM assembly looks at a different region of the sky. The solar panels, shown deployed in the figure, are along the Pitch axes, and they point towards the Sun. Sun avoidance angles for all payloads are ensured during pointed observations as well as during maneuvers from one source to another. In addition to these five payloads, a Charged Particle Monitor \cite[CPM;][]{Rao-etal2017} is also accommodated on the spacecraft to measure the flux of the highly energetic particles.

\subsection{Payloads onboard \textit{AstroSat}}\label{sec2:2.1}

\textit{AstroSat} successfully completed its initial design life of five years on 28 September 2020, and continues to operate effectively to date (with operational constrains), contributing significantly to scientific research \cite[]{Seetha-Kasturirangan2021,Ghosh-etal2021,Bhattacharyya-etal2021,Yadav-etal2021,Sarwade-etal2021}. The primary scientific objectives of \textit{AstroSat} are to detect X-ray transients, observe various astronomical sources across multiple bands of the electromagnetic spectrum, and study their temporal variability \cite[]{Agrawal2006}. To achieve these goals, five scientific payloads have been configured. A brief description of each payload is given below.

\begin{figure*}
    \begin{center}
    \includegraphics[width=0.75\textwidth]{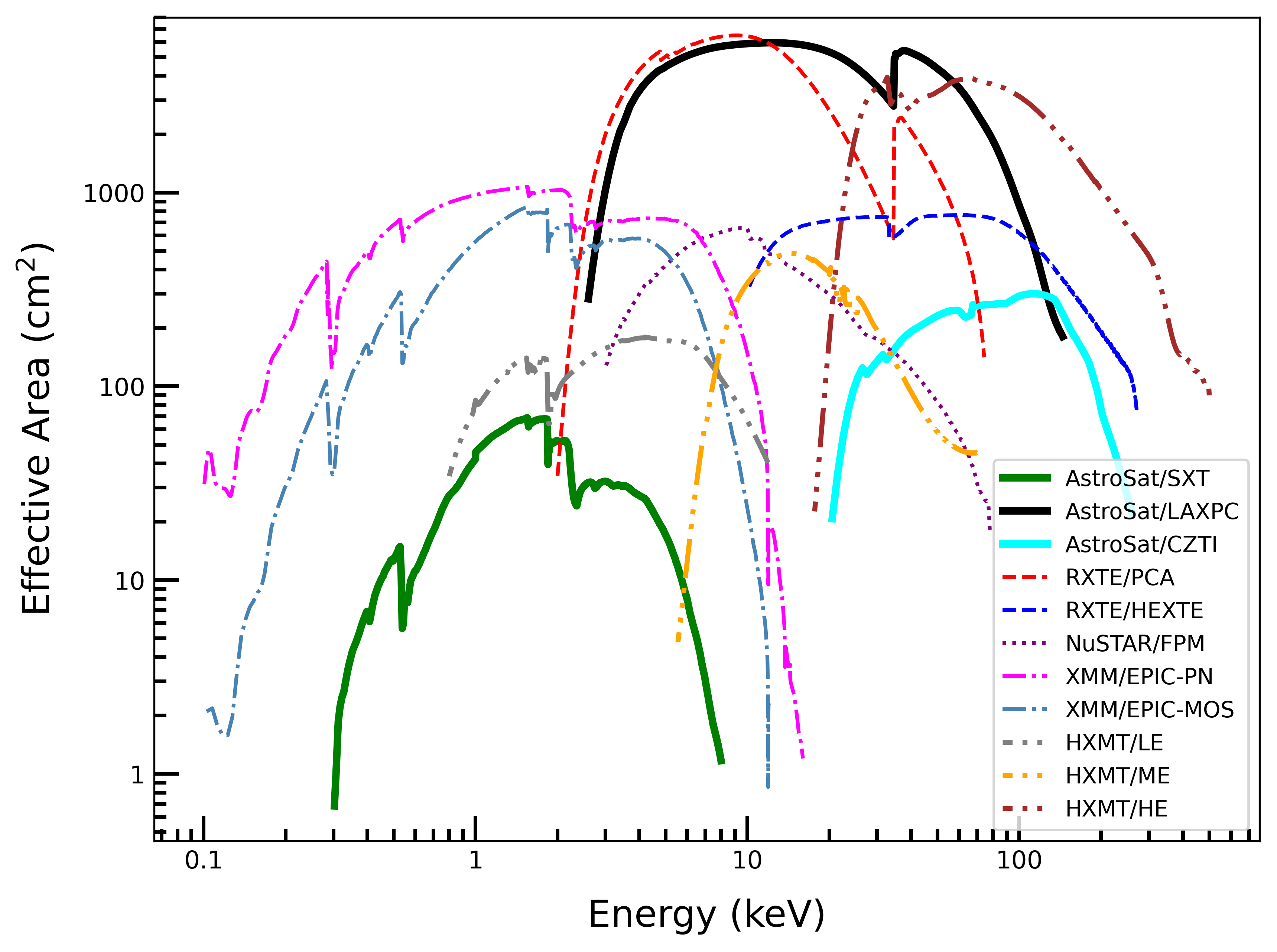}
    \end{center}
    \caption{Effective area curves of the scientific instruments onboard \textit{AstroSat}, \textit{RXTE}, \textit{NuSTAR}, \textit{XMM-Newton} and \textit{HXMT} in $0.1-500.0$ keV energy range. Different line styles correspond to different missions, while colors with identical line styles represent individual instruments of a mission. The specific instruments for each mission are mentioned in the legend. 
    }
    \label{fig:fig2}
\end{figure*}

\begin{itemize}

    \item \textbf{UVIT} \cite[]{Subramaniam-etal2016,Tandon-etal2017a,Tandon-etal2017b} comprises two identical telescopes (see Fig. \ref{fig:fig1}), each with a $375$ mm aperture, designed to observe in the ultraviolet ($130-300$ nm) and visible ($320-550$ nm) wavelength bands. It provides high angular resolution of approximately $1.8$ arcseconds and a circular field of view (FoV) of $28$ arcminutes. This combination of features makes UVIT a powerful instrument for studying hot stars and star formation in nearby galaxies.

    \item \textbf{SXT} \cite[]{Singh-etal2016} is one of the instruments onboard \textit{AstroSat}, as shown in Fig. \ref{fig:fig1}. It has a FoV of approximately $40$ arcminutes and is designed for moderate resolution of $\sim 4$ arcsecond square per pixel for the imaging \cite[]{Singh-etal2017} of X-ray sources. The effective area of SXT is $\sim 40$ cm$^{2}$ at around $1$ keV (see Fig. \ref{fig:fig2}). The CCD detector offers an energy resolution of approximately $130$ eV at $6$ keV, making SXT a powerful instrument for spectroscopic studies in the soft X-ray band.

    \item \textbf{LAXPC} \cite[]{Yadav-etal2016b} is designed for high time resolution and moderate spectroscopic resolution ($10-15\%$ energy resolution at $20$ keV) over a broad energy range ($3-80$ keV). It has a relatively large FoV with a Full Width at Half Maximum (FWHM) of about $1^{\circ} \times 1^{\circ}$. The instrument comprises three similar detectors, namely LAXPC10, LAXPC20, and LAXPC30, as shown in Fig. \ref{fig:fig1}. The combined effective area of the three LAXPCs is approximately $6000$ cm$^{2}$ below $20$ keV \cite[]{Antia-etal2017}. It offers the largest effective area in the $20-60$ keV range among all spaced based missions to date, as illustrated in Fig. \ref{fig:fig2}. With a timing resolution of $10~\mu$s, LAXPC is well-suited for in-depth timing studies across a wide energy band.    

    \item \textbf{CZTI} \cite[]{Vadawale-etal2016} is an instrument onboard \textit{AstroSat} (see Fig. \ref{fig:fig1}) designed for high-resolution spectroscopy in the $20-150$ keV energy band (see Fig. \ref{fig:fig2}). In addition, at energies above $100$ keV, CZTI functions as an open detector, enabling it to monitor transient events such as Gamma-ray Bursts (GRBs) and to perform polarimetric measurements in the extended energy range of $100-300$ keV. The instrument has a FoV of $4.6^{\circ} \times 4.6^{\circ}$ in the $10-100$ keV range, which widens to $17^{\circ} \times 17^{\circ}$ at energies beyond $100$ keV. CZTI offers a time resolution of $20~\mu$s and an energy resolution of approximately 11\% at $60$ keV \cite[]{Bhalerao-etal2017}.

    \item{SSM} \cite[]{Ramadevi-etal2017} consists of three X-ray cameras mounted on a rotating platform, operating in the $2.5-10.0$ keV band. Each camera uses a gas-filled proportional counter with one-dimensional coded-mask optics. The two edge cameras are canted by $45^\circ$ from the central camera plane and further tilted by $\pm12^\circ$ to provide partially overlapping sky coverage. A stepper-motor rotation system enables scan and reverse operation from $5^\circ$ to $355^\circ$ in stare-and-step mode, with a typical stare time of 10 min and a step size of $10^\circ$. The edge cameras have a FoV of $26.8^\circ \times 100.0^\circ$, while the central camera covers $22.1^\circ \times 100.0^\circ$. The angular resolution is $\sim10-13$ arcminute along the short axis and $\sim2.5^\circ$ along the long axis with a timing accuracy of $100\,\mu$s.

\end{itemize}

\textit{AstroSat} stands out as a unique mission in modern astronomy owing to its true multi-wavelength observational capability. Its broad spectral coverage, extending from the optical and ultraviolet (UV) to high-energy $\gamma$-rays through instruments such as UVIT, SXT, LAXPC, and CZTI, renders \textit{AstroSat} unparalleled among contemporary missions. While previous missions have pursued multi-wavelength studies via coordinated observations across multiple observatories, \textit{AstroSat}’s integrated platform effectively eliminates the logistical complexities inherent to such operations. Furthermore, its substantial effective area in the $\sim 20-60$ keV range makes it exceptionally suited for probing rapid variability in BH–XRBs. A comparative plot of the effective area as a function of energy for \textit{AstroSat}, along with decommissioned and currently operational missions, is shown in Fig.~\ref{fig:fig2}, with the respective instruments indicated in the legend. When compared to \textit{RXTE} ($2-200$ keV; PCA and HEXTE), \textit{NuSTAR} ($3-79$ keV; FPMA and FPMB), and \textit{XMM-Newton} ($0.1-15.0$ keV; EPIC-PN and EPIC-MOS), \textit{AstroSat} provides broader X-ray energy coverage spanning $0.3-150.0$ keV through SXT, LAXPC, and CZTI. However, \textit{HXMT} ($1-250$ keV) offers the widest X-ray energy range with significant effective area only in the hard X-ray band. It is worth mentioning the LAXPC instrument offers the largest effective area, approximately four times greater than that of \textit{RXTE}/PCA beyond $20$ keV\footnote{\url{https://web.iucaa.in/~astrosat/AstroSat_handbook.pdf}}, among all presently operational X-ray astronomy missions. This makes it particularly advantageous for high time-resolution studies in the hard X-rays. The enhanced effective area in $20-60$ keV band thus enables the detection and detailed characterization of fast temporal variabilities as well as wide-band spectral decomposition studies.

\begin{figure*}
    \begin{center}
    \includegraphics[width=0.8\textwidth]{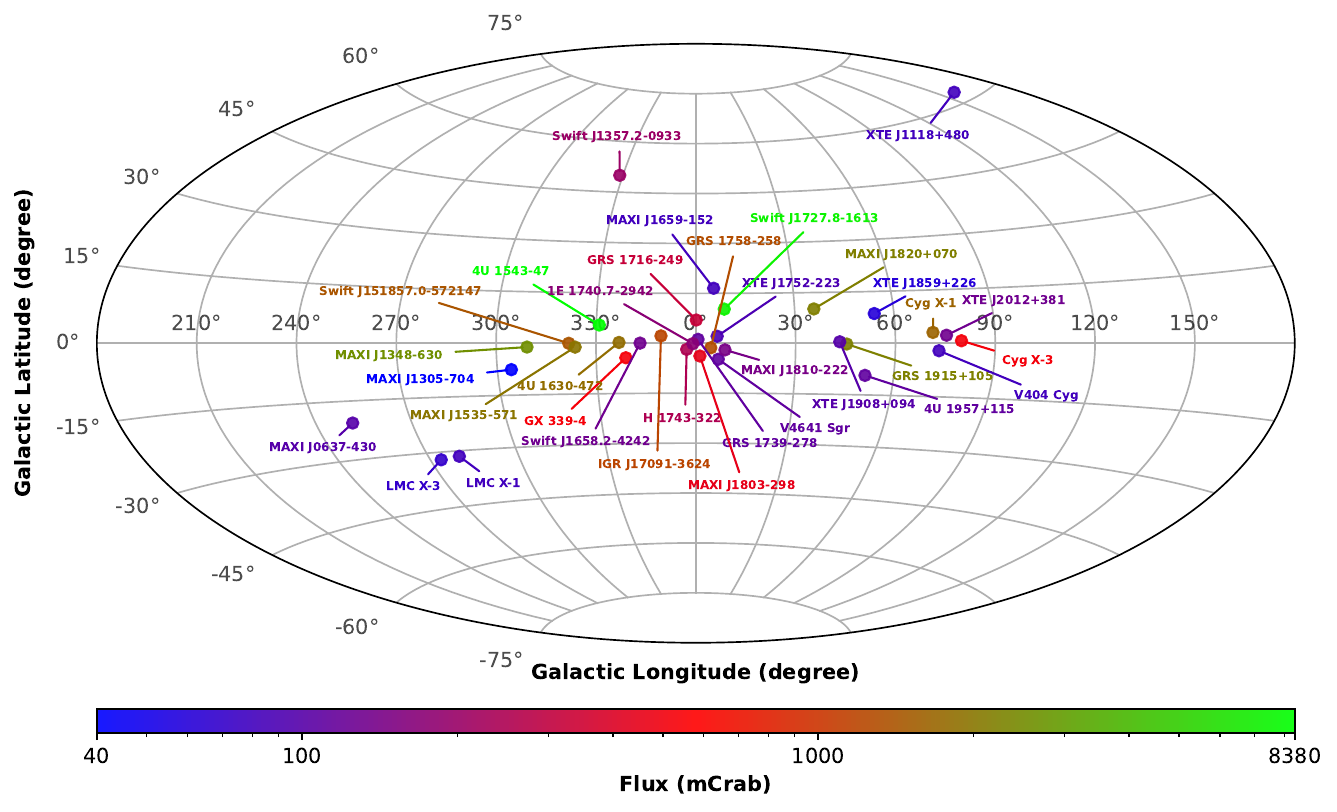}
    \end{center}
    \caption{Distribution of $34$ BH-XRBs observed by the X-ray payloads onboard \textit{AstroSat} during $2015-2025$, shown in Galactic coordinates \citep[Aitoff projection:][]{Aitoff1889}. Color map indicates peak flux of each source in mCrab ranging from $\sim 40$ mCrab to $8380$ mCrab within $2-20$ keV energy band. 
    }
    \label{fig:fig3}
\end{figure*}

\begin{figure*}
    \begin{center}
    \includegraphics[width=0.88\textwidth]{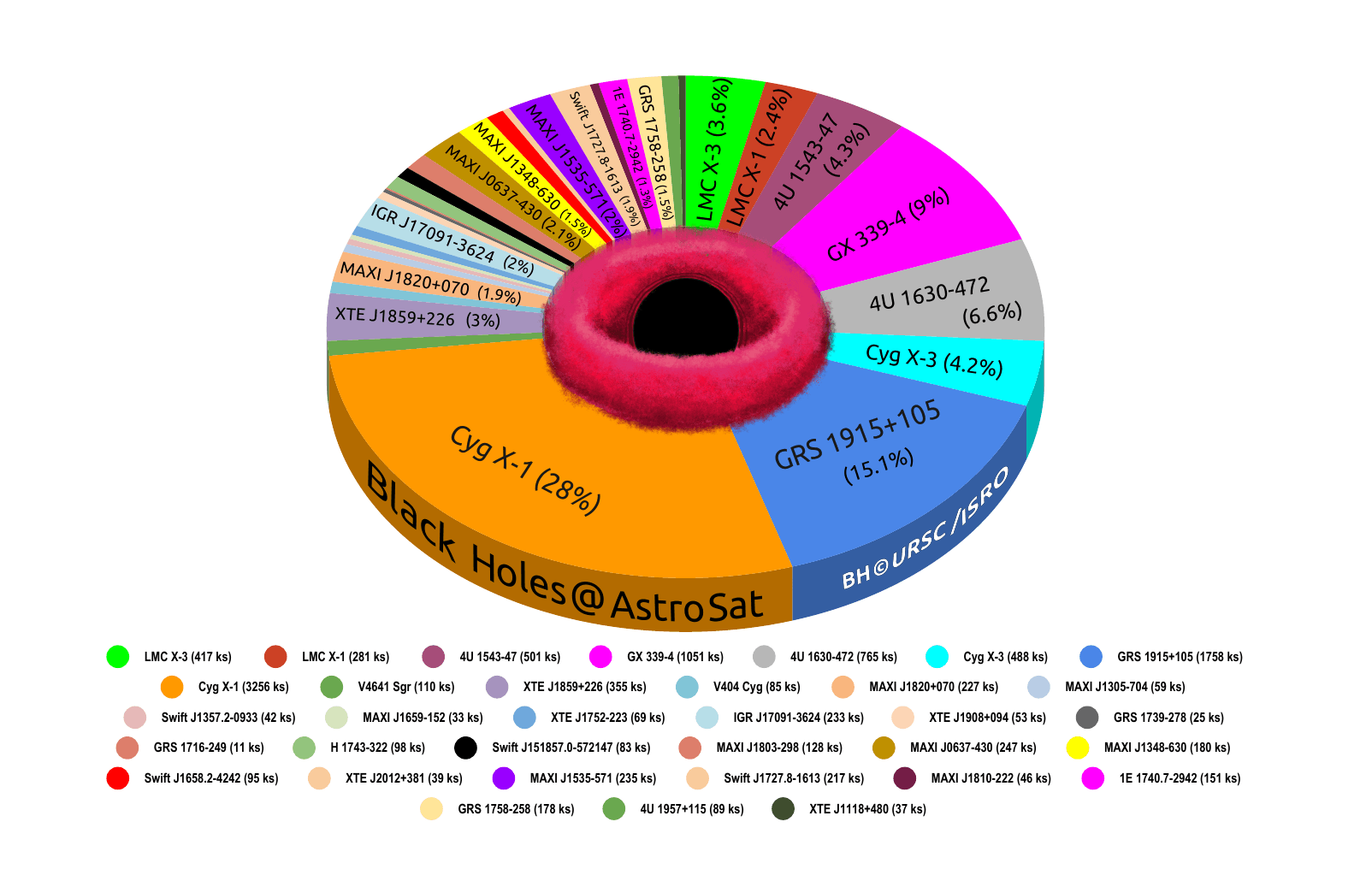}
    \end{center}
    \caption{Graphical illustration (Pie chart) showing the exposure time distribution of $34$ black hole sources observed with \textit{AstroSat} over a total of $12$ Ms during the past decade. Source names and corresponding exposure times are given in the legend. Different colors represent individual sources. For clarity, $17$ sources with exposure time $\geq 1.5\%$ of the total observation time are labelled in the figure.
    } 
    \label{fig:fig4}
\end{figure*}

\subsection{Galactic \& Extragalactic Stellar Mass BHs Observed with \textit{AstroSat}}\label{sec2:2.2}

\begin{table*}
\centering
\caption{Summary of $34$ BH-XRBs observed by \textit{AstroSat} during $2015$ to $2025$, including source name with source type (PcV, P and T), sky coordinates (RA, Dec), and key parameters (mass, spin, distance and inclination). The maximum \textit{MAXI}/GSC intensity in the $2-20$ keV band during the \textit{AstroSat} observation period and total exposure are also listed. {\red The final column indicates the BH identification (confirmed or candidate)}. Rows shaded in red indicate sources with limited observations during \textit{AstroSat} observations. See text for details.
}
\label{tab:Table-01}
\resizebox{\textwidth}{!}{
\begin{tabular}{l l l l l l l l l l l l} \hline \hline
     
Sl. & Source Name & RA (degree)  & Dec (degree) & Mass ($M_{\rm BH}$) & Spin ($a_{\rm k}$)  & Distance & Inclination & Max. Intensity & Exposure & \red BH Identification\\ 

No.  & (Type: PcV, P, T)   &  (J2000) & (J2000)   & (M$_{\odot}$) & & (kpc) & (degree) & (Crab) & (ks)  & \red (Confirmed/Candidate) \\
    \hline

1 & GRS 1915$+$105 (PcV) & 288.798  &  $+10.946$  &  $12.4_{-1.8}^{+2.0}$ (20)  &  $0.990-0.997$ (21)  &  $9.4\pm0.6$ (18)  &  $64\pm4$ (18) &  2.205  &  1758 & Confirmed \\

2 & Cyg X$-$1 (PcV) & 299.591  &  $+35.202$  &  $21.2\pm2.2$ (22)  &  $> 0.9985$ (23)  &  $2.22_{-0.17}^{+0.18}$ (22)  &  $27.51_{-0.57}^{+0.77}$ (22) &  1.617 &  3256 & Confirmed  \\

3 & Cyg X$-$3 (PcV) & 308.107  &  $+40.958$  &  $2.4_{-1.1}^{+2.1}$ (17)  &  $-$  &  $8.95\pm0.96$ (18)  &  $29.5\pm1.2$ (19) &  0.520  &  488 & Candidate \\

4 & LMC X$-$1 (P) & 84.912  &  $-69.743$   &  $10.91\pm1.41$ (4)  &  $0.92_{-0.07}^{+0.05}$ (5)  &  $48.1\pm2.2$ (4)  &  $36.38\pm1.92$ (4) &  0.083 &  281 & Confirmed \\ 

5 & LMC X$-$3 (P) & 84.734  &  $-64.082$   &  $6.98\pm0.56$ (1)  &  $0.22-0.41$ (2)  &  $48.1\pm2.2$ (3)  &  $69.24\pm0.72$ (1) &  0.069 &  417 & Confirmed \\ 

6 & GRS 1758$-$258 (P) & 270.302  &  $-25.743$  &  $>8.6$ (82)  &  $-$  & $8\pm1$ (80)  &  $61\pm2$ (81) &  1.280  &  178 & Candidate \\ 

7 & 1E 1740.7$-$2942 (P) & 265.970  &  $-29.740$  &  $\sim5$ (79) & $\gtrsim0.5$ (79)  &  $8\pm1$ (80)  &  $55\pm1$ (81) &  0.170  &  151 & Candidate \\ 

8 & 4U 1957$+$115 (P) & 299.850  &  $+11.709$  &  $<6.2$ (83)  &  $\sim 0.99$ (84)  &  $20-40$ (83)  &  $13-78$ (83) &  0.110 &  89 & Candidate \\ 

9 & IGR J17091$-$3624 (T) & 257.282  &  $-36.407$  &  $8.7-15.6$ (45)  & $-$ &  $11-17$ (46)  &  $30-40$ (47) &  1.180  &  233 & Candidate \\ 

10 & GX 339$-$4 (T) & 255.706  &  $-48.790$   &  $2.3-9.5$ (10)  &  $0.95_{-0.09}^{+0.08}$ (11)  &  $8.4\pm 0.9$ (11)  &  $40-60$ (12) &  0.527  &  1051 & Confirmed \\

11 & Swift J1658.2$-$4242 (T) & 254.552  &  $-42.699$  &  $13-14.2$ (67)  &  $>0.96$ (68)  &  $> 3$ (69)  &  $64_{-3}^{+2}$ (68) &  0.133  &  95 & Candidate \\ 

12 & 4U 1630$-$472 (T) & 248.504  &  $-47.393$  &  $3-9$ (13)  &  $0.817\pm 0.014$ (14)  &  $11.5\pm0.3$ (15)  &  $64^{+2}_{-3}$ (16) &  1.796 &  765 & Candidate \\

13 & 4U 1543$-$47 (T) & 236.786  &  $-47.669$  &  $9.4\pm1.0$ (6)  &  $0.46\pm0.12$ (7)  &  $7.5\pm0.5$ (8)  &  $20.7\pm1.5$ (9) &  8.378  &  501 & Confirmed \\ 

14 & MAXI J1535$-$571 (T) & 233.832  &  $-57.230$  &  $5.14-7.83$ (72)  &   $0.67_{-0.04}^{+0.16}$ (73)  &  $4.1_{-0.5}^{+0.6}$ (74)  &  $67.4\pm0.8$ (75) &  1.982  &  235 & Candidate \\ 

15 & MAXI J1820$+$070 (T) & 275.091  &  $+7.186$  &  $8.48_{-0.72}^{+0.79}$ (31)  &  $0.77\pm0.21$ (32)  &  $2.96\pm0.33$ (33)  &  $63\pm3$ (33) &  2.166 &  227 & Confirmed \\ 

16 & MAXI J1348$-$630 (T) & 207.054  &  $-63.274$  &  $11\pm2$ (64)  &  $0.79\pm0.13$ (65)  &  $3.4\pm0.3$ (64)  &  $65\pm7$ (66) &  2.575 &  180 & Candidate \\ 

17 & H 1743$-$322 (T) & 266.565  &  $-32.234$  &  $11.21_{-1.96}^{+1.65}$ (56,57)  &  $-0.3-0.7$ (58)  &  $8.5\pm0.8$ (58)  &  $75\pm3$ (58) &  0.245 &  98 & Candidate \\

18 & Swift J1727.8$-$1613 (T) & 261.937  &  $-16.212$   &  $>3.12\pm0.10$ (87)  &  $0.98_{-0.07}^{+0.02}$ (76) &  $2.7\pm0.3$ (77)  & $40.0_{-0.8}^{+1.2}$ (76) &  7.270 &  217 & Confirmed \\ 

19 & MAXI J0637$-$430 (T) & 99.098  &  $-42.868$  &  $5.1\pm1.6$ (62)  &  $<0.25$ (62)  &  $8.7\pm 2.3$ (62)  &  $46.1_{-5.3}^{+4.0}$ (63) &  0.096  &  247 & Candidate \\ 

20 & MAXI J1803$-$298 (T) & 270.762  &  $-29.831$  &  $3-10$ (60)  &  $>0.991$ (61)  &  $-$  &  $>65$ (60) &  0.447  &  128 & Candidate \\ 

21 & GRS 1716$-$249 (T) & 259.904  &  $-25.018$  &  $6.4_{-2.0}^{+3.2}$ (54)  &  $0.464_{-0.007}^{+0.016}$ (55)  & $6.9\pm1.1$ (54)  &  $61\pm15$ (54) &  0.323 &  11 & Candidate \\ 

22 & Swift J151857.0$-$572147 (T) & 229.736  &  $-57.362$  &  $3.67-8.07$ (59)  &  $0.84_{-0.26}^{+0.17}$ (59)  &  $5.8\pm2.5$ (59)  &  $21.1_{-3.6}^{+4.5}$ (59) &  1.410  &  83 & Candidate \\ 

23 & GRS 1739$-$278 (T) & 265.670  &  $-27.760$  &  $<9.5$ (51)  &  $0.994_{-0.003}^{+0.004}$ (52)  &  $6-8.5$ (53)  &  $54_{-4}^{+8}$ (52) &  0.112  &  25 & Candidate \\

24 & XTE J2012$+$381 (T) & 303.158  &  $+38.184$  &  $\gtrsim11.8\pm4.6$ (70)  &  $0.988_{-0.030}^{+0.008}$ (70)  &  $5.4\pm2.1$ (71)  &  $68_{-11}^{+6}$ (70) &  0.127 &  39 & Candidate \\ 

\rowcolor{red!20}
25 & XTE J1859$+$226 (T) & 284.673  &  $+22.658$  &  $7.8\pm1.9$ (25)  &  $0.986_{-0.004}^{+0.001}$ (26)  &  $8\pm3$ (27)  &  $66.6\pm4.3$ (25) &  0.061 &  355 & Confirmed \\ 

\rowcolor{red!20}
26 & V4641 Sgr (T) & 274.840  &  $-25.407$  &  $6.4\pm0.6$ (24)  &  $-$  &  $6.2\pm0.7$ (24)  &  $72.3\pm4.1$ (24) &  0.105  &  110 & Confirmed \\

\rowcolor{red!20}
27 & V404 Cyg (T) & 306.016  &  $+33.868$  &  $9.0_{-0.6}^{+0.2}$ (28)  &  $> 0.92$ (29)  &  $2.39\pm0.14$ (30)  &  $67_{-1}^{+3}$ (28) &  0.075 &  85 & Confirmed \\ 

\rowcolor{red!20}
28 & MAXI J1305$-$704 (T) & 196.735  &  $-70.451$  &  $8.9_{-1.0}^{+1.6}$ (34)  &  $0.87_{-0.13}^{+0.07}$ (35)  &  $7.5_{-1.4}^{+1.8}$ (34)  &  $72_{-8}^{+5}$ (34) &  0.039  &  59 & Confirmed \\

\rowcolor{red!20}
29 & Swift J1357.2$-$0933 (T) & 209.320  &  $-9.544$   &  $12.4\pm3.6$ (36)  &  $-$  &  $\geq6$ (37)  &  $87.4_{-5.6}^{+2.6}$ (38) &  0.196 &  42 & Candidate \\ 

\rowcolor{red!20}
30 & MAXI J1659$-$152 (T) & 254.757  &  $-15.258$  &  $4.9-5.7$ (39)  &  $-1-0.44$ (40)  &  $8.6\pm3.7$ (41)  &  $65-80$ (41) &  0.085 &  33 & Candidate \\  

\rowcolor{red!20}
31 & XTE J1752$-$223 (T) & 268.063  &  $-22.342$  &  $12_{-0.68}^{+0.77}$ (42)  &  $0.52\pm0.11$ (43)  &  $7.11_{-0.25}^{+0.27}$ (42)  &  $35\pm4$ (44) &  0.095 &  69 & Candidate \\ 

\rowcolor{red!20}
32 & XTE J1908$+$094 (T) & 287.221  &  $+9.385$  &  $6.5_{-0.7}^{+0.5}$ (48)  &  $0.55_{-0.45}^{+0.29}$ (49)  &  $3-10$ (50)  &  $27_{-3}^{+2}$ (49) &  0.104  &  53 & Candidate \\

\rowcolor{red!20}
33 & MAXI J1810$-$222 (T) & 272.500  &  $-22.200$  &  $-$  &  $-$  &  $\gtrsim6$ (78)  &  $-$ &  0.138  &  46 & Candidate \\ 

\rowcolor{red!20}
34 & XTE J1118$+$480 (T) & 169.545  &  $+48.037$  &  $7.06-7.24$ (85)  &  $-$  &  $1.72\pm0.10$ (86)  &  $74\pm4$ (85) &  0.076 &  37 & Confirmed \\ 

\hline \hline

\end{tabular}
}
\justify Note: References for source mass, spin, distance and inclination are provided within the parenthesis.    
\justify
	\noindent{\bf References:} 
    1: \cite{Orosz-etal2014}, 2: \cite{Bhuvana-etal2021}, 3: \cite{Straub-etal2011}, 4: \cite{Orosz-etal2009}, 5: \cite{Gou-etal2009}, 6: \cite{Russell-etal2006}, 7: \cite{Chen-Wang-2024}, 8: \cite{Jonker-etal2004}, 9: \cite{Orosz-etal2003}, 10: \cite{Heida-etal-2017}, 11: \cite{Parker-etal2016}, 12: \cite{Zdziarski-etal2019}, 13: \cite{Baby-etal2020}, 14: \cite{Liu-etal2022a}, 15: \cite{Kalemci-etal2018}, 16: \cite{King-etal2014}, 17: \cite{Zdziarski-etal2013}, 18: \cite{Reid-etal2023}, 19: \cite{Antokhin-etal2022}, 20: \cite{Reid-etal2014}, 21: \cite{Sreehari-etal2020},  22: \cite{Miller-Jones-etal2021}, 23: \cite{Zhao-etal2021}, 24: \cite{MacDonald-etal2014},  25: \cite{Yanes-Rizo-etal2022}, 26: \cite{Mall-etal2024}, 27: \cite{Tetarenko-etal2016},  28: \cite{Khargharia-etal2010}, 29: \cite{Walton-etal2017}, 30: \cite{Miller-jones-etal2009},    31: \cite{Torres-etal2020}, 32: \cite{Banerjee-etal2024}, 33: \cite{Atri-etal2020} 34: \cite{MataSanchez-etal2021}, 35: \cite{Feng-etal2023},  36: \cite{Casares2016}, 37: \cite{Charles-etal2019},  38: \cite{Casares-etal2022},  39: \cite{Torres-etal2021}, 40: \cite{Feng-etal2022a}, 41: \cite{Kuulkers-etal2013},  42: \cite{Abdulghani-etal2024}, 43: \cite{Reis-etal2011}, 44: \cite{Garcia-etal2018},    45: \cite{Iyer-etal2015}, 46: \cite{Rodriguez-etal2011}, 47: \cite{Xu-etal2017},    48: \cite{Chatterjee-etal2021}, 49: \cite{Draghis-etal2021}, 50: \cite{Chaty-etal2006}, 51: \cite{Wang-etal2018a}, 52: \cite{Zhao-etal2026}, 53: \cite{Greiner-etal1996}    54: \cite{Casares-etal2023}, 55: \cite{Zhao-etal2024},   56: \cite{Molla-etal2017}, 57: \cite{Tursunov-Kolos2018}, 58: \cite{Steiner-etal2012a},   59: \cite{Peng-etal2024b},    60: \cite{Mata-etal2022}, 61: \cite{Feng-etal2022b},   62: \cite{Soria-etal2022}, 63: \cite{Jia-etal2023},  64: \cite{Lamer-etal2021}, 65: \cite{Guan-etal2024}, 66: \cite{Titarchuk-Seifina2023},    67: \cite{Mondal-Jithesh2023}, 68: \cite{Xu-etal2018}, 69: \cite{Russell-etal2018},    70: \cite{Draghis-etal2023b}, 71: \cite{Gaia-Collab2016},    72: \cite{Sreehari-etal2019b}, 73: \cite{Sridhar-etal2019}, 74: \cite{Chauhan-etal2019}, 75: \cite{Miller-etal2018},    76: \cite{Peng-etal2024a}, 77: \cite{Mata-etal2024}, 78: \cite{Russel-etal2022},
    79: \cite{Stecchini-etal2020}, 80: \cite{Tetarenko-etal2020}, 81: \cite{Bhuvana-etal2023}, 82: \cite{Keck-etal2001}, 83: \cite{Gomez-etal2015}, 84: \cite{Barillier-etal2023}, 85: \cite{Cherepashchuk-etal2019}, 86: \cite{Gelino-etal2006}, 87: \cite{MataSanchez-etal2025}

\end{table*}   

Until now, 80 stellar-mass black holes (Galactic and extra-Galactic) are known to exist, among which 72 are transients\footnote{BlackCAT website: \url{https://www.astro.puc.cl/BlackCAT/transients.php}} and 8 are persistent sources \cite[]{Corral-Santana-etal2016,Tetarenko-etal2016}. Over the past decade, \textit{AstroSat} has observed 34 stellar-mass black hole sources, as listed in Table~\ref{tab:Table-01}. Of these, 26 are transients (T), 5 are persistent (P), and 3 are persistent-cum-variable (PcV) sources. For instance, GRS 1915$+$105, Cyg X$-$1 and Cyg X$-$3 exhibit aperiodic persistence on long timescales. Among the 5 persistent sources observed by \textit{AstroSat}, 2 are dynamically confirmed black holes (LMC X$-$1 and LMC X$-$3), while among the 26 transients, {\red 17} are black hole candidates and {\red 9} are confirmed black holes. Furthermore, out of the 3 persistent-cum-variable sources, 2 are dynamically confirmed black holes (Cyg X$-$1 and GRS 1915$+$105).

Table~\ref{tab:Table-01} provides source names along with their classification (PcV, P, T), Right Ascension (RA), Declination (Dec), mass, spin, distance, inclination, maximum intensity (\textit{MAXI}), and total exposure (\textit{AstroSat}), respectively. Fig. \ref{fig:fig3} displays the sky distribution of the $34$ black hole sources observed by \textit{AstroSat} during $2015-2025$ in galactic coordinates. The color code represents the maximum flux in the $2-20$ keV band detected by \textit{MAXI}/GSC during the \textit{AstroSat} campaign. These sources exhibit peak fluxes ranging from approximately $40$ mCrab\footnote{The average \textit{MAXI}/GSC count rate for the Crab in the $2-20$ keV band is $3.74$ photons/s/cm$^{2}$. The maximum GSC counts for all sources were converted to Crab units accordingly.} (MAXI J1305$-$704) to over $8380$ mCrab (4U 1543$-$475). To generate the skymap, the SIMBAD astronomical database\footnote{\url{https://simbad.u-strasbg.fr/simbad/sim-basicIdent=m33&submit=SIMBAD+search}} was queried for the equatorial coordinates (J2000 frame) of these sources. These RA and Dec values were converted to galactic longitude and latitude, and plotted using Aitoff sky projection.

Almost all of the 34 sources lie within the Milky Way Galaxy, except for LMC X$-$1 and LMC X$-$3, which belong to the nearby dwarf galaxy, the Large Magellanic Cloud (LMC). Among the Galactic sources, most are confined to the Galactic plane, with a few exceptions, such as Swift J1357.2$-$0933, MAXI J0637$-$430, and XTE J1118$+$480, which are located at higher Galactic latitudes within the Galactic halo, away from the plane. Among these, XTE J1118$+$480 was the first ever black hole discovered in the Galactic halo located at a high Galactic latitude of $\sim + 62^{\circ}$ \cite[]{Remillard-etal2000}.

\textit{AstroSat} has observed these $34$ black hole sources for a total of $\sim 12$ Ms during the last decade. The exposure times for each source are listed in Table~\ref{tab:Table-01}. Fig.~\ref{fig:fig4} presents a pie chart illustrating the distribution of exposure times among the different sources observed by \textit{AstroSat}. Each color is used to denote a given source with the source names and their respective exposure times (in ks) shown in the legend. 
Moreover, the names of sources for which significant coordinated observations ($\geq 1.5\%$ of $12$ Ms) are carried out by \textit{AstroSat} are labelled on the corresponding ‘slices’ of the pie chart. Note that the black hole source Cyg X$-$1 was observed most extensively with a total exposure of $\sim 3$ Ms accounting for $\sim 25\%$ of the total observation time of BH-XRBs. 

\section{Brief Overview of the Observed Black Hole Sources}\label{sec3:BH_overview}

This section presents a brief overview of the BH-XRBs observed by \textit{AstroSat}. {\red BH-XRBs are commonly classified as persistent or transient based on their long-term X-ray behaviour \citep{Tanaka-Shibazaki1996,Chen-etal1997b,Corral-Santana-etal2016,Tetarenko-etal2016}. However, this scheme does not adequately describe systems that remain continuously active while exhibiting large amplitude flux ($F$) and spectral variability. We therefore introduce the term \textit{persistent-cum-variable} for such sources. GRS~1915$+$105 and Cyg~X$-$1 are representative examples, displaying substantial long term flux variations with $F_{\max}/F_{\min}\sim83$ and $\sim6-7$, respectively, together with recurrent spectral state transitions and pronounced spectro-temporal evolution \citep{Grinberg-etal2014,Kushwaha-etal2021,Athulya-etal2023,Broadbent-Phillipson2024,Majumder-etal2025a}. In contrast, the persistent BH-XRBs considered here exhibit more modest variability, with $F_{\max}/F_{\min}\sim2-5$, while maintaining sustained X-ray activity. For each source, we summarize its key physical parameters and observational properties, and classify it as persistent-cum-variable (PcV), persistent (P), or transient (T) according to its long-term X-ray emission characteristics. A comprehensive summary of the source properties is provided in Table~\ref{tab:Table-01}.}

{\red Indeed, black hole mass and spin are the two fundamental physical parameters that govern the nature of a black hole and its interaction with the surrounding environment. Reliable determination of these quantities is therefore essential for understanding black hole accretion processes, jet formation, and the overall evolution of black hole systems. However, both parameters are often subject to substantial observational and methodological uncertainties. It is noteworthy that black hole mass estimates derived from X-ray spectral modelling are often affected by significant model degeneracies and systematic uncertainties, and therefore do not carry the same level of reliability as dynamical measurements based on binary orbital parameters \cite[]{Casares-Jonker2015,Lopez-Jonker2019}. Likewise, black hole spin estimates are highly model dependent and subject to considerable systematic uncertainties \cite[see][for details]{Zdziarski-etal2026}. At present, spin measurements are primarily obtained using thermal continuum fitting \cite[]{McClintock-etal2014} and relativistic reflection modelling \cite[]{Reynolds2014}, both of which rely on assumptions regarding the accretion geometry, disc structure, and radiative properties of the system.

In this work, we have re-evaluated the mass estimates for all BH-XRBs and explicitly distinguished dynamically constrained masses from those obtained through indirect, model-dependent methods. For sources lacking dynamical confirmation, the quoted masses are retained but clearly identified as spectral model estimates. Furthermore, we introduce a classification in Table~\ref{tab:Table-01} that differentiates dynamically confirmed and candidate systems, thereby clarifying the varying levels of observational confidence associated with the reported mass values. We also indicate, wherever available in the literature, the method used to estimate the black hole spin for each source in the following subsections, enabling readers to assess the corresponding level of confidence and possible systematic uncertainties associated with the reported spin values \cite[see also][]{McClintock-etal2014, Shashank-etal2025, Garcia-etal2026}.}

\subsection{Persistent-cum-Variable Sources}\label{sec3:pcv}

BH-XRBs exhibiting persistent yet aperiodic variability are herein categorized as persistent-cum-variable systems. These sources maintain a persistently bright X-ray emission on relatively short timescales (days to weeks), while manifesting transient like variations over longer durations (months). \textit{AstroSat} has observed three such BH systems, such as GRS 1915$+$105, Cyg X$-$1, and Cyg X$-$3, which are described below.

\begin{itemize}
    \item \textbf{GRS 1915$+$105} is a well-known Galactic microquasar \cite[]{Mirabel-Rodrigue1994}, and one of the brightest BH-XRBs, discovered in 1992 by the {\it WATCH} instrument onboard \textit{GRANAT} \cite[]{Castro-Tirado-etal1992}. The system comprises a K–M III-type giant companion star with a mass of $\sim1.4~{\rm M}_\odot$ and a black hole as the primary compact object, with an orbital period of 34 days \cite[]{Greiner-etal2001a}. Based on its rapid X-ray variability, the black hole mass was initially estimated to lie in the range $10-33~{\rm M}_\odot$ \cite[]{Morgan-etal1997}. Subsequent radial velocity studies provided a more robust constraint yielding a mass of $\sim 14~{\rm M}_\odot$ \cite[]{Greiner-etal2001b}. 
    Using parallax measurements, \citet{Reid-etal2014} further {\red dynamically constraint} the black hole mass to be $ 12.4^{+2.0}_{-1.8}~{\rm M}_\odot$ and refined the distance to $8.6$ kpc.
    However, a more recent and robust estimation suggests the distance {\red and inclination} of the source to be $9.4\pm0.6$ kpc {\red and $64^{\circ}\pm4^{\circ}$ respectively} \cite[]{Reid-etal2023}. Several indirect measurements indicate that GRS 1915$+$105 hosts a rapidly rotating black hole \cite[]{Middleton-etal2006, Miller-etal2013, Mills-etal2021, Wang-etal2023b} with the binary system viewed at a moderate inclination \cite[]{Fender-etal1999, Reid-etal2014}. Note that, wide-band spectral modelling of {\it AstroSat} observations constrained the black hole mass as $12.4-13.1~{\rm M}_\odot$ and the spin parameter as $a_{\rm k}=0.990-0.997$ {\red using continuum fitting method} \cite[]{Sreehari-etal2020}. GRS 1915$+$105 also exhibits variable radio fluxes in different spectral states \cite[]{Mirabel-Rodrigue1994, Fender-etal1999, Vadawale-etal2003,Fender-etal2004}. Most importantly, a superluminal relativistic jet ejection with an apparent velocity of $\sim 1.25$ c also reported for GRS 1915$+$105, which provides key insights into the disc–jet connection in BH-XRBs \cite[]{Mirabel-Rodrigue1994}. In general, radio quiet states of the source manifest very weak emission ($\lesssim 5$ mJy), whereas plateau and flaring states produce strong radio fluxes often exceeding $100$ mJy \cite[]{Fender-etal1999, Rodriguez-Mirabel1999}.

    Interestingly, GRS 1915$+$105 exhibits a remarkably broad range of peculiar and repeated structured rapid X-ray variability on timescales of seconds to minutes. Based on the morphology of its lightcurves and the corresponding patterns in color$-$color diagrams (CCDs), these variability patterns have been classified into $15$ distinct classes ($\alpha$, $\beta$, $\gamma$, $\delta$, $\theta$, $\kappa$, $\lambda$, $\mu$, $\nu$, $\rho$, $\phi$, $\chi$, $\omega$, $\xi$, and $\tau$), as observed with {\it RXTE}, {\it IXAE}, {\it INTEGRAL} and {\it AstroSat} \cite[]{Paul-etal1997, Paul-etal1998, Yadav-etal1999, Muno-etal1999, Belloni-etal2000, Naik-etal2002, Klein-wolt-etal2002, Hannikainen-etal2005, Pahari-etal2009, Athulya-etal2022, Majumder-etal2022, Dhaka-etal2025,Majumder-etal2026}.

    Over nearly three decades of activity, GRS 1915$+$105 has exhibited diverse timing features, including both LFQPOs \cite[]{Chen-etal1997a, Paul-etal1997, Nandi-etal2001a, Vadawale-etal2001a, Belloni-etal2012} and HFQPOs \cite[]{Morgan-etal1997, Strohmayer2001a, Belloni-etal2006, Belloni-Altamirano2013} observed with {\it RXTE}. Among these, stable HFQPOs in $63-71$ Hz range have been detected across several variability classes \cite[$\kappa$, $\gamma$, $\mu$, $\delta$, $\omega$, $\rho$, and $\nu$;][]{Belloni-Altamirano2013} along with simultaneous features at $\sim 34$ Hz and $\sim 41$ Hz reported in a few instances \cite[]{Strohmayer2001b}.

    Meanwhile, both soft ($\sim \red 0.1-7.4$ ms) and hard ($\sim \red 0.1-80.0$ ms) time-lags are observed at the LFQPO frequency ($\red 0.5-10.0$ Hz) of GRS 1915$+$105 with {\it RXTE} {\red (reference band: $2-5$ keV)} and {\it AstroSat} {\red (reference band: $3-6$ keV)} \cite[]{Reig-etal2000, Qu-etal2010, Pahari-etal2013, van-den-Eijnden-etal2016, Yadav-etal2016b,Dutta-etal2018, Rawat-etal2019, Zhang-etal2020b, Belloni-etal2024, Majumder-etal2026b}. {\red Note, however, that previous studies have used different energy band combinations for time-lag measurements. Using {\it RXTE}, \citet{Reig-etal2000}, \citet{Pahari-etal2013}, and \citet{van-den-Eijnden-etal2016} measured time-lags in the $5-13$ keV, $6-9$ keV, and $7-13$ keV bands relative to $2-5$ keV, $2-5$ keV, and $2-7$ keV reference bands, respectively. Using {\it AstroSat}, \citet{Yadav-etal2016a} and \citet{Rawat-etal2019} investigated time-lags across multiple energy bands relative to $3-4$ keV reference band, whereas \citet{Majumder-etal2024,Majumder-etal2025a} employed $6-25$ keV band relative to $3-6$ keV reference band. More recently, \citet{Majumder-etal2026b} measured time-lags in the $6-20$ keV band using the same $3-6$ keV reference band.} The time-lag at LFQPOs often exhibits sign reversal, indicating towards the physical origin of the lag behaviour \cite[]{Qu-etal2010}. Furthermore, \cite{Majumder-etal2024} reported the first detection of a soft lag at the HFQPO frequency ($\sim 70$ Hz), with a maximum delay of $\sim 3$ ms. Notably, the presence of a hard lag at the HFQPO of GRS 1915$+$105 was identified in earlier studies during the {\it RXTE} era \cite[]{Mendez-etal2013}. In addition to the timing characteristics, the wide-band spectral properties of GRS 1915$+$105 have been extensively studied using data from several missions. For example, simultaneous modelling of {\it RXTE} and {\it CGRO/OSSE} wide-band ($3-600$ keV) spectra revealed a power-law like spectral shape without any break in $\chi$, $\alpha$, $\omega$, $\nu$, $\rho$ and $\kappa$ classes \cite[]{Zdziarski-etal2001, Zdziarski-etal2005}. Furthermore, wide-band energy spectra ($3-300$ keV) with \textit{RXTE} and \textit{INTEGRAL} in different variability classes ($\nu$, $\lambda$, $\rho$, $\beta$ and $\chi$) are also studied and found to be dominated by thermal and/or non-thermal Comptonizations at higher energies \cite[]{Hannikainen-etal2005, Rodriguez-etal2008, Droulans-Jourdain2009}.

    \item \textbf{Cyg X$-$1}, one of the brightest HMXBs in the sky, was discovered in X-rays by the \textit{Uhuru} satellite in $1971$ \cite[]{Tananbaum-etal1972}. The system comprises a dynamically confirmed black hole \cite[]{Bolton1972,Webster-Murdin1972} with a mass of $21.2 \pm 2.2~{\rm M}_{\odot}$ \cite[]{Miller-Jones-etal2021}, and a massive O9.7 Iab-type supergiant companion HDE 226868 \cite[]{Walborn1973}. The distance and inclination of the system are estimated to be $2.22^{+0.18}_{-0.17}$ kpc and $27.51^{\circ+0.77}_{~-0.17}$, respectively \cite[]{Miller-Jones-etal2021}. The spin of the black hole has been estimated as $a_{\rm k} > 0.98$ \cite[]{Tomsick-etal2014,Gou-etal2014,Kushwaha-etal2021, Zhao-etal2021} {\red using continuum fitting method}, although \citet{Zdziarski-etal2024} noted that continuum fitting spin estimates can be highly model dependent. 

    Over the years, numerous spectral studies have shown that Cyg X$-$1 predominantly remains in LHS, however, it occasionally transits to HSS for durations from months to years \cite[]{Grinberg-etal2014,Kushwaha-etal2021}. The LHS spectra are well described by thermal Comptonization models \cite[]{Ling-etal1983,Parker-etal2015} while a multicolor disc blackbody component becomes dominant in HSS due to the strong accretion disc emission. \citet{Kushwaha-etal2021} reported an `extremely' soft state during \textit{AstroSat} observations with a disc temperature of $\sim 0.5$ keV and $\sim 45 \%$ of fractional disc flux contribution. In the prolonged LHS, Cyg X$-$1 shows weak yet persistent radio emission of approximately $5-30$ mJy \cite[]{Stirling-etal2001,Zdziarski-etal2020b}. In comparison, compact jets with fluxes of about $0.5-10.0$ mJy are also observed during the softer states \cite[]{Rushton-etal2012,Zdziarski-etal2020b}. While the radio emission in the LHS correlates with both hard and soft X‑rays, in the soft state the radio signal lags the soft X‑rays by roughly $\sim 100$ days \cite[]{Zdziarski-etal2020b}.
     
    The short timescale variability of this source has been extensively studied using \textit{RXTE} and \textit{AstroSat} data \cite[]{Pottschmidt-etal2003,Grinberg-etal2014,Misra-etal2017}. The PDS typically show multiple broad components which are satisfactorily modelled with \texttt{Lorentzian} profiles. An exception occurs during the HSS, where a single \texttt{Lorentzian} combined with a \texttt{powerlaw} component adequately describes the PDS \cite[]{Churazov-etal2001,Axelsson-etal2005}. The centroid frequencies of these \texttt{Lorentzian} components shift towards higher values during transitions from LHS to HSS, but do not exceed $\sim 25$ Hz \cite[]{Axelsson-etal2005}. Moreover, the total rms variability and flux (count rate) for Cyg X$-$1 is found to be linear across all its spectral states, including LHS, IMS and HSS. However, the slope of this relation is generally steeper in the hard state compared to the soft and intermediate states \cite[]{Gleissner-etal2004}. In addition, \cite{Kushwaha-etal2021} found that the total rms during HSS ($\sim 25\%$) is comparable to that during LHS. Interestingly, significant polarized emission has been detected for this source in both LHS and HSS \cite[]{Krawczynski-etal2022, Majumder-etal2025b}. Recently, \cite{Rout-etal2025} reported a narrow dip in coherence at $0.005$ Hz possibly resulting in QPO signature using \textit{AstroSat} observations.
    
    \item \textbf{Cyg X$-$3} is an HMXB and its companion is a Wolf–Rayet star \cite[]{van-Kerkwijk-etal1992}. Despite extensive studies, the nature of the compact object remains uncertain. \citet{Bhargava-etal2017} reported an orbital period of 4.8 h, showing remarkable stability over more than four decades. The estimated mass of the compact object is $2.4^{+2.1}_{-1.1}~{\rm M}_{\odot}$ \cite[]{Zdziarski-etal2013}, although \cite{Koljonen-etal2017} reported its mass to be $\le 5~{\rm M}_{\odot}$. This renders the characteristics of the compact object enigmatic, as it resides within the mass gap of $2-5 ~{\rm M}_\odot$ \cite[]{Zdziarski-etal2013}. Recently \cite{Miura-etal2025} suggested a mass range of $\sim 1.3-5.1~{\rm M}_\odot$ for $i\sim 25^{\circ}$, which is consistent with the previous mass estimates, however for $i\sim 35^{\circ}$, source mass is estimated as $\sim 1.3-24.0~{\rm M}_\odot$. This is based on \textit{XRISM/Resolve} high-resolution spectroscopy, where orbital phase-resolved spectral analysis is performed to obtain the velocity amplitude which is further used to estimate the mass of the central compact object. On the other hand, \cite{Antokhin-etal2022} suggests a mass of $\sim 7.2~{\rm M}_\odot$ and inclination $\sim 29.5^{\circ} \pm 1.2^{\circ}$ based on modelling X-ray and infrared lightcurves. Recent analysis suggests the distance of the source to be $8.95\pm0.96$ kpc \cite[]{Reid-etal2023}. Compared to other X-ray binaries, Cyg X$-$3 is exceptionally radio loud possibly due to the interaction between its relativistic jet and the dense stellar wind of the companion \cite[]{Zdziarski-etal2016}. Cyg X$-$3 has exhibited giant radio flares of flux $\sim 20$ Jy on several occasions \cite[]{Schalinski-etal1995, Mioduszewski-etal2001}. The jets during $1997$, $2000$ and $2001$ were observed to be moderately relativistic in nature with velocity $\sim 0.48-0.81$ c \cite[]{Miller-etal2004}.  During the \textit{AstroSat} era, the radio peak flux was observed to be $\sim 20$ Jy using RATAN-600 telescope at $4.6$, $8.2$ and $11.2$ GHz \cite[]{Trushkin-etal2017}. 
    
    LFQPOs are detected in the range of $\nu_{\rm QPO} \sim 5-8$ mHz, $12-14$ mHz, and $18-24$ mHz, which were reported during the rising phase of the flaring hard state and are observed to disappear during the decay phase \cite[]{Pasham-etal-2013,Pahari-etal2017}. These variations are attributed to changes in the accretion flow driven by the stellar wind. Corresponding soft-lags and an increase in rms variability with energy were also detected \cite[]{Pahari-etal2017}. {\red Notably, the energy dependent time-lag analysis is performed using the $2-5$, $5-10$, $10-14$, and $14-30$ keV energy bands, with $10-14$ keV band considered as the reference band.} Earlier, \cite{Rajeev-etal1994} studied the spectral variability of this source. \cite{Koljonen-etal2010} suggested six spectral states for this source based on X-ray and radio observations. Hypersoft state (equivalently HSS) is characterized by softer spectrum along with low or quenched radio emission, whereas flaring soft X-rays (FSXR; equivalently SIMS) shows relatively harder spectrum with strong radio emission. \cite{Pahari-etal2018a} observed a rapid transition from the hypersoft state to FSXR with a sharp rise in radio flux indicating possible jet ejection. \citet{Zdziarski-etal2016} reported a complex radio/X-ray correlation in the hard state, with soft X-rays positively and hard X-rays negatively correlated with radio flux, indicating an intrinsic intermediate spectral state affected by strong local absorption.

    The high polarization fraction of $\sim 20\%$ suggests the presence of collimated outflow forming a funnel like geometry of opening angle $<15^{\circ}$ in the inner disc region of Cyg X$-$3. This outflow produces the geometrical beaming of radiation and results in the apparent high luminosity of the source lying in the ultra-luminous regime \cite[]{Veledina-etal2024, Majumder-etal2025b}. Notably, the presence of an optically thick outflow covering the line of sight at a high inclination in SS $433$ is coarsely comparable to the case of Cyg X$-$3, which is observed at a much smaller inclination \cite[]{Veledina-etal2024}.

    {\red
    Given the long standing uncertainty regarding the nature of the compact object in Cyg X$-$3, we mention that the source exhibits powerful radio flaring and relativistic jet activity that are strongly coupled to its X-ray spectral evolution \citep{Szostek-etal-2008,Koljonen-etal2010,Zdziarski-etal-2013}, and its radio luminosity is reported to be at least $\sim 30$ times higher than that typically observed in neutron star X-ray binaries \citep{Migliari-etal-2006}. In addition, infrared and X-ray lightcurve modelling yields a compact object mass of $\sim 7.2\,M_{\odot}$, favouring a black hole interpretation \citep{Antokhin-etal-2022}, while the absence of detectable X-ray pulsations provides no direct evidence for a neutron star nature \citep{Kitamoto-etal1992,Zdziarski-etal-2013}. X-ray polarimetric measurements reveal a polarization degree higher than that typically observed in black hole systems, although these results do not uniquely constrain the nature of the compact object \citep{Veledina-etal2024,Majumder-etal2026}. Moreover, recent orbital phase-resolved XRISM spectroscopy yields compact object mass ranges that are consistent with either a neutron star or a black hole, depending on the assumed orbital inclination \citep{Miura-etal2025}. Therefore, no single observation currently provides a definitive identification of the compact object. Nevertheless, the collective evidence from its X-ray spectral behaviour, extreme radio jet activity, absence of pulsations, and available mass estimates weighs more strongly in favour of a black hole than a neutron star. Accordingly, Cyg X$-$3 is best regarded as a \textit{black hole candidate}, although its compact object nature remains unconfirmed. 
    }
    
\subsection{Persistent Sources}\label{sec3:persist}

BH-XRBs accreting matter at quasi-stable rate are generally categorized as persistent systems, and characterized by their steady and relatively bright X-ray emission over extended periods. These sources primarily accrete matter from their companion stars through stellar winds. A total of eight such persistent systems are known, among which \textit{AstroSat} has observed five sources. Notably, LMC X$-$1 and LMC X$-$3 are dynamically confirmed extragalactic black hole systems. The observational and physical characteristics of all five persistent sources are described below.

    \item \textbf{LMC X$-$1}, located in the neighbouring LMC galaxy at a distance of approximately $48$ kpc, is the first BH-XRB discovered beyond our Galaxy \cite[]{Mark-etal1969}. It is classified as an HMXB, where accretion occurs through stellar winds from its massive B5 type supergiant companion. The {\red dynamically estimated} system parameters are well constrained with an orbital inclination of $36.38^{\circ} \pm 1.92^{\circ}$, a companion mass of $31.79 \pm 3.48~{\rm M}_{\odot}$, and a black hole mass of $10.91 \pm 1.41~{\rm M}_{\odot}$ \cite[]{Orosz-etal2009}. {\red Using continuum fitting method, \cite{Gou-etal2009} estimated source spin as $0.92^{+0.05}_{-0.07}$, while the Fe line fitting analysis yielded a spin $\sim 0.97$ \cite[]{Steiner-etal2012b}. Recent measurements using wideband \textit{AstroSat} and \textit{NuSTAR} observations reported spin estimates as $0.82-0.92$ (continuum fitting) and $0.93-0.94$ (Fe-line fitting), respectively \cite[]{Bhuvana-etal2021, Bhuvana-etal2022}.}

    LMC X$-$1 is a persistent X-ray source, which remains steadily bright and is consistently observed in the soft spectral state by various X-ray missions \cite[]{Haardt-etal2001, Nowak-etal2001}. Its soft state spectrum is typically characterized by a multi-temperature disc blackbody with a temperature of around 1 keV and a steep power-law tail ($\Gamma > 2.0$). The hydrogen column density ($n_{\rm H}$) along the line of sight is estimated to be $\sim 1 \times 10^{22}~{\rm cm}^{-2}$ \cite[]{Hanke-etal2010}. During the soft state, QPOs have been detected at $\sim 26-29$ mHz, $75$ mHz, and $142$ mHz \cite[]{Ebisawa-etal1989, Alam-etal2014}. The PDS of LMC X$-$1 approximately follows a $f^{-1}$ dependence with low fractional rms, which is consistent with that of canonical soft state of other BH-XRBs. Over long timescales spanning several years, LMC X$-$1 exhibits stable flux emission and hence, has less variance. However, short-term variability is found to be relatively higher in this source.
    
    The accretion disc of LMC X$-$1 is known to deviate from the canonical $L \propto T^{4}$ relation \cite[]{Ruhlen-etal2011}, suggesting a complex interplay between the disc structure and the corona.  Furthermore, Imaging X-ray Polarimetry Explorer (\textit{IXPE}) \cite[]{Weisskopf-etal2022} observations revealed no significant X-ray polarization with an upper limit of $1.1\%$ at the $99\%$ confidence level \cite[]{Podgorny-etal2023, Majumder-etal2025b}.

    \item \textbf{LMC X$-$3}, located in LMC at a distance of $48.1$ kpc \cite[]{Straub-etal2011}, is an X-ray binary whose companion is a B3 V-type star with a mass of $3.63 \pm 0.57~{\rm M}_{\odot}$. The system belongs to LMXB class, where accretion occurs via Roche lobe overflow. The source mass has been dynamically confirmed to be $6.98 \pm 0.56~{\rm M}_{\odot}$ with an inclination angle of $69.24^{\circ}\pm0.72^{\circ}$ \cite[]{Orosz-etal2014}. {\red Using the continuum fitting method in stable spectral state yielded black hole spin parameter of $a_{\rm k} = 0.25^{+0.20}_{-0.29}$ \cite[]{Steiner-etal2014}. These results were later corroborated by \cite{Bhuvana-etal2021,Bhuvana-etal2022} by constraining the spin in range $0.22-0.41$, based on \textit{AstroSat} observations.}

    LMC X$-$3 is the brightest X-ray source in LMC and exhibits remarkable variability in X-rays. This variability extends over timescales ranging from days to years and is often associated with transitions between different spectral states. Although the source spends the majority of its time in soft spectral state, similar to LMC X$-$1, it occasionally transits into a hard state. The soft state spectrum of LMC X$-$3 is dominated by a thermal disc blackbody component with a temperature of $\sim 1$ keV accompanied by a steep ($\Gamma > 2$) high energy power-law tail. In contrast, the hard state is characterized by a cooler disc ($T < 0.5$ keV) and a dominant hard ($\Gamma < 2$) power-law component. During the hard state, the PDS displays strong wide-band ($0.01-100.00$ Hz) variability with an rms amplitude of $\sim 40\%$ \cite[]{Boyd-etal2000}, along with a QPO at $\sim 0.46$ Hz having an rms amplitude of $\sim 14\%$. The source is also known to exhibit an Anomalous Low/Hard State (ALS) distinguished by an extremely low luminosity ($L \sim 10^{33}$ erg s$^{-1}$ in $0.6-5.0$ keV range) and a purely absorbed power-law spectrum with $\Gamma \sim 1.4$ \cite[]{Torpin-etal2017}. Despite its pronounced long term variability, the source shows relatively weak short-term variability in contrast to LMC X$-$1. Note that LMC X$-$3 maintains a stable inner disc radius, suggesting that the accretion disc remains fixed at the innermost stable circular orbit ($R_{\rm ISCO}$) \cite[]{Steiner-etal2010}. Recent measurements with \textit{IXPE} detected a significant polarization degree of $3.04 \pm 0.40 \%$ at a polarization angle of $-44.24^{\circ} \pm 3.77^{\circ}$ in the $2-8$ keV energy range \cite[]{MajumderSeshadri-etal2024,Garg-etal2024}, with the inferred spin value consistent with previous estimates.

    Interestingly, LMC X$-$3 has been extensively observed across multiple wavelengths \cite[]{Brocksopp-etal2001, Nowak-etal2001, Page-etal2003, Torpin-etal2017}. Its X-ray emission primarily arises from the inner accretion disk, while the UV emission can originate from several components including the accretion disk, ionized stellar wind, X-ray irradiated surface of the B-star companion, and possibly a hot spot on the outer disk. Simultaneous X-ray, UV, and optical observations help disentangle these contributions \cite[]{Torpin-etal2017}. Note that the optical/UV emission leads the X-ray variability by approximately $\sim 8$ days \cite[]{Brocksopp-etal2001, Torpin-etal2017}.

    \item \textbf{GRS 1758$-$258} (alternatively known as 1RXS J180112.6$-$254435, Swift J1801.1$-$2544) is one of the brightest BH-XRBs in our Galaxy, located close to the Galactic Centre at an angular distance of approximately $5^{\circ}$ away \cite[]{Sunyaev-etal1991}. Owing to its proximity to the Galactic Centre, the high optical extinction has made the detection of its companion star challenging. Nevertheless, \citet{Chen-etal1994} ruled out the presence of a massive O-type or red supergiant companion through optical, near-infrared, and radio observations. From the mass–luminosity relation, \citet{Keck-etal2001} estimated the black hole mass to be greater than $8.6~{\rm M}_{\odot}$. \cite{Tetarenko-etal2020} estimated an average jet power of $\sim (1.1-5.7) \times 10^{36}$ erg s$^{-1}$ over $0.12-0.31$ Myr by studying the radio jet-ISM interactions {\red and constrained the distance as $8\pm 1$ kpc. Reflection modelling using the wideband \textit{AstroSat} spectrum yielded an inclination angle of $61^{\circ}\pm2^{\circ}$ \cite[]{Bhuvana-etal2023}.} The source was identified as a microquasar exhibiting two extended and symmetric radio lobes, resembling relativistic jets, in deep $6$ cm \textit{VLA} observations \cite[]{Rodriguez-etal1992}. These jets are observed during its hard spectral state and are quenched during the soft state \cite[]{Soria-etal2011}.
    
    The source predominantly remains in the hard spectral state, showing bright flux of $\sim 2 \times10^{37}$ erg s$^{-1}$ in the $4-300$ keV range \cite[]{Mereghetti-etal1994}. Based on \textit{ASCA} observations, \citet{Mereghetti-etal1997} estimated the interstellar hydrogen column density along its line of sight to be $\sim 1.5 \times10^{22}~{\rm cm}^{-2}$. Latter, the hard spectral state of the source is characterized by a power-law spectrum with a photon index of $\sim 1.5$ and a high energy cut-off above $100$ keV \cite[]{Lin-etal2000}. The hard state energy spectrum exhibits a broad Fe K$\alpha$ emission line at $6.4$ keV and a high energy relativistic reflection component \cite[]{Sidoli-Mereghetti2002,Jana-etal2022b,Bhuvana-etal2023}. During the hard state, PDS display flat-topped noise that breaks into a power-law at frequencies between $0.1$ and $0.8$ Hz, accompanied by a high fractional rms variability \cite[]{Main-etal1999}. Although the source primarily persists in the hard state, it occasionally undergoes transitions into intermediate and soft states \cite[]{Smith-etal2001b,Pottschmidt-etal2006}. While no disc component is detected during the hard state, both intermediate and soft states reveal a prominent thermal disc component \cite[]{Soria-etal2011,Bhuvana-etal2023}. A distinctive feature of the soft state in this source is the absence of a luminosity increase, contrary to the behaviour observed in most BH-XRBs. Instead, the soft state luminosity is found to be lower by a factor of $\sim 3$ compared to that of the hard state, leading to its classification as the `dim-soft' state \cite[]{Smith-etal2001a,Miller-etal2002}.

    \item \textbf{1E 1740.7$-$2942} (alternatively known as The Great Annihilator) was first detected as a persistent hard X-ray source by the \textit{Einstein} observatory \cite[]{Hertz-Grindlay1984}. It lies in a densely populated region near the Galactic Centre, approximately $5^{\circ}$ away from GRS 1758$-$258. Similar to GRS 1758$-$258, 1E 1740.7$-$2942 exhibits double sided radio jets and is therefore classified as a microquasar \cite[]{Mirabel-etal1992}. {\red Wideband spectral reflection modelling results in a moderately rotating ($a_{\rm k} \gtrsim 0.5$) black hole of mass $\sim 5~{\rm M}_{\odot}$ \cite[]{Stecchini-etal2020} with inclination angle $\geq 50^{\circ}$. A similar inclination angle of $ 55^{\circ}\pm1^{\circ}$ is reported using \textit{AstroSat} data \cite[]{Bhuvana-etal2023}. The distance of this source is constraint as $8\pm1$ kpc \cite[]{Tetarenko-etal2020}.} The source suffers from a high hydrogen column density ($n_{\rm H} \sim 12 \times 10^{22}$ atoms cm$^{-2}$), attributed to a dense molecular cloud along the line of sight, which has impeded the identification of its companion star \cite[]{Sheth-etal1996}. Based on \textit{VLT} observations, \cite{Marti-etal2000} suggested that the companion is likely of spectral type between B8 V and G5 III, indicating that 1E 1740.7$-$2942 belongs to the LMXB class. Radio observations by \textit{VLA} have revealed a double sided jet structure, where the jet has been fitted with the precessing jet model \cite[]{Luque-Escamilla-etal2015}. \cite{Tetarenko-etal2020} estimated an average jet power of $\sim (0.7-3.5)\times 10^{37}$ erg s$^{-1}$ over $0.10-0.26$ Myr by studying the radio jet-ISM interactions. 

    The source displays spectral and temporal characteristics closely resembling GRS 1758$-$258. It predominantly remains in the hard state, emitting strongly in hard X-rays up to $\sim 200$ keV, with a spectral cut-off corresponding to an electron temperature of $kT_{\rm e} \sim 100$ keV \cite[]{Heindl-etal1993, delSanto-etal2005}. The hard state spectra exhibit canonical features, including an Fe K$\alpha$ emission line, a `Compton reflection hump', and a Comptonized continuum \cite[]{Castro-etal2014, Stecchini-etal2020}. Interestingly, unlike GRS 1758$-$258, the hard state spectra also reveal a soft thermal component with a temperature of $\sim 0.4$ keV \cite[]{Reynolds-Miller2010, Natalucci-etal2014}. The source occasionally transits between intermediate and hard states \cite[]{Smith-etal2002, delSanto-etal2005}. A notable feature of 1E 1740.7$-$2942 is the occurrence of an unusual `dim–soft' state, characterized by a disc blackbody with $kT_{\rm in} \sim 0.5$ keV, a steep power-law component ($\Gamma > 2$), and an overall luminosity lower than that of the hard state \cite[]{delSanto-etal2005}. 
    
    \item \textbf{4U 1957$+$115} is a persistent X-ray source discovered in 1974 \cite[]{Giacconi-etal1974}. Its inclination angle {\red and mass} is not well constrained and has been proposed to lie in the range $\sim13^{\circ}-78^{\circ}$ {\red and $\rm<6.2~M_{\odot}$ respectively} \cite[]{Gomez-etal2015, Maitra-etal2014}. However, using a relativistic disc model, \citet{Mudambi-etal2022} constrained the source inclination to $\sim 72^{\circ}$ assuming a distance of $\sim 10$ kpc and a black hole mass of $6.0~{\rm M}_{\odot}$. The source distance has also been estimated to be in the range of $\sim 20-40$ kpc \cite[]{Gomez-etal2015}. The system possibly hosts a maximally rotating black hole {\red and its spin is estimated as $a_{\rm k} \sim 0.99$ using both continuum fitting method and reflection modelling \cite[]{Barillier-etal2023}. Evidence of a secondary star is not yet detected at any wavelength \cite[]{Hakala-etal-2014}.} 
    
    The source is observed to be in disc dominated state \cite[]{Mudambi-etal2022, Barillier-etal2023}. A positive correlation has been observed between the mass accretion rate and the inner disc radius, with a Spearman rank correlation coefficient of $\sim 0.93$. The spectral variability of the source was also  studied earlier by \cite{Singh-etal1994}. Polarimetric constraints in the soft state of the source confirm the presence of returning radiation, indicating a rapidly rotating black hole at the center \cite[]{Marra-etal2024, Majumder-etal2025b}.

\subsection{Transient Sources}\label{sec3:transi}

BH-XRBs accreting at relatively low rates are generally observed in a quiescent phase (QP) during which their X-ray luminosity remains extremely faint, typically below the detection limit of $\sim 15$ milliCrab based on one day average of \textit{MAXI}/GSC data \cite[]{Sugizaki-etal2011}. When the accreted matter from the companion star overflows Roche lobe, a sudden increase in the mass accretion rate triggers an outburst, leading to a substantial rise in X-ray brightness \cite[]{Lasota2001,Frank-etal2002}. Such systems are classified as transient sources which are characterized by episodic outbursts separated by long intervals of quiescence. Generally, the source outburst profiles are of two types, namely FRSD (fast rise $\sim 1-5$ days and slow decay $\sim 50-100$ days), and SRSD (slow rise $\sim 10-50$ days and slow decay $\sim 200-300$ days) \cite[][and references therein]{Chen-etal1997b,Nandi-etal2012,Corral-Santana-etal2016,Tetarenko-etal2016,Nandi-etal2018,Sreehari-etal2018,Shidatsu-etal2019}. To date, $72$ transient BH-XRBs have been detected, of which \textit{AstroSat} has observed 26 sources. Among these, seven systems are dynamically confirmed black hole binaries. The observational and physical characteristics of all 26 transient sources observed by \textit{AstroSat} are summarized below.

    \item \textbf{IGR J17091$-$3624} is a Galactic BH-XRB transient, first identified in $2003$ with {\it INTEGRAL/IBIS} \citep{Kuulkers-etal2003}. The system exhibits recurrent outbursts, typically separated by quiescent intervals of approximately four years \cite[]{Capitanio-etal2009, Krimm-etal2011b, Miller-etal2016, Katoch-etal2021}. Despite multiple outbursts over the past two decades, the fundamental system parameters, such as the black hole mass, inclination, and distance remain poorly constrained due to the non-detection of its optical counterpart. Nevertheless, indirect estimates based on X-ray reflection spectroscopy suggest an inclination in the range of $\sim 30^{\circ}-40^{\circ}$ \cite[]{Xu-etal2017, Wang-etal2018b}, whereas spectro-temporal modelling and correlation studies indicate a black hole mass in the range of $8.7-15.6~{\rm M}_{\odot}$ \citep{Iyer-etal2015}. Luminosity based arguments further suggest a source distance of $\sim 11-17$ kpc \cite[]{Rodriguez-etal2011}. The source is found to exhibit compact radio jets during the hard states and weak radio emission during state transitions \cite[]{Rodriguez-etal2011, Gatuzz-etal2020}. 

    IGR J17091$-$3624 exhibits a wide range of complex X-ray variability signatures remarkably similar to those seen in GRS 1915$+$105, including `structured' classes such as $\rho$, $\nu$, $\alpha$, $\lambda$, and $\beta$ \cite[]{Altamirano-etal2011, Capitanio-etal2012, Radhika-etal2018, Katoch-etal2021, Wang-etal2024}. These classes are characterized by distinct combinations of quasi-periodic flares, dips, and transitions between hard and soft states on timescales of seconds to minutes. A comprehensive classification of the nine variability classes \cite[]{Court-etal2017} identifies IGR J17091$-$3624 as a unique laboratory that bridges the gap between canonical black hole transients and highly variable GRS 1915$+$105. HFQPO has also been detected in IGR J17091$-$3624 at $\sim 66$ Hz, indicating strong similarities with the HFQPOs observed in GRS 1915$+$105 \cite[]{Altamirano-Belloni2012}. Spectral studies of IGR J17091$-$3624 reveal rapid state transitions generally observed for black hole transients \cite[]{Altamirano-etal2011, Capitanio-etal2012}. The source shows a typical Comptonized spectrum in the hard state and a dominant disc component in the soft state \citep{Radhika-etal2018}. {\it AstroSat} observations reveal the spectro-temporal evolution of the source during its 2016 outburst \cite[]{Katoch-etal2021}. Furthermore, exceptionally high polarization degree of $\sim 9\%$ is observed in the LHS of IGR J17091$-$3624 with {\it IXPE} observations \cite[]{Ewing-etal2025, Majumder-etal2025b}.

    \item \textbf{GX 339$-$4} was discovered in 1973 by the {\it OSO$-$7} satellite \cite[]{Markert-etal1973}, and the source undergoes regular outbursts with a recurrence period of approximately $2-3$ years \cite[]{Sreehari-etal2018}. Since its discovery, GX 339$-$4 has been extensively studied across multiple wavelengths \cite[]{Mendez-etal1997, Corbel-etal2003, Hynes-etal2003, Zdziarski-etal2004, Belloni-etal2005, Shaposhnikov-Titarchuk2009, Nandi-etal2012}. Near-infrared absorption-line studies \cite[]{Heida-etal-2017} indicate the presence of a K-type giant companion star feeding a central black hole through Roche-lobe overflow, with an orbital period of $\rm P_{orb}=1.76$ d \cite[]{Hynes-etal2003, Heida-etal-2017}. {\red The dynamically constrained black hole mass is computed as $\sim 2.3-9.5$ M$_{\odot}$ \cite[]{Heida-etal-2017}. In addition, using the continuum fitting method,} the source mass is estimated to lie in the range $8.3-11.9~{\rm M}_{\odot}$ \cite[]{Sreehari-etal2019a}, with a distance of $8.4 \pm 0.9$ kpc \cite[]{Parker-etal2016}. Although earlier {\red relativistic reflection} studies suggested a high spin parameter of $a_{\rm k}=0.95^{+0.08}_{-0.09}$ \cite[]{Garcia-etal2015,Parker-etal2016}, \citet{Zdziarski-etal2025} recently reported a negative spin value. {\red Further, \cite{Zdziarski-etal2019} constrained the inclination angle as $40^{\circ}-60^{\circ}$.} This source exhibited strong radio flares during the HIMS$-$SIMS transition with radio flux $\sim 20-50$ mJy \cite[]{Gallo-etal2003, Corbel-etal2013, Islam-Zdziarski2018, Corbel-etal2021}. During $2002$ outburst, the optically thin jet exhibited an apparent velocity of $\sim 0.9\,{\rm c}$. 

    During outbursts, GX 339$-$4 exhibits rapid spectral state transitions, tracing all four canonical spectral states \cite[LHS, HIMS, SIMS and HSS:][and references therein]{Belloni-etal2005, Debnath-etal2010, Nandi-etal2012,Aneesha-etal2024}. The PDS of the source shows a wide range of timing features including both broad-band noise and narrow QPOs. The presence of type-A, type-B, and type-C QPOs along with harmonic features were reported earlier \cite[]{Belloni-etal2005,Motta-etal2011,Aneesha-etal2024, Sharma-etal2026}, whereas energy dependent wide-band features were systematically investigated using {\it AstroSat} observations \cite[]{Chand-etal2024}. Furthermore, {\it IXPE} observations reveal a polarization degree of $1.3 \pm 0.3\%$ and a polarization angle of $74^{\circ} \pm 7^{\circ}$ in the $3-8$ keV energy band \cite[]{Mastroserio-etal2025, Majumder-etal2025b}.

    \item \textbf{Swift J1658.2$-$4242} underwent an outburst in $2018$ \cite[]{Grebenev-etal2018}. The system seems to host a mid- to late–K type companion star and has an orbital period of approximately $4.0 \pm 0.4$ hours \cite[]{Mondal-Jithesh2023}. Source mass, inclination, and distance are estimated to be $ 13.0-14.2~{\rm M}_{\odot}$, $64^{\circ+2^\circ}_{~-3^\circ}$, and $> 3$ kpc, respectively \cite[]{Xu-etal2018, Russell-etal2018,Mondal-Jithesh2023}, while the spin is constrained to $a_{\rm k} > 0.96$ {\red using reflection modelling} \cite[]{Xu-etal2018} . It has exhibited strong radio flare with flux $\sim 60$ mJy during its 2018 outburst \cite[]{Bogensberger-etal2020}. 

    During the outburst, observations with \textit{NuSTAR} and \textit{Swift} confirmed a hard spectral state with $\Gamma = 1.6$ and $kT_{\rm e} \sim 22$ keV. The presence of a broad Fe K$\alpha$ line and a `Compton hump' indicates strong relativistic reflection feature, while an absorption line at $\sim 7.03$ keV suggests possible signature of outflows. Type-C QPOs in the $0.14-0.21$ Hz range were also reported \cite[]{Xu-etal2018}. The decrease of flux as $\sim 45\%$ was observed by \textit{NuSTAR}, \textit{Swift}, and \textit{XMM-Newton} along with the detection of $6-7$ Hz QPOs and an increase in rms from $2\%$ to $10\%$ \cite[]{Xu-etal2019}. In \textit{AstroSat} data, type-C QPOs were observed in the $1.5-6.6$ Hz range \cite[]{Jithesh-etal2019}. During the state transition, type-A QPOs appeared at $\nu_{\rm QPO} \sim 7$ Hz and type-B QPOs at $\nu_{\rm QPO} \sim 4$ Hz \cite[]{Choudhury-etal2025}. Notably, type-C and type-A QPOs exhibited soft-lags of $\le 10.0$ ms and $\sim 5.0$ ms, respectively, whereas type-B QPOs showed a hard-lag of $\sim 10.0$ ms \cite[]{Choudhury-etal2025}, {\red considering subject and reference bands as $6-15$ keV and $2-6$ keV, respectively}. The source also exhibited $15$ `flip-flop' events during the outburst, where the brighter states were on average $\sim 77\%$ more luminous than the dim states \cite[]{Bogensberger-etal2023}.

    \item \textbf{4U 1630$-$472} was first observed in 1972 by \textit{Uhuru} \cite[]{Jones-etal1976}. Source distance is estimated to be $11.5\pm0.3$ kpc \cite[]{Kalemci-etal-2018}. {\red The black hole mass is constrained through spectral modelling as $3-9$ M$_{\odot}$ \cite[]{Baby-etal2020}, whereas scaling method yielded source mass as $\sim 10.0 \pm 0.1~{\rm M}_{\odot}$ \cite[]{Seifina-etal2014}}. The system is highly inclined with angle $\red 64^{\circ+2}_{~-3}$ \cite[]{King-etal2014}. The spin of the source is estimated as $\red a_{\rm k} = 0.817\pm0.014$ \cite[]{Liu-etal2022a}, {\red using relativistic reflection fitting, whereas continuum spectral modelling resulted in $a_{\rm k} = 0.92 \pm 0.04$ \cite[]{Pahari-etal2018b}.} Although no optical companion has been detected, \cite{Augusteijn-etal2001} reported a possible infrared counterpart, which could be an early-type or Be star. 
    
    Till date, the source has exhibited more than $25$ outbursts since its discovery. Among them, a few ‘super-outbursts’, recurring on typical timescales of $12-14$ years, lasted for $\sim 2$ years with intensities $\ge 0.8$ Crab \cite[]{Parmar-etal1997, Share-etal1978, Tomsick-etal2005,Baby-etal2020,Chatterjee-etal2022,Parra-etal2025}, while numerous ‘mini-outbursts’, occurring every $\sim 600$ days, persisted for $\sim 100-200$ days with peak intensities of $\sim0.5$ Crab \cite[]{Baby-etal2020}. \cite{Pahari-etal2018b} reported two moderately blue shifted Fe lines at $6.705 \pm 0.002$ keV and $6.974^{+0.004}_{-0.003}$ keV, attributed to low velocity ionized disc winds. The source is generally observed in a disc dominated state, with a disc flux fraction of $\sim 0.97$.  
    The source exhibited similar outburst profiles in $2016$ and $2018$, characterized by a `C'-shaped pattern in HID \cite[]{Baby-etal2020}. Based on the HID, the outbursts display three spectral states (LHS, IMS and HSS). No QPOs were detected in the PDS using \textit{AstroSat} observations. Using \textit{NICER} observations, \citet{Chopra-etal2025} analysed multiple outbursts ($2018$, $2020$, and $2021$) and detected $27$ type-C QPOs that show a strong correlation between $\nu_{\rm QPO}$ and photon index ($\Gamma$). A change in the slope of the fitted linear relation at $\nu_{\rm QPO} \sim 2.3$ Hz is observed that suggests a possible transition in the coronal properties. During the $2022$ and $2023$ outbursts, the flux–temperature relation followed $F \propto T^{3.92}$ and $F \propto T^{4.91}$, respectively \cite[]{Peng-etal2024c}. Interestingly, an exceptionally high polarization degree up to $\sim 8.5\%$ has been observed in the thermal and steep power-law states of 4U 1630-472 \cite[]{Kushwaha-etal2023, Rodriguez-etal2023, Ratheesh-etal2024, Majumder-etal2025b}.

    \item \textbf{4U 1543$-$47} (alternatively known as IL Lup) was first discovered during an outburst in $1971$ by \textit{Uhuru} \cite[]{Matilsky-etal1972}. {\red The system hosts a dynamically confirmed black hole of mass $9.4 \pm 1.0~{\rm M}_\odot$ and a companion star of mass $2.45 \pm 0.15~{\rm M}_\odot$ \cite[]{Russell-etal2006} with an estimated distance of $7.5 \pm 0.5$ kpc \cite[]{Jonker-etal2004}. Reflection modelling of \textit{NuSTAR} data yielded black hole spin of $a_{\rm k} = 0.98^{+0.01}_{-0.02}$ and an inclination angle of $68^{\circ+3^\circ}_{~-4^\circ}$ degree \cite[]{Draghis-etal2023a}. Recently, \cite{Chen-Wang-2024, Choudhury-etal2025} estimated the source spin as $\sim 0.46\pm0.12$ using continuum fitting method. \cite{Orosz-etal2003} obtained the inclination angle as $\sim 20.7^{\circ}\pm1.5^{\circ}$ based on optical lightcurve modelling.} During the $2002$ outburst, the Giant Metrewave Radio Telescope (GMRT) detected multiple episodes of radio flaring at different phases of the event \cite[]{Park-etal2004}. Using multi-wavelength data (X-ray, optical, infrared, and radio), \cite{Russell-etal2020} confirmed the presence of a compact jet during SIMS, although jets seem to contribute negligibly to the total luminosity.    
    
    Since discovery, it has exhibited recurrent X-ray outbursts approximately every decade and most recently in $2021$ after nearly $19$ years of quiescence \cite[]{Negoro-etal2021a}. The $2021$ outburst was particularly remarkable as the X-ray brightness reached  $11$ Crab in the $2-4$ keV band as observed by \textit{MAXI}/GSC \cite[]{Negoro-etal2021b}, making it one of the brightest known BH-XRBs. In a recent study, \cite{Prabhakar-etal2023} reported that the source reached super-Eddington luminosity during the peak of the outburst while remaining in the soft spectral state with inverse-Comptonized photons contributing less than $3\%$ of the total emission. Note that a dynamic absorption feature between $\sim 8-11$ keV energy range was detected consistently throughout the outburst, which is the first such detection in any X-ray binary system. This feature is attributed to absorption by highly ionized, relativistic ($\sim 0.3\,$c) disc winds. Furthermore, \cite{Ram-etal2024} reported rapid rms fluctuations and sudden flux transitions during the late stages of the outburst.
   
    \item \textbf{MAXI J1535$-$571} was independently detected by \textit{MAXI} and \textit{Swift/BAT} on $2017$ September $2$ \cite[]{Negoro2017, Kennea2017}. The source hosts a black hole with a mass in the range $5.14-7.83~{\rm M}_{\odot}$ \cite[]{Sreehari-etal2019b} and the possible estimate of inclination of {\red $67.4^{\circ} \pm 0.8^{\circ}$ \cite[]{Miller-etal2018} obtained through spectral modelling}. While \cite{Miller-etal2018} inferred a nearly maximal spin from a single NICER spectrum using the \texttt{relxill} model, multiple {\it AstroSat} observations suggest a moderate spin of $a_{\rm k} = 0.67^{+0.16}_{-0.04}$ \cite[]{Sridhar-etal2019}. The source distance is estimated to be $4.1^{+0.6}_{-0.5}$ kpc using H I absorption spectra from the Australian Square kilometer Array Pathfinder (ASKAP) with an upper bound of $6.7$ kpc \cite[]{Chauhan-etal2019}. Murchison Widefield Array (MWA) and More of Karoo Array Telescope (MeerKAT) observations of this source indicate the presence of transient jets that show an apparent motion with jet velocity $\sim 0.69\,$c \cite[]{Russell-etal2019a, Chauhan-etal2021b}. A radio jet was detected prior to the transition into the soft state, which coincided with the change from type-C to type-B QPOs \cite[]{Russell-etal2019a}. \cite{Rawat2023} suggested that the source may host two distinct coronae: one corresponding to the time-averaged spectrum and another possibly associated with the jet.

    LFQPOs ($2.6-11.4$ Hz) are detected during HIMS/SIMS of $2017$ outburst of this source observed with {\it HXMT} and the source is found to exhibit soft phase-lag of $0.1-0.3$ rad \citep{Huang2018} at QPO frequency, {\red considering subject and refernec bands as $3-7$ keV and $1-3$ keV, respectively.}. The centroid frequency of the LFQPOs ($\sim 1.7-2.7$ Hz) observed with \textit{AstroSat} are found to be correlated with $\Gamma$, accretion rate, inner disc radius and coronal heating rate of the system \cite[]{Bhargava2019,Garg-etal2022}, whereas the dynamic power spectra indicate the persistent QPO features during the observations \cite[]{Chatterjee2021}. Unlike typical black hole transients, the source displayed four consecutive outbursts (each lasting $30-40$ days) with each flare exhibiting a `q'-diagram and state transitions similar to those observed during the main outburst \cite[]{Cuneo2020}.
    
    \item \textbf{MAXI J1820$+$070} is a bright, low mass BH-XRB first discovered on $2018$ in the optical band as ASASSN-18ey by the All-Sky Automated Survey for SuperNovae (\textit{ASAS-SN}) \cite[]{Denisenko2018}, followed by \textit{MAXI}/GSC X-ray detection six days later \cite[]{Kawamuro-etal2018}. The primary outburst persisted until the end of 2018 and was followed by several re-brightening episodes up to June 2020 with additional weak activity observed in early 2021. The system has a high orbital inclination of $63^\circ \pm 3^\circ$ \cite[]{Atri-etal2020} and the black hole mass is {\red dynamically constrained} to be $8.48^{+0.79}_{-0.72}~{\rm M}_\odot$ \cite[]{Torres-etal2020}. Using radio parallax, \citet{Atri-etal2020} estimated the source distance to be approximately $2.96 \pm 0.33$ kpc. From optical spectroscopy, \citet{Torres-etal2019} determined an orbital period of $16.4518 \pm 0.0002$ hours. {\red The spin of the source is not yet well constrained, with estimates ranging from $a_{\rm k}\sim0.2$ to $0.8$ based on relativistic precession modelling and continuum fitting methods \cite[]{Guan-etal2021,Bhargava-etal2021}. Using the continuum fitting method, \cite{Banerjee-etal2024} further estimated the spin to be $a_{\rm k}=0.77\pm0.21$.} \cite{Bright-etal2020} reported the signature of strong radio flare ($\sim 120$ mJy) that exhibits superluminal motions with apparent velocities $\sim 1.7$ c and $0.7$ c. Furthermore, the radio luminosity is also observed to correlate positively with the X-ray and optical luminosities \cite[]{Bright-etal2025}.
    
    The evolution of the coronal geometry also remains elusive for this source. \cite{Kara-etal2019} inferred a decrease in the vertical height of the corona based on the reduced reverberation time-lags, even though significant changes are not observed in the Fe K$\alpha$ line profile. Similarly, \citet{Buisson-etal2019} suggested a vertically contracting corona accompanied by a truncated disc with the inner disc radius evolving from $\sim 1000\,r_{\rm g}$ to $\le 10\,r_{\rm g}$ (where $r_{\rm g}$ is the gravitational radius), based on the shift in the Fe line peak energy ($6.5-7.0$ keV) and the rms variability. Type-C QPOs are detected during the LHS \cite[]{Mereminskiy-etal2018, Zampieri-etal2018, Fiori-etal2018, Ma-etal2021}, whereas type-B QPOs emerge during the transition from HIMS to SIMS and are followed by a bright radio flare \cite[]{Homan-etal2020}. 

    \item \textbf{MAXI J1348$-$630} was discovered in 2019 by \textit{MAXI}/GSC \cite[]{Yatabe-etal2019}. Through indirect method the mass, distance, and inclination of the source have been constrained as $11 \pm 2~{\rm M}_{\odot}$, $3.4 \pm 0.3$~kpc \cite[]{Lamer-etal2021}, and $65^{\circ} \pm 7^{\circ}$ \cite[]{Titarchuk-Seifina2023}, respectively. {\red The source is suggested to host a rapidly spinning black hole, with $a_{\rm k}>0.97$ inferred from reflection modelling and continuum fitting methods \citep{Mall-etal2023}. In contrast, \citet{Guan-etal2024} obtained a lower spin estimate of $a_{\rm k}=0.79\pm0.13$ using the same approaches.} The source exhibits relativistic radio jet with flux $\ge 100$ mJy by MeerKAT at $1.3$ GHz, and jet velocity $\sim 0.69$ c  during its $2019-2020$ outburst \cite[]{Carotuneto-etal2021}.
    
    {\red Using \textit{NICER} observations, \citet{Belloni-etal2020} detected a type-B QPO at $\nu_{\rm QPO}\sim4.5$ Hz and reported energy dependent hard-lags consistent with a Comptonization origin. The time-lags were estimated across $18$ energy bands covering $0.8-12.0$ keV relative to a $2.0-3.0$ keV reference band.} A type-A QPO at $\nu_{\rm QPO} \sim 7$~Hz was also observed using \textit{NICER} data, which is suggested to originate from a vertically extended Comptonization region \cite[]{Zhang-etal2023}. The type-B QPO in this source seems to be associated with a precessing jet, while the type-C QPO is possibly linked to a corona or jet \cite[]{Liu-etal2022b}. The source exhibited a re-flare following the major outburst \cite[]{Dai-etal2023}, during which the corona was found to dynamically reconfigure in reflaring phases as well as in HIMS, correlating with QPO evolution, accretion state transitions, and jet activity \cite[]{Alabarta-etal2025}. The second outburst remained in the hard state ($\Gamma \sim 1.59-1.68$) dominated by non-thermal emission with detection of type-C QPOs ($0.95-0.98$ Hz) \cite[]{Bhowmick-etal2022}. 
    
    \item \textbf{H 1743$-$322} (also known as IGR J17464$-$3213 or XTE J17464$-$3213) was discovered in 1977 by the \textit{Ariel~V} satellite \cite[]{Kaluzienski-Holt1977}. The source is located at a distance of $8.5\pm0.8$~kpc \cite[]{Steiner-etal2012a}. The companion star is a late-type main-sequence star \cite[]{Chaty-etal2015} with estimated period $P=15.07$~h \cite[]{Aneesha-Mandal2020}, and the black hole mass is estimated as $11.2^{+1.7}_{-2.0}~{\rm M}_{\odot}$ \cite[]{Molla-etal2017, Tursunov-Kolos2018}. {\red \citet{Steiner-etal2012a} estimated the orbital inclination to be $75^{\circ}\pm3^{\circ}$ from \textit{VLA} radio observations and constrained the black hole spin to $-0.3<a_{\rm k}<0.7$ using the thermal continuum fitting method applied to \textit{RXTE} data.} 
    During \textit{RXTE} era, it has exhibited strong radio flare during the HIMS to SIMS transition. During its $2002-2003$ outburst it exhibits a relativistic jet with $\beta \sim 0.8$ \cite[]{McClintock-etal2009}, whereas during $2009$ outburst the velocity is constrained as $\sim 0.2\,$c \cite[]{Miller-etal2012}. 
    
    The brightest outburst of H 1743$-$322 was observed in 2003 (peak flux of $\sim 100$~cts~s$^{-1}$ in \textit{RXTE/ASM}) by \textit{RXTE} and \textit{INTEGRAL} \cite[]{Steiner-etal2012a, Homan-etal2005a, Remillard-McClintock2006, Kalemci-etal2006}. Over the past two decades, the source exhibited multiple outbursts, including both complete and failed spectral state transitions, as well as episodes of strong thermal–radiative winds \cite[]{Shidatsu-Done2019} and jet activity \cite[and references therein]{Choudhury-etal2025}. Following the 2003 outburst, H 1743$-$322 displayed outbursts with a recurrence timescale of $0.5-1$~yr \cite[]{Stiele-Yu2016,Aneesha-Mandal2020}, showing a characteristic hysteresis pattern in HID while traversing through all canonical spectral states. The source showed QPOs across a wide range of frequencies such as a HFQPO at $240$~Hz during the $2003$ outburst \cite[]{Homan-etal2003}, and a millihertz QPO at $\sim 11$~mHz during the $2010$ outburst \cite[]{Altamirano-Strohmayer2012}. Furthermore, during the $2010$ outburst, the type-C QPO frequency evolved from $0.92$ Hz to $4.79$ Hz in the rising phase and decreased from $6.42$ Hz to $79$ mHz during the declining phase \cite[]{Debnath-etal2013}. In the subsequent outbursts, type-C QPOs were observed to evolve between $0.09-2.64$ Hz with rms amplitudes of $5.53-2.88\%$, while type-B QPOs appeared at $\sim 5.37-4.43$ Hz with rms values of $2.08\%-3.68\%$ \cite[]{Husain-etal2023, Aneesha-etal2024}.
        
    \item \textbf{Swift J1727.8$-$1613} is detected on August 2023, following a rapid rise in X-ray flux that peaked at $\sim 7.6$ Crab in the $15-50$ keV energy range \citep{Palmer-etal2023}. Subsequent multi-wavelength follow up observations established its nature as a typical black hole transient with detections across X-ray \citep{Nakajima-etal2023, Draghis-etal2023c, Peng-etal2024a}, optical \citep{Castro-Tirado-etal2023, Wang-etal2023a}, and radio bands \citep{Miller-Jones-etal2023}. The source exhibited pronounced variability in radio flux density during HIMS and SIMS reaching a peak flux of $\sim 350$ mJy as observed with \textit{VLITE} at 338 MHz \citep{Peters-etal2023, Nandi-etal2024}. {\red Recent studies constrained the distance, spin, and inclination of the source as $ 2.7\pm0.3$ kpc, $0.98^{+0.02}_{-0.07}$, and $40.0^{\circ+1.2^\circ}_{~-0.8^\circ}$, respectively, with the spin inferred from reflection modelling \citep{Mata-etal2024, Peng-etal2024a}. The mass has also been dynamically constrained to exceed $\rm 3.12\pm0.10~M_{\odot}$ \cite{MataSanchez-etal2025}.} Interestingly, \cite{Wood-etal2024} reported the largest resolved radio jet in Swift J1727.8$-$1613, which exhibited relativistic outflows during two major radio flares with estimated velocities of approximately $0.70-0.85$ c and $0.55-0.73$ c, respectively \cite[see][]{Choudhury-etal2025}.

    \cite{Nandi-etal2024} reported the detection of evolving type-C LFQPOs in hard X-rays (up to $\sim 100$ keV) using {\it AstroSat}, likely associated with the inner Comptonizing corona. However, {\it NICER} observations suggest a vertically elongated, jet-like corona based on the evolution of type-C QPOs \cite[]{Liao-etal2024}. Similarly, LFQPOs are detected in hard X-rays with {\it Insight-HXMT} observations \cite[]{Yu-etal2024, Chatterjee-etal2024, Bollemeijer-etal2025}. Moreover, \cite{Veledina-etal2023} reported the first significant detection of X-ray polarization for this source in HIMS. Following this, \cite{Ingram-etal2024} studied the evolution of X-ray polarization through different spectral states of the source during its outburst. However, a dramatic drop in the polarization degree is observed during the soft state of the source \cite[]{Svoboda-etal2024b}. Further, it has been found that the X-ray polarization angle, detected in the HIMS of the source, remains aligned along the radio jet of the source. This suggests a radial corona in the equatorial plane \cite[]{Ingram-etal2024, Majumder-etal2025b}.
    
    \item \textbf{MAXI J0637$-$430} was discovered by \textit{MAXI}/GSC during an outburst in 2019 \cite[]{Negoro-etal2019}. The outburst lasted for approximately six months and was monitored across multiple wavelengths by several instruments, including \textit{AstroSat} \cite[]{Thomas-etal2019}. The mass of the black hole in MAXI J0637$-$430 was initially estimated using the continuum fitting method to lie within the range $3.0-19.0$ M$_{\odot}$ with a distance of less than $15$ kpc \cite[]{Baby-etal2021}. This broad range reflects the uncertainty in the system parameters for this newly discovered source. Subsequent X-ray and optical studies refined the mass {\red and distance} estimate to $5.1 \pm 1.6~{\rm M}_{\odot}$ {\red and $8.7\pm2.3$ kpc respectively, while constraining} the black hole spin to $a_{\rm k} < 0.25$ \cite[]{Soria-etal2022} {\red using irradiated accretion disk model}. These studies also estimated a binary orbital period of $\sim$2.2 h, making it the shortest known orbital period among Galactic BH-XRBs. Furthermore, strong Fe~K$\alpha$ emission was detected, and relativistic reflection modelling indicated a black hole spin of $a_{\rm k} > 0.72$ and an inclination angle of $\red 46.1^{\circ+4.0^\circ}_{~-5.3^\circ}$ \cite[]{Jia-etal2023}. Note that independent model estimates yielded the mass of the source as $\sim 8.0~{\rm M}_{\odot}$ \cite[]{Chatterjee-etal2023}. The radio jet observed in this source is very weak with flux $\sim 60~\mu$Jy and the nature of the jet could not be determined \cite[]{Russell-etal2019b}. 
    
    Based on observations from \textit{NICER}, \textit{AstroSat}, and \textit{Swift-XRT}, \citet{Baby-etal2021} reported that the source exhibited a distinctive ‘c’-shaped track in HID in contrast to the typical ‘q’-shaped pattern seen in most transient BH-XRBs. The source was observed to evolve through multiple spectral states during the course of the outburst \cite[]{Tetarenko-etal2021, Jana-etal2021b}. Interestingly, its thermal, disc dominated spectrum showed peculiar features that could not be well described by the standard disc-corona model. Instead, an additional soft thermal component was required, which is perhaps originated from a region within the ISCO \cite[]{Lazar-etal2021, Mummery-etal2024}.
 
    \item \textbf{MAXI J1803$-$298} was first discovered in 2021 by \citet{Serino-etal2021}. Observations from \textit{MAXI}/GSC and \textit{Swift}/BAT revealed a transition from hard to soft spectral state, accompanied by a Compton thick outflow that scattered the X-ray flux \cite[]{Megumi-etal2022}. {\red The source mass and inclination have been estimated as $3-10~{\rm M}_{\odot}$ and $>65^{\circ}$, respectively \citep{Mata-etal2022}. Reflection modelling further suggests a rapidly spinning black hole with $a_{\rm k}>0.991$} \cite[]{Feng-etal2022b}.  MAXI J1803$-$298 has exhibited discrete jet during SIMS with an apparent velocity of $\sim 1.5$ c \cite[]{Wood-etal2023}. {\red While the source distance is yet to be constrained, a value of $\sim 8$ kpc (equivalent distance of the Galactic Centre) has been commonly assumed \cite[]{Jana-etal2022a, 58_chand_2022_maxij1803, Wood-etal2023}.}
    
    Simultaneous \textit{Insight-HXMT} and \textit{NICER} observations detected type-C QPOs in the frequency range $\nu_{\rm QPO} \sim 0.16-2.60$ Hz, which are attributed to Lense–Thirring precession. In contrast, the origin of type-B QPOs is linked to a dual-corona model \cite[]{Zhu-etal2023}. The source also exhibited lightcurve variability characterized by periodic dips and non-dip intervals with a period of $\sim 7$ hours \cite[]{Jana-etal2022a, Adegoke2025}. These dips are likely caused by the presence of an ionized absorber located in the outer accretion disk.

    \item \textbf{GRS 1716$-$249} was discovered by \textit{GRANAT} in 1993 \citep{Ballet-etal1993}. {\red Dynamical measurements constrain its mass, distance, and inclination to $6.4^{+3.2}_{-2.0}~{\rm M}_{\odot}$, $6.9 \pm 1.1$ kpc, and $61^{\circ} \pm 15^{\circ}$, respectively \citep{Casares-etal2023}. A high spin, $a_{\rm k} \ge 0.92$, was inferred from continuum fitting and reflection modelling by \cite{Tao-etal2019}, whereas \cite{Zhao-etal2024} obtained a significantly lower value of $\sim 0.464^{+0.016}_{-0.007}$ through joint continuum and reflection modelling with \texttt{kerrbb2} and \texttt{relxill}. The source has an orbital period of approximately 7 h \citep{Jana-etal2022a} and lies on the radio quiet branch of the radio$-$X-ray correlation plane \citep{Bassi-etal2019}.}

    The source underwent a renewed outburst in 2016 \cite[]{Negoro-etal2016}. During this episode, it transitioned to the soft spectral state on three occasions but never attained the canonical soft state, suggesting a failed outburst \cite[]{Bassi-etal2019}. Spectral analysis by \citet{Bharali-etal2019} revealed that the emission was dominated by thermal Comptonization accompanied by a weak disc blackbody component, a broad Fe K$\alpha$ emission line, and a prominent `Compton hump' around $\sim 30$ keV, respectively. The PDS exhibited LFQPOs in both \textit{Swift/XRT} and \textit{AstroSat} observations \cite[]{Chatterjee-etal2021}. The excess infrared emission suggests its origin from synchrotron emission in jets \cite[]{Rout-etal2023}.

    \item \textbf{Swift J151857.0$-$572147} is a newly discovered BH–XRB, first detected by \textit{Swift/BAT} on March $2024$. {\red Using indirect method, \citet{Peng-etal2024b} constrained the black hole mass, inclination, and distance to $3.7-8.1~{\rm M}_{\odot}$, $21.1^{\circ+4.5^\circ}_{~-3.6^\circ}$, and $5.8 \pm 2.5$ kpc, respectively, while reflection modelling yielded a spin parameter of $0.84^{+0.17}_{-0.26}$.}
    
    Initial observation using \textit{Swift} (XRT and BAT) indicated a harder power-law spectrum with a photon index of $1.78 \pm 0.02$ \cite[]{Kennea-etal2024}. Subsequently, \cite{Chatterjee-etal2025} performed a comprehensive study of $2024$ outburst using {\it Insight-HXMT}. As the outburst evolved, the source transits from a harder to a softer spectral state with an increase in the photon index ($\Gamma \sim 2.4-3.5$). During this period, type-C QPOs with frequencies in $3-9$ Hz range were detected at photon energies below $48$ keV \cite[]{Chatterjee-etal2025}. Around the same period, an exceptionally bright radio flare was detected by {\it ATCA} on $9$ March $2024$, which evidently suggest the spectral state transition \cite[]{Carotenuto-Russell2024}. 

    \item \textbf{GRS 1739$-$278} was discovered by SIGMA/\textit{GRANAT} in 1996 \cite[]{Paul-etal1996}. {\red The black hole mass is estimated to be below $9.5~{\rm M}{\odot}$ \cite[]{Wang-etal2018a}, and the source is located at a distance of $6.0-8.5$ kpc \cite[]{Greiner-etal1996}. \cite{Miller-etal2015} estimated the spin and inclination of the source as $a_{\rm k} \sim 0.8\pm0.2$ and $32.5^{\circ}$, respectively. In a recent analysis, reflection based modelling favoured a nearly maximally spinning black hole with $a_{\rm k}\sim0.994^{+0.004}_{-0.003}$ and an inclination of $54^{\circ+8^\circ}_{~-4^\circ}$ \cite[]{Zhao-etal2026}.}

    The $1996$ outburst exhibited strong spectral evolution consistent with the canonical outburst sequence. The source shows evolving QPO from $\sim 5.0$ Hz to $\sim 5.6$ Hz during IMS of the decay phase of the outburst. The QPO rms increased from $1.30\%$ to $5.01\%$ and the corresponding phase-lag rises from $0.003$ to $0.220$ rad as the energy increases \cite[]{Borozdin-etal2000,Wijnads-etal2001}. {\red Here, reference band is considered as $4.3-5.0$ keV for time-lag measurements across multiple subject energy bands.} After $18$ years of quiescence, the source underwent another outburst in $2014$, during which QPOs in the range $0.3-0.7$ Hz were detected. The QPO frequency was found to correlate with the source flux that increases as the spectrum softened \cite[]{Mereminskiy-etal2019}.

    \item \textbf{XTE J2012$+$381} was discovered in $1998$ with the ASM on board \textit{RXTE} \citep{Remillard-etal1998} and the source exhibited two outbursts in $1998$ and $2022$. The mass, distance, inclination, and spin of the source are estimated to be $\gtrsim 11.8 \pm 4.6~{\rm M_\odot}$ \cite[]{Campana-etal2002, Gaia-Collab2016, Draghis-etal2023b}, $5.4 \pm 2.1$ kpc \cite[]{Gaia-Collab2016}, $68^{\circ +6^\circ}_{~-11^\circ}$, and $a_{\rm k} \sim 0.988_{-0.030}^{+0.008}$ \cite[]{Draghis-etal2023b}, respectively{\red, with the spin estimated through relativistic reflection modelling}.

    During $1998$ outburst, the lightcurve displayed a FRED like profile \cite[]{Chen-etal1997b} with secondary maxima \cite[]{Vasiliev-etal2000}. The X-ray spectrum was well described by a multicolor disc blackbody with an inner disc temperature of $\sim 0.76$ keV plus a power-law component with a photon index of $\sim 2.9$ \cite[]{White-etal1998}. Timing analysis revealed a dominant Very Low Frequency Noise (VLFN) component, approximated by a power-law model ($P \propto f^{\alpha}$) with a slope $\alpha = -1.08 \pm 0.08$ in the $0.001-0.500$ Hz range \cite[]{Vasiliev-etal2000}. Detailed spectral studies of the $1998$ outburst using \textit{RXTE} and \textit{BeppoSAX} showed that the disc temperature evolved with the outburst profile \citep{Vasiliev-etal2000, Campana-etal2002}, a trend also observed during the $2022$ outburst \cite[]{Draghis-etal2023b}.

    \item \textbf{XTE J1859$+$226} was discovered by \textit{RXTE/ASM} in 1999 \cite[]{Wood-etal1999}. This binary system hosts a {\red dynamically confirmed} black hole of mass $7.8 \pm 1.9~{\rm M}_{\odot}$ \cite[]{Yanes-Rizo-etal2022}, orbiting a companion star of spectral type K5$-$7 V \cite[]{Corral-Santana-etal2011} with an orbital period of $9.2$ hr \cite[]{Garnavich-Quinn2000, Sanchez-Fernandez-etal2000}. The source is located at a distance of $8 \pm 3$ kpc \cite[]{Tetarenko-etal2016} and inclination of $66.6^{\circ} \pm 4.3^{\circ}$ \cite[]{Yanes-Rizo-etal2022}. {\red Using reflection modelling, the source spin is estimated as $a_{\rm k}=0.986^{+0.001}_{-0.004}$ \cite[]{Mall-etal2024}}. During the same outburst, multiple radio flares with flux densities of $\sim 10-100$ mJy were observed during SIMS that coincides with the appearance of type-B/C* QPOs \cite[]{Brocksopp-etal2002, Radhika-Nandi2014, Radhika-etal2016a,Choudhury-etal2025} indicating a strong coupling between the accretion disc and the jet. 

    Shortly after its discovery, \textit{RXTE/PCA} observations revealed a $0.45$ Hz QPO with the X-ray spectrum being dominated by a hard power-law component during the $1999$ outburst \cite[]{Markwardt-etal1999b}. Subsequently, a series of soft X-ray flares were detected, showing QPOs in the $6-7$ Hz range and HFQPOs between $82$ Hz and $187$ Hz \cite[]{Cui-etal2000}. Furthermore, the QPO frequency was observed to increase from $\sim 0.6$ Hz to $\sim 6$ Hz, suggesting a reduction of the coronal region \cite[]{Nandi-etal2018}. Recently, \cite{Yamaoka-etal2025} reported that the inner disc temperature increased from $\sim 0.4$ keV to $\sim 0.9$ keV, while the inner disc radius decreased from $\sim 300$ km to $\sim 60$ km. Interestingly, during the decay phase, the disc temperature declined to $\sim 0.3$ keV, although the inner disc radius remained nearly constant at $\sim 60$ km.

    \item \textbf{V4641 Sgr} (also known as SAX J1819.3$-$2525) was discovered in $1999$ by \textit{BeppoSAX} \cite[]{IntZand-etal1999} and \textit{RXTE} \cite[]{Markwardt-etal1999a}. The system hosts a {\red dynamically confirmed} black hole of mass $6.4 \pm 0.6$ M$_{\odot}$, orbiting a B9III-type companion star of mass $2.9 \pm 0.4~{\rm M}_{\odot}$ at a distance of $6.2 \pm 0.7$ kpc, with an orbital inclination of $72.3^{\circ} \pm 4.1^{\circ}$ \cite[]{MacDonald-etal2014}. The orbital period of the binary is estimated to be $2.82$ days \cite[]{Orosz-etal2001}. V4641 Sgr exhibited strong transient radio jets ($\sim 0.4$ Jy) with an estimated velocity $\sim 0.9$ c \cite[]{Hjellming-etal2000}, whereas its subsequent outbursts generally yield comparatively weaker radio jets. During 2002 outburst, a strong jet with flux $\sim 133$ mJy was detected \cite[]{Campbell-etal2002}, and observations with the GMRT revealed radio flux of $\sim 100$ mJy at $610$ and $244$ MHz \cite[]{Ishwara-etal2005}.

    V4641 Sgr exhibits outburst characteristics that are highly atypical compared to other LMXBs. In September 1999, it underwent an exceptionally bright and short-lived outburst that reached a peak intensity of $\sim 12.2$ Crab and dropped to $\sim 0.1$ Crab within two hours \cite[]{hjellming2000}. Such rapid and intense variability is rarely observed in LMXBs. Since then, the source has displayed recurrent outbursts with a median recurrence interval of $\sim 220$ days \cite[]{Tetarenko-etal2016}, although none have attained the extreme luminosity of the 1999 event. Interestingly, V4641 Sgr is seen to maintain a soft, disc-dominated spectral state even at unusually low Eddington fractions, as evidenced by soft X-ray spectra observed during the 2014 and 2015 outbursts at luminosities of $L_{X} \sim 0.01 \, {\rm L}_{\mathrm{Edd}}$ and $L_{X} \sim 6 \times 10^{-4}\,{\rm L}_{\mathrm{Edd}}$, respectively \cite[]{Pahari-etal2015,Bahramian-etal2015}.
     
    \item \textbf{V404 Cyg} (also known as GS 2023$+$338) was discovered by \textit{GINGA} during its $1989$ outburst \cite[]{Makino1989}. The system hosts a {\red dynamically confirmed} black hole of mass $9.0^{+0.2}_{-0.6}~{\rm M}_{\odot}$ \cite[]{Khargharia-etal2010} located at a distance of $2.39 \pm 0.14$ kpc \cite[]{Miller-jones-etal2009} with an orbital inclination of $67^{\circ+3^\circ}_{~-1^\circ}$ \cite[]{Khargharia-etal2010}. The black hole spin is estimated to be $a_{\rm k}>0.92$ {\red using relativistic reflection modelling} \cite[]{Walton-etal2017} . During $1989$ outburst, transient jet with radio flux $\sim 0.05-0.15$ Jy is detected in \textit{VLA} observations \cite[]{Hjellming-etal1989}, whereas multiple radio flares \cite[]{Mooley-etal2015} including a strong one of flux $\sim 1$ Jy is seen in $2015$ outburst monitored by \textit{VLA} \cite[]{Hughes-etal2023, Fender-etal2023}. In addition, V404 Cyg displays persistent radio emission, attributed to a self-absorbed synchrotron jet \cite[]{Gallo-etal2005, Miller-Jones-etal2008}. 

    This source is well known for its high luminosity, however most observations during the $1989$ outburst suggested the absence of a soft spectral component \cite[]{Brocksopp-etal2004}. This interpretation was later reconsidered by \citet{Zycki-etal1999} with evidence indicating that the source may have briefly entered a soft spectral state. The source underwent another major outburst in $2015$, during which it exhibited various spectral state transitions, where spectral parameters evolved in a manner distinct from those of typical outbursting BH–XRBs \cite[]{Radhika-etal2016b}. Moreover, numerous X-ray flares were detected during this outburst, displaying rapid spectral and timing variabilities \cite[]{Jana-etal2021a}.
    
    \item \textbf{MAXI J1305$-$704} was discovered in $2012$ hosting a {\red dynamically confirmed} black hole of mass $8.9^{+1.6}_{-1.0}\,{\rm M_\odot}$ with an inclination in the range $72^{\circ+5^\circ}_{~-8^\circ}$ and a distance of $7.5^{+1.8}_{-1.4}$ kpc \cite[]{MataSanchez-etal2021}. The spin of the black hole is estimated to be $a_{\rm k}=0.87^{+0.07}_{-0.13}$ using the continuum fitting method \cite[]{Feng-etal2023}. A faint radio signal with a flux density of $0.46 \pm 0.02$ mJy at $5.5$ GHz is also reported \cite[]{Coriat-etal2012}.

    The $2012$ outburst exhibited both LHS and HSS along with recurrent absorption dips occurring at intervals of $\sim 9.7$ hr \cite[]{Shidatsu-etal2013}. The absorption spectra indicate the presence of a failed wind \cite[]{Miller-etal2014}. During the soft state, continuum fitting of the X-ray spectrum yielded an inner disc temperature of $\sim 0.8$ keV, while the power density spectrum showed a total fractional rms of $\sim 1\%$ without any variability features \cite[]{Feng-etal2023}.

    \item \textbf{Swift J1357.2$-$0933} was discovered in $2011$ by \textit{Swift/BAT} \cite[]{Krimm-etal2011a} and exhibited three outbursts to date. The black hole mass is {\red estimated as $12.4 \pm 3.6\,{\rm M_\odot}$ using indirect method} \cite[]{Casares2016}, and the system has a short orbital period of $2.8$ h \cite[]{Corral-Santana-etal2013}. The distance is constrained to be $>6$ kpc \cite[]{Charles-etal2019}. Recently, the source inclination is estimated to be $87.4^{\circ+2.6^\circ}_{~-5.6^\circ}$ \cite[]{Casares-etal2022}.

    Millihertz QPOs are observed from this source during both $2011$ outburst \cite[$\sim 6$ mHz;][]{Armas-Padilla-etal2013} and $2019$ outburst \cite[$\sim 1$ and $5$ mHz;][]{Beri-etal2023}. Spectral analysis indicates a peak luminosity of $\sim 10^{35}-10^{36}$ erg s$^{-1}$, placing this system among the very faint X-ray transients \cite[]{Beri-etal2023}.

    \item \textbf{MAXI J1659$-$152} exhibited a single outburst in $2010$ detected by \textit{Swift/BAT} \cite[]{Mangano-etal2010} and \textit{MAXI}/GSC \cite[]{Negoro-etal2010}. The binary consists of a black hole with a mass of $4.9-5.7\,{\rm M_\odot}$ {\red estimated using optical spectroscopy} \cite[]{Torres-etal2021}, an inclination in the range $65^{\circ}-80^{\circ}$, and a distance of $8.6 \pm 3.7$ kpc \cite[]{Kuulkers-etal2013}. Continuum fitting of the X-ray spectra suggests that the black hole may possess an extreme retrograde spin, with $-1.00 < a_{\rm k} \lesssim 0.44$ \cite[]{Feng-etal2022a}.

    During the $2010$ outburst, timing analysis revealed both type-C ($1.6-7.3$ Hz) and type-B ($1.6-4.1$ Hz) QPOs \cite[]{Yamaoka-etal2012}. Spectral modelling indicates the presence of a multicolor accretion disc accompanied by a Comptonizing medium. The inner disc temperature and photon index increase during the rising phase before declining. The Comptonized flux appears to account for the observed type-C QPOs \cite[]{Munoz-Darias-etal2011}. Furthermore, the inner disc radius decreases as the type-C QPO frequency increases, whereas the innermost radius remains constant during the type-B QPO intervals \cite[]{Yamaoka-etal2012}.

    \item \textbf{XTE J1752$-$223} underwent an outburst in $2009-2010$ \cite[]{Markwardt-etal2009}, which was monitored by \textit{RXTE}. The source is located at a distance of $7.1^{+0.3}_{-0.2}$ kpc and hosts a black hole with a mass of $\rm 12.0^{+0.8}_{-0.7}~M_{\odot}$ and an inclination angle of $ 35^{\circ}\pm4^{\circ}$ \cite[]{Miller-etal2011, Shaposhnikov-etal2010, Garcia-etal2018, Debnath-etal2021, Abdulghani-etal2024}. {\red The system hosts a moderately rotating black hole with the spin parameter constrained to be $a_{\rm k}=0.52\pm0.11$ from reflection modelling and continuum fitting methods \cite[]{Reis-etal2011, Choudhury-etal2025}.} Multiple radio flares were detected during the SIMS, HSS, and the decay phase associated with the SIMS to HIMS transition, occurring (quasi) simultaneously with enhanced X-ray flux \cite[]{Brocksopp-etal2013}. During the third radio flare in SIMS, the jet was moderately relativistic with a velocity of $\sim 0.7$ c \cite[]{Miller-etal2011}. 

    The source underwent a canonical outburst and reached a peak X-ray flux of $\sim 5 \times 10^{-8}$ erg s$^{-1}$ cm$^{-2}$ in the $1-100$ keV range. The disc flux fraction was $<5\%$ during the LHS, increased to $\sim 10-13\%$ during the HIMS, further reached $\sim 42\%$ in the SIMS, and finally attained $\sim 69-92\%$ in the HSS \cite[]{Chatterjee-etal2020}. Type-C QPOs were detected with frequencies increasing from $2-6$ Hz and QPO rms decreasing from $7\%-2\%$. A type-B QPO was observed near the first radio flare with a frequency of $\sim 3.5$ Hz and rms of $\sim 5\%$. A type-A QPO was observed near the second radio flare at a frequency of $\sim 2.1$ Hz and rms of $\sim 2.3\%$. Type-C and type-B QPOs exhibited hard-lags of $\sim 12-15$ ms and $\sim 10$ ms, {\red respectively, with $6-15$ keV and $2-6$ keV as subject and reference bands, respectively,} whereas type-A QPO displayed a soft-lag of $\sim 2$ ms \citep{Choudhury-etal2025}. 

    \item \textbf{XTE J1908$+$094} was discovered serendipitously in $2002$ during \textit{RXTE} observations of the soft gamma-ray repeater SGR 1900$+$14 \citep{Woods-etal2002}. Since then, it has exhibited three outbursts in $2003$, $2013$, and $2019$ \cite[]{Draghis-etal2021}. The estimated mass, distance and inclination of the source are $6.5^{+0.5}_{-0.7}~{\rm M_\odot}$ \cite[]{Debjit-Chatterjee-etal2021}, $3-10$ kpc \cite[]{Chaty-etal2006}, and $27^{\circ +2^\circ}_{~-3^\circ}$ \cite[]{Draghis-etal2021}. {\red Using continuum fitting method, spin of the source is estimated as $0.55^{+0.29}_{-0.45}$ \cite[]{Draghis-etal2021}}. An expanding radio jet with a flux density of about $15$ mJy was also reported \cite[]{Rushton-etal2017}.

    During the $2013$ outburst, the source displayed a characteristic outburst profile commonly observed in black hole transients. The \textit{NuSTAR} lightcurve revealed a $\sim 40$ ks flare \cite[]{Tao-etal2015, Zhang-etal2015}, accompanied by spectral softening and an increase in the power-law flux \cite[]{Zhang-etal2015}. The spectrum further showed a broad Fe K$\alpha$ line and a prominent disc reflection component \cite[]{Tao-etal2015, Zhang-etal2015}. \cite{Zhang-etal2015} suggested the flare could be associated with the relativistic jets which was later confirmed by \cite{Rushton-etal2017}. During the $2019$ outburst, the source remained predominantly in the SIMS \cite[]{Debjit-Chatterjee-etal2021}.

    \item \textbf{MAXI J1810$-$222} was discovered in $2018$ by \textit{MAXI}/GSC \cite[]{Negoro-etal2018}. The source distance is estimated to be $\gtrsim 6$ kpc \citep{Russel-etal2022} although its mass, inclination, and spin remain unconstrained. Radio observations from $2019$ to $2021$ reported flux densities in the range $\sim 70-570~\mu$Jy \cite[]{Russel-etal2022}. During this period, an outflow of $0.05-0.15$ c is observed which could be due to the radiation pressure or most likely magnetically driven winds \cite[]{Del-Santo-etal2023}.
    
    In $2018$ outburst, \textit{NuSTAR} observed the source in the soft state with no evidence of X-ray pulsations or bursts \cite[]{Oeda-etal2019}. Subsequently, in August $2019$, the \textit{Swift}/BAT survey recorded an increase in the hard X-ray flux, revealing a peculiar outburst evolution. Unlike typical transients, the source does not follow the canonical HID, repeatedly transiting between the hard and soft states \cite[]{Russel-etal2022}. The energy spectra show a multicolor disc blackbody along with a Comptonizing medium. In the soft state, Comptonization was not required, and the inner disc temperature was $kT_{\rm in} \sim 0.5$ keV. In the hard state, $kT_{\rm in}$ decreased to $\sim 0.3$ keV. The photon index during the IMS and LHS varied between $1.4$ and $2.1$. Wide-band PDS analysis ($0.06-1000.00$ Hz) displayed only Poisson noise with a total rms $<2\%$ in the soft state, whereas a flat-top noise without any QPO feature (rms$_{\rm total} \sim 25\%$) was observed in the hard state \cite[]{Russel-etal2022}.

    \item \textbf{XTE J1118$+$480} was discovered in $2000$ with ASM onboard \textit{RXTE} \cite[]{Remillard-etal2000} and exhibited two outbursts so far, in $2000$ and $2005$. The source hosts a {\red dynamically confirmed} black hole with a mass of $7.06-7.24~{\rm M_\odot}$ inclined at $74^\circ \pm 4^\circ$ \cite[]{Cherepashchuk-etal2019}, and located at a distance of $1.72 \pm 0.10$ kpc \cite[]{Gelino-etal2006}. Notably, the spin of the source is yet to be constrained.

    The first outburst in $2000$ lasted for approximately seven months, showing an initial decay after the first peak followed by a re-brightening into a `plateau state' that persisted for nearly five months during which the source remained in the LHS \cite[]{Chaty-etal2003, Brocksopp-etal2010}. \cite{Frontera-etal2003} analysed the \textit{BeppoSAX} spectra during this outburst and the spectra are modelled with a combination of blackbody emission and thermal Comptonization. The blackbody luminosity was found to decrease with time, whereas the electron temperature of the Comptonizing plasma showed no significant variation. Furthermore, \cite{Reis-etal2009} reported thermal disc emission with a temperature of $0.21$ keV based on \textit{Chandra} observations. During the second outburst in $2005$, the source exhibited a FRED type lightcurve and again remained in the LHS \cite[]{Swank-etal2005, Zurita-etal2006}. However, in contrast to the first outburst, the disc blackbody emission was more prominent accompanied by short lived jet ejections \cite[]{Brocksopp-etal2010}. In $2000$ and $2005$, the source underwent failed outbursts. During the $2000$ event, it exhibited a strong, compact radio jet while in the LHS \cite[]{Fender-etal2001}. In $2005$ outburst, the maximum radio flux reached $\sim 12$ mJy at $4.7$ GHz as measured with the \textit{VLA} \cite[]{Brocksopp-etal2010}, and a compact jet was present during this phase with its peak flux occurring simultaneously with the outburst maximum.
    
\end{itemize}

\cprotect\section{Observations with \textit{AstroSat}} \label{sec4:obs}

In this section, we present the observational details of the $34$ BH–XRBs analyzed using \textit{AstroSat} data. In total, \textit{AstroSat} carried out $261$ pointed observations of the BH–XRBs under consideration, with a cumulative exposure time of approximately $12$ Ms. We analyse only those observations with effective exposure times exceeding $\sim0.5$ ks to ensure a minimum signal-to-noise ratio (SNR) of $10$. The SNR is defined as $S t/\Sigma$, where $S$ is the source count rate, $t$ is the exposure time, and $\Sigma$ is the total noise expressed as $\Sigma=\sqrt{S t + B t + pR_n^{2}}$, with $B$ denoting the background count rate, $p$ the number of pixels in the source region, and $R_n$ the read noise \citep{Fano1947,Newberry1991}. As the contribution of $R_n$ is negligible for both SXT and LAXPC, we approximate the SNR as $\mathrm{SNR}=St/\sqrt{(S+B)t}$. All observations considered in this work are publicly available at the Indian Space Science Data Centre (ISSDC)\footnote{\url{https://webapps.issdc.gov.in/astro_archive/archive/Home.jsp}}. 

\begin{figure*}
    \begin{center}
    \includegraphics[width=0.46\textwidth]{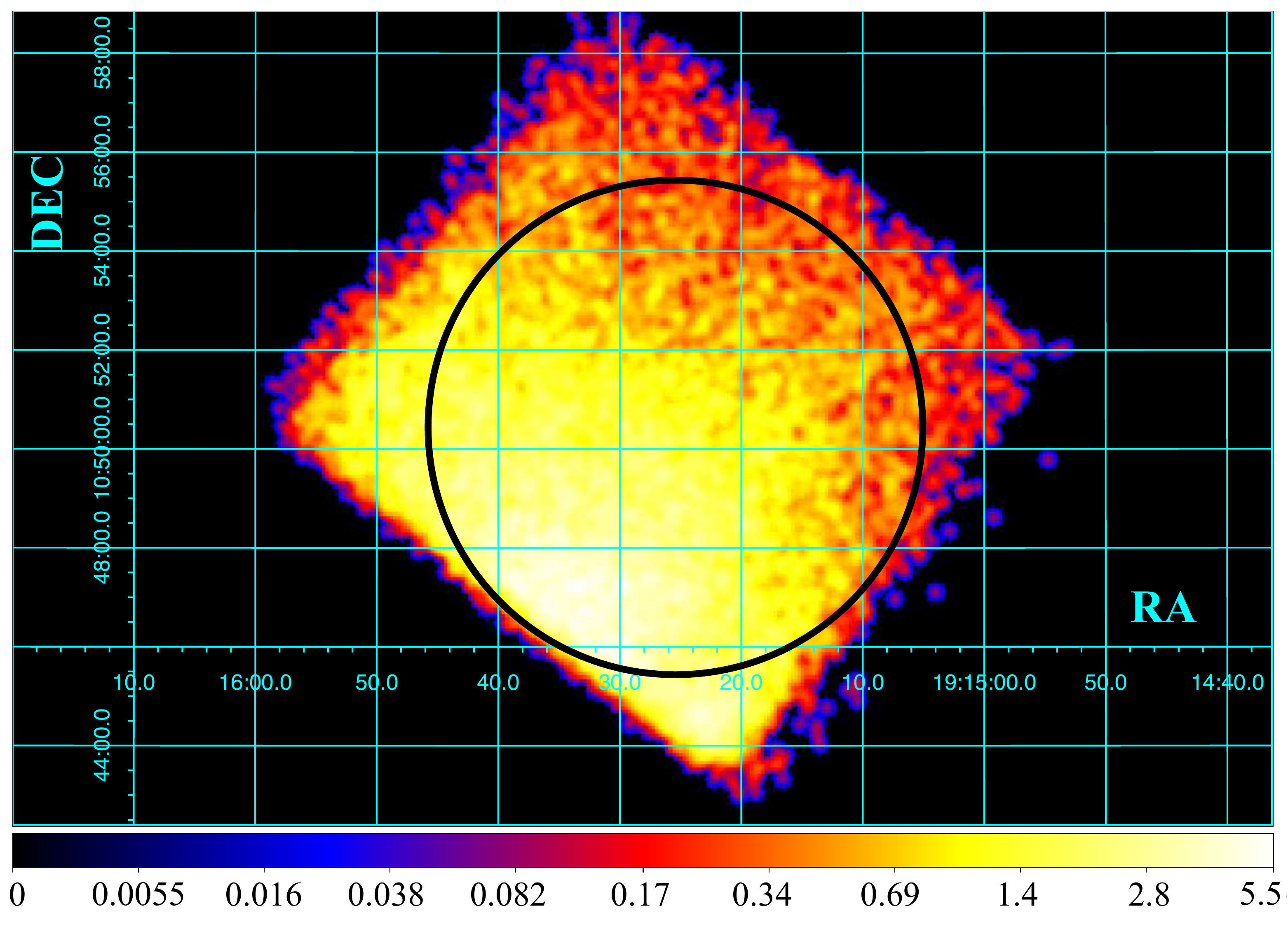}
    \includegraphics[width=0.45\textwidth]{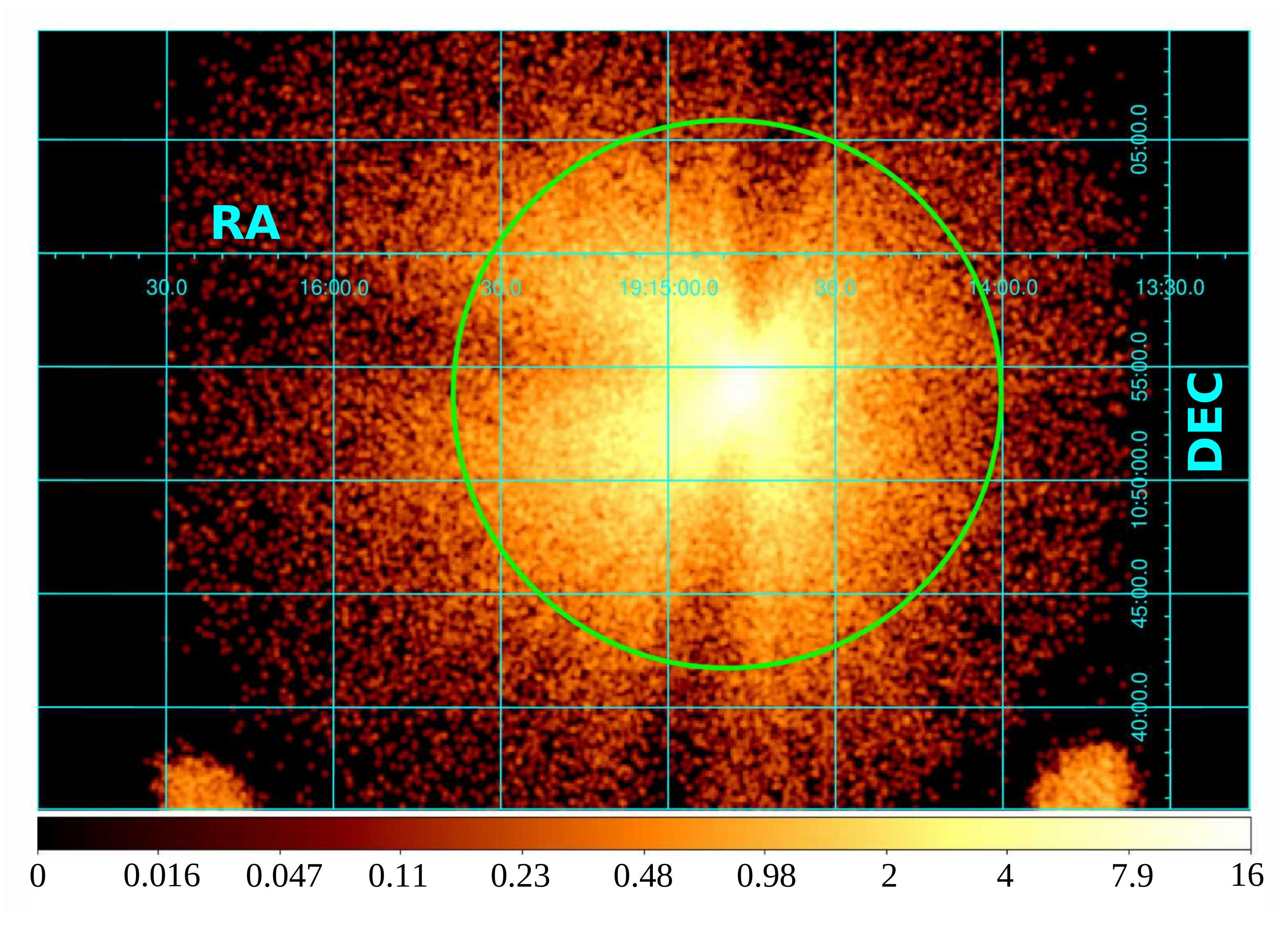}
    \end{center}
    \caption{{\it Left}: {\it AstroSat}/SXT image of GRS 1915$+$105 observed during $\kappa$ class (AS18) in Photon Counting (PC) mode. The source region is marked by a green circle with a radius of $12$ arcminute. The colorbar represents the intensity distribution of the source. The bright point sources visible in the bottom corners correspond to two of the four calibration sources of the SXT instrument. {\it Right}: {\it AstroSat}/SXT image of the source observed with Fast Window (FW) mode in $\delta$ class (AS8). The black circle is the $5$ arcminute region from which the source events are extracted. See text for details.}
    \label{fig:fig5}
\end{figure*}

\subsection{Persistent-cum-Variable (PcV)}\label{sec4:4.1}

GRS 1915$+$105 was the first BH–XRB observed by the SSM, merely two days after the commissioning of the payload. Among all the instruments onboard \textit{AstroSat}, the SSM was the first to point at this source. The initial observation was carried out with the SSM1 and SSM2 slanted cameras at a fixed platform angle of $\sim 5^{\circ}$ on 14 October 2015 covering approximately $15$ consecutive spacecraft orbits. After accounting for Earth occultation and South Atlantic Anomaly (SAA) passages, the effective good time interval (GTI) for GRS 1915$+$105 was $\sim 530$~min. On the following day, 15 October 2015, the source was observed again using the SSM3 camera at a similar platform angle of $\sim 5^{\circ}$. This observation spanned about $10$ orbits, during which the source was detected for an effective exposure of $\sim 240$~min.

\begin{figure*}
    \begin{center}
    \includegraphics[width=\textwidth]{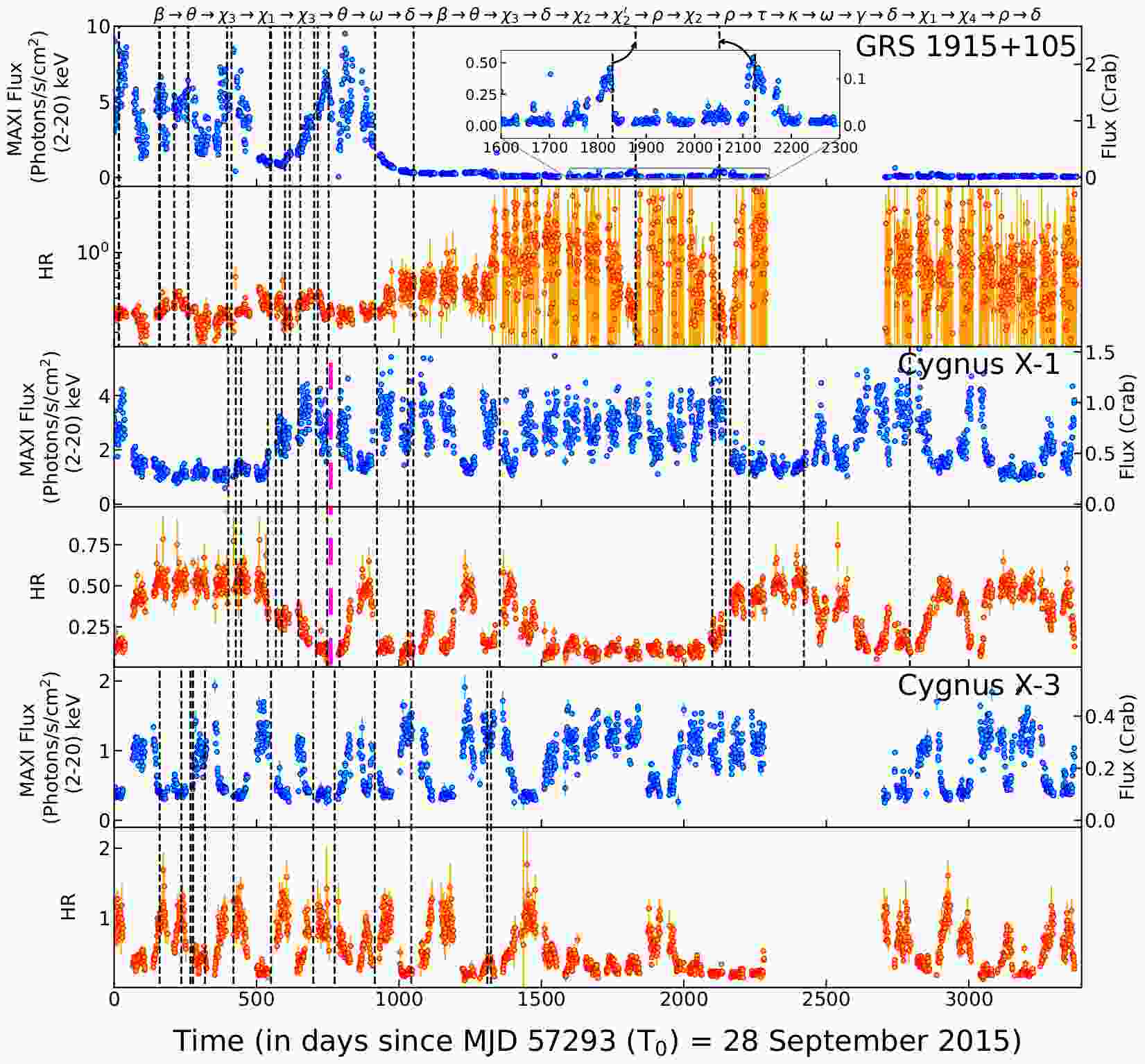}    
    \end{center}
    \caption{\textit{MAXI}/GSC monitoring of the persistent-cum-variable sources (GRS 1915$+$105, Cyg X$-$1 and Cyg X$-$3) in $2-20$ keV energy band during the \textit{AstroSat} observation period considered in this work. Variation of the hardness ratio (HR) observed with \textit{MAXI}/GSC is plotted for all sources. The vertical black dashed lines indicate the epochs of {\it AstroSat} observations and vertical magenta big-dashed line denotes the softest state of Cyg X$-1$ observed. At the top of the figure, we indicate all possible class transitions of GRS 1915$+$105 observed during the complete {\it AstroSat} observational period and show the re-brightening phase of the source in the inset. The lightcurve and hardness ratio are obtained from $1$-day averaged \textit{MAXI}/GSC data from the launch date of \textit{AstroSat}. See text for details.
    }
    \label{fig:fig6}
\end{figure*}

Subsequently, \textit{AstroSat} carried out $47$ pointed observations of GRS 1915$+$105 comprising $38$ Guaranteed Time (GT) observations, four Announcement of Opportunity (AO) cycle observations, and five Target of Opportunity (ToO) observations using the \textit{LAXPC} and \textit{SXT} instruments onboard \textit{AstroSat}. To avoid redundancy arising from multiple observations of the same variability class, we select $25$ representative observations for this study to examine the spectro-temporal variability of GRS 1915$+$105 across distinct variability classes. The details of these selected observations, categorized by their variability classes, are listed in Table~\ref{tab:Table-02}. Fig. \ref{fig:fig6} (top panel) presents the long-term \textit{MAXI}/GSC monitoring of the source in the $2-20$ keV energy band with the epochs of the \textit{AstroSat} observations and the corresponding class transitions marked for reference.

Among all BH–XRBs, \textit{AstroSat} observed Cyg X$-$1 maximum time with a total exposure of $\sim 3$ Ms. The source has been observed in $54$ pointed observations, comprising $41$ GT, $4$ ToO, and $9$ AO observations. The first observation was performed during the post-launch verification phase on $15$ November $2015$ (MJD $57341$) followed by the first regular observation on $8$ January $2016$ (MJD $57395$). In this work, we analyse $21$ observations of Cyg X$-$1 selected from a total of $48$ combined SXT and LAXPC observations obtained between January $2016$ and May $2023$ in order to avoid repetition of similar spectral states. The log of the selected observations is presented in Table \ref{tab:Table-03}. 

Cyg X–3 is a persistent-cum-variable source frequently observed by \textit{AstroSat} between $2015$ and $2019$. A total of $15$ pointed observations were carried out including $11$ GT and $3$ ToO observations. The $2015$ observation could not be used in the present analysis due to the unavailability of a reliable background spectrum and hence, only $14$ observations from $2016-2019$ are considered. The details of these \textit{AstroSat} observations used for the spectro-temporal analysis are listed in Table~\ref{tab:Table-03}.

\begin{table*}
\centering
	\caption{Details of the \textit{AstroSat} observations of GRS 1915$+$105. The notations and symbols have their usual meaning. See \S \ref{sec4:obs}, \S \ref{sec6:6.1_timing} and \S \ref{sec7:7.1_pcv} for details. 
    }
	\renewcommand{\arraystretch}{1.4}
	\resizebox{1.0\textwidth}{!}{%
	\begin{tabular}{c c c c c c c c c c c c}
	\hline
	Epoch & ObsID & Orbit & Date & MJD & Exposure$^\dagger$ & $r_{\rm det}$ & $r_{\rm in}$ & HR1 & HR2 & Variability \\
			
	& & & (dd-mm-yyyy) & &  (ks) & (cts/s) & (cts/s) & (B/A)$^*$ & (C/A)$^*$ & Class \\	\hline

    AS0$^\ddagger$ & P01\_122T01\_9000000026 & 00242 & 14-10-2015 & 57309.3 & 31.8 & $-$ & $-$ & $-$ & $-$ &  $\beta$ \\
            
	AS1 & T01\_030T01\_9000000358 & 02351 & 04-03-2016 & 57451.9 & 3.5 &  7360 & 8725 & 0.66 &  0.08 &  $\theta$ \\

    AS2 & T01\_030T01\_9000000358 & 02360 & 05-03-2016 & 57452.5 & 2.0 & 6829  & 8049  & 0.68  & 0.13   & $\chi^{}_3$ \\
			
	AS3 & T01\_030T01\_9000000358 & 02365 & 05-03-2016 & 57452.8 & 3.1 & 3940 & 4300  & 0.75 & 0.13 &   $\chi^{}_1$ \\

     AS4 & T01\_030T01\_9000000358 & 02367 & 05-03-2016 & 57452.9 & 4.6 & 5458  & 6181  & 0.69  & 0.14  & $\chi^{}_3$ \\
			
	AS5 & T01\_030T01\_9000000358 & 02373 & 04-03-2016 & 57453.4 & 2.5 & 9518 & 11932  & 0.61 & 0.07 &  $\theta$ \\
			
	AS6 & G05\_214T01\_9000000428 & 03124 &  26-04-2016 & 57504.0 & 2.5 & 7697  & 9202 & 0.70 & 0.04 & $\omega$ \\
			
	AS7 & G05\_189T01\_9000000492 & 03841 & 13-06-2016 & 57552.6 & 3.1 & 9762 & 12317 & 0.54 & 0.03 & $\delta$ \\
			
	AS8 & G05\_189T01\_9000000492 & 03860 & 14-06-2016 & 57553.9 & 2.5 & 7974 & 9600  & 0.70 & 0.05 & $\delta$ \\
			
    AS9 & G06\_033T01\_9000000760 & 05862 & 28-10-2016 & 57689.1 & 2.7 & 5830 & 6655  & 0.54 & 0.06 &  $\beta$ \\
			
	AS10 & G06\_033T01\_9000000760 & 05870 & 28-10-2016 & 57689.5 & 2.2 & 6507 & 7552  & 0.63 & 0.06 & $\theta$ \\
			
	AS11 & G06\_033T01\_9000000792 & 06102 & 13-11-2016 & 57705.2 & 3.6 & 6820 & 7977  & 0.61 & 0.03 &  $\delta$ \\
            
    AS12 & G06\_033T01\_9000001116 & 08108 & 28-03-2017 & 57840.9 & 4.0 & 1714 & 1779  & 1.00  & 0.37  & $\chi^{}_2$ \\

	AS13 & G06\_033T01\_9000001116 & 08110 &	29-03-2017 & 57841.4  & 4.3  & 1785  & 1856 & 0.91 & 0.31 & $\chi^{\prime}_2$ \\

    AS14 & G06\_033T01\_9000001116 & 08120  & 	28-03-2017 & 57841.7 & 2.6 & 1859 & 1929  & 0.80  & 0.24 & $\rho$ \\

	AS15 & G07\_046T01\_9000001124 & 08158 & 01-04-2017 & 57844.0 & 2.5 & 1763 & 1831  & 0.77 & 0.20 & $\chi^{}_2$ \\
			
	AS16 & G07\_046T01\_9000001236 & 08876 & 19-05-2017 & 57892.8 & 3.6 & 1395 & 1437 & 0.63 &  0.17 &  $\rho$ \\
			
    AS17 & G07\_028T01\_9000001272 & 09130 & 05-06-2017 & 57909.8 & 3.0 & 2235 & 2346 & 0.62 & 0.04 & $\tau$ \\

	AS18 & G07\_046T01\_9000001374 & 09670 & 12-07-2017 & 57946.3 & 2.7 &  3231 & 3469  & 0.76 & 0.07 & $\kappa$ \\
			
	AS19 & G07\_028T01\_9000001500 & 10394 & 30-08-2017 & 57995.3 & 3.0 & 5726 & 6520  & 0.79 & 0.06 & $\omega$ \\
			
	AS20 & G07\_046T01\_9000001534 & 10583 &  12-09-2017 & 58008.0 & 1.9 & 7392 & 8770  & 0.90 & 0.06 &  $\gamma$ \\
					
	AS21 & A04\_180T01\_9000001622 & 11147 & 20-10-2017 & 58046.0 & 3.1 & 8432 & 10274  & 0.58 & 0.03 & $\delta$ \\
			
	AS22 & A04\_180T01\_9000002000 & 13559 & 01-04-2018 & 58209.1 & 2.6 & 2761 & 2933  & 0.69 & 0.15 & $\chi^{}_1$ \\

    AS23 & G08\_028T01\_9000002306 & 15562 & 14-08-2018 &  58344.4 & 3.6 & 410 & 414 & 0.89 & 0.47 &  $\chi^{}_4$ \\

    AS24 & T03\_231T01\_9000003910 & 27094 & 01-10-2020 &  59123.5 & 3.8 & 218 & 220 & 0.53 & 0.26 &  $\rho$ \\

	AS25 & T04\_029T01\_9000004582 & 31451 & 22-07-2021 &  59417.9 & 3.1 & 336 & 338 & 0.47 & 0.06  & $\delta$ \\

	\hline
\end{tabular}%
}
\begin{tablenotes}
    \item $^*$A, B and C are the count rates in $3-6$ keV, $6-15$ keV and $15-60$ keV energy ranges, respectively.
    \item $^\dagger$Refers effective exposure, not the stare time of the instruments.
    \item {\red $^\ddagger$SSM observation.}
\end{tablenotes}

\label{tab:Table-02}

\end{table*}

\begin{table*}
    \caption{Details of the \textit{AstroSat} observations of persistent-cum-variable (PcV) source Cyg X$-$1 and Cyg X$-$3 considered in this study. The notations and symbols have their usual meaning. See \S \ref{sec4:obs}, \S \ref{sec6:6.1_timing} and \S \ref{sec7:7.1_pcv} for details.}
    \renewcommand{\arraystretch}{1.4}
    \resizebox{1.0\textwidth}{!}{%
    \begin{tabular}{c c c c c c c c c c}
    \hline
    Source & Epoch & ObsID & Orbit & Date & MJD & Exposure$^\dagger$ & $r_{\rm det}$ & $r_{\rm in}$ & $F_{\rm var}$  \\

    & & & & (dd-mm-yyyy) & &  (ks) & (cts/s) & (cts/s) & (\%) \\ \hline

    Cyg X$-$1 & AS1  &  G06\_028T01\_9000000768  & 05931 & 01-11-2016 & 57693.7 & 3.1 & 2699 & 2863 & 18.4 \\

    &  AS2 & G06\_028T01\_9000000834  & 06324 & 28-11-2016 & 57720.3 & 4.3 & 2817 & 2997 & 20.9 \\

    &  AS3 & G06\_028T01\_9000000890  & 06597 & 16-12-2016 & 57738.7 & 2.5 & 3548 & 3837 & 17.2 \\

    &  AS4 & G06\_028T01\_9000001094 & 07990 & 20-03-2017 & 57832.8 & 4.2 & 2579 & 2729 & 10.8 \\

    &  AS5 & G07\_027T01\_9000001180 & 08403 & 17-04-2017 &  57860.9 &  6.7 & 3013  &  3223 & 11.1  \\

    &  AS6  & G07\_027T01\_9000001210 & 08708 & 08-05-2017 &  57881.4 & 2.8 & 2964 & 3164 & 9.4 \\

    & AS7  & G07\_027T01\_9000001358 & 09569 & 05-07-2017 & 57939.5  & 4.0 & 2903 & 3093 & 11.5 \\

    &  AS8  & G07\_027T01\_9000001516  & 10485 & 05-09-2017 & 58001.5  & 4.9 & 3697 & 4012 & 29.1 \\

    &  AS9  & G08\_075T01\_9000001616  & 11068 & 14-10-2017 & 58040.9  & 2.6 & 3505 & 3788 & 24.5 \\

    &  AS10  & G08\_075T01\_9000001636 & 11212 & 24-10-2017 & 58050.6 & 3.0 & 1444 & 1489 & 26.4 \\

    &  AS11  & G08\_075T01\_9000001726 & 11717 & 27-11-2017 & 58084.6 & 4.5 & 3133 & 3373 & 24.9 \\

    &  AS12  & G08\_075T01\_9000002024 & 13657 & 07-04-2018 & 58215.7 & 6.7 & 1864 & 1941 & 30.7 \\

    & AS13  & T02\_073T01\_9000002252 & 15250 & 24-07-2018 & 58323.4 & 3.7 & 1958 & 2046 & 20.0 \\

    & AS14  & G08\_030T01\_9000002302 & 15548 & 13-08-2018 & 58343.5  & 4.6 & 1274 & 1310 & 20.4 \\

    &  AS15  & A05\_180T01\_9000002986 & 20036 & 12-06-2019 & 58646.7 & 4.6 & 3920 & 4277 & 10.8 \\

    &  AS16  & A10\_109T01\_9000004492 & 31080 & 27-06-2021 & 59392.8 & 3.9 & 2271 & 2386 & 7.6 \\

    &  AS17  & A10\_109T01\_9000004638 & 31775 & 13-08-2021 & 59439.7 & 3.4 & 2839 & 3021  & 7.8 \\

    &  AS18  & A10\_109T01\_9000004678 & 32013 & 29-08-2021 & 59455.7 & 7.8 & 2172 & 2277 & 10.0 \\

    & AS19  & A11\_080T01\_9000004756 & 33001 & 04-11-2021 & 59522.5  & 6.0 & 1984 & 2072 & 8.9 \\

    &  AS20  & A11\_080T01\_9000005146 & 35840 & 15-05-2022 & 59714.3 & 2.4 & 2742 & 2912 & 13.6 \\

    &  AS21  & T05\_105T01\_9000005658 & 41345 & 22-05-2023 & 60086.1 & 3.2 & 2638 & 2795 & 26.2 \\

    \hline
Cyg X$-$3   & AS1 & T01\_028T01\_9000000360& 02376$-$02392 & 06-03-2016 & 57453.0 & 18.9 & 1361 & 1443 & 47.8 \\

    & AS2 & G05\_192T01\_9000000466& 03498$-$03504 & 21-05-2016 & 57529.0 & 26.5 & 2297 & 2542 & 30.5 \\
    
    & AS3 & G05\_170T01\_9000000510& 03975 &  22-06-2016 & 57561.0 & 7.7 & 705 & 726 & 20.8 \\
    
    & AS4 & G05\_192T01\_9000000522& 04094$-$04099 & 30-06-2016 & 57569.0 & 7.39 & 1157 & 1216 & 60.3 \\
    
    & AS5 & G05\_192T01\_9000000526& 04111$-$04112 & 01-07-2016 & 57570.0 & 6.9 & 1451 & 1545 & 13.9 \\  
    
    & AS6 & G05\_192T01\_9000000594& 04728$-$04736 & 12-08-2016 & $57612.0$ & $20.5$ & $2050$ & $2243$&  $64.5$ \\
    
    & AS7 & G06\_103T01\_9000000812& 06206$-$06222 & 20-11-2016 & $57712.0$ & $38.6$ & $1494$ & $1594$ & $44.0$ \\
    
    & AS8 & G07\_043T01\_9000001126& 08163$-$08178 & 01-04-2017 & $57844.0$ & $16.9$ & $1049$ &  $1097$& $17.3$ \\
    
    & AS9 & G07\_048T01\_9000001494& 10362 & 27-08-2017 & $	57992.0$ & $1.8$ & $1616$ & $1734$&  $38.7$ \\
    
    & AS10 & G08\_032T01\_9000001680& 11463$-$11468 & 10-11-2017 & $58067.0$ & $11.6$ & $1280$ &  $1352$& $56.0$ \\
    
    & AS11 & G08\_032T01\_9000001998& 13553 & 31-03-2018 & $58208.0$ & $1.9$ & $1121$ & $1176$ & $21.7$ \\
    
    & AS12 & G08\_032T01\_9000002280& 15448 & 06-08-2018& $58336.9$ & $1.4$ & $1238$ & $1306$&  $50.4$ \\   
    
    & AS13 & T03\_105T01\_9000002876& 19400 & 30-04-2019& $58603.9$ & $4.5$ & $1629$ & $1749$ & $45.5$ \\     
    
    & AS14 & T03\_115T01\_9000002914& 19400 & 13-05-2019& $58616.0$ & $41.2$ &$1112$ &  $1166$ & $37.8$ \\ 
 \hline

\end{tabular}%
}
\begin{tablenotes}
    \item $^\dagger$Refers effective exposure, not the stare time of the instruments.
\end{tablenotes}

\label{tab:Table-03}
\end{table*}

\subsection{Persistent (P)}\label{sec4:4.2}

The persistent sources LMC X$-$1, LMC X$-$3, GRS 1758$-$258, 1E 1740.7$-$2942, and 4U 1957+115 are studied using archival {\it AstroSat} observations spanning the period $2016-2023$. Details of all the observations considered in this study are provided in Table \ref{tab:Table-04}.

LMC X$-$1 was observed during $2016$, $2017$, and $2020$ comprising four pointed observations, all obtained under the AO cycle. LMC X$-$3 was observed in $2016$, $2017$, $2018$, and $2020$ with a total of $9$ pointed observations. Of these, $2$ observations from 2016 correspond to GT data, while the remaining are from the AO cycles. GRS 1758$-$258 was monitored between $2016$ and $2018$ through four pointed observations including three AO cycle data sets and one GT observation. Similarly, 1E 1740.7$-$2942 was observed between $2016$ and $2017$ with five pointed observations consisting of three GT and two AO data sets. The source 4U 1957$+$115 was observed by {\it AstroSat} in $2016$, $2017$, and $2018$ with $7$ pointed observations in total as $6$ GT and $1$ AO data set. 

\subsection{Transient (T)}\label{sec4:4.3}

{\it AstroSat} has observed $26$ transient BH–XRBs over multiple epochs throughout its operational period to date. {\red Among these, $16$ sources exhibit variability in both spectral and timing properties. The remaining $10$ sources (XTE J1859$+$226, V4641 Sgr, V404 Cyg, MAXI J1305$-$704, Swift J1357.2$-$0933, MAXI J1810$-$222, MAXI J1659$-$152, XTE J1908$+$094, XTE J1752$-$223 and XTE J1118$+$480) are excluded owing to insufficient photon statistics for comprehensive study.} 

{\it AstroSat} monitored Swift J1727.8$-$1613 over $8$ epochs between $2-14$ September 2023, including $1$ slew mode and $7$ ToO observations. During these observations, the LAXPC20 detector was switched off for safety reasons, and data were acquired solely using the LAXPC10 detector operated in low gain mode \citep{Nandi-etal2024}. For timing analysis, we utilize LAXPC10 data in the $20-100$ keV energy band. The source was simultaneously observed by {\it Insight–HXMT} during several of the {\it AstroSat} epochs, and we employ $4$ {\it HXMT} observations for spectral analysis. MAXI J1535$-$571 was observed through ToO observations conducted between $12-17$ September $2017$ as highlighted in the outburst profile and hardness ratio plots shown in Fig. \ref{fig:fig21}. We analyze all available observations of the source, however, to avoid repetition of similar results from closely spaced epochs exhibiting only marginal spectro-temporal variability, we present the results for eleven representative epochs (see Table \ref{tab:Table-05}). Swift J1658.2$-$4242 was observed during its $2018$ outburst with three ToO observations of {\it AstroSat}.

GX 339$-$4 was observed extensively with {\it AstroSat} across multiple outbursts spanning $2016-2024$. In this study, we consider $23$ epochs of {\it AstroSat} observations to investigate the detailed spectro-temporal variability properties of the source. This consists of $17$ ToO data, $5$ AO data and $1$ GT data. Similarly, for H 1743$-$322, we use $4$ observations obtained during its $2016-2017$ outburst, where $1$ observation corresponds to ToO data and the rest are AO data. 4U 1630$-$472 exhibited several outbursts to date, with {\it AstroSat} observations available from the $2016$, $2018$, $2020$, and $2023$ outbursts. All together, there are $10$ ToO data, $4$ AO data and $3$ GT data. In Fig. \ref{fig:fig25}, we show the outburst profiles from {\it MAXI}/GSC for 4U 1630$-$472, GX 339$-$4 and H 1743$-$322 with the epochs of {\it AstroSat} observations marked. 

{\it AstroSat} observed 4U 1543$-$47 over $6$ epochs during its $2021$ outburst, marked in the {\it MAXI}/GSC daily-averaged lightcurve shown in Fig. \ref{fig:fig29}. The first two observations were performed with a $40$ arcmin offset owing to the high brightness of the source. In addition, $3$ {\it AstroSat} observations were carried out during the $2018$ outburst of MAXI J1820+070, as indicated in the {\it MAXI}/GSC lightcurve of the outburst (Fig. \ref{fig:fig29}).  The transient sources MAXI J0637$-$430 and GRS 1716$-$249 were observed by \textit{AstroSat} under ToO mode during their $2019$ outbursts, with four and three observations respectively, the latter exhibiting a failed outburst. The source MAXI J1348$-$630 was also observed three times by \textit{AstroSat} during its $2019$ outburst. MAXI J1803$-$298 was observed in ToO mode with five observations during its $2021$ outburst. {\red Swift J151857.0$-$572147 was observed during the first outburst from 16th March $2024$ to 18th March $2024$. Similarly, GRS 1739$-$278 and XTE J2012$+$381 was observed by \textit{AstroSat} during July $2021$ and March $2023$ respectively.} A summary of all \textit{AstroSat} observations analyzed in this study for the respective transient sources is presented in Table~\ref{tab:Table-05}. 

\begin{table*}
\caption{Details of the \textit{AstroSat} observations of persistent sources LMC X$-$1, LMC X$-$3, GRS 1758$-$258, 1E 1740.7$-$2942 and 4U 1957$+$115. The notations and symbols have their usual meaning. See \S \ref{sec4:obs}, \S \ref{sec6:6.1_timing} and \S \ref{sec7:7.2_persis} for details.}
\renewcommand{\arraystretch}{1.3}
\resizebox{1.0\textwidth}{!}{%
\begin{tabular}{l c c c c c c c c c} \hline
Source & Epoch & ObsID & Orbit & Date & MJD & Exposure$^\dagger$ & $r_{\rm det}$ & $r_{\rm in}$ & $F_{\rm var}$ \\

& & & & (dd-mm-yyyy) & &  (ks) & (cts/s) & (cts/s) & (\%) \\ \hline
LMC X$-$1  & AS1 & A02\_118T01\_9000000826 & 23205$-$23222  & 25-11-2016 & 57717  & 62.2 & $40.1$ &  40.2 & $6.5$  \\
         & AS2 & A03\_119T02\_9000001496 & 10371$-$10375  & 28-08-2017 & 57993  & 14.5 & $41.3$ &  41.4 & $7.7$  \\
         & AS3 & A08\_008T01\_9000003414 & 23113$-$23128  & 06-01-2020 &  58854 & 62.1 & $44.9$ &  45.0 & $5.3$  \\
         & AS4 & A08\_008T01\_9000003596 & 24373$-$24374  & 31-03-2020 &  58939 & 53.5 & $50.2$ &  50.3 & $14.9$  \\
\hline
LMC X$-$3  & AS1 & G05\_212T01\_9000000600 & 04761$-$04767  & 14-08-2016 & 57614  & 25.7 & $49.9$ & 50.0 & $2.7$  \\
         & AS2 & G05\_212T01\_9000000672 & 05295$-$05301  & 19-09-2016 &  57650 & 21.4 & $89.8$ &  90.1 & $5.8$  \\
         & AS3 & A03\_106T01\_9000001130 & 08190$-$08195  & 03-04-2017 & 57846  & 19.4 & $74.9$ &  75.1 & $6.4$  \\
         & AS4 & A03\_106T01\_9000001190 & 08432$-$08439  & 19-04-2017 & 57862  & 28.6 & $218.0$ &  220.0 & $10.8$  \\
         & AS5 & A04\_112T01\_9000001778 & 11565$-$11574  & 17-11-2017 & 58074  & 28.9 & $92.6$ &  93.0 & $2.4$  \\
         & AS6 & A04\_112T01\_9000001698 & 12014$-$12025  & 17-12-2017 & 58104  & 42.9 & $83.7$ &  84.0 & $2.7$  \\
         & AS7 & A04\_112T01\_9000001884 & 12760$-$12768  & 06-02-2018 & 58155  & 32.5 & $98.3$ &  98.7 & $2.7$  \\
         & AS8 & A04\_112T01\_9000001908 & 12966$-$12974  & 20-02-2018 & 58169  & 32.7 & $66.4$ &  66.6 & $2.8$  \\
         & AS9 & A08\_008T02\_9000003430 & 23205$-$23222  & 01-12-2020 & 59184  & 56.9 & $241.1$ &  243.6 & $4.6$  \\
         
\hline
GRS 1758-258$^\ddagger$ & AS1  & A02\_077T01\_9000000732 &  05671  & 17-10-2016 & 57678  & 58.6  & $4.3$ & 4.3 &$13.8$  \\
        & AS2 & A03\_053T01\_9000001410 &  09910  & 28-07-2017 &  57962 & 43.4 & $2.6$ & 2.6  &$6.6$  \\
        & AS3 & A03\_053T01\_9000001542 &  10708  & 19-09-2017 & 58015  & 48.2 & $3.2$ &3.2  &$6.1$  \\
        & AS4 & G08\_070T01\_9000002028 &  13673  & 08-04-2018 & 58216  & 7.5 & $2.0$ &  2.0&$13.2$  \\
             
\hline

1E 1740.7-2942$^{\ddagger}$ & AS1 & G05\_158T01\_9000000630 & 04965$-$04981   & 28-06-2016 & 57567  & 88.6 & $0.3$ & 0.3 &$11.4$  \\
        & AS2 & A02\_086T01\_9000000714 &  05547$-$05561  & 06-10-2016 & 57667  & 70.0 & $0.5$ & 0.5 &$15.2$  \\
        & AS3 & G08\_045T01\_9000001920 &  13047$-$13060  & 05-06-2018 & 58274  & 77.2 & $0.6$ &  0.6 &$10.1 $  \\
        & AS4 & G08\_045T01\_9000002092 &  14153$-$14164  & 11-05-2018 & 58249  & 68.6 & $0.5$ & 0.5 &$10.0$  \\
        & AS5 & A04\_229T01\_9000002096 &  14174$-$14187  & 12-05-2021 & 59346  & 94.5 & $0.5$ & 0.5 &$6.5$  \\

\hline
4U 1957+115 & AS1 & G05\_194T01\_9000000490 & 03802  &  10-06-2016 & $57549$ & $3.8$ & $946.0$ & $985.0$ & $3.5$\\

        & AS2 & G05\_194T01\_9000000660 & 05206  & 13-09-2016 & $57644$ & $3.1$ & $802.0$ & $829.0$ &$4.4$ \\
        
        & AS3 & G07\_052T01\_9000001270 & 09124  & 05-06-2017 & $57909$ & $2.0$ & $820.0$ & $849.0$ & $3.6$\\   
        
        & AS4 & G07\_052T01\_9000001362 & 09586  & 06-07-2017 & $57940$ & $10.5$ & $911.0$ &  $947.0$ &$4.0$ \\        
        & AS5 & A03\_117T06\_9000001404 & 09856$-$09891 & 24-07-2017 & $57958$ & $8.8$ & $685.0$ & $705.0$ &$10.2$ \\
        
        & AS6 & G08\_044T01\_9000001658 &  11328 & 01-11-2017 & $58058$ & $10.4$ & $689.0$ & $709.0$ & $2.8$\\
        
        & AS7 & G08\_044T01\_9000002014 & 13615$-$13620  & 05-04-2018 & $58213$ & $9.0$ & $482.0$ & $491.0$ &$4.5$ \\ \hline

\end{tabular}%
}
\begin{tablenotes}
    \item $^\dagger$Refers effective exposure, not the stare time of the instruments.
    \item $^{\ddagger}$The estimated parameters are expected to be affected due to contamination from nearby sources within the FoV of LAXPC.
\end{tablenotes}

\label{tab:Table-04}
\end{table*}

\begin{table*}
 \caption{Details of the \textit{AstroSat} observations of transient sources under considerations. The notations and symbols have their usual meaning. See \S \ref{sec4:obs}, \S \ref{sec6:6.1_timing} and \S \ref{sec7:7.3_transi} for details. 
 }
 \renewcommand{\arraystretch}{1.4}
 \resizebox{1.0\textwidth}{!}{%
 \begin{tabular}{l c c c c c c c c c}
 \hline
 Source & Epoch & ObsID & Orbit & Date & MJD & Exposure$^\dagger$ & $r_{\rm det}$ & $r_{\rm in}$ & $F_{\rm var}$ \\

 & & & & (dd-mm-yyyy) & &  (ks) & (cts/s) & (cts/s) & ($\%$) \\  \hline
 IGR J17091-3624$^{\ddagger}$  & AS1 & T01\_118T01\_9000000430 & 3128  & 26-04-2016 & 57504.4 & 2.3 & 2145 & 2248 &  15.5  \\

    & AS2 & T05\_020T01\_9000005050 & 35195  & 01-04-2022 & 59670.8 & 2.9 & 1428 & 1472 & 12.8  \\ \hline

Swift J1658.2-4242 & AS1 & T02\_004T01\_9000001910 &  12977 & 20-02-2018  &  $58170.0$  & $9.2$ & $959$  & $999$ &$6.6$\\

    & AS2 & T02\_004T01\_9000001910 &  12981 &  20-02-2018 &  $58170.4$  & $3.1$ & $997$  & $1040$ & $6.4$ \\
    
    & AS3 & T02\_011T01\_9000001940 &  13138 &  03-03-2018 &  $58180.9$  & $2.5$ & $1236$ &  $1304$ &$4.4$\\
    
    & AS4 & T02\_011T01\_9000001940 &  13143 &  03-03-2018  &  $58181.1$  & $54.8$& $1500$ & $1601$  &$15.1$ \\
    
    & AS5 & T02\_020T01\_9000001990 &  13509 &  28-03-2018  &  $58205.9$  & $0.3$  & $510$  &  $521$ &$54.9$ \\
    
    & AS6 & T02\_020T01\_9000001990 &  13516 &  28-03-2018  &  $58206.3$  & $8.6$ & $1164$ &  $1224$ &$4.1$ \\
    
    & AS7 & T02\_020T01\_9000001990 &  13517 &  28-03-2018  &  $58206.5$  & $6.7$ & $1251$ & $1320$  &$20.2$ \\
    
    & AS8 & T02\_020T01\_9000001990 &  13520 & 28-03-2018  &  $58206.6$  &  $2.3$& $987$  &  $1030$ & $4.3$\\
\hline

4U 1630-472$^{\ddagger}$ & AS1 &G05\_157T01\_9000000626  & 04944 & 27-08-2016   & $57627.0$     & $5.0$  & $546$ & $559$  & $18.2$ \\

    & AS2 &G05\_157T01\_9000000626  & 04955 & 27-08-2016   & $57627.7$     & $5.0$  & $546$ & $559$  & $18.2$ \\

    & AS3 &G05\_132T01\_9000000686  & 05416$-$05433 & 28-09-2016   & $57659.0$     & $3.4$  & $4458$ &  $5485$  & $5.6$\\
   
    & AS4 &T02\_076T01\_9000002274	& 15413$-$15419 & 04-08-2018  & $58334.4$     & $11.3$ & $1104$ & $1158$  &$5.2$  \\
    
    & AS5 &T02\_077T01\_9000002282	& 15454 &  07-08-2018  & $58337.3$     & $9.6$  & $1468$ & $1564$  &  $5.3$\\
    
    & AS6 &T02\_078T01\_9000002294	& 15497$-$15501 & 10-08-2018  & $58340.2$     & $6.8$  & $1361$ &  $14431$  & $3.3$ \\
    
    & AS7 &T02\_079T01\_9000002298	& 15522$-$15526 & 11-08-2018  & $58341.8$     & $9.1$  & $1356$  & $1438$ &$2.2$ \\
    
    & AS8 &T02\_083T01\_9000002304	& 15559 &  14-08-2018  & $58344.3$     & $3.5$  & $1380$  & $1465$ & $1.9$\\
    
    & AS9 &T02\_084T01\_9000002318	& 15656$-$15689 &  20-08-2018  & $58350.9$     & $4.3$  & $1280$  & $1353$ & $2.2$\\
    
    & AS10 &T02\_097T01\_9000002340	& 15803$-$15808 &  30-08-2018  & $58360.8$     & $1.4$  & $7654$  & $11280$ &$4.3$ \\
    
    & AS11 &T02\_100T01\_9000002354	& 15978 & 11-09-2018  & $58372.6$     & $2.5$  & $1559$  & $1668$ & $5.3$ \\
    
    & AS12 &T02\_101T01\_9000002372	& 16060$-$16061 &  17-09-2018  & $58378.1$     & $7.9$  & $1681$  & $1809$ & $5.6$\\
    
    & AS13 &T03\_188T01\_9000003612	& 24587$-$24592 &  15-04-2020  & $58954.1$     & $8.3$  & $1316$  & $1393$ & $3.9$ \\
                         
    & AS14 &A12\_056T02\_9000005538	& 40285$-$40291 & 11-03-2023   & $60014.0$     & $24.1$  & $2991$  & $3192$ & $15.2$ \\
    
    & AS15 &A12\_056T02\_9000005538	& 40345$-$40350 & 15-03-2023   & $60018.0$     & $16.2$  & $2134$  & $2234$ & $47.9$ \\
    
    & AS16 &A12\_056T02\_9000005538	& 40356$-$40358 & 16-03-2023   & $60019.0$     & $16.2$  & $2137$  & $2237$ & $32.8$ \\
    
    & AS17 &A12\_056T02\_9000005538	& 40400$-$40408 & 19-03-2023   & $60022.0$     & $26.7$  & $1935$  & $2017$ & $26.0$ \\

\hline
4U 1543-47 & AS1 & T04\_018T01\_9000004494 & 31126 & 01-07-2021 & 59396.0 & 7.9 & 1433 & 1841 & $14.7$ \\
	
    & AS2 & T04\_021T01\_9000004526 & 31265 & 10-07-2021 & 59405.4 & 7.7  & 1076 & 1406  & $6.9$ \\
	
    & AS3 & T04\_030T01\_9000004588 & 31499 & 26-07-2021 & 59421.2 & 11.9  & 3012 & 4052 & $0.8$ \\
		      
    & AS4 & T04\_035T01\_9000004622 & 31638 & 04-08-2021 & 59430.6 & 10.3  & 2696 & 3627 & $4.1$ \\
	
    & AS5 & T04\_046T01\_9000004680 & 32029 & 31-08-2021 & 59457.1 & 15.7  & 2377 & 3217 & $1.4$ \\
	
    & AS6 & T04\_051T01\_9000004686 & 32089 & 04-09-2021 & 59461.1 & 51.2 & 2172 & 2952  & $1.3$\\
\hline

MAXI J1535-571 & AS1 & T01\_191T01\_9000001536 & 10585 & 12-09-2017 & 58008.2 & 4.9  & 4479 & 5532 & $10.1$    \\
			
    & AS2 & T01\_191T01\_9000001536 & 10589 & 12-09-2017 & 58008.5 & 4.6  & 5337 & 6902 & 2.9    \\
			
    & AS3 & T01\_191T01\_9000001536 & 10594 & 12-09-2017 & 58008.8 & 3.3  & 5398 & 7005 & 3.2    \\
	
    & AS4 & T01\_191T01\_9000001536 & 10604 & 13-09-2017 & 58009.5 & 3.7  & 5593 & 7337 & 2.1   \\
	
    & AS5 & T01\_191T01\_9000001536 & 10618 & 14-09-2017 & 58010.5 & 4.1  & 5708 & 7536 & 2.9    \\
	
    & AS6 & T01\_191T01\_9000001536 & 10625 & 14-09-2017 & 58010.9 & 3.4  & 5910 & 7892 & 2.0    \\
	
    & AS7 & T01\_191T01\_9000001536 & 10628 & 15-09-2017 & 58011.1 & 4.5  & 5923 & 7915 & 2.1   \\
	 
    \hline

\end{tabular}%
}
\begin{tablenotes}
    \item $^\dagger$Refers effective exposure, not the stare time of the instruments.
    \item $^{\ddagger}$Estimated parameters are expected to be affected due to contamination from nearby sources within the FoV of LAXPC.
\end{tablenotes}

\label{tab:Table-05}
\end{table*}

\begin{table*}
\ContinuedFloat
\caption{Continued.}
\renewcommand{\arraystretch}{1.3}
\resizebox{1.0\textwidth}{!}{%
\begin{tabular}{l c c c c c c c c c} \hline
Source  & Epoch & ObsID & Orbit & Date       & MJD & Exposure$^\dagger$ & $r_{\rm det}$ & $r_{\rm in}$ & $F_{\rm var}$ \\
        &       &       &       & (dd-mm-yyyy) &     &  (ks)     & (cts/s)       & (cts/s)      & ($\%$)  \\ \hline
    MAXI J1535-571   & AS8 & T01\_191T01\_9000001536 & 10636 & 15-09-2017 & 58011.7 & 3.5  & 6011 & 8073 & 1.2    \\
    
    & AS9 & T01\_191T01\_9000001536 & 10644 & 16-09-2017 & 58012.2 & 4.9  & 6189 & 8397 & 2.1    \\
			
    & AS10 & T01\_191T01\_9000001536 & 10651 & 16-09-2017 & 58012.7 & 3.4  & 6282 & 8570 & - \\ 
    
    & AS11 & T01\_191T01\_9000001536 & 10656 & 17-09-2017 & 58013.1 & 3.7  & 6338 & 8674 & -\\
    
    \hline
    
MAXI J1820$+$070 & AS1 & T02\_038T01\_9000001994 & 13533 & 30-03-2018 & 58208.0 & 2.4  & 4424 & 4760 & $3.0$\\
	
    & AS2 & T02\_060T01\_9000002216 & 15038 & 10-07-2018 & 58309.0 & 2.5  & 2516 & 3284 & $4.6$\\
		
    & AS3 & T02\_066T01\_9000002324 & 15724 & 25-08-2018 & 58355.0 & 7.9  & 955 & 1336 & $8.7$ \\ \hline
                   
MAXI J1348$-$630  & AS1 & T03\_083T01\_9000002722   &	18368 & 19-02-2019 &  $58542.6$ & $4.7$ & $4001$ & $4809$ & $20.6$ \\
                  
    & AS2 & T03\_083T01\_9000002742   &   18495 &  28-02-2019 & $58612.2$  & $19.9$& $3453$ & $4039$ &  $4.9$ \\
    
    & AS3 & T03\_120T01\_9000002990   &   20071 &  14-06-2019 &  $58649.1$ & $4.6$ & $2027$ & $2216$ & $11.9$ \\ \hline                  

MAXI J0637$-$430  & AS1 & T03\_147T01\_9000003290 & 22231$-$22247   & 08-11-2019 & 58795.0  &  5.6 & $154$ & 155 &$7.4$ \\
    
    & AS2 & T03\_147T01\_9000003306 & 22334$-$22349   & 15-11-2019 &  58802.0 & 8.0  & $92$ & 92.4 &$9.0$ \\
    
    & AS3 & T03\_149T01\_9000003314 & 22412   & 20-11-2019 & 58807.0  & 5.2 & $212$ & 214 &$20.6$ \\
    
    & AS4 & T03\_159T01\_9000003328 & 22426$-$22437  & 21-11-2019 & 58808.0  & 6.0 & $96$ & 96.4 &$4.3$ \\ \hline
    
MAXI J1803$-$298 & AS1 &T04\_003T01\_9000004368  & 30377 &  11-05-2021   & $59345.0$   & $9.0$ & $1083$ & $1135$  &  $5.2$  \\
    
    & AS2 &T04\_006T01\_9000004370	& 30381 & 11-05-2021    & $59345.6$   & $3.6$ & $1126$ &  $1182$  & $5.4$  \\
    & AS3 &T04\_006T01\_9000004370	& 30385 & 11-05-2021    & $59345.9$   & $4.1$ & $842$ &  $857$  & $8.6$  \\
     & AS4 &T04\_006T01\_9000004370	& 30393 & 12-05-2021    & $59346.5$   & $2.4$ & $844$ &  $859$  & $4.9$  \\          
    & AS5 &T04\_028T01\_9000004584	& 31477$-$31478 &  24-07-2021   & $59419.0$     & $6.0$  & $705$  & $726.5$ & $4.7$  \\ 
    & AS6 &T04\_028T01\_9000004584	& 31484 &  25-07-2021   & $59420.1$     & $1.3$  & $418$  & $421$ & $8.4$  \\ \hline

GRS 1716$-$249$^{\ddagger}$ & AS1 & T01\_156T01\_9000001034 &  07488 &  15-02-2017 & $57799.0$  & $1.6$ & $1811$ & $1906$ & $12.7$ \\

    & AS2 & T01\_164T01\_9000001140 &  08238 &  06-04-2017 & $57849.0$  & $3.1$ & $1925$ & $2094$  & $7.6$ \\
    
    & AS3 & T01\_176T01\_9000001378 &  09686 &  13-07-2017 & $57947.0$  & $2.1$ & $1053$ & $1102$ & $15.6$\\
    
     \hline

Swift J1727.8$-$1613  & AS1 & SLEW\_01234\_9000005829 & $-$ & 02-09-2023 & 60189.6 & 0.011  & $-$  & $-$ & $-$ \\
    
    & AS2 & T05\_145T01\_9000005836 & 42961$-$42971 & 08-09-2023 & 60195.1 & 31.8  & 1443  &  1537& 10.2 \\
    
    & AS3 & T05\_148T01\_9000005838 & 42975$-$42986 & 09-09-2023 & 60196.0 &  30.8 & 1397  &  1486 & 9.8 \\
    
    & AS4 & T05\_148T01\_9000005838 & 42995$-$43002 & 10-09-2023 & 60196.9 & 32.6  &  1301 & 1377 & 9.4 \\
    
    & AS5 & T05\_148T01\_9000005838 & 43005$-$43016 & 11-09-2023 & 60198.0 &  30.9 & 1174  & 1236 & 8.3 \\
    
    & AS6 & T05\_148T01\_9000005838 & 43019$-$43033 & 12-09-2023 & 60199.0 & 30.6  & 1089 & 1142 & 7.3 \\
    
    & AS7 & T05\_148T01\_9000005838 & 43034$-$43047 & 13-09-2023 & 60199.9 & 33.1  & 1070 & 1121 & 7.2 \\
    
    & AS8 & T05\_148T01\_9000005838 & 43048$-$43054 & 14-09-2023 & 60200.9 &  20.2 & 1141  & 1199 & 7.5 \\ \hline

\rowcolor{lightgray}	

    GX 339-4 & AS1 & T01\_031T01\_9000000368 & 2449$-$2452 & 11-03-2016  & $57458.5$ &  $14.2$ & $38$  &$54$  & $6.2$ \\

\rowcolor{lightgray}
 	
    & AS2 & G05\_253T01\_9000000560 & 4445$-$4471 & 24-07-2016  & $57593.4$ &  $75.0$ & $40$  & $67$  & $2.7$ \\
 				
 	& AS3 &A04\_109T01\_9000001578 & 10916$-$10926 & 04-10-2017  & $58030.9 $ &  $35.2$ &$61$  &$77$  & $8.4$ \\
 	
    & AS4 &A05\_166T01\_9000003192 & 21544$-$21557 & 22-09-2019  & $58748.7$&  $42.8$ &$140$  &$162$  & $7.8$ \\
 	
    & AS5 &T03\_275T01\_9000004180 & 29084$-$29096 & 13-02-2021  & $59258.0$&  $38.3$ &$440$  &$486$  & $4.1$ \\
 	
    & AS6 &T03\_279T01\_9000004218 & 29335$-$29378 & 02-03-2021  & $59275.0$&  $124.0$ &$516 $ &$569$  &  $4.1$ \\
 	
    & AS7 &T03\_280T01\_9000004222 & 29379$-$29385 & 05-03-2021  & $59278.1$&  $13.3$ & $447$ &$496$  & $4.6$ \\
 	
    & AS8 &T03\_281T01\_9000004224 & 29385$-$29386 & 05-03-2021  & $59278.4$&  $2.5$ & $473$ &$523$  &$9.8$ \\
 	
    & AS9 &T03\_282T01\_9000004226 & 29386$-$29387 & 05-03-2021  & $59278.4$&  $2.4$ & $498$ &$550$  & $9.0$\\
 	
    & AS10 &T03\_291T01\_9000004278 & 29750$-$29764 & 30-03-2021 & $59303.1 $&  $36.5$ & $727$ & $915$ & $3.8$  \\
 	
    & AS11 & T03\_291T01\_9000004278 & 29765$-$29779 & 31-03-2021  & $59304.0$&  $42.4$ &$678$  &$873$  & $3.1$  \\
 	
    & AS12 &T03\_291T01\_9000004278 & 29784$-$29794 & 01-04-2021  & $59305.0$&  $40.9$ &$612$  &$802$  &$4.8$  \\
 	
    & AS13 &T03\_291T01\_9000004278 & 29799$-$29814 & 02-04-2021  & $59306.0$&  $47.5$ &$611$  &$803$  & $4.0$\\
\hline
\end{tabular}%
}
\begin{tablenotes}
    \item $^\dagger$Refers effective exposure, not the stare time of the instruments.
    \item Observations of GX 339$-$4 in QP are highlighted in light gray.
\end{tablenotes}

\label{tab:Table-05}
\end{table*}

\begin{table*}
\ContinuedFloat
\caption{Continued.}
\renewcommand{\arraystretch}{1.3}
\resizebox{1.0\textwidth}{!}{%
\begin{tabular}{l c c c c c c c c c} \hline
Source  & Epoch & ObsID & Orbit & Date       & MJD & Exposure$^\dagger$ & $r_{\rm det}$ & $r_{\rm in}$ & $F_{\rm var}$ \\
        &       &       &       & (dd-mm-yyyy) &     &  (ks)     & (cts/s)       & (cts/s)      & ($\%$)  \\ \hline
GX 339$-$4 & AS14 &T03\_291T01\_9000004278 & 29816$-$29823 & 03-04-2021  & $59307.5$&  $33.1$ &$605$  &$797$  & $5.9$ \\
 	
    & AS15 &T03\_291T01\_9000004278 & 29828$-$29838 & 04-04-2021 & $59308.0$&  $42.2$ &$592$  & $785$ &  $3.9$ \\
 	
    & AS16 &T03\_291T01\_9000004278 & 29843$-$29859 & 05-04-2021 & $59309.0$&  $51.1$&$584$&$780$& $3.1$  \\
 	
    & AS17 &T03\_291T01\_9000004278 & 29860$-$29866 & 06-04-2021  & $59310.3$&  $21.8$ & $565$ & $773$ & $1.9$ \\
 	
    & AS18 &T05\_051T01\_9000005334 & 37541$-$37558 & 07-09-2022 & $59829.5$&  $41.4$ &$141$  &$164$  & $3.3$   \\
    & AS19 & T05\_053T01\_9000005338 & 37572$-$37582 & 09-09-2022 & $59831.5$ & $90.3$ & $148$ & $172$  & $3.3$    \\
 	\rowcolor{lightgray}
 	& AS20 & A12\_099T01\_9000005716 & 41889$-$41914 & 28-06-2023 & $60123.0$ & $50.8$ & $41$ & $56$ & $9.9$   \\
 							
 	& AS21 & A05\_166T01\_9000006070 & 45315$-$45343 & 14-02-2024 & $60354.7$ & $28.6$ & $660$ & $817$ & $4.5$   \\
 	
    & AS22 & A13\_028T01\_9000006122 & 45694$-$45723 & 10-03-2024 & $60379.4$ & $84.3$ & $160$ &$220$ & $1.3$  \\
 	
    & AS23 & T05\_187T01\_9000006132 & 45854$-$45864 & 21-03-2024 & $60390.0$ & $34.0$ & $110$ & $158$ & $2.9$   \\  \hline

H 1743$-$322 & AS1 & T01\_045T01\_9000000364 & 2417$-$2423 & 09-03-2016 & $57456.4$ & $15.4$ & $262$ & $289$ & $9.3$\\
 			
    \rowcolor{lightgray}
 	& AS2 & G05\_253T03\_9000000562 & 4489$-$4498 & 26-07-2016 & $57595.1$ & $66.9$ & $27$ & $69$ & $3.7$ \\
 				
    \rowcolor{lightgray}
 	& AS3 & G05\_144T01\_9000000612 & 4818$-$4828 & 18-08-2016 & $57618.6$ & $22.9$ & $51$ & $68$ & $7.6$ \\
 				
 	& AS4 & G07\_039T01\_9000001444 & 10067$-$10075 & 08-08-2017 & $57973.3$ & $8.5$ & $252$ & $277$ & $7.6$\\ \hline

{\red  Swift J151857.0$-$572147} & AS1 & T05\_088T01\_9000006126 & 45803$-$45809 & 17-03-2024 & 60386.7 & 14.9 & 529 & 535 &  8.9  \\ 

& AS2 & T05\_088T01\_9000006126 & 45810$-$45817 & 18-03-2024 & 60387.4 & 22.9 & 422 & 428 &  4.2  \\ \hline

{\red  GRS 1739$-$278} & AS1 & A10\_119T01\_9000004590 & 31507$-$31524 & 27-07-2021 & 59422.3 & 23.0 & 533 & 539 &  14.0  \\  \hline

{\red  XTE J2012$+$381} & AS1 & T05\_070T01\_9000005544 & 40517$-$40536 & 27-03-2023 & 60030.2 & 33.2 & 516 & 522 & 17.3   \\ \hline
    
\end{tabular}%
}
\begin{tablenotes}
    \item $^\dagger$Refers effective exposure, not the stare time of the instruments.
    \item Observations of GX 339$-$4 and H 1743$-$322 in QP are highlighted in light gray.
\end{tablenotes}

\label{tab:Table-05}
\end{table*}

\subsection{Observation with UVIT onboard \textit{AstroSat}}\label{sec4:4.4}

UVIT observed seven galactic BH-XRBs (LMC X$-$1, LMC X$-$3, GX 339$-$4, GRS 1758$-$258, IGR J17091$-$3624, MAXI J0637$-$430, and MAXI J1820$+$070) in both imaging and spectroscopic modes over the course of its $10$ years of operation. The observations, conducted with a variety of filters, provide valuable insights into the UV characteristics of these sources. A summary of all the observations is given in Table \ref{uvit_summary}. Among these sources, GX 339$-$4, GRS 1758$-$258, IGR J17091$-$3624, and MAXI J0637$-$430 resulted in non-detections. For MAXI J0637$-$430, \cite{Thomas-etal2022} reported that the UVIT spectra from individual pointing could not be combined due to a change in the instrument's effective area. On the other hand, sources like LMC X$-$1 and LMC X$-$3 showed consistent detections across multiple observations including in both FUV and NUV filters.  These sources were observed over several years with no major variations in their flux values. MAXI J1820+070 was observed with UVIT in the spectroscopic mode which uses a grating to perform low-resolution slitless spectroscopy with a spectral resolution of $17$ \AA \space in FUV and $33$ \AA \space in NUV bands \citep{Tandon-etal2017a,Tandon-etal2017b}. 

\section{Methodology of Data Reduction}\label{sec5:methods}

In this section, we describe the comprehensive data reduction methodology employed to extract and analyze BH-XRB data from various instruments onboard {\it AstroSat}. In addition, we also use {\it MAXI} data for the long term monitoring of the black hole sources under consideration.

\subsection{SXT}\label{sec5:5.1_sxt}

We use the Level$-2$ SXT data to extract source events for all observations. The SXT data reduction and analysis are carried out following the standard procedures and guidelines provided by the SXT instrument team\footnote{\url{https://www.tifr.res.in/~astrosat_sxt/index.html}}. The SXT data are available in both Fast Window (FW) and Photon Counting (PC) modes \citep{Singh-etal2017}. For a given observation, the Level$-2$ data are read into \texttt{XSELECT} within the {\tt HEASOFTv6.35.2} package. Following \citet{Singh-etal2017, Majumder-etal2022, Majumder-etal2024, Pahari-etal2024}, we adopt a circular extraction region of radius $12$ arcmin for source spectra in PC mode (see Fig.~\ref{fig:fig5}). This choice is justified as $\sim 90\%$ of the source photons are found to lie within this region \cite[]{Chand-etal2021, Baby-etal2020, Katoch-etal2021, Athulya-etal2022, Majumder-etal2022, Sharma-etal2023, Aneesha-etal2024}. For FW mode data, the source is often found to be offset in the CCD frame, and hence a smaller circular extraction region of $5$ arcmin radius is used (see Fig.~\ref{fig:fig5}). The SXT data are generally found to be free from pile-up for most sources, and thus no pile-up correction is applied \citep{Baby-etal2020, Sreehari-etal2020, Majumder-etal2022, Majumder-etal2024}. In particular, the source count rate remains below $40$ counts s$^{-1}$, effectively ruling out the possible pile-up effects\footnote{\url{https://web.iucaa.in/~astrosat/AstroSat_handbook.pdf}} \cite[]{Majumder-etal2022}. However, the pointed observations (AS1 and AS2) of 4U 1543-47 and most of the observations of Cyg X$-$1 are found to be affected by pile-up. Accordingly, annular regions of $2-15$ arcmin for 4U 1543$-$47 and $3-12$ arcmin for Cyg X$-$1 are used for source flux extraction, respectively. Similarly, MAXI J1820+070 is found to be extremely bright during HSS, and the last two epochs (MJD $58309$ and MJD $58355$) show significant pile-up \cite[]{Prabhakar-etal2022}. For these observations, an annular extraction region with inner and outer radii of $3$ arcmin and $16$ arcmin are adopted. Since the pile-up affects the spectra beyond $4$ keV, only Grade-0 events ($i.e.$, pure X-ray events) are selected following the recommendations of \cite{Singh-etal2021}. The standard background files, along with instrument response and ancillary files provided by the SXT team, are used for spectral modelling.

\subsection{LAXPC}\label{sec5:5.2_laxpc}

The LAXPC data are processed using the software package \texttt{LAXPCSOFT v3.4.5} released on 16 January 2025 to convert the Level$-1$ data into Level$-2$ data format following the standard procedures described in \cite{Antia-etal2017, Antia-etal2021}. During Level$-1$ processing, the software automatically generates GTIs by excluding time intervals affected by SAA. The data within the GTIs are subsequently used for scientific analysis. For the present analysis, the longest continuous observation within each orbit is considered. 

Following \cite{Sreehari-etal2019b, Sreehari-etal2020, Majumder-etal2022}, we generate background subtracted LAXPC10 and LAXPC20 combined lightcurves in different energy ranges using the corresponding LAXPC channels inside the software routine \citep{Antia-etal2017}. For this, the individual lightcurves of the LAXPC detector units are combined using the {\tt lcmath} tool available in {\tt HEASOFTv6.35.2}. From these lightcurves, we compute the incident count rates ($r_{\rm in}$), deadtime corrected incident count rates ($r_{\rm det}$), and hardness ratios (HR1 and HR2; see \S \ref{sec6:6.1_timing} for details). It is worth mentioning that we extract the source spectra from LAXPC20 data only, due to its stable gain throughout the entire observational period \cite[]{Antia-etal2021, Majumder-etal2022}. The appropriate energy response files of the LAXPC20 instrument are used for spectral modelling. Notably, 4U 1543$-$47 has been observed with $40$ arcmin offset by \textit{AstroSat} due to high brightness of the source. In the case of offset observations, offset response files are used in the spectral modelling. Further, we also notice the presence of the Xenon edge feature near $\sim 29-34$ keV in the LAXPC20 spectrum \cite[]{Antia-etal2017, Sreehari-etal2019b, Sreehari-etal2020, Majumder-etal2022}, which is accounted for during the spectral modelling.

\subsection{SSM}\label{sec5:5.3_ssm}

The SSM cameras onboard {\it AstroSat} record the events at a time resolution of $100~\mu$s. We use the SSM data to study the variability properties of GRS 1915$+$105, providing the first glimpse of any BH-XRB source observed with {\it AstroSat}. For this, the event lists were used to generate lightcurves in three energy bands: $2-6$ keV, $6-12$ keV and $2-12$ keV, each with a time bin of $3$ s. Further, in order to remove the outliers, Z-score threshold based \citep{Barnett-etal1994} filtering for a rolling window was applied. For each rolling window, the mean ($\bar{x}$) and standard deviation ($s$) is computed, and outliers are identified for $|(x_i - \bar{x})/s| > t$, where $t$ is the threshold set. A window size of $6$ bins and a threshold $t$ of $0.5$ yielded optimum results. The hardness ratios (HR) are computed on these filtered lightcurves of $3$ s time bins and the uncertainties of the lightcurves are propagated while computing the ratios. These propagated error values has a distinguishable pattern for charged particle events, which is evident from event lists near SAA region, and these HR uncertainty values are further used to filter the lightcurves as well as the HR values.

\subsection{UVIT}\label{sec5:5.4_unit}

We analyze UVIT imaging data (using NUV and FUV filters) in Photon Counting (PC) mode with Level-2 products generated by the UVIT Pipeline available from the \textit{AstroSat} data archive\footnote{\url{https://webapps.issdc.gov.in/astro_archive/archive/Home.jsp}}. The pipeline provides flat-fielded, drift-corrected astrometric images for each filter. The images for various orbits in an observation are coadded to obtain the final image for the respective observations. An aperture of $5$ pixels radius size is selected to do photometry which encompasses $\sim 95$\% of the source pixels. Thereafter, we convert the extracted counts into flux densities for each filter using the unit conversion (UC) as suggested by \cite{Tandon-etal2020}. The obtained flux densities are corrected for foreground extinction in Milky Way. Adopting the ratio of total-to-selective extinction as $R_V=3.1$ \citep{Whitford-1958} for Milky Way, the extinction coefficient in V band ($A_V$) is calculated for LMC X$-$3. This $A_V$ is used to determine the extinction for all the filters using the reddening relation of \cite{Cardelli-etal1989}. The flux for FUV F1 filter ranges from $1.03-1.28$ erg s$^{-1}$ cm$^{-2}$ \AA$^{-1}$, while for NUV F2 and F5 flux is around $0.91$ and $0.58-0.80$ erg s$^{-1}$ cm$^{-2}$ \AA$^{-1}$, respectively. The UVIT color composite image of LMC X$-$3 is shown in Fig. \ref{fig:fig19}. With \textit{AstroSat}, LMC X$-$3 was observed simultaneously with SXT and LAXPC (see \S \ref{sec5:5.1_sxt} and \S \ref{sec5:5.2_laxpc} for data extraction). 

\subsection{MAXI/GSC and Insight-HXMT}\label{sec5:5.5_maxi_hxmt}

In this work, we utilize the on-demand daily monitoring data from the {\it MAXI}/GSC \citep{Matsuoka-etal2009} processed in the $2-20$ keV energy band. The data are accessed through the user interface provided by the instrument team\footnote{\url{http://maxi.riken.jp/mxondem/}}. We extract background subtracted lightcurves in different energy bands by specifying the source coordinates and the desired energy ranges. For this study, we consider {\it MAXI}/GSC lightcurves with a $1-$day time resolution in the $2-20$ keV range to investigate the long term evolution of the sources under considerations. The hardness ratio (HR) is calculated using the definition $HR = C_1 / C_2$, where $C_1$ and $C_2$ denote the photon count rates in the $6-20$ keV and $2-6$ keV energy bands, respectively.

For the {\it Insight-HXMT} data reduction, we use the dedicated software package \texttt{HXMTDAS} and follow the standard procedures outlined by the instrument team\footnote{\url{http://hxmten.ihep.ac.cn/software.jhtml}}. This process involves generating energy spectra, background spectra, and response files for each observation considered in this study.

\subsection{Procedure to Remove Contamination}\label{sec5:5.6_remove_contamination}

\begin{figure*}
    \begin{center}
    \includegraphics[width=0.4\textwidth]{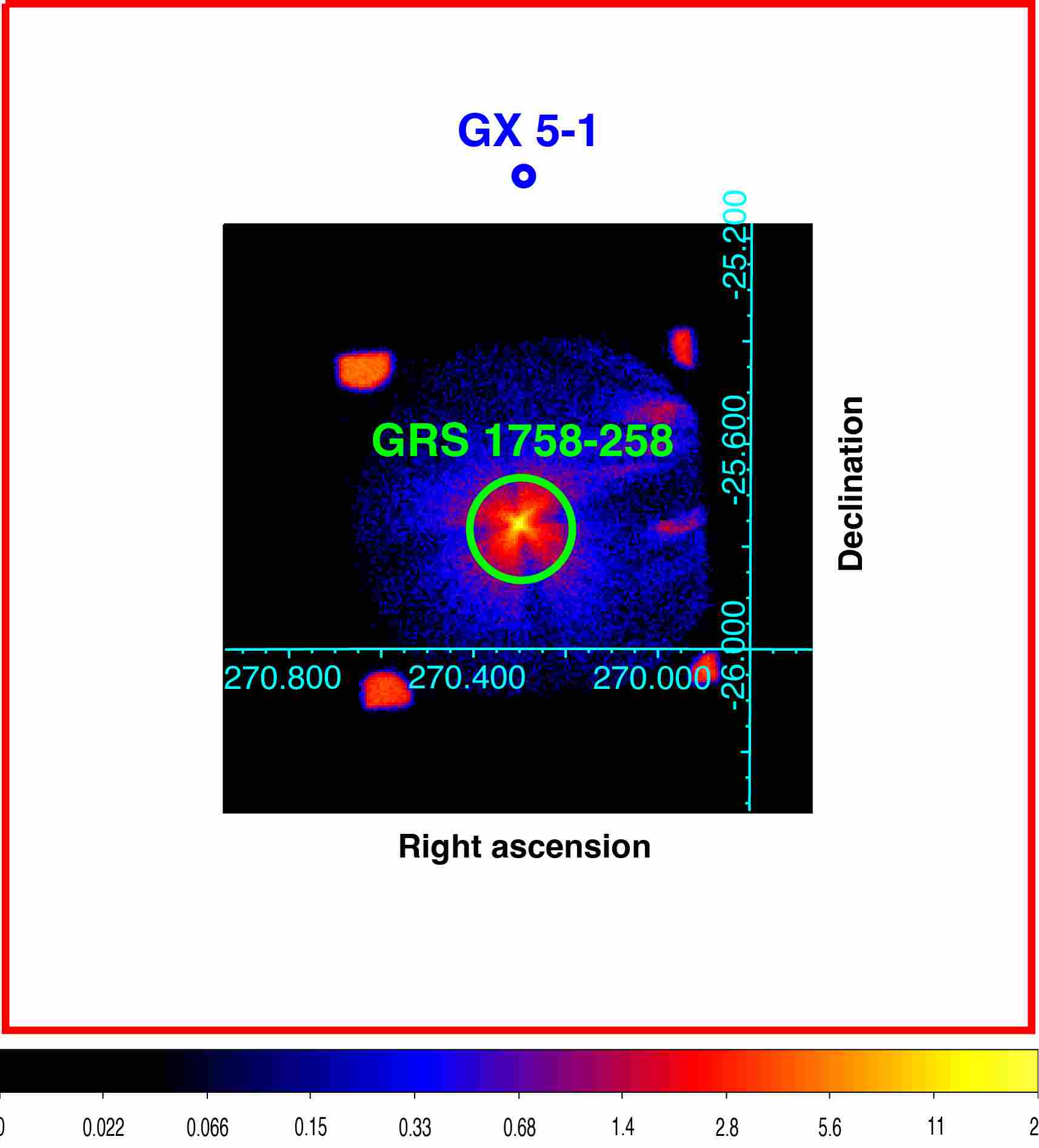}
    $\qquad$
    \includegraphics[width=0.4\textwidth]{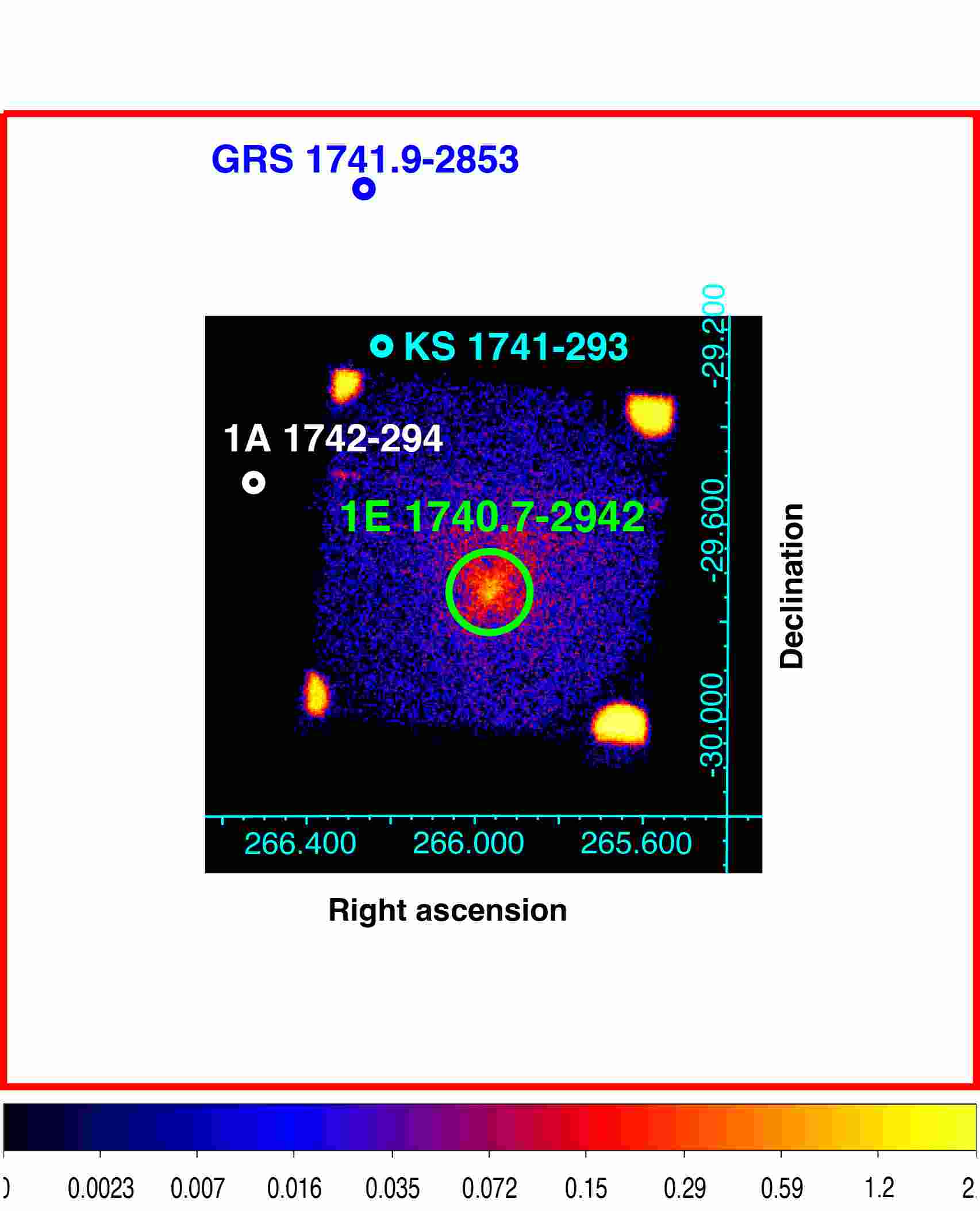}
    \end{center}
    \caption{SXT images of GRS 1758$-$258 (left) and 1E 1740.7$-$2942 (right) with the \textit{LAXPC} field of view indicated by the red box. In addition to the target sources, nearby X-ray sources that may contribute to contamination in the LAXPC observations are also marked. Specifically, the bright low-mass X-ray binary GX 5$-$1 lies within the LAXPC FoV of GRS 1758$-$258, while sources KS 1741$-$293, 1A 1742$-$294 and GRS 1741.9$-$2853 are located near 1E 1740.7$-$2942. See text for details.}
    \label{fig:fig7}
\end{figure*}

Among the $34$ BH–XRBs under consideration, $4$ sources (GRS 1758$-$258, 1E 1740.7$-$2942, IGR 17091$-$3624, and 4U 1630$-$472) are found to suffer contaminated LAXPC observations due to nearby objects. GRS 1758$-$258 and 1E 1740.7$-$2942 are located in close proximity to the Galactic Centre, a region known for the site of high density crowded X-ray sources. As a result, there is a significant likelihood of contamination from nearby sources within the FoV of the target observations. While SXT observations mitigate this issue by restricting the extraction region, LAXPC data are more susceptible to such contamination. For GRS 1758$-$258, the neutron star binary GX 5$-$1 lies at an angular separation of $\sim 40$ arcmin within the LAXPC FoV (see Fig. \ref{fig:fig7}). Similarly, for 1E 1740.7$-$2942, three neutron star systems, namely KS 1741$-$293, 1A 1742$-$292, and GRS 1741.9$-$2853, are located at angular separations of $\sim 27$ arcmin, $31$ arcmin, and $50$ arcmin (see Fig. \ref{fig:fig7}), respectively contributing to contamination in the LAXPC data \cite[see][for details]{Bhuvana-etal2023}.

For 4U 1630$-$472, the LAXPC spectrum is contaminated by the pulsar IGR J16320$-$4751, located at an angular separation of $\sim 34.8$ arcmin. To mitigate this, we apply a spectral subtraction technique prior to conducting spectral analysis \cite[see][for details]{Baby-etal2020}. This technique involves simulating offset spectra for the contaminating sources based on assumed spectral states and incorporating them as additional background spectra along with the standard LAXPC background. Similarly, off-axis response files are employed in the spectral analysis of IGR 17091$-$3624 observations with LAXPC due to contamination from nearby objects \cite[see][for details]{Katoch-etal2021}.

\section{Analysis and Modelling}\label{sec6:analysis_modelling}

In this section, we outline the general methodology adopted for the analysis and modelling of the {\it AstroSat} data for sources under consideration. The analysis involves both timing and spectral investigations comprising the modelling of (a) power spectra to characterize variability features and (b) energy spectra to probe the underlying emission processes. From these analyses, we extract the model fitted parameters to quantify the temporal and spectral properties of the sources.

\subsection{Temporal Analysis}\label{sec6:6.1_timing}

We first generate $1$s binned combined lightcurves from the LAXPC10 and LAXPC20 detectors in the $3-60$ keV energy band for each source to investigate their variability properties. Considering a detector deadtime of $\tau_{\rm d} = 42.5~\mu$s for each LAXPC unit, we estimate the incident photon count rate ($r_{\rm in}$) from the detected count rate ($r_{\rm det}$) as \cite[]{Yadav-etal2016, Sreehari-etal2019b, Sreehari-etal2020},
\begin{equation}
r_{\text{in}} = \frac{r_{\text{det}}}{1 - \tau_{\rm d} r_{\text{det}} / N},
\end{equation}
where $N$ is the number of LAXPC detectors used. In case of GRS 1915+105, we further compute the soft color ($HR1$) and hard color ($HR2$) from lightcurves extracted in different energy bands. For this purpose, lightcurves are generated in the $3-6$ keV, $6-15$ keV, and $15-60$ keV energy ranges. Following \citet{Sreehari-etal2020, Majumder-etal2022}, we define the soft and hard colors as $HR1 = B/A$ and $HR2 = C/A$, where $A$, $B$, and $C$ correspond to the photon count rates in the $3-6$ keV, $6-15$ keV, and $15-60$ keV bands, respectively. The color–color diagrams (CCDs) are then generated by plotting $HR1$ against $HR2$. The long term variability of the sources is examined using the daily binned {\it MAXI} lightcurves in the $2-20$ keV band. The fractional variability of each lightcurve, spanning the period $2016-2025$, is estimated using $F_{\rm var}=\frac{\sqrt{S^2-\overline{\sigma^2_{\rm err}}}}{\overline{x}}$ \cite[see][for details]{Vaughan-etal2003}, where $S^{2}$ is the total variance of the lightcurve, $\overline{\sigma_{\rm err}^{2}}$ is the mean squared measurement error and $\overline{x}$ is the mean count rate.

We further investigate the power spectral properties of the sources using {\it AstroSat} observations by constructing $1$~ms resolution lightcurves from the combined data of LAXPC10 and LAXPC20. Using these lightcurves, we compute PDS for each observation consistent with the Nyquist frequency of $500$ Hz. Each PDS is generated using $32,768$ bins per interval and subsequently averaged to obtain the final PDS. A geometric rebinning factor of $1.03$ is applied during the analysis to rebin the data in frequency space. The resulting power spectra are corrected for instrumental dead-time effects following the methodology described in \citet{Sreehari-etal2019b, Sreehari-etal2020, Majumder-etal2022}. Accordingly, we correct for the dead-time effects on the normalized rms power following the relation \cite[]{Zhang-etal1995, Bachetti-etal2015, Sreehari-etal2019b} as given by,
\begin{equation}
\mathrm{rms}_{\mathrm{in}} = \mathrm{rms}_{\mathrm{det}} (1 + \tau_{\rm d} \, r_{\mathrm{in}}/N) = \frac{\mathrm{rms}_{\mathrm{det}}}{1 - \tau_{\rm d} \, r_{\mathrm{det}}/N},
\end{equation}
where $\mathrm{rms}_{\mathrm{in}}$ is the dead-time corrected rms and $\mathrm{rms}_{\mathrm{det}}$ is the measured rms by the instrument. It is important to note that the correction for dead-time effects becomes particularly crucial when studying HFQPO features, as these typically feature close to the Poisson noise level in the PDS \citep{vanderKlis1989b, Zhang-etal1995, Morgan-etal1997, Sreehari-etal2020}. Each PDS is modelled using a combination of a \texttt{constant} component representing the Poisson noise, and multiple \texttt{Lorentzian} components corresponding to the variability features observed at distinct frequencies. In particular, broad-band limited noise (BLN) components at relatively lower frequency in the PDS are generally fitted with broad and zero-centered Lorentzians, whereas coherent peaks are modelled using narrow Lorentzian profiles \cite[see][for details]{VanderKlis1989a,Nowak-etal2000, Belloni-etal2002, Nandi-etal2024}. In contrast, a simple power-law distribution is sufficient to describe the PDS at higher frequencies.

To characterize and interpret the QPO features in the PDS, we adopt the standard definitions of the quality factor ($Q$–factor $= \nu_{\rm QPO}/FWHM$, where $\nu_{\rm QPO}$ and $FWHM$ represent the centroid frequency and the full width at half maximum of the QPO feature) and significance ($\sigma_{\rm QPO} = norm/err_{\rm neg}$, where $norm$ refers normalisation and $err_{\rm neg}$ denotes the negative error in $norm$) as detailed in \cite{vanderKlis1989b, Casella-etal2005, Belloni-Altamirano2013, Sreehari-etal2020, Majumder-etal2022}. The fractional $rms$ amplitude of each QPO is computed as the square root of the integrated power of the best fitted \texttt{Lorentzian} function. Based on centroid frequencies and characteristic properties, QPOs observed in BH-XRBs are classified into different types as summarized in Table \ref{tab:Table-06}. Furthermore, the total $rms$ amplitude of the wide-band PDS is calculated over the $0.01-500$ Hz frequency range for each observation. 

\begin{table*}
	\caption{Consolidated properties of LFQPOs (type-A, type-B, type-C, type-C* and type-B Cathedral) and HFQPOs, namely QPO frequency ($\nu_{\rm QPO}$), Q-factor ($\nu_{\rm QPO}/{\rm FWHM}$), significance ($\sigma$), percentage rms (rms$_{\rm QPO}\%$), Noise types, percentage rms total (rms$_{\rm total}\%$), time-lag ($t_d$), spectral states and ratio of Comptonised to bolometric fluxes in $1-100$ keV ($f_{\rm th}=F_{\rm Comp}/F_{\rm bol}$), respectively, obatined from \textit{RXTE} and \textit{AstroSat} observations.
    }	
\begin{tabular}{l c c c c c c c}
\hline

Parameter & \multicolumn{5}{c}{LFQPO}& & HFQPO$^{\#}$ \\
\cline{2-6}  
            & Type-C   & Type-B & Type-A & Type-C* & Type-B Cathedral&  & \\ \hline
$\nu_{\rm QPO}$ (Hz)            & 0.1$-$30.0                          & 1$-$7$^{\dagger}$ & 5$-$8   & {\red 5}$-$8 & 5$-$6 & & 65$-$75 \\
Q-factor                        & $\sim 10$                          & 3$-$6           & $\leq 3$ & 3$-$5 & 3$-$6 & & 3$-$5 \\
Significance ($\sigma$)         & 3$-$20                            & $\ge 3$ & 2.5$-$3.0 & $\ge 3$ & $\ge 3$& & {\red 3$-$11} \\
rms$_{\rm QPO}\%$               & 5$-$20                            & 2$-$5 & $\leq 4$ & 4$-$6 & 2$-$5 & & 1$-${\red 6} \\
Noise type         in PDS       & Flat-top noise $+$ Red noise      & Red noise & Red noise & Red noise & Red noise& &  White noise\\
rms$_{\rm total}$\%$^{\boxdot}$ & 10$-$35                           & 4$-$6 & $\leq 5$ & 5$-$12 & 4$-$6  & & 10$-${\red 50}\\
Power-law noise (\%)            & 6$-$17                            & 2$-$4 & $\le 4$  & 5$-$7  & $\le 4$ & & --- \\
Harmonics                       & Sub, $1^{\rm st}$, $2^{\rm nd}$   & $1^{\rm st}$  & None & Sub, $1^{\rm st}$  &  Sub& & $1^{\rm st}$ \\
Energy band (keV)                     & 2$-$30$^{\ddagger}$         & 2$-$25 & $\sim$ 6$-$10 & 2$-$30 & 2$-$25 & & {\red 6}$-$25\\
Time-lag ($i<50^{\circ}$)       & hard                             & hard & soft & --- & --- & & --- \\
Time-lag ($i\ge60^{\circ}$)     & hard/soft                         & hard/soft & soft & soft & hard & & hard/soft \\
Spectral state                  & LHS, HIMS                         & SIMS & SIMS & SIMS& SIMS& & SIMS, HSS \\
$f_{\rm th}$                    & 0.50$-$0.85                       & $\leq 0.50$ & $\leq 0.38$ & $\leq 0.50$ & $\leq 0.45$ & & $0.05-0.25$ \\ \hline

\end{tabular}
\label{tab:Table-06}
\begin{tablenotes}
    \item $^{\#}$For GRS 1915$+$105 .
    \item $^\dagger$Type-B QPOs are also observed at $\sim16$ Hz in 4U 1543$-$47 and XTE J1650$-$500 \cite[]{Choudhury-etal2025}. 
    \item $^{\boxdot}$rms$_{\rm total}\%$ is calculated in $0.01-62.50$ Hz range with Nyquist frequency of $62.50$ Hz derived from an $8$ ms lightcurve. For HFQPO, the frequency range is considered as $0.1-500.0$ Hz.
    \item $^{\ddagger}$Type-C QPO is observed above $30$ keV in few sources \cite[see][for details]{Ma-etal2021,Zhu-etal2023,Nandi-etal2024}.
    \item LFQPOs references: \citealt[and references therein]{Wijnands-etal1999, Casella-etal2004, Casella-etal2005, Belloni-etal2005, Motta-etal2009, Motta-etal2011, Nandi-etal2012,Choudhury-etal2025}. 
    \item HFQPOs references: \citealt[and references therein]{Belloni-Altamirano2013, Majumder-etal2022,Majumder-etal2025a,Majumder-etal2025b}.

\end{tablenotes}

\end{table*}

To further investigate the temporal properties of the sources, we compute the time-lags between different energy bands. The time-lag is obtained from the argument of the complex cross-spectrum \cite[]{Vaughan-Nowak1997}, which is defined as $C(j) = X^{*}_{1}(j) X_{2}(j)$, where $X_{1}(j)$ and $X_{2}(j)$ are the Fourier coefficients at frequency $\nu(j)$ corresponding to two simultaneous lightcurves $x_{1}(k)$ and $x_{2}(k)$ at time $t_{k}$ in two different energy bands \cite[]{van_der_klis1987}. The Fourier transform is defined as,
\begin{equation}
    X(j) = \sum_{k=0}^{m-1} x(k)e^{2\pi i \nu_{j} t_{k}}.
\end{equation}
Here, $x(k)$ is the $k^{\rm th}$ count in the lightcurve which consists of $m$ evenly spaced time bin of length $\Delta t$ in seconds. $X(j)$ is the complex Fourier transform of $x(k)$ at frequency $\nu_{j}=j/(m\Delta t)$, where $j \in [-m/2, m/2]$ \cite[see][for details]{vanderKlis1989b}. $X^{*}(j)$ is the complex conjugate of $X(j)$. The phase lag is defined as $\phi(j)=arg[C(j)]$ and the corresponding time-lag is defined as, $\delta t(j)=\phi(j)/2\pi\nu_{j}$ \cite[see][for details]{Vaughan-Nowak1997}.

We compute the cross-spectrum between the $6-20$ keV and $3-6$ keV energy bands using the \texttt{LAXPCsoftware}, based on data from LAXPC10 and LAXPC20. The frequency resolution and Nyquist frequency are considered to be $0.1$ Hz and $500$ Hz, respectively. Following \cite{Reig-etal2000}, the time-lag is averaged over the FWHM of the QPO feature. {\red It is important to note that, the current version of \texttt{LAXPCsoftware} does not incorporate corrections for detector dead-time effects in the time-lag calculations. However, such effects can introduce artificial correlations between different energy channels, which results in a negative offset in the real part of the cross-spectrum, particularly at high frequencies, where Poisson noise dominates over intrinsic source variability (see \citealt[]{vanderKlis-1997} for details). To mitigate the influence of dead-time on the measured time-lags, we follow the correction procedure implemented in the GHATS\footnote{\url{http://astrosat-ssc.iucaa.in/uploads/ghats_home.html}} software package. We first compute the cross-spectra between the light curves extracted in the $3-6$ keV and $6-20$ keV energy bands. The dead-time induced offset is then estimated by averaging the real part of the cross-spectrum over a high-frequency interval ($>100$ Hz), where Poisson noise dominates. This estimated offset is subsequently subtracted from the real part of the cross-spectrum at all Fourier frequencies, while the imaginary part remains unchanged. Subsequently, the corrected cross-spectra are used to calculate the phase-lag and corresponding time-lag spectra, from which the time-lag at QPO frequencies are calculated for the respective sources. This correction is applied only to HFQPOs, as dead-time effects on LFQPO lags are insignificant.}

\subsection{Spectral Analysis}\label{sec6:6.2_spec}

We investigate the wide-band ($0.7-50.0$ keV) spectral properties of the sources observed with {\it AstroSat} by modelling the SXT and LAXPC20 spectra using \texttt{XSPECv12.15.0} \citep{Arnaud-1996}. Specifically, the SXT data are analyzed over the $0.7-7$ keV band, while the LAXPC20 spectra are considered in the $3-50$ keV range. In general, BH–XRB spectra exhibit two major components such as thermal emission originating from the accretion disc and non-thermal emission arising from the Comptonizing corona. The thermal component is modelled using the multi-temperature accretion disc model \texttt{diskbb} \citep{Makishima-etal1986}, whereas a simple \texttt{power-law} component provides a phenomenological description of the non-thermal emission at higher energies. Several physical models have been developed over the years to describe the Comptonized emission observed at high energies in BH-XRB spectra including \texttt{compbb} \citep{Nishimura-etal1986}, \texttt{compTT} \citep{Titarchuk1994}, \texttt{compPS} \citep{Poutanen-etal1996}, \texttt{nthcomp} \citep{Zdziarski-etal1996}, and \texttt{thcomp} \citep{Zdziarski-etal2020a}. These models differ primarily in their assumptions regarding the geometry and physical properties of the Comptonizing medium. In addition, strong reflection features are occasionally present in BH-XRB spectra and are typically modelled using variants of the \texttt{relxill} family of models \cite[see][and references therein]{Garcia-etal2014}. Meanwhile, variety of disc, corona, and reflection models were employed to characterize the spectra of BH–XRBs observed with {\it AstroSat} \cite[]{Sreehari-etal2020,Bogensberger-etal2020,Garg-etal2022,Prabhakar-etal2023,Bhuvana-etal2023,Dhaka-etal2023,Belloni-etal2024,Majumder-etal2025a}. However, in this work, we adopt the recently developed convolution model (\texttt{thcomp}) to explain the Comptonized emission component for the purpose of uniformity in analyzing the spectral data from all sources, unless stated otherwise.

To model the energy spectra for each observation, we employ the composite model combination \texttt{constant$\times$tbabs$\times$(thcomp$\otimes$diskbb)}. In this model prescription, \texttt{tbabs} \citep{Wilms-etal2000} accounts for photoelectric absorption from both interstellar medium and the source, while \texttt{constant} compensates for cross-calibration differences between the SXT and LAXPC20 instruments. This model provides a consistent and physically motivated description of the wide-band spectra for most BH–XRBs observed with {\it AstroSat}. However, for few sources, additional components are required to capture source specific features. In particular, for Cyg X–3, we adopt the model combination \texttt{constant$\times$tbabs$\times$pcfabs$\times$(thcomp$\otimes$diskbb)}, where the inclusion of \texttt{pcfabs} accounts for partial covering absorption caused by the dense stellar wind from the companion star. The specific cases requiring such modifications are discussed individually in the respective sections.

Apart from this, we include the \texttt{smedge} \cite[]{Ross-etal1996} absorption component in the spectral modelling to fit the absorption feature near $7-9$ keV in the LAXPC20 spectra depending on its strength varying across different observations of various sources \cite[]{Majumder-etal2022, Harikesh-etal2025}. In addition, a \texttt{Gaussian} component is used to model the strong Fe K$\alpha$ line emission present in the spectra around $\sim 6.4$ keV wherever applicable \cite[]{Tanaka-etal1995b}. Further, we use gain correction with a fixed slope of $1$ using the gain fit command to flatten the residuals at low energies $\sim 1.8$ keV and $\sim 2.2$ keV \cite[]{Singh-etal2017, Sreehari-etal2019b}. The gain command shifts the energies on which the response matrix is defined and shifts the effective area curve to match. In addition to \textit{gain fit}, the \texttt{edge} component is also used to model these instrumental features, depending on their strength. Similarly, the Xenon edge feature, which generally appears around $\sim 29-34$ keV in the LAXPC spectra \cite[]{Antia-etal2017}, is modelled using either an \texttt{edge} \cite[]{Sreehari-etal2019b, Athulya-etal2022, Majumder-etal2022} or a \texttt{Gaussian} component \cite[]{Sharma-etal2020, Beri-etal2021} wherever applicable. A systematic error of $2\%$ is applied in the spectral modelling \cite[]{Antia-etal2017, Sreehari-etal2019b}.

We note that the hydrogen column density for GRS 1915$+$105 is fixed at $n_{\rm H} = 6 \times 10^{22}$ atoms cm$^{-2}$ as it remains unconstrained during the spectral modelling. However, for the remaining sources, the column density is allowed to vary freely during the spectral modelling, unless stated otherwise. Furthermore, we find that the aforementioned model combination fails to reproduce the higher energy part of the \texttt{Insight-HXMT} spectra of Swift J1727.8$-$1613. Instead, the model combination \texttt{constant$\times$tbabs$\times$(thcomp$\otimes$diskbb + cutoffpl)} provides a satisfactory description of the wide-band spectra of the source across different epochs. Similarly, to model the Comptonized emission, the \texttt{nthcomp} model is employed for LMC X$-$1, LMC X$-$3, GRS 1758$-$258, 4U 1543$-$47, 1E 1740.7$-$2942, and MAXI J1820$+$070, where \texttt{thcomp} model alone is found to be inadequate in fitting the spectra. It is worth noting that, in most observations, statistically significant spectral data are limited to $20-30$ keV range. Beyond this energy range, background counts dominate over the source counts owing to the decreasing effective area of LAXPC at higher energies. In addition, for these sources, few hard state observations exhibit Fe-line emission around $\sim 6.4$ keV, which is modelled using a \texttt{Gaussian} component. For MAXI J0637$-$430, the \texttt{laor} model \cite[]{Laor1991} is used to account for the relativistically broadened Fe-line emission.

It is worth mentioning that the relativistic disc model \texttt{kerrbb} \cite[]{Li-etal2005} is employed to fit the soft state spectra of LMC X$-$1 and LMC X$-$3 to constrain the physical parameters of the system. However, for GRS 1758$-$258 and 1E 1740.7$-$2942, the relativistic reflection model \texttt{relxillCp} \cite[]{Garcia-etal2010, Dauser-etal2010, Dauser-etal2016} is required to obtain an acceptable fit. Strong reflection features are also detected in 4U 1543$-$47 and MAXI J1820$+$070, modelled using different variants of \texttt{relxill}. These models, namely \texttt{kerrbb}, \texttt{relxill}, and \texttt{relxillCp} require several parameters including source distance ($D$), black hole mass ($M_{\rm BH}$), and disc inclination angle ($i$) to constrain the spin. However, as the inclination angles of GRS 1758$-$258 and 1E 1740.7$-$2942 are unknown, their spins cannot be reliably estimated. For MAXI J0637$-$430, the iron abundance ($A_{\rm Fe}$) in the reflection models is fixed at the solar value during spectral fitting.

Next, we compute the flux corresponding to each spectral component, namely the disc flux ($F_{\rm disc}$), Comptonized flux ($F_{\rm Comp}$), and the total bolometric flux ($F_{\rm bol}$) in the $1-100$ keV energy range using the \texttt{cflux} model \cite[\cite{Arnaud-1996}, see also][and references therein]{Majumder-etal2022}. In this procedure, the \texttt{cflux} component is applied after the absorption model \texttt{tbabs}. We further compute the ratio of disc to Comptonized fluxes ($F_{\rm ratio} = F_{\rm disc}/F_{\rm Comp}$) to investigate the relative contribution of the emission components. We estimate the percentage bolometric luminosity as $L_{\rm bol} = (4\pi D^{2} F_{\rm bol}/{\rm L}_{\rm Edd}) \times 100$, where $D$ is the source distance and ${\rm L}_{\rm Edd}$ refers to the Eddington luminosity\footnote{$\rm L_{Edd} = 1.26\times10^{38} \left(\frac{M_{\rm BH}}{{\rm M}_\odot}\right)$ erg s$^{-1}$ (\citealp[see][]{Frank-etal2002} for details).}. Thereafter, following \cite{Walpole-etal2012}, we calculate the Pearson correlation coefficient ($P_{\rm CC}$) to examine possible correlations among the spectro-temporal model fitted parameters.

\section{Results}\label{sec7:results}

In this section, we present the results of the spectro-temporal analyses of $34$ Galactic BH–XRBs observed with {\it AstroSat}. Results are derived from systematic analyses and modelling of wide-band power and energy spectra to investigate the evolution of temporal and spectral properties among the various spectral states governed by the accretion dynamics.

\subsection{Persistent-cum-Variable Sources}\label{sec7:7.1_pcv}

\subsubsection{GRS 1915$+$105}\label{sec7:7.1.1}

\subsubsection*{(a) First `Light' from a Black Hole with \textit{AstroSat}}

GRS 1915$+$105 was the first black hole source observed by {\it AstroSat} with the SSM instrument. Owing to the large FoV of SSM cameras, multiple X-ray sources were present in the field in addition to the targeted source during these observations. While reconstructing the sky image, contributions from other bright sources such as Ser X$-1$, MAXI J1910$-$057, Aql X$-$1, Cyg X$-$1, Cyg X$-$2, and Cyg X$-$3 were taken into account. The cleaned GTI event lists from these observations were subsequently used for analysis. Although all sources contributed to the total event list, the emission from GRS 1915$+$105 dominated, as it was the brightest source in the field during the SSM1 and SSM2 observations on $14$ October $2015$. Given the coarse spectral resolution of the detectors, the analysis was restricted to the $2-12$ keV energy range, and HR was defined as the ratio of the $6-12$ keV to $2-6$ keV fluxes.

\begin{figure}
    \begin{center}
        \includegraphics[width=\columnwidth]{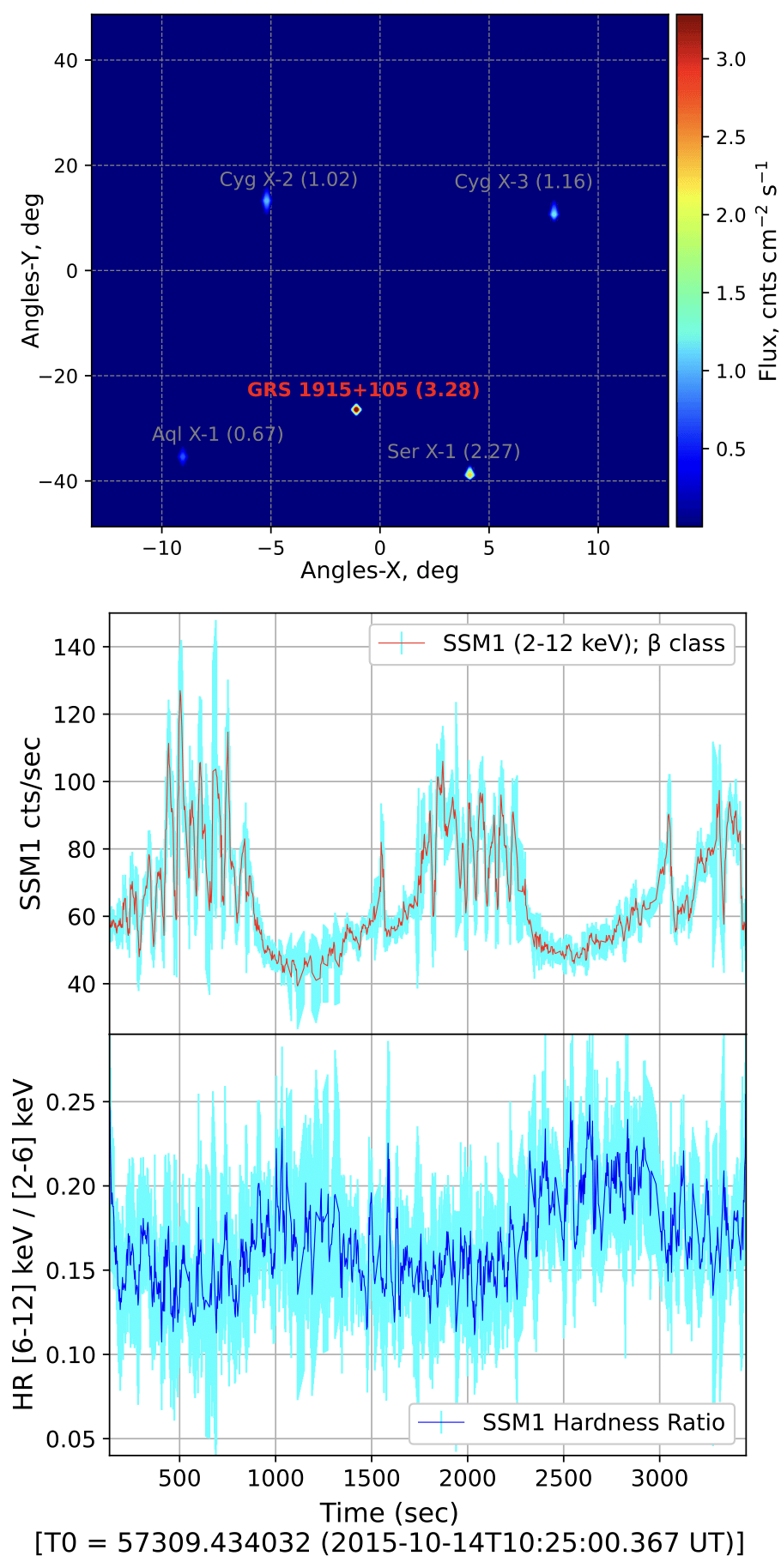}
    \caption{SSM1 onboard \textit{AstroSat} observation of GRS 1915$+$105 exhibiting $\beta$ class. Top panel shows the reconstructed sky image in camera coordinates, where GRS 1915$+$105 is detected at ($-0.75^\circ, -24.90^\circ$) on $14$ October $2015$ along with other sources. Detected fluxes (in brackets) indicate that GRS 1915$+$105 was the brightest source throughout these observations. Middle panel displays $2-12$ keV lightcurve (Orbit $00242$) for $\beta$ class variability. Bottom panel represents the corresponding HR varying consistently with $\beta$ class pattern. See text for details.
    }
    \label{fig:fig8}
\end{center}
\end{figure}

During the SSM1/SSM2 observations, GRS 1915$+$105 exhibited $\beta$ class variability \cite[see][]{Ramadevi-etal2015}, characterized by HR variations in the range $\sim 0.10-0.25$ which closely follow the source flux variations. Fig. \ref{fig:fig8} shows the lightcurve variation corresponding to MJD $57309.434032$ for a duration of $\sim 3.3$ ks (Orbit $00242$) on $14$ October $2015$. The reconstructed sky image (upper panel of Fig. \ref{fig:fig8}) revealed the detected sources in order of increasing flux (in counts~cm$^{-2}$~s$^{-1}$) as Aql X$-$1 ($0.67$), Cyg X$-$2 ($1.02$), Cyg X$-$3 ($1.16$), Ser X$-$1 ($2.27$), and GRS 1915$+$105 ($3.28$). GRS 1915$+$105, detected at camera coordinates (Angle-X, Angle-Y) = ($-0.75^\circ$, $-24.90^\circ$), remained the brightest in the field throughout these observations. A follow-up observation using SSM3 also detected GRS 1915$+$105 on the next day only for a short duration, during which the $\beta$ class variability was again evident.

\subsubsection*{(b) Temporal Variability Properties}

\begin{figure*}
    \begin{center}
    \includegraphics[width=0.9\textwidth]{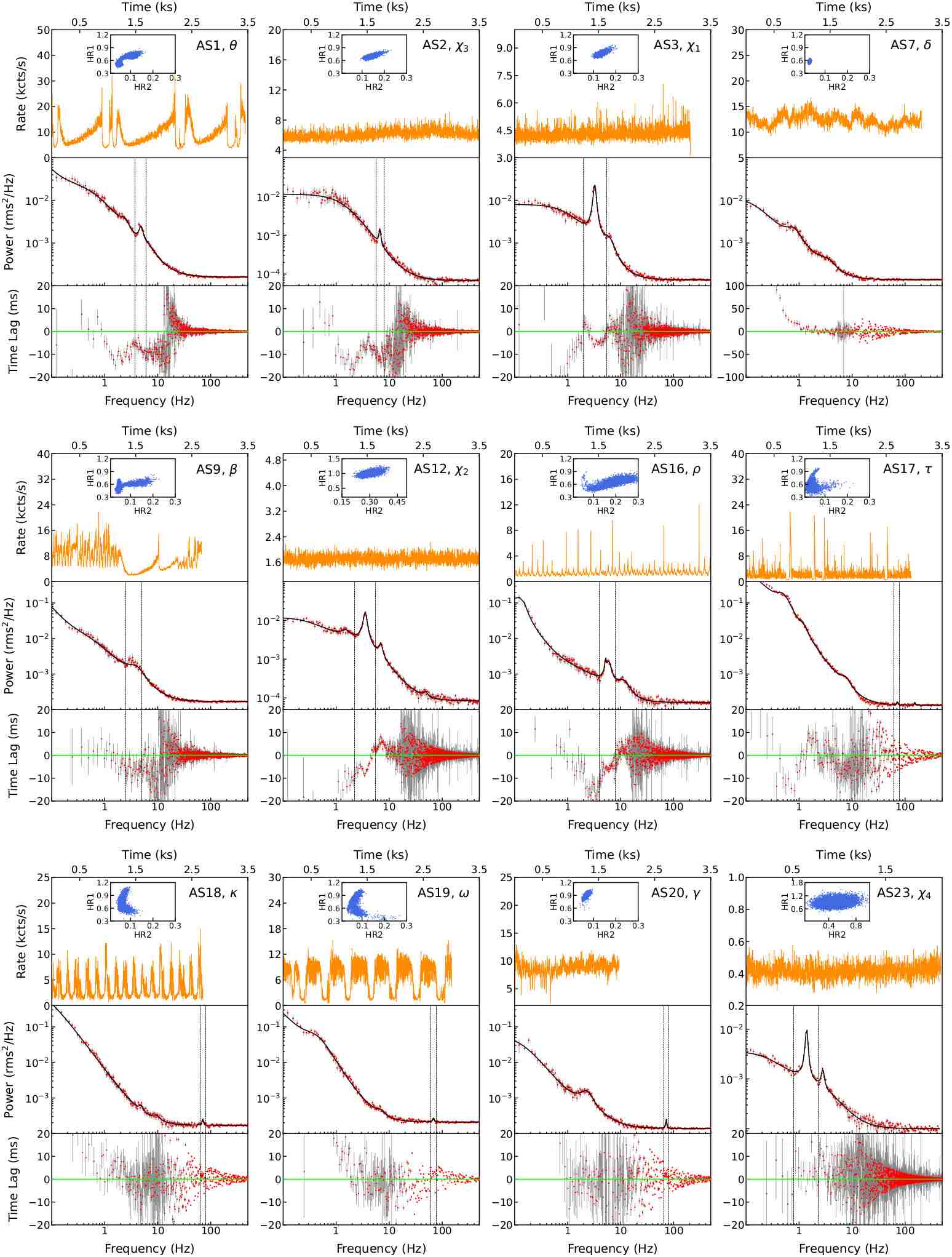}
    \end{center}
    \caption{Temporal variabilities of GRS 1915$+$105 across $12$ variability classes observed with \textit{AstroSat} during $2016-2021$. Each panel shows the lightcurve with CCD, corresponding PDS, and lag–frequency spectrum for the respective variability class. See text for details.
    }
    \label{fig:fig9}
\end{figure*}

\textit{AstroSat} observations of GRS 1915$+$105 over a span of ten years reveal nine distinct variability classes ($\beta$, $\theta$, $\chi$, $\omega$, $\delta$, $\rho$, $\tau$, $\kappa$, $\gamma$) along with several intra-class variants ($\chi^{}_1$, $\chi^{}_2$, $\chi_2^{\prime}$, $\chi^{}_3$, $\chi^{}_4$) \cite[]{Yadav-etal2016, Rawat-etal2019, Sreehari-etal2020, Athulya-etal2022, Majumder-etal2022,Majumder-etal2026b}. The source exhibited sequential class transition during the \textit{AstroSat} campaign as $\beta \rightarrow \theta \rightarrow \chi^{}_3 \rightarrow \chi^{}_1 \rightarrow \chi^{}_3 \rightarrow \theta \rightarrow \omega \rightarrow \delta \rightarrow \beta \rightarrow \theta \rightarrow \delta \rightarrow \chi^{}_2 \rightarrow \chi_2^{\prime} \rightarrow \rho \rightarrow \chi_2 \rightarrow \chi \rightarrow \rho \rightarrow \tau \rightarrow \kappa \rightarrow \omega \rightarrow \gamma \rightarrow \delta \rightarrow \chi^{}_1 \rightarrow \chi^{}_4 \rightarrow \rho \rightarrow \delta$. Among these variability classes, $\beta$, $\theta$, $\omega$, $\delta$, $\rho$, $\tau$, $\kappa$, and $\gamma$ correspond to the typical `softer’ variability classes, whereas $\chi$ represents the characteristic `harder’ variability state. Note that, from $2019$ onward, GRS 1915$+$105 entered an obscured phase with an average flux level of $\sim 0.1$ Crab followed by multiple rebrightening episodes in X-ray intensity \cite[see Fig. \ref{fig:fig6} (upper panel);][]{Athulya-etal2022}. To investigate the variability properties of GRS 1915$+$105, in Fig. \ref{fig:fig9}, we depict the lightcurve in $3-60$ keV energy range along with the corresponding CCDs. The CCDs are generated from energy resolved lightcurves across different energy bands \cite[see][for details]{Sreehari-etal2020, Majumder-etal2022}. 

\begin{table*}
\caption{Temporal properties of GRS 1915$+$105 observed with {\it AstroSat} in different variability classes. The notations and symbols have their usual meanings. See \S \ref{sec6:6.1_timing} and \S \ref{sec7:7.1_pcv} for details.}
\renewcommand{\arraystretch}{1.4}
\resizebox{\textwidth}{!}{
\begin{tabular}{c c c c c c c c c c c}\hline

	Epoch & $\nu_{\rm QPO}$ & FWHM & Significance & Q-factor & rms$_{\rm QPO}\%$ & time-lag  & rms$_{\rm total}\%$ & $\chi^2\big/$dof & QPO-Type & Class \\
			
	& (Hz) & (Hz) & ($\sigma$) &  & & (ms)  & & & (LF/HF) & \\ \hline
			
	AS1 & $4.85_{-0.05}^{+0.05}$ & $0.98$ & 6.2 & 4.9 & $6.09\pm2.61$ & $-7.9\pm0.5$  & $35.16\pm12.32$ & 163/227 & LF  & $\theta$ \\

    AS2 &  6.69$_{-0.08}^{+0.08}$  & 0.59 & 4.5 & 11.3 & $3.57\pm0.87$ & $-6.5\pm1.2$  & $24.58\pm 2.29$ & 176/237  & LF & $\chi_3$\\
            
	AS3 &  $3.25_{-0.01}^{+0.01}$ & $0.46$ & 31.5 & 7.1 & $13.43\pm1.88$ & $-5.3\pm0.4$  & $31.87\pm2.43$ & 202/234 & LF & $\chi_1$ \\

    AS4 & 4.81$_{-0.02}^{+0.02}$  & 0.65 & 13.9 & 7.4 & 7.46 $\pm$ 0.27 & $-9.5\pm0.3$  & $26.11\pm 0.04$ & 422/345  & LF & $\chi_3$ \\
			
	AS5 &  $6.15_{-0.11}^{+0.11}$  & $2.08$ & 4.8  & 2.9 & $6.71\pm0.93$ & $-12.3\pm0.7$  & $34.78\pm3.19$ & 153/233 & LF & $\theta$ \\

	AS6 &  $71.56_{-0.23}^{+0.25}$ & $1.68$ & 3.6 & 42.6 & $4.48\pm0.72$ & $\red 1.4\pm0.1$ & $35.92\pm5.79$ & 159/229& HF & $\omega$\\
			
	AS7 &  $-$ & $-$ & $-$ & $-$ & $-$ & $-$ & $27.34\pm1.36$ & 171/232 & $-$ & $\delta$ \\
			
	AS8 &  $69.28_{-0.27}^{+0.13}$ & $1.75$ & 7.3  & 39.6 & $5.54\pm0.51$ & $\red 2.9\pm0.2$  & $27.52\pm 1.63$  & 137/230 & HF & $\delta$ \\
			
    AS9 &  $3.33_{-0.05}^{+0.06}$ & $0.61$ & $3.9$ & $5.5$ & $2.80\pm0.54$ & $-4.4\pm0.7$ & $16.53\pm0.73$ & 131/232 & LF & $\beta$ \\

	AS10 &  $5.98_{-0.12}^{+0.11}$  & $2.18$ & $6.5$ & $2.8$ & $7.19\pm0.76$ & $-5.4\pm0.2$ & $29.46\pm3.82$ & 179/235 & LF & $\theta$ \\

    AS11 & $-$ & $-$ & $-$ & $-$ & $-$ & $-$ & $28.22\pm2.11$ & 158/234 & $-$ & $\delta$ \\

    AS12 &  3.64$_{-0.01}^{+0.01}$ & 0.87  & 25.2  & 4.2 & $11.26\pm0.22$ & $-6.8\pm0.3$  & $28.72\pm0.06$  & 227/178 & LF & $\chi_2$ \\

    AS13 &  $4.30_{-0.02}^{+0.02}$ & $1.30$ & 22.0  & 3.3 & $10.49\pm0.21$ & $-7.4\pm0.3$  &  $28.99\pm0.40$ & 217/173 & LF & $\chi_2'$ \\

    AS14 & $5.32_{-0.02}^{+0.02}$ & $1.50$ & 36.5 & 3.6 & $8.59\pm0.15$ & $-7.4\pm0.3$ & $26.44\pm0.04$ & 221/176 & LF & $\rho$ \\
			
	AS15 &  $4.58_{-0.04}^{+0.03}$ & $1.19$  & 26.0 & 3.9 & $11.51\pm2.36$ & $-6.5\pm0.3$  & $33.26\pm3.22$ & 123/226 & LF & $\chi_2$ \\

	AS16 &  $5.94_{-0.08}^{+0.10}$ & $1.47$ & 6.4  & 4.1 & $7.13\pm1.59$ & $-6.1\pm0.3$  & $35.79\pm17.43$  & 143/226 & LF & $\rho$ \\

    AS17$^\boxtimes$ &  $71.15_{-1.05}^{+1.01}$ & $5.33$ & $3.0$ & $13.4$ & $4.69\pm0.78$ & $\red -1.0\pm0.3$ & $57.78\pm7.06$ & 220/224 & HF & $\tau$ \\

    &  $^{\dagger}$$152.30_{-2.06}^{+1.55}$ & $7.23$ & $4.7$ & $21.1$ & $5.75\pm0.81$ & $\red 0.4\pm0.1$ & $57.78\pm7.06$ & 220/224 & HF & $\tau$ \\
			
	AS18 & $70.36_{-0.59}^{+0.61}$ & $5.34$ & 6.0 & 13.2 & $6.29\pm0.93$ & $\red 1.5\pm0.3$ & $44.36\pm8.28$ & 118/231 & HF & $\kappa$ \\
			
	AS19 & $68.18_{-0.63}^{+0.69}$ & $5.09$ & 5.2 & 13.4 & $5.21\pm0.65$ & $\red 2.4\pm0.1$  & $44.02\pm8.06$& 172/231  & HF  & $\omega$ \\
            
	AS20 & $72.31_{-0.21}^{+0.24}$ & $3.67$ & 10.3 & 19.7 & $5.14\pm0.68$ & $\red 1.7\pm0.1$ & $29.03\pm2.94$ & 140/232 & HF & $\gamma$  \\
			
	AS21 &  $-$ & $-$ & $-$ & $-$ & $-$ & $-$ & $27.38\pm2.41$ &156/232  & $-$ & $\delta$ \\
            
	AS22 &  $3.25_{-0.02}^{+0.02}$ & $0.79$ & 11.4  & 4.1 & $15.04\pm2.78$ & $-7.3\pm0.2$ & $34.67\pm3.74$ & 132/229 & LF & $\chi_1$ \\

    AS23 &   $1.38_{-0.01}^{+0.01}$ &  $0.15$ &  $21.9$ &  $9.2$ &  $4.68\pm0.11$ & $-2.9\pm1.8$ & $13.56\pm0.04$ & 256/240 &  LF & $\chi_4$ \\

    AS24$^\otimes$ &   $0.02\ddagger$ &  $0.002\ddagger$ &  $5.5$ &  $10.0$ &  $6.27\pm0.36$ & $-$ & $8.65\pm0.62$ & 153/136 &  LF & $\rho$ \\
			
	AS25$^\otimes$ & $0.17_{-0.01}^{+0.07}$ & $0.06$ & $5.5$ & $2.8$ & $9.46\pm0.36$ & $-$ & $13.37\pm0.51$  & 133/116 & LF & $\delta$ \\

	\hline
\end{tabular}
}
\begin{tablenotes}
    \item $^\dagger$Possible detection of HFQPO harmonic in GRS 1915$+$105 in new variability class \cite[see also][]{Majumder-etal2026}.
    \item $^\boxtimes$HFQPO exhibits negligible time-lag yielding error bar larger than the measured value. The error scales as the inverse square root of the number of segments set by the total exposure time divided by the segment length.
    \item $^\ddagger$Errors are insignificant. 
    \item $^\otimes$Time-lag is not reliably constrained because of poor statistics. This is primarily due to the limited number of segments available for averaging, which results from the use of a large segment length and hence fewer independent measurements.
\end{tablenotes}

\label{tab:Table-07}
\end{table*}

The `softer’ classes of GRS 1915$+$105 exhibit distinct variability signatures. The $\kappa$ and $\omega$ classes are generally characterized by alternating low-count (`dip’) and high-count (`non-dip’) segments lasting a few tens of seconds with their CCDs tracing a $C$-shaped pattern \cite[see also][]{Majumder-etal2022}. A similar, though more irregular, pattern appears in the $\tau$ class, where aperiodic dips and spike like structures dominate the lightcurve \cite[]{Majumder-etal2025b}. The $\tau$ class displays unique variability, distinct from all previously known classes of the source, and has therefore been identified as a new class. The $\rho$ class shows heartbeat like variability with regular spike features and a CCD extending to higher HR2 values. The $\gamma$ and $\delta$ classes remain comparatively less variable, exhibiting red-noise like fluctuations; however, short, small amplitude dips in the $\gamma$ class distinguish it from the $\delta$ class \cite[]{Majumder-etal2022, Majumder-etal2024, Majumder-etal2025a}. Their CCDs are confined to lower HR2 and higher HR1 regions with only marginal differences. During the $\theta$ class, the lightcurve displays an `M’-shaped pattern lasting a few hundred seconds with the CCD forming distinct regions at low HR1 and high HR2 values. The $\beta$ class shows highly irregular variability, consisting of low-count dips followed by prolonged low-flux intervals, and a CCD with a pronounced tail toward higher HR2. In contrast, the `harder’ $\chi$ class exhibits relatively steady count rates and a CCD dominated by high energy photons. Further analysis allows intra-class divisions into $\chi^{}_1$, $\chi^{}_2$, $\chi^{\prime}_2$, $\chi^{}_3$, and $\chi^{}_4$ \cite[]{Majumder-etal2026b}, based on HR1 and HR2 values \cite[]{Belloni-etal2000}. Among all the $\chi$ variants, the hardness of $\chi_1$ and $\chi_3$ are relatively lower, while hardness of $\chi_4$ is found to be the highest.

We further investigate the power spectral properties of GRS 1915$+$105 for each variability class observed with {\it AstroSat}, following the analysis procedures described in \S\ref{sec6:6.1_timing}. The best fit PDS corresponding to different classes are shown in Fig. \ref{fig:fig9}. We detect LFQPOs in the frequency range of $1.38$$-7.55$ Hz, with rms amplitudes varying between $4.68-15.04\%$ for $\theta$, $\chi$, and $\rho$ classes (see Table \ref{tab:Table-07}). Similar LFQPO features of GRS 1915$+$105 have been reported using \textit{AstroSat} observations in the previous studies \cite[]{Yadav-etal2016, Belloni-etal2024,Athulya-etal2022,Majumder-etal2026b}. Moreover, \cite{Majumder-etal2026b} identified distinct wide-band PDS features among different $\chi$ variants. For example, $\chi^{}_3$ exhibits higher QPO frequency without harmonics, whereas $\chi^{}_1$ and $\chi^{}_2$ predominantly show QPOs with harmonic and sub-harmonic features, though occasionally only one component is detected. In contrast, $\chi^{}_4$ consistently displays a sharp low-frequency QPO accompanied by a strong harmonic.

{\red Time-lag analysis of all LFQPO observations obtained with {\it AstroSat} reveals the presence of soft-lags (see Fig. \ref{fig:fig9} and Table \ref{tab:Table-07}) ranging $2.9-12.3$ ms \cite[see][for details]{Majumder-etal2026b}. The minimum soft-lag is measured as $2.9$ ms for $\chi^{}_4$ class, while the maximum ($6.5-12.3$ ms) is observed for $\chi^{}_3$ and $\theta$ classes. For $\chi^{}_4$ observation (AS23), a soft-lag is detected even at an LFQPO of frequency $1.4$ Hz, in contrast to the {\it RXTE} results \cite[]{Zhang-etal2020b}, where hard-lags were reported for LFQPOs below $2.2$ Hz. Moreover, \cite{Majumder-etal2026b} observed the rms$_{\rm QPO}$ to increase with QPO frequency upto $\sim 3.4$ Hz and then decrease further. A similar trend is also observed for \textit{RXTE} observations with maximum rms$_{\rm QPO}$ at $\sim 2.2$ Hz \cite[]{Zhang-etal2020b}. In addition, \citet{Majumder-etal2026b} investigated the energy dependence of both rms$_{\rm QPO}$ and time-lag, finding that both quantities increase with photon energy.}

\subsubsection*{(c) Deeper view of HFQPOs in GRS 1915$+$105}

\begin{figure}[h]
    \begin{center}
    \includegraphics[width=\columnwidth]{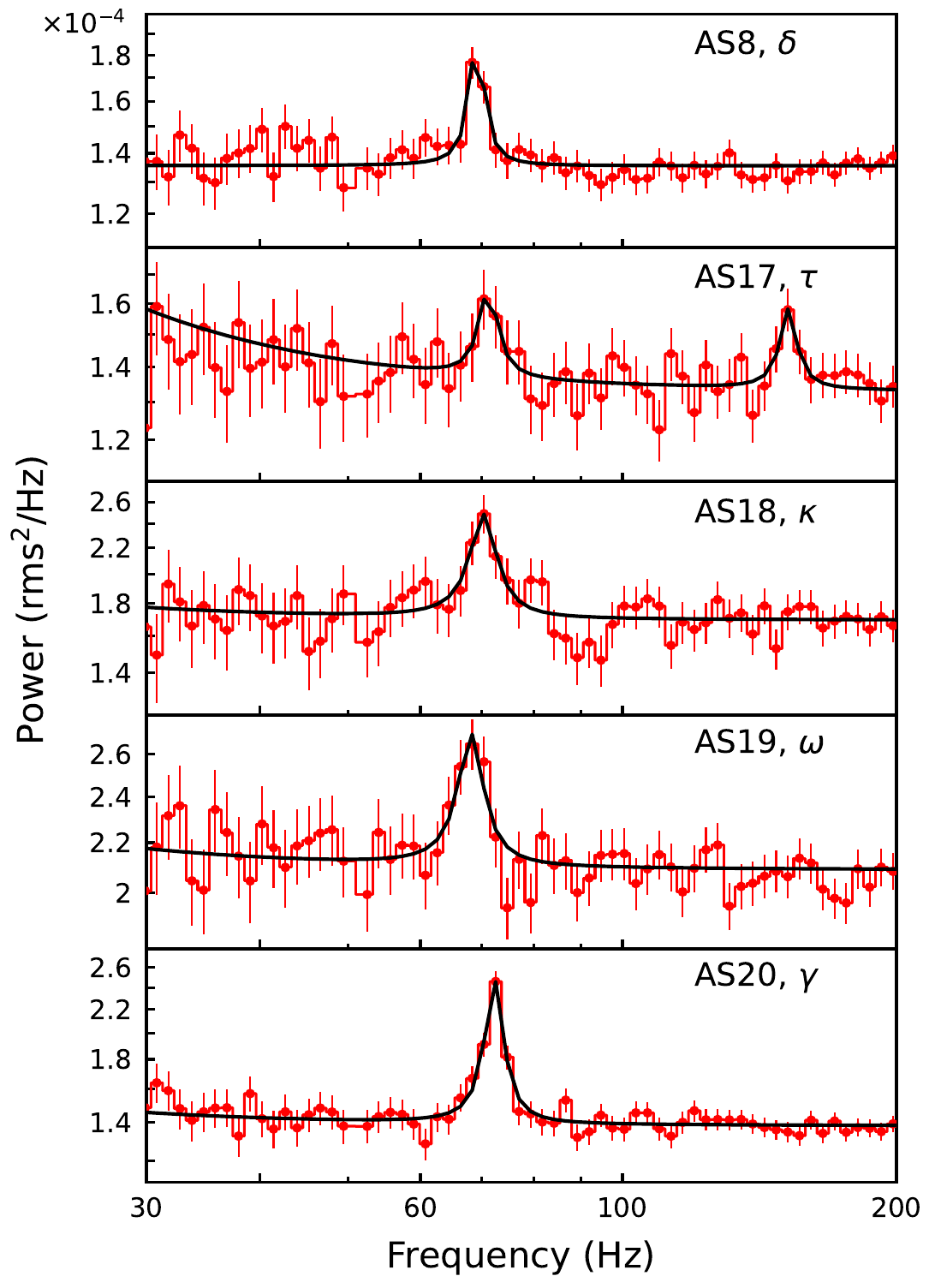}
    \end{center}
    \caption{HFQPOs of GRS 1915$+$105 observed with AstroSat in five variability classes ($\delta$, $\tau$, $\kappa$, $\omega$, $\gamma$) in the $3-60$ keV energy range. For each class, the best fit PDS is shown over $30-200$ Hz frequency band for illustration. See text for details.
    }
    \label{fig:fig10}
\end{figure}

\begin{figure}[h]
    \begin{center}
    \includegraphics[width=\columnwidth]{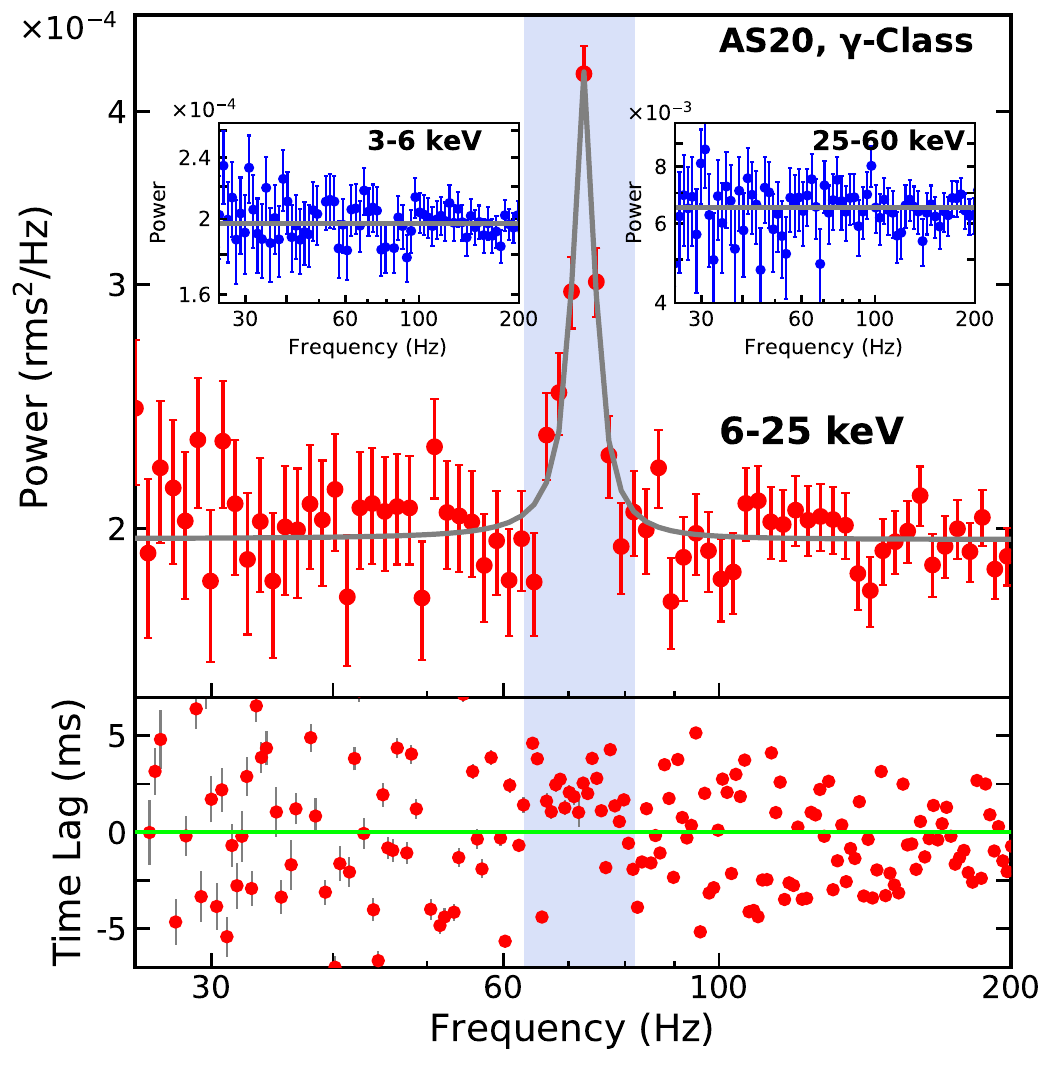}
    \end{center}
    \caption{{\it Top:} Energy dependent PDS of of GRS 1915$+$105 in the $\gamma$ variability class (epoch AS20) observed with \textit{AstroSat}. 
    {\it Bottom:} Variation of time-lag in the high frequency domain of the PDS. See text for details.
    }
    \label{fig:fig11}
\end{figure}

In this section, we focus on the high frequency domain of the PDS where HFQPOs at $\sim 70$ Hz is detected in five `softer’ variability classes, namely $\delta$, $\gamma$, $\kappa$, $\omega$, and $\tau$. We also detect an additional peak at $\sim 152$ Hz in the $\tau$ class observation, which, together with the stable $\sim 70$ Hz feature, represents the first possible evidence of a twin peak HFQPO in GRS 1915$+$105. Interestingly, \cite{Belloni-etal2006} also reported a broad and marginally significant ($\sigma_{\rm QPO}\lesssim 3$) peak at $\sim 170$ Hz in $\theta$ variability class of \textit{RXTE} observations. The best fitted and derived HFQPO parameters are listed in Table \ref{tab:Table-07}. In Fig. \ref{fig:fig10}, we present the best fitted PDS representing the HFQPO feature detected in these five variability classes. In particular, \cite{Sreehari-etal2020,Majumder-etal2022} demonstrated that HFQPOs are strongly energy dependent, appearing only in $6-25$ keV range (see Fig.~\ref{fig:fig11}), where thermal Comptonization dominates the emission. Furthermore, \citet{Majumder-etal2022,Harikesh-etal2025} showed that HFQPOs preferentially occur during high-count (`non-dip’) intervals of the lightcurve and disappears during low-count (`dip’) phases. {\red
Recently, \citet{Majumder-etal2024} reported soft-lags at the HFQPO frequency of $\sim 70$ Hz. However, in the present study we find hard time-lags in the range of $\sim 1.4-2.9$ ms across all four variability classes ($\delta$, $\omega$, $\kappa$, and $\gamma$) except $\tau$ class, where soft-lag ($-1.0$ ms) is observed. The difference between the two results may arise from the treatment of instrumental dead-time effects. The analysis of \citet{Majumder-etal2024} was performed using \texttt{LAXPCsoftware}, which does not include a correction for detector dead-time in time-lag calculations, whereas our analysis incorporates a dead-time correction in the cross-spectrum prior to deriving the phase- and time-lag spectra (see \S\ref{sec6:6.1_timing} for details). 
}

\subsubsection*{(d) Spectral Variability Properties}

\begin{figure}
    \begin{center}
    \includegraphics[width=\columnwidth]{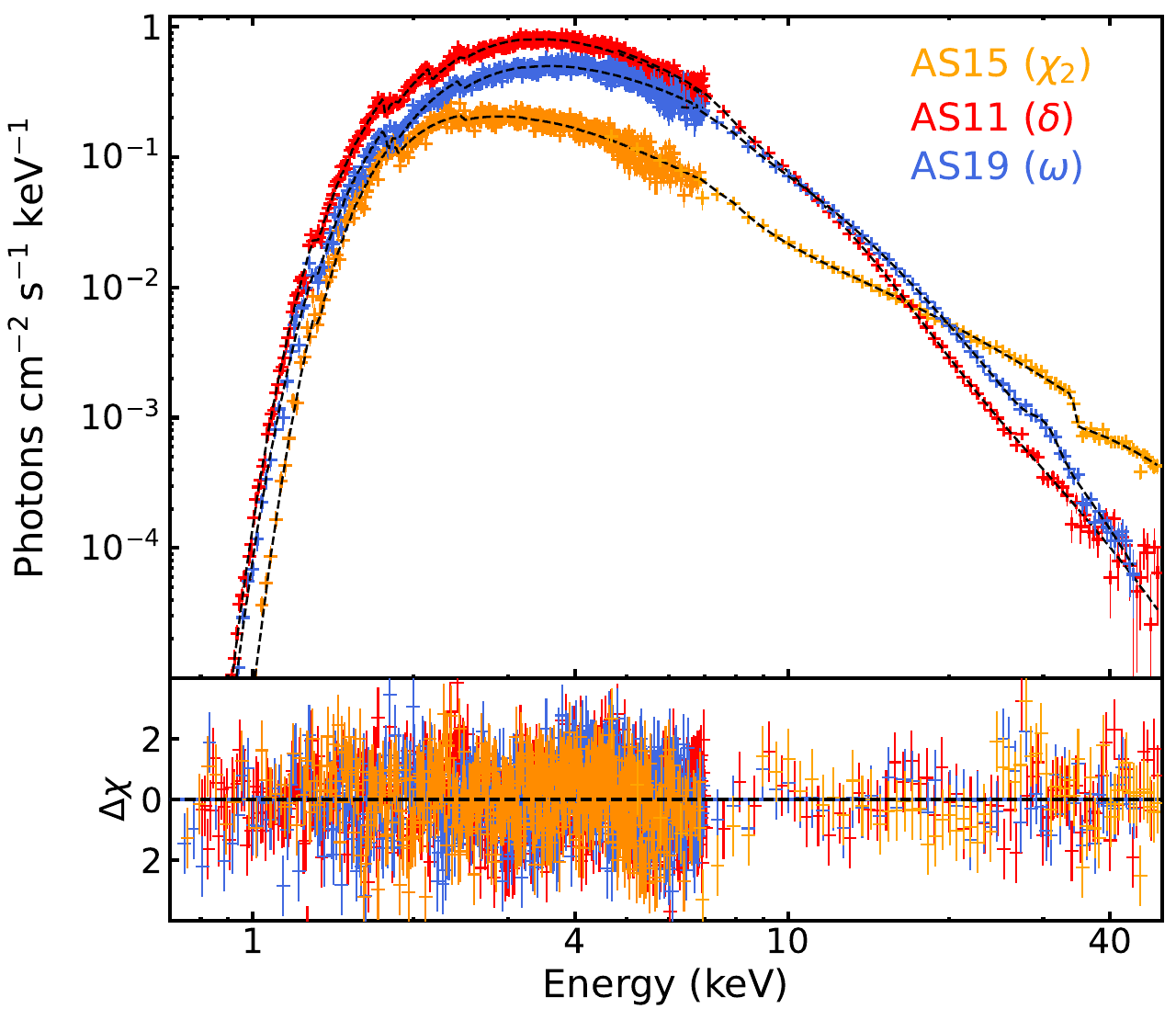}
    \end{center}
    \caption{{\it Top:} Wide-band energy spectra ($0.7-50$ keV) of GRS 1915$+$105 in the $\chi^{}_2$, $\delta$, and $\omega$ variability classes, obtained using SXT and LAXPC observations. Each spectrum is modelled with the combination \texttt{constant$\times$tbabs$\times$(thcomp$\otimes$diskbb)}. {\it Bottom}: Corresponding best fit residuals for each spectrum. See text for details.
    }
    \label{fig:fig12}
\end{figure}

We investigate the wide-band ($0.7-50.0$ keV) energy spectral properties of GRS 1915$+$105 using combined SXT and LAXPC20 data onboard {\it AstroSat}. The details of the spectral modelling and analysis are described in \S\ref{sec6:6.2_spec}, and the best fitted and derived spectral parameters are listed in Table \ref{tab:Table-08}. The inner disc temperature ($kT_{\rm in}$) and electron temperature ($kT_{\rm e}$) are found to vary in the ranges $0.25$$-2.38$ keV and $2.78$$- 30.21$ keV, respectively, across all variability classes. The photon index ($\Gamma_{\rm th}$) lies between $1.71$ and $3.09$, corresponding to intermediate and softer spectral states. These findings suggest that most of the observed states during the {\it AstroSat} campaign correspond to the `softer' variability classes of GRS 1915$+$105. Conversely, harder variability classes ($\chi$ variants) are found to exhibit higher $kT_{\rm e}$ $\sim 13-30$ keV along with photon index ($\Gamma_{\rm th}$) in the range of $\sim 1.71-2.57$. The bolometric luminosity is estimated to be within $2-44\%~{\rm L}_{\rm Edd}$ indicating that the source accretes in a sub-Eddington regime. Fig. \ref{fig:fig12} shows representative spectra for the $\delta$, $\chi^{}_2$, and $\omega$ classes.

During the LFQPO observations, the spectra are dominated by thermal Comptonized emission, which contributes up to $\sim 60\%$ of the total flux. The electron temperature and photon index vary in the ranges $10.00-30.21$~keV and $1.71-2.57$, respectively. The covering fraction (cov$_{\rm frac}$) spans $0.10-0.87$, implying that a substantial fraction of the seed photons undergo Comptonization in the corona. Recently \cite{Majumder-etal2026b} reported that the fraction of Comptonized flux correlates with the QPO frequency, fractional rms amplitude, and time-lag properties. Specifically, the Comptonized flux fraction increases as the QPO frequency decreases, while the rms amplitude grows with increasing Comptonized flux but declines beyond $\sim 45\%$ of Comptonized flux. This trend indicates that LFQPOs possibly originate due to the modulation of the Comptonizing region (see \S\ref{sec10p5:QPO-Lag} for details). 

\begin{table*}
\centering
	\caption{Wide-band spectral properties of GRS 1915$+$105 obtained using {\it AstroSat} data, modelled with \texttt{constant$\times$tbabs$\times$(thcomp$\otimes$diskbb)}. The parameters $kT_{\rm in}$, $kT_{\rm e}$, $\Gamma_{\rm th}$, and cov$_{\rm frac}$ correspond to the inner disc temperature, electron temperature, photon index, and covering fraction, respectively. The flux components $F_{\rm disc}$, $F_{\rm Comp}$, and $F_{\rm bol}$ in $1-100$ keV represent the disc, Comptonized, and total bolometric fluxes, while $L_{\rm bol}$ denotes the bolometric luminosity. During spectral fitting, the hydrogen column density along the line of sight was fixed at $n_{\rm H}=6\times10^{22}$ atoms cm$^{-2}$. Errors  are calculated with $90\%$ confidence in model fitted parameters. See text for details.}	
	\renewcommand{\arraystretch}{1.4}
    \begin{tabular}{l c c c c c c c c c c c} \hline

	Epoch  & $kT_{\rm in}$ & $kT_{\rm e}$ & $\Gamma_{\rm th}$ & cov$_{\rm frac}$ & $\chi^2\big/$dof  & $F_{\rm disc}$ & $F_{\rm Comp}$ & $F_{\rm bol}$ & $L_{\rm bol}$ & Class \\

	& (keV) & (keV) & &  &  & \multicolumn{3}{c}{($\times10^{-8}$ erg s$^{-1}$ cm$^{-2}$)} & ($\%{\rm L}_{\rm Edd}$) & \\ \hline
			
	AS1 & $1.40_{-0.05}^{+0.04}$ & $19.02_{-3.73}^{+7.94}$ & $2.57_{-0.07}^{+0.09}$ & $0.52_{-0.07}^{+0.11}$ & 605/551 & 3.54 & 1.59 & 5.14 & 29 & $\theta$ \\

    AS2 & $1.38_{-0.06}^{+0.06}$ & $19.88_{-5.96}^{+6.23}$ & $2.57_{-0.11}^{+0.12}$ & $0.57_{-0.11}^{+0.12}$ & 516/472 & 3.66 & 1.82 & 5.48 & 31 & $\chi_3$ \\
        
	AS3 & $0.84_{-0.03}^{+0.03}$ & $20.92_{-2.74}^{+3.75}$ & $2.52_{-0.02}^{+0.02}$ & $0.87$* & 613/543 & 2.96 & 3.23 & 6.19 & 35 & $\chi_1$ \\

    AS4 & $1.32_{-0.04}^{+0.03}$ & $13.06_{-1.59}^{+2.29}$ & $2.43_{-0.06}^{+0.06}$ & $0.50_{-0.05}^{+0.07}$ & 635/532 & 2.94 & 1.59 & 4.53 & 25 & $\chi_3$ \\
		
	AS5 & $1.43_{-0.07}^{+0.07}$ & $14.73_{-2.54}^{+4.43}$ & $2.57_{-0.09}^{+0.09}$ & $0.50_{-0.07}^{+0.10}$ & 296/297 & 4.40  & 1.91 & 6.31  &  35 & $\theta$ \\
		
	AS6 & $1.18_{-0.02}^{+0.02}$ & $4.15_{-0.14}^{+0.16}$ & $2.77_{-0.05}^{+0.05}$ & $0.60^{*}$ & 851/622 & 5.36  & 0.93 & 6.29  & 31 & $\omega$ \\
		
	AS7 & $1.62_{-0.01}^{+0.02}$ & $6.50^{*}$ & $3.04_{-0.07}^{+0.08}$ & $0.26_{-0.02}^{+0.02}$ & 678/535  & 7.14  & 0.65 & 7.82  & 44 & $\delta$ \\
		
	AS8 & $1.44_{-0.01}^{+0.01}$ & $5.19_{-0.19}^{+0.22}$ & $2.95_{-0.04}^{+0.05}$ & $0.70^{*}$ & 554/468 & 4.72  & 0.94 & 5.64 & 28 & $\delta$\\
			
	AS9 & $1.47_{-0.05}^{+0.06}$ & $10.00^{*}$ & $2.55_{-0.04}^{+0.05}$ & $0.28_{-0.02}^{+0.03}$ & 537/440 & 3.26 & 0.77 & 4.05  & 23 & $\beta$ \\	
			
	AS10 & $1.76_{-0.06}^{+0.06}$ & $10.00^{*}$ & $2.42_{-0.04}^{+0.04}$ & $0.37_{-0.04}^{+0.04}$ & 716/577 & 5.57 & 1.85 & 7.42 & 42 & $\theta$ \\
			
	AS11 & $2.02_{-0.04}^{+0.05}$ & $10.00^{*}$ & $3.09_{-0.12}^{+0.15}$ & $0.12_{-0.02}^{+0.02}$ & 760/676 & 5.45 & 0.28 & 5.81 & 33 & $\delta$ \\	

    AS12 & $1.03_{-0.04}^{+0.05}$ & $28.22_{-4.47}^{+6.38}$ & $2.12_{-0.02}^{+0.02}$ & $0.70^{*}$ & 605/548 & 1.38 & 1.48 & 2.86 & 16 & $\chi_2$ \\

    AS13 & $1.18_{-0.06}^{+0.05}$ & $29.00^{*}$ & $2.15_{-0.03}^{+0.03}$ & $0.56_{-0.07}^{+0.08}$ & 591/537 & 1.47 & 1.49 & 2.88 & 16 & $\chi_2^{'}$ \\

    AS14 & $1.32_{-0.06}^{+0.05}$ & $29.00^{*}$ & $2.15_{-0.04}^{+0.05}$ & $0.44_{-0.06}^{+0.09}$ & 627/576 & 1.77 & 1.31 & 3.01 & 17 & $\rho$ \\
            
	AS15 & $0.27_{-0.01}^{+0.01}$ & $20.00^{*}$ & $2.24_{-0.02}^{+0.02}$ & $0.14_{-0.02}^{+0.02}$ & 640/544 & 1.61 & 2.03 & 3.64 & 21 & $\chi_2$ \\	

    AS16 & $1.39_{-0.02}^{+0.02}$ & $19.30^{+7.24}_{-3.66}$ & $1.87_{-0.05}^{+0.05}$ & $0.16_{-0.02}^{+0.03}$ & 653/633 &  1.33 & 0.60 & 1.93  & 11 & $\rho$ \\

    AS17 & $1.06_{-0.02}^{+0.02}$ & $20.00^{*}$ & $2.88_{-0.02}^{+0.04}$ & $0.60^{*}$ & 758/644 & 1.63 & 0.66 & 2.34 & 13 & $\tau$ \\
        		
	AS18 & $2.33_{-0.08}^{+0.07}$ & $8.00^{*}$ & $2.78_{-0.11}^{+0.13}$ & $0.16_{-0.06}^{+0.09}$ & 765/622 & 2.05 & 0.17 & 2.23 & 13 & $\kappa$ \\	
			
	AS19 & $2.38_{-0.02}^{+0.04}$ & $6.45_{-0.25}^{+0.32}$ & $2.83_{-0.03}^{+0.02}$ & $0.29^{*}$ & 728/611 & 3.42 & 0.27 & 3.71 & 19 & $\omega$ \\	
			
	AS20 & $1.13_{-0.06}^{+0.05}$ & $4.81_{-0.17}^{+0.18}$ & $2.86_{-0.05}^{+0.05}$ & $1.00^{\boxdot}$ & 683/570 & 4.32 & 1.68 & 6.02 & 30 & $\gamma$ \\
		
	AS21 & $1.18_{-0.02}^{+0.02}$ & $3.67_{-0.16}^{+0.19}$ & $2.89_{-0.06}^{+0.07}$ & $0.50^{*}$ & 651/510 & 4.96 & 1.37 & 6.32 & 36 & $\delta$ \\	

	AS22 & $0.25_{-0.02}^{+0.02}$ & $30.21_{-6.22}^{+5.78}$ & $2.25_{-0.06}^{+0.05}$ & $0.10_{-0.04}^{+0.04}$ & 307/359 & 3.13 & $3.52$ & $6.66$ & $38$ & $\chi_1$ \\

    AS23 & $2.26_{-0.25}^{+0.26}$ & $16.39_{-2.25}^{+3.67}$ & $1.71_{-0.07}^{+0.07}$ & $0.64_{-0.11}^{+0.15}$ & 635/574 & 0.35 & 0.63 & 0.98 & 5 & $\chi_4$ \\

    AS24 & $1.18_{-0.14}^{+0.11}$ & $20.00^*$ & $2.07_{-0.10}^{+0.15}$ & $0.18_{-0.04}^{+0.06}$ & 188/188 & 0.29 & 0.11 & 0.40 & 2 & $\rho$ \\

     AS25 & $0.70_{-0.11}^{+0.19}$ & $2.78_{-0.23}^{+0.25}$ & $2.64_{-0.21}^{+0.19}$ & $0.50^{*}$ & 361/382 & 0.63 & 0.31 & 0.94 & 5 & $\delta$ \\
    \hline
	\end{tabular}%
    \begin{tablenotes}
        \item $^{*}$Frozen values.
        \item $^{\boxdot}$Pegged at hard limit.
    \end{tablenotes}

    \label{tab:Table-08}
\end{table*}

\subsubsection{Cyg X$-$1}

\subsubsection*{(a) Timing Variability Properties}

The long term \textit{MAXI} lightcurve of Cyg X$-$1 in $2-20$ keV energy range along with HR variation is shown in Fig. \ref{fig:fig6}. The vertical dashed lines denote the \textit{AstroSat} observations analysed in this work. The source is considered to be in the LHS when the flux is $<0.5$ Crab and the HR is $>0.5$, whereas observations with flux in the range $0.50-0.75$ Crab and HR between $0.25-0.50$ correspond to the IMS. The HSS corresponds to observations with a flux of $\sim 1$ Crab and HR $<0.25$. The HR variation clearly indicates transitions between spectral states. The source remained in the LHS at the beginning of $2016$ and reached the HSS by the end of $2017$ via the IMS. 

During the LHS, $F_{\rm var}$ is obtained in the range $8.9-20.9\%$, while during the IMS it decreases to $7.6-11.5\%$ (see Table~\ref{tab:Table-03}). In contrast, during the HSS, $F_{\rm var}$ is significantly higher, lying in the range $20.0-30.7\%$. The PDS also shows distinct morphological changes across spectral states. To model the PDS for all observations, broad \texttt{Lorentzian} components are required along with BLN components (see \S\ref{sec6:6.1_timing} for details). In LHS, three broad \texttt{Lorentzian} features are present, whereas two relatively strong \texttt{Lorentzian} components appear in the IMS (see left panel of Fig.~\ref{fig:fig13}). Conversely, the PDS in HSS is featureless which is modelled solely with BLN components. The model fitted parameters of the \texttt{Lorentzian} components with the maximum {\red rms$_{\rm BL}$} (merely broad features not representing QPOs) for each observation are listed in Table~\ref{tab:Table-09}. Interestingly, the total fractional rms (rms$_{\rm total}$), calculated in the $0.01-50$ Hz band, varies systematically with spectral state. During the LHS, rms$_{\rm total}$ ranges as $21.85-28.14\%$, while in IMS it lies between $15.29\%$ and $24.14\%$. In contrast, the highest rms$_{\rm total}$ values are found in HSS, where it varies between $18.10-33.39\%$.

\subsubsection*{(b) Spectral Properties}

We study wide-band ($0.7-70$ keV) energy spectral characteristics of Cyg X$-$1 using \textit{AstroSat} observations under consideration. The details of the spectral modelling are described in \S\ref{sec6:6.2_spec}, and the best fitted spectral parameters along with the derived quantities are listed in Table \ref{tab:Table-10}. The spectral parameters exhibit distinct characteristics across the different spectral states. In LHS, the inner disc temperature ($kT_{\rm in}$) and the photon index ($\Gamma$) are comparatively lower, lying in the ranges $0.11-0.31$ keV and $1.62-1.80$, respectively. In contrast, the electron temperature ($kT_{\rm e}$) is higher varying between $49.45$ and $76.63$ keV. During this state, the Comptonized component contributes $\sim 96-98\%$ of the total flux.

In IMS, both $kT_{\rm in}$ and $\Gamma$ increase in the ranges $0.23-0.41$ keV and $1.74-2.00$, respectively. The electron temperature varies as $\sim 22-50$ keV, and the Comptonized flux fraction lies between $\sim 65-92\%$. Relatively higher values of $kT_{\rm in}$ ($0.39-0.56$ keV) and $\Gamma$ ($2.21-2.70$) are observed in HSS, while $kT_{\rm e}$ remains within $20-40$ keV. As expected, the Comptonized flux fraction remains lowest in HSS ranging from $\sim 26-57\%$.

\begin{figure*}
\centering
    \includegraphics[width=0.49\textwidth]{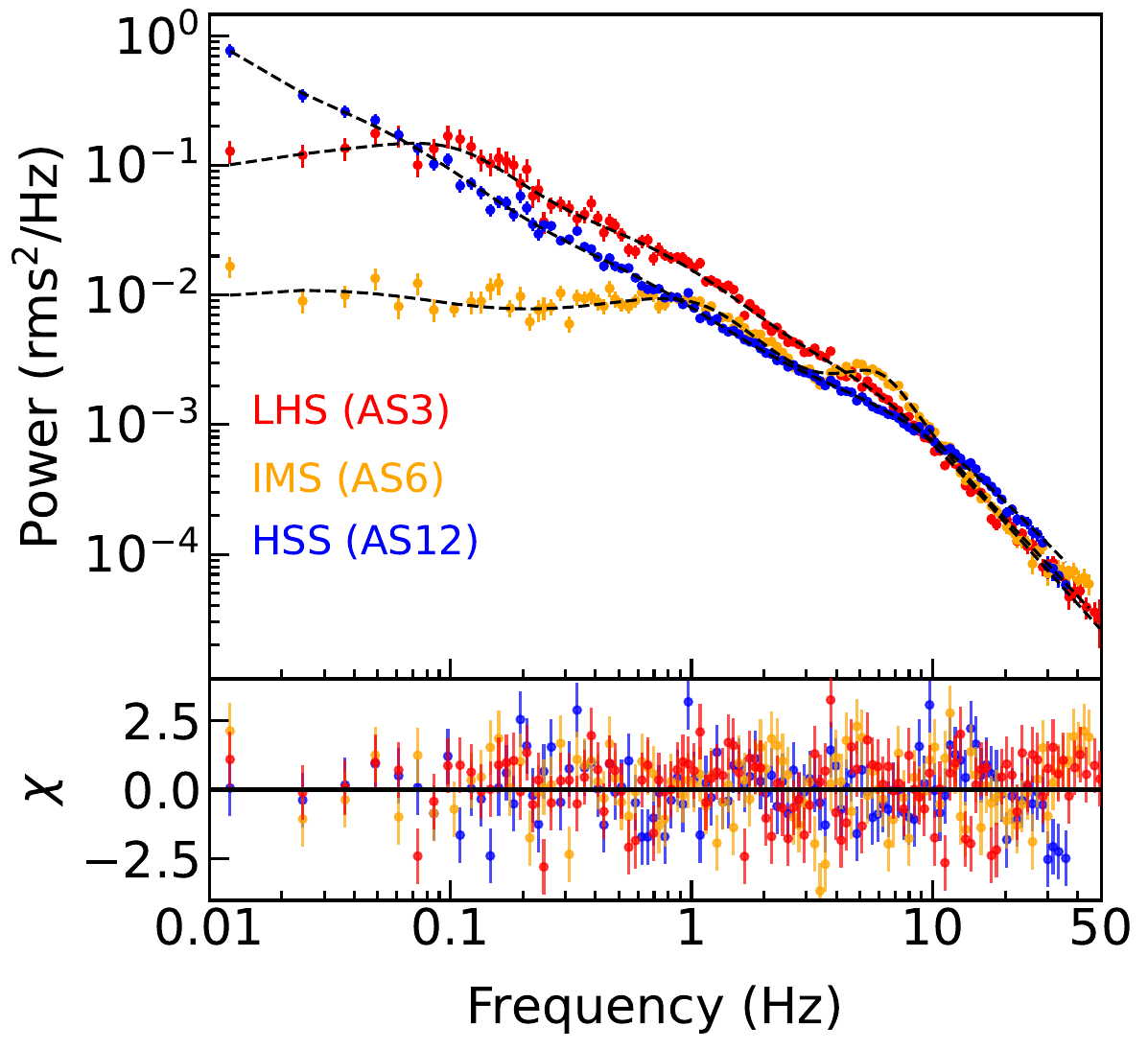}
    \includegraphics[width=0.49\textwidth]{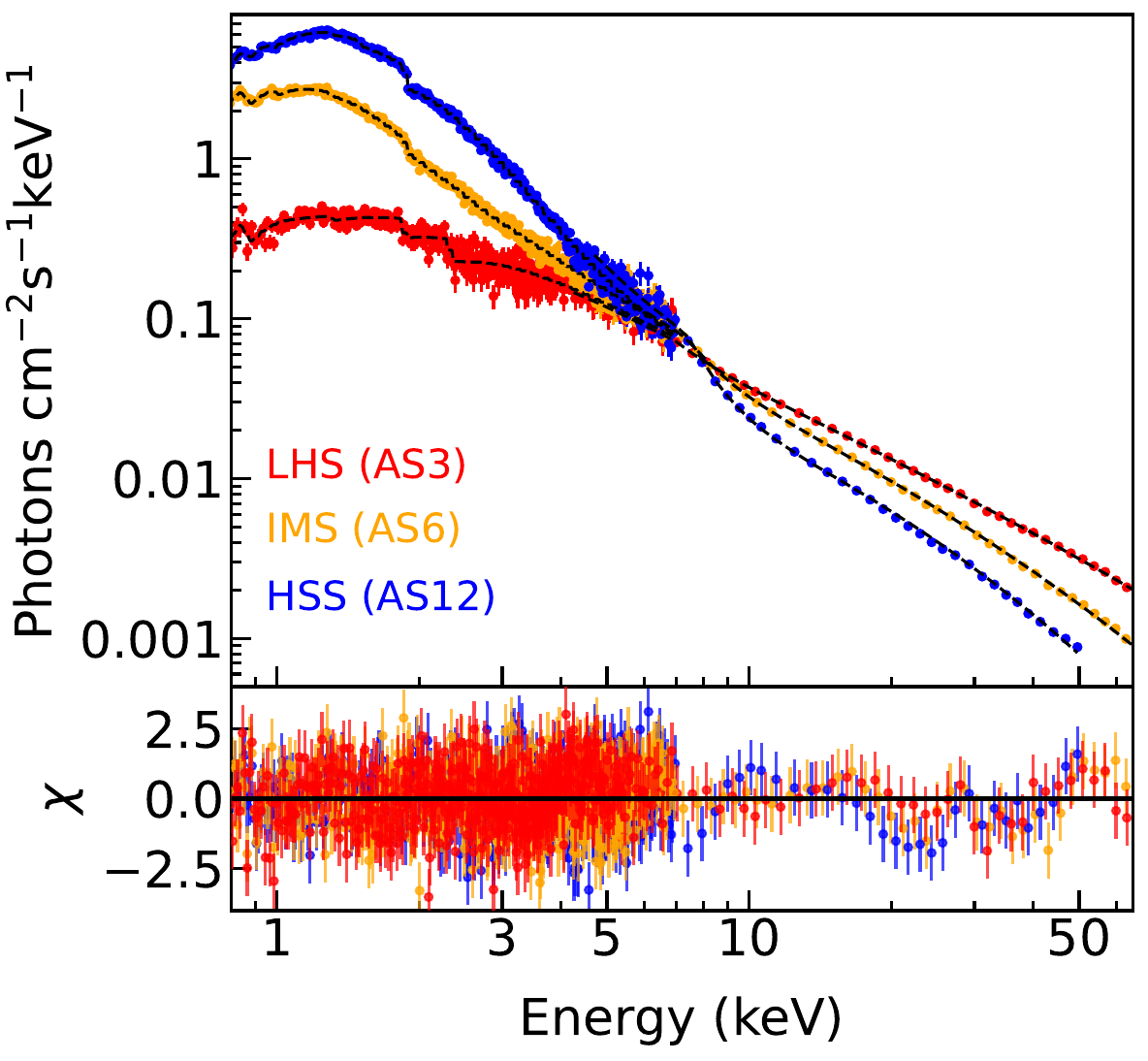}
    \caption{\textit{Left:} Model fitted PDS of Cyg X$-$1 in $0.01-50$ Hz frequency range, using multiple broad \texttt{Lorentzians}. The LHS, IMS, and HSS data are shown in red, orange, and blue, respectively.
    \textit{Right:} Energy spectra of Cyg X$-$1 observed with \textit{AstroSat} is fitted using \texttt{constant$\times$tbabs$\times$(thcomp$\otimes$diskbb)}. Red, orange and blue data points represent the spectra during LHS, IMS and HSS observations, respectively.
    }
    \label{fig:fig13}
\end{figure*}

The representative energy spectra for LHS, IMS, and HSS are shown in the right panel of Fig. \ref{fig:fig13}. It is evident that as the source transits from LHS to HSS via IMS, the low energy flux increases as the emission from the accretion disc dominates the energy spectra. This clearly demonstrates the dynamic nature of the accretion flow in Cyg X$-$1, which has persisted for over five decades \cite[]{Ling-etal1983,Gou-etal2014,Jiang2024}. Note that \cite{Kushwaha-etal2021} reported the `softest spectral state' of Cyg X$-$1 during an \textit{AstroSat} observation (AS10 in Table~\ref{tab:Table-03} and Fig. \ref{fig:fig6}), where the HR value was lower than that measured in the previously known soft state observed with \textit{Suzaku} \cite[]{Kawano-etal2017}. Furthermore, detailed LAXPC spectral modelling using a disc component and a broken power-law revealed a higher value of the low energy photon index ($\Gamma_1 = 4.13$) compared to the earlier soft state ($\Gamma_1 \sim 4.00$; \cite{Kawano-etal2017}). This evidently confirms the `softest spectral state' of this source observed with \textit{AstroSat} \cite[see][for details]{Kushwaha-etal2021}. However, present analysis with Comptonization model yields a photon index of $\Gamma_{\rm th} = 2.70$ (see Table \ref{tab:Table-10}) for this observation, which eventually indicates the `softest' spectral characteristic of the source during the \textit{AstroSat} era. Applying the continuum fitting method to this `softest state' observation, \cite{Kushwaha-etal2021} estimated black hole spin to be $a_{\rm k} > 0.998$, which is in agreement with earlier study \cite[]{Gou-etal2014}.

\subsubsection{Cyg X$-$3}

\subsubsection*{(a) Timing Variability Properties}

The \textit{MAXI} lightcurve of Cyg X$-$3, shown in Fig. \ref{fig:fig6}, exhibits pronounced variability in both flux and HR. It is observed that when HR reaches its maximum ($>1.0$), the flux drops to its minimum ($\le 0.2$ Crab), indicating that the source belongs to LHS. Conversely, when the HR is minimum $\le 0.5$, the flux reaches its peak value $\sim 0.4$ Crab, suggesting that the source is in HSS. The source is also found in IMS with intermediate HR and flux values. Overall, $F_{\rm var}$ is found to lie in the range $\sim 13-60\%$ during \textit{AstroSat} observations. Moreover, the source is observed to trace different spectral states, namely LHS (HR $\ge1.0$), IMS (HR$\sim0.3-0.6$), and HSS (HR$\le 0.2$).

Previous studies \cite[]{Pahari-etal2017} reported QPOs in $5-24$ mHz range, however, we detect two observations exhibiting higher QPO frequencies at $\sim 0.3$ Hz and $\sim 4.1$ Hz, each with a significance $\ge 5 \sigma$ corresponding to observations AS2 and AS7 (see Table \ref{tab:Table-09}). Fig. \ref{fig:fig6} evidently indicates that AS2 and AS7 belong to LHS (HR$\sim 1.02$) and HIMS (HR$\sim 0.75$), respectively. However, the rms$_{\rm QPO}$ for AS2 and AS7 are observed to be low as $\sim0.5\%$ and $1.2\%$ in $3-60$ keV energy band. The $\mathrm{rms}_{\rm total}$ for these observations lies in the range $\sim 11-13\%$, whereas $\mathrm{rms}_{\rm total}$ is relatively less ($\sim 1.2-3.5\%$) when QPOs are not detected. In AS2 and AS7, detected QPOs are of type-C, where we were unable to constrain the time-lag due to poor statistics with limited number of segments. The model fitted PDS corresponding to these observations are presented in the left panel of Fig. \ref{fig:fig14}. The observation corresponding to AS12 belongs to HSS (HR$\sim0.25$) without any detection of QPO.

\subsubsection*{(b) Spectral Variability Properties}

The wide-band ($0.7-50.0$ keV) energy spectral properties of Cyg X$-$3 is investigated using \textit{AstroSat} observations. The best fitted model parameters are listed in Table \ref{tab:table_spectra_Cyg} following the model prescription described in \S\ref{sec6:6.2_spec}. The spectral analysis indicates that during LHS, the inner disc temperature is relatively low ($kT_{\rm in}\sim 0.39-0.60$ keV), and the Comptonized flux ($F_{\rm Comp}$) strongly dominates over the disc flux ($F_{\rm disc}$). The electron temperature remains in the range $kT_{\rm e}\sim 7-9$ keV, while the covering fraction is cov$_{\rm frac} \sim 0.20-0.30$. The bolometric luminosity is estimated to be $\red \sim 15-21\%\,{\rm L}_{\rm Edd}$. During IMS, the disc becomes hotter with $kT_{\rm in} \sim 0.61-0.99$ keV, and the electron temperature is found to span $kT_{\rm e} \sim 7-18$ keV. In this state, the disc and Comptonized fluxes are comparable. The bolometric luminosity increases to $\red \sim 14-49\%\,{\rm L}_{\rm Edd}$. In HSS, the inner disc temperature further rises to $kT_{\rm in} \sim 1.1-1.8$ keV, while $kT_{\rm e}$ is obtained as $5-10$ keV, although in some observations $kT_{\rm e}$ could not be constrained and therefore it is kept fixed to best fitted value. As expected, the disc flux becomes dominant over the Comptonized component in this state, and the bolometric luminosity reaches $\red \rm \sim 25-61\%~L_{Edd}$. {\red It is noteworthy that the large uncertainty in the source mass ($2.4^{+2.1}_{-1.1}\,M_{\odot}$), together with the distance uncertainty ($8.95\pm0.96$ kpc), yields a poorly constrained bolometric luminosity of $\sim11-257\%\,{\rm L}_{\rm Edd}$.}

\begin{figure*}
	\centering
    \includegraphics[width=0.9\textwidth]{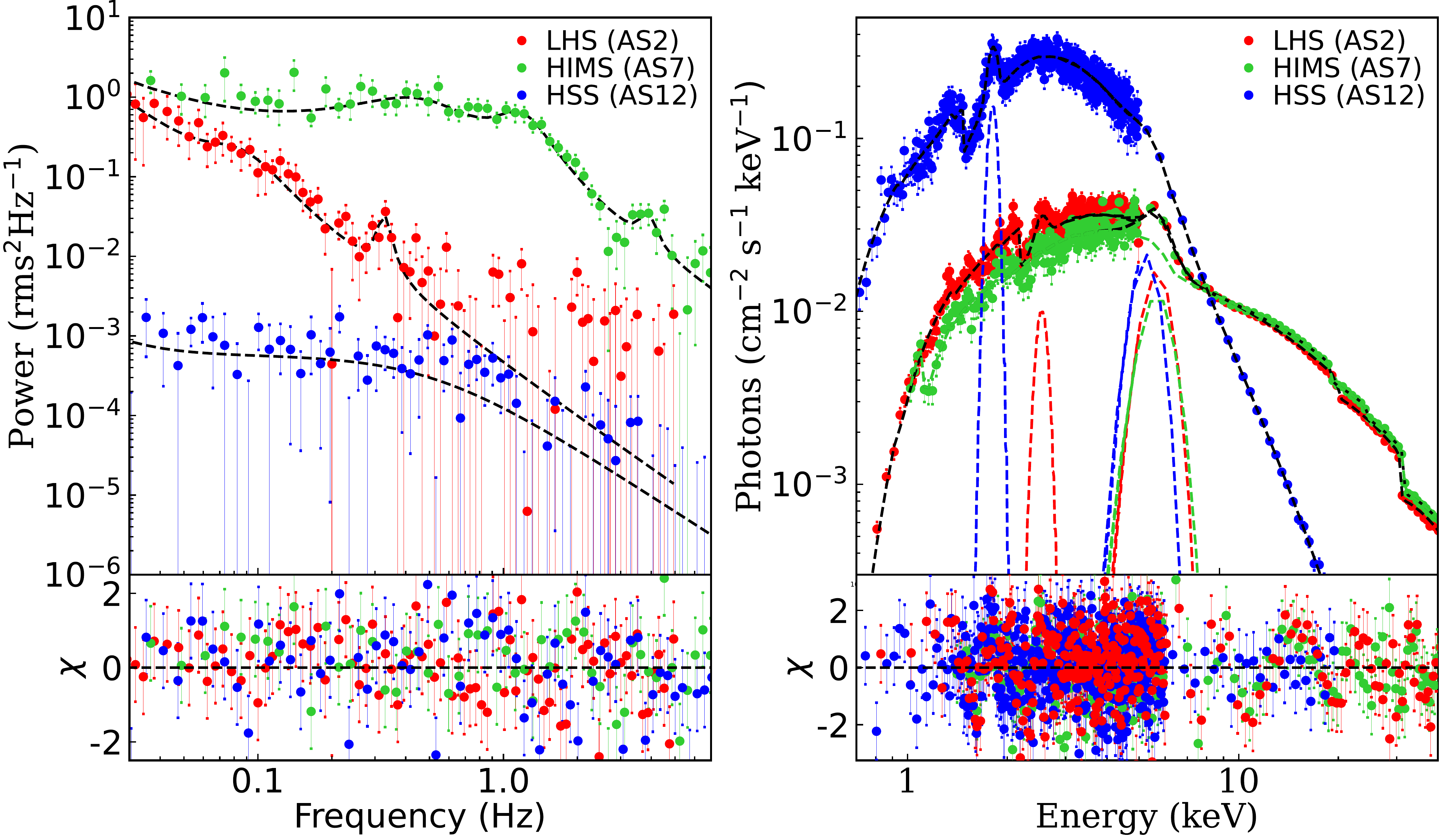}
    \caption{{\it Left:} Model fitted PDS of Cyg X$-$3 in $0.01-10$ Hz frequency range, where QPOs are observed at $\sim 0.33$ Hz (red) and $\sim 4.10$ Hz (green), without any detection of QPO in HSS (AS12). {\it Right:} Energy spectra of Cyg X$-3$, where AS2 (red) and AS7 (green) correspond to the observations with low flux in \textit{MAXI}/GSC and AS12 (blue) exhibits the highest flux in \textit{MAXI}/GSC.
    }
    \label{fig:fig14}
\end{figure*}

The hydrogen column density and partial covering fraction associated with the \texttt{pcfabs} component are obtained as $n_{\rm H} \sim 12-20$ atoms cm$^{-2}$ and $0.5-0.9$, respectively. Emission features at $\sim 6.5$ keV (Fe complex line) and $\sim 2.3$ keV (S K$\alpha$ line) are modelled using a \texttt{gaussian} component.  The right panel of Fig. \ref{fig:fig14} shows three different energy spectra of the source. For observations exhibiting QPOs (AS2 and AS7), the spectral fit yields  $\Gamma_{\rm th}\sim1.58-1.66$, $kT_{\rm in}\sim 0.4-0.5$ keV and $kT_{\rm e}\sim 8$ keV, with a bolometric luminosity of $\red \sim 15-25\%~{\rm L}_{\rm Edd}$. In both cases, the Comptonized flux dominates over the disc flux. These results are consistent with the spectral states inferred from the HR values derived using \textit{MAXI} data. The model fitted parameters are tabulated in Table \ref{tab:table_spectra_Cyg}.

\begin{table*}
\caption{Temporal properties of Cyg X$-$1 and Cyg X$-$3 observed with {\it AstroSat}. The notations and symbols have their usual meanings. See \S \ref{sec6:6.1_timing} and \S \ref{sec7:7.1_pcv} for details.
}
\renewcommand{\arraystretch}{1.4}
\resizebox{\textwidth}{!}{%
\begin{tabular}{c c c c c c c c c c c c} \hline

Source & Epoch &  $\red \nu_{\rm BL}$ & FWHM & Significance & Q-factor & rms$_{\rm \red BL}\%$ & time-lag  & rms$_{\rm total}\%$ & $\chi^2\big/$dof & QPO-Type \\

& & (Hz) & (Hz) & ($\sigma$) & &  & (ms) &   &  &  \\ \hline

Cyg X$-1^{\dagger}$  & AS1 & $2.00_{-0.53}^{+0.53}$ & $8.68$ & 10.1 & 0.2 & $15.47\pm0.76$ & $-$ & $26.48\pm0.17$ & 108/109 & $-$ \\

    & AS2 & $0.92_{-0.33}^{+0.32}$ & $7.41$ & 16.0 & 0.1 & $18.14\pm0.56$ & $-$ & $28.14\pm0.15$ & 169/112 & $-$ \\

    & AS3 & $0.07_{-0.03}^{+0.02}$ & $0.18$ & 5.7 & 0.4 & $15.36\pm1.38$ & $-$ & $25.83\pm0.16$ & 155/111 & $-$ \\

    & AS4 & $0.29_{-0.04}^{+0.08}$ & $0.88$ & 4.2 & 0.3 & $17.56\pm2.09$ & $-$ & $23.83\pm0.16$ & 163/110 & $-$ \\

     & AS5 & $0.35_{-0.07}^{+0.13}$ & $0.93$ & 2.1 & 0.3 & $15.83\pm1.43$ & $-$ & $22.42\pm0.23$ & 174/106 & $-$ \\

    & AS6 & $0.75_{-0.06}^{+0.08}$ & $1.86$ & 13.9 & 0.4 & $13.23\pm0.47$ & $-$ & $19.38\pm0.14$ & 161/105 & $-$ \\

    &  AS7  & $0.11_{-0.04}^{+0.05}$ & $0.98$ & 13.6 & 0.1 & $17.29\pm0.63$ & $-$ & $24.14\pm0.15$ & 128/108 & $-$ \\

    & AS8 & $7.71_{-2.01}^{+1.93}$ & $14.85$ & 3.4 & 0.5 & $6.26\pm0.93$ & $-$ & $20.04\pm0.09$ & 184/103 & $-$ \\

    & AS9 & $-$ & $-$ & $-$ & $-$ & $-$ & $-$ & $22.03\pm0.01$ & 107/102 & $-$ \\

    & AS10 & $-$ & $-$ & $-$ & $-$ & $-$ & $-$ & $18.10\pm0.09$ & 110/88 & $-$ \\

    & AS11 & $-$ & $-$ & $-$ & $-$ & $-$ & $-$ & $25.40\pm0.26$ & 147/102 & $-$ \\

    & AS12 & $-$ & $-$ & $-$ & $-$ & $-$ & $-$ & $21.06\pm0.15$ & 140/105 & $-$ \\

    & AS13 & $-$ & $-$ & $-$ & $-$ & $-$ & $-$ & $25.09\pm0.59$ & 87/92 & $-$ \\

    & AS14 & $-$ & $-$ & $-$ & $-$ & $-$ & $-$ & $26.80\pm0.21$ & 131/86 & $-$ \\

    & AS15 & $6.42_{-1.47}^{+0.48}$ & $6.54$ & $4.2$ & $1.0$ &  $6.19\pm0.74$ & $-$ & $17.91\pm0.05$  & 91/90 & $-$ \\

    & AS16 & $1.03_{-0.16}^{+0.14}$ & $3.61$ & $14.0$ & $0.3$ &  $11.22\pm0.40$ & $-$ & $15.29\pm0.04$  & 116/97 & $-$ \\

    & AS17 & $3.96_{-0.25}^{+0.25}$ & $7.34$ & $14.4$ & $0.5$ &  $12.95\pm0.44$ & $-$ & $18.33\pm0.09$  & 103/99 & $-$ \\

    & AS18 & $0.13_{-0.03}^{+0.03}$ & $0.79$ & $21.0$ & $0.2$ & $16.09\pm0.38$ & $-$ & $22.00\pm0.08$  & 121/108 & $-$ \\

    &  AS19  & $3.12_{-0.19}^{+0.19}$ & $6.85$ & 17.5 & 0.5 & $15.51\pm0.44$ & $-$ & $21.85\pm0.03$ & 137/102 & $-$ \\

    & AS20 & $1.23_{-0.23}^{+0.23}$ & $1.97$ & $3.1$ & $0.6$ & $10.90\pm1.75$ & $-$ & $23.68\pm0.16$ & 117/108 & $-$ \\

    & AS21 & $0.97_{-0.21}^{+0.12}$ & $2.19$ & $7.5$ & $0.4$ &  $19.01\pm1.26$ & $-$ & $33.39\pm$0.14  & 158/85 & $-$ \\

    \hline

    Source & Epoch &  $\nu_{\rm QPO}$ & FWHM & Significance & Q-factor & rms$_{\rm QPO}\%$ & time-lag  & rms$_{\rm total}\%$ & $\chi^2\big/$dof & QPO-Type \\

& & (Hz) & (Hz) & ($\sigma$) & &  & (ms) &   &  &  \\ \hline

Cyg X$-3$ & AS1 & $-$ & $-$ & $-$ & $-$ & $-$ & $-$ & $1.40\pm0.10$ & $-$ &  \\
& AS2$^{\ddagger}$ & $0.33\pm0.01$ & $0.03\pm0.01$  & $9.3$ & $4.2$ & $0.51\pm0.02$ & $-$ & $11.31\pm0.23$ & $104/100$ & C \\

    & AS3-AS6  & $-$ & $-$ & $-$ & $-$ & $-$ & $-$ & $1.40-3.50$ & $-$ &  \\

    & AS7$^{\ddagger}$ & $4.14\pm0.18$ & $0.74$  & $5.6$ & $3.0$ & $1.18\pm0.02$ & $-$ & $13.55\pm0.23$ & $96/98$ &  C \\

    & AS8-AS15 & $-$ & $-$ & $-$ & $-$ & $-$ & $-$ & $1.10-3.20$ & $-$ & \\

\hline
\end{tabular}
}
\begin{tablenotes}
    \item $^\dagger$Timing parameters corresponding to broad \texttt{Lorentzian} features ($\red \nu_{\rm BL}$) with maximum rms$_{\rm QPO}$. Time-lag is not constrained due to broad feature (${\rm FWHM > \red \nu_{BL}}$) of Cyg X$-1$ following \cite{Reig-etal2000}.
    \item $^{\ddagger}$Time-lag is not reliably constrained due to poor statistics, which resulted from the limited number of segments available for averaging caused by the use of a large segment length that reduces the number of independent measurements.
\end{tablenotes}
	
    \label{tab:Table-09}
\end{table*}

\begin{table*}
\caption{Wide-band spectral properties of Cyg X$-$1 and Cyg X$-$3 using {\it AstroSat} data, modelled with \texttt{constant$\times$tbabs$\times$(thcomp$\otimes$diskbb)}. The parameters $kT_{\rm in}$, $kT_{\rm e}$, $\Gamma_{\rm th}$, and cov$_{\rm frac}$ correspond to the inner disc temperature, electron temperature, photon index, and covering fraction, respectively. The flux components $F_{\rm disc}$ and $F_{\rm Comp}$ in $1-100$ keV represent the disc and Comptonized fluxes, while $L_{\rm bol}$ denotes the bolometric luminosity. During spectral fitting, the hydrogen column density is kept fixed as $n_{\rm H}=0.7\times10^{22}$ atoms cm$^{-2}$ for Cyg X$-1$ and $n_{\rm H}=2.0\times10^{22}$ atoms cm$^{-2}$ for Cyg X$-3$, respectively. Errors in model fitted parameters are calculated with $90\%$ confidence. See text for details.}
\centering
\renewcommand{\arraystretch}{1.4}
	\label{tab:table_spectra_Cyg}
	\resizebox{\textwidth}{!}{%
\begin{tabular}{lccccccccc}
\hline
Source &    Epoch  & $kT_{\rm in}$ & $kT_{\rm e}$  & $\Gamma_{\rm th}$  & cov$_{\rm frac}$ &  $\chi^2\big/$dof & $F_{\rm disc}$ & $F_{\rm Comp}$ &   $L_{\rm bol}$ \\
&  & (keV) & (keV) & (keV) & & & \multicolumn{2}{c}{($\times10^{-8}$ erg s$^{-1}$ cm$^{-2}$)} & ($\%{\rm L}_{\rm Edd}$) \\ \hline

Cyg X$-1$ & AS1 & $0.108_{-0.002}^{+0.003}$ & $76.63_{-34.37}^{\boxdot}$ &  $1.63_{-0.01}^{+0.01}$  & $0.011_{-0.002}^{+0.002}$ & 674/570 & 0.05 &  3.06 & 0.7 \\

    & AS2 & $0.16_{-0.01}^{+0.01}$ & $49.60_{-13.31}^{+49.98}$ & $1.62_{-0.01}^{+0.01}$	& $0.22_{-0.08}^{+0.10}$  & 401/418 & 0.02 & 1.95 & 0.4 \\

    & AS3 & $0.11_{-0.004}^{+0.005}$  & $68.28_{-24.67}^{\boxdot}$ & $1.63_{-0.01}^{+0.01}$	& $0.03_{-0.01}^{+0.01}$	& 493/451 & 0.05 & 3.68 & 0.8 \\

    & AS4 & $0.29_{-0.01}^{+0.02}$ & $50.00^*$ & $1.77_{-0.01}^{+0.01}$ & $0.46_{+0.04}^{-0.03}$ & 595/496 & 0.17 & 2.13  & 0.5  \\

    & AS5 & $0.23_{-0.01}^{+0.01}$ & $50.00^*$ & $1.82_{-0.01}^{+0.01}$ & $0.21_{-0.02}^{+0.04}$ & 638/570 & 0.33 & 2.91  & 0.7 \\

    & AS6 & $0.32_{-0.01}^{+0.01}$ & $45.00^*$ & $1.92_{-0.01}^{+0.01}$ & $0.28_{+0.02}^{-0.02}$ & 594/559 & 0.72 & 2.64 & 0.7 \\

    &  AS7  & $0.27_{-0.01}^{+0.01}$ & $40.00^*$ & $1.74_{-0.01}^{+0.01}$ & $0.45_{-0.04}^{+0.05}$ & 586/516 & 0.15 & 2.35 & 0.6 \\

    & AS8 & $0.39_{-0.02}^{+0.02}$ & $37.63_{-7.89}^{+15.38}$  &  $2.25_{-0.02}^{+0.02}$	& $0.69_{-0.06}^{+0.08}$ & 701/588	& 1.36  & 4.06 & 1.1 \\

    & AS9 & $0.51_{-0.02}^{+0.03}$ & $32.98_{-3.48}^{+20.65}$  &  $2.53_{-0.02}^{+0.05}$ & $0.76_{-0.05}^{+0.14}$ & 695/558 & 1.79 & 2.45   & 0.9 \\

    & AS10 & $0.54_{-0.01}^{+0.01}$ & $20.00^*$ & $2.70_{-0.02}^{+0.02}$   & $0.24_{-0.02}^{+0.02}$ & 569/536 & 2.32 & 0.82  &  0.7 \\

    & AS11 & $0.45_{-0.02}^{+0.02}$ & $20.00^{*}$ & $2.45_{-0.01}^{+0.01}$ & $0.53_{-0.02}^{+0.02}$ & 630/567 & 1.59 & 2.18  & 0.8 \\

    & AS12 & $0.52_{-0.1}^{+0.1}$ & $40.00^*$  & $2.22_{-0.01}^{+0.01}$	& $0.21_{-0.01}^{+0.01}$  & 566/561 & 2.85  & 1.83 & 1.0 \\

    & AS13 & $0.44_{-0.01}^{+0.01}$ & $50.00^{*}$ & $2.50_{-0.01}^{+0.01}$ & $0.49_{-0.02}^{+0.01}$ & 599/565 & 1.58 &  1.80 & 0.7 \\

    &  AS14 & $0.48_{-0.01}^{+0.01}$ & $30.00^*$ & $2.60_{-0.02}^{+0.01}$  & $0.33_{-0.02}^{+0.01}$ & 652/517 & 1.96 &  1.04 & 0.7 \\

    & AS15 & $0.34_{-0.01}^{+0.01}$ & $22.26_{-1.95}^{+2.59}$  & $1.87_{-0.02}^{+0.02}$  & $0.48_{-0.02}^{+0.03}$ & 772/631 & $1.03$ & $7.27$  & $1.7$ \\

    & AS16 & $0.41_{-0.01}^{+0.01}$ & $30.00^*$  & $2.00_{-0.01}^{+0.01}$  & $0.26_{-0.01}^{+0.01}$ & 565/548 & $3.15$ & $5.95$ & $1.9$ \\

    & AS17 & $0.37_{-0.02}^{+0.02}$ & $33.53_{-5.68}^{+10.60}$ & $1.87_{-0.02}^{+0.02}$  & $0.47_{-0.03}^{+0.04}$ & 635/540  & $1.00$  & $5.62$ & $1.4$ \\

    & AS18 & $0.20_{-0.01}^{+0.01}$  & $49.45_{-11.74}^{+33.74}$  & $1.71_{-0.01}^{+0.01}$ 	& $0.32_{-0.05}^{+0.06}$ & 689/568  & $0.19$ & $5.87$ & $1.3$ \\

    & AS19 & $0.31_{-0.01}^{+0.02}$ & $37.26_{-6.18}^{+11.55}$ & $1.80_{-0.01}^{+0.01}$ & $0.46_{-0.03}^{+0.04}$ & 587/517  & 0.49 & 4.54   & 1.1 \\

    & AS20 & $0.15_{-0.01}^{+0.01}$ & $59.76_{-18.24}^{\boxdot}$ & $1.65_{-0.01}^{+0.01}$ 	& $0.16_{-0.05}^{+0.06}$ & 439/428 & $0.04$ & $2.61$ & $0.6$ \\

    & AS21 & $0.56_{-0.03}^{+0.04}$ & $21.68_{-3.58}^{+6.31}$ & $2.21_{-0.03}^{+0.03}$  & $0.39_{-0.04}^{+0.05}$ & 476/424 & $8.92$ & $10.60$  & $4.0$ \\

\hline

Cyg X$-3^{\dagger}$ & AS1  &  $0.96\pm0.08$ & $13.28\pm0.85$  &	$2.18\pm0.04$  &	$0.28\pm0.02$&	$111/75$ &$0.59$ & $0.22$ &	$\red 13.6$ \\

    &AS2&$0.52\pm0.02$  & $8.32\pm0.19$&   $1.69\pm0.03$ &	$0.27\pm0.01$&	$853/568$ &$0.25$   & $0.65$	&	$\red 15.2$\\

    &AS3&	$0.52\pm 0.05$  & $9.18\pm0.61$ &	$1.90\pm	0.04$	&$0.26\pm0.06$	&$638/572$  &$0.49$	&$0.75$	&	$\red 20.9$\\

    &AS4&	$0.62\pm	0.05$ &	$8.18\pm0.61$ &	$2.10\pm	0.04$	&$0.26\pm0.06$	&$638/573$ &$0.49$	&$0.75$	&	$\red 20.9$\\

    &AS5&	$0.98\pm0.07$ &	$16.83\pm1.3$ &	$2.43\pm0.06$ &	$0.12\pm0.04$&	$89/74$  &$1.12$ &	$0.76$	&$\red 31.7$ \\

    &AS6&	$0.39\pm0.04$  &  $7.97\pm0.19$ &	$1.58\pm0.02$ &	$0.10^*$&	$583/519$ &$0.35$ &$0.69$	&	$\red 17.5$ \\

    &AS7&	$1.38\pm0.02$  &	$10.00^*$ &	$1.82\pm0.06$&	$0.025\pm0.004$&	$808/581$  &	$1.38$ &	$0.10$	&$\red 24.9$ \\

    &AS8&	$0.61\pm0.04$ &$7.82\pm0.46$ &	$1.64\pm0.03$	&	$0.10^*$&	$246/168$   &$1.10$ &$0.92$	&	$\red 34.0$ \\

    &AS9&	$0.60\pm0.03$  &	$7.61\pm0.23$ &	$1.61\pm0.02$	&$0.21\pm0.02$	&$659/555$  &	$0.33$  &$0.62$  	&$\red 16.0$ \\
    
    &AS10&	$0.74\pm0.05$  &	$18.09\pm2.78$  &$2.18\pm0.05$	&$0.15^*$	&$300/292$	&$1.91$  &$1.02$	&$\red 49.3$ \\
    
    &AS11&	$1.13\pm0.03$ &	$5.00^*$  &	$2.62\pm0.28$	&$0.08\pm0.02$ & $598/512$ &	$3.42$ &	$0.18$  &	$\red 60.6$ \\

    &AS12&	$1.82\pm0.05$  &	$12.87\pm1.55$ &$2.04\pm0.02$ &	$0.05$&	$657/540$ &$2.76$  &	$0.22$	&	$\red 49.7$ \\

    &AS13&	$1.20\pm0.04$ &	$16.80\pm2.75$  &$2.21\pm0.03$&	$0.20\pm0.03$	&$724/578$  &$1.93$ &	$0.63$	&	$\red 43.1$ \\

\hline
\end{tabular}
}
\begin{tablenotes}
    \item $^{*}$Frozen values. 
    \item $^\boxdot$Pegged at hard limit. 
    \item {\red $^{\dagger}$The bolometric luminosity is estimated assuming source mass of $4.5\,{\rm M}_{\odot}$ and distance of $8.95$ kpc for Cyg X$-$3.}
\end{tablenotes}

\label{tab:Table-10}
\end{table*}

\subsection{Persistent Sources}\label{sec7:7.2_persis}

\subsubsection*{(a) Timing Variability Properties}

\begin{figure*}
    \begin{center}
    \includegraphics[width=\textwidth]{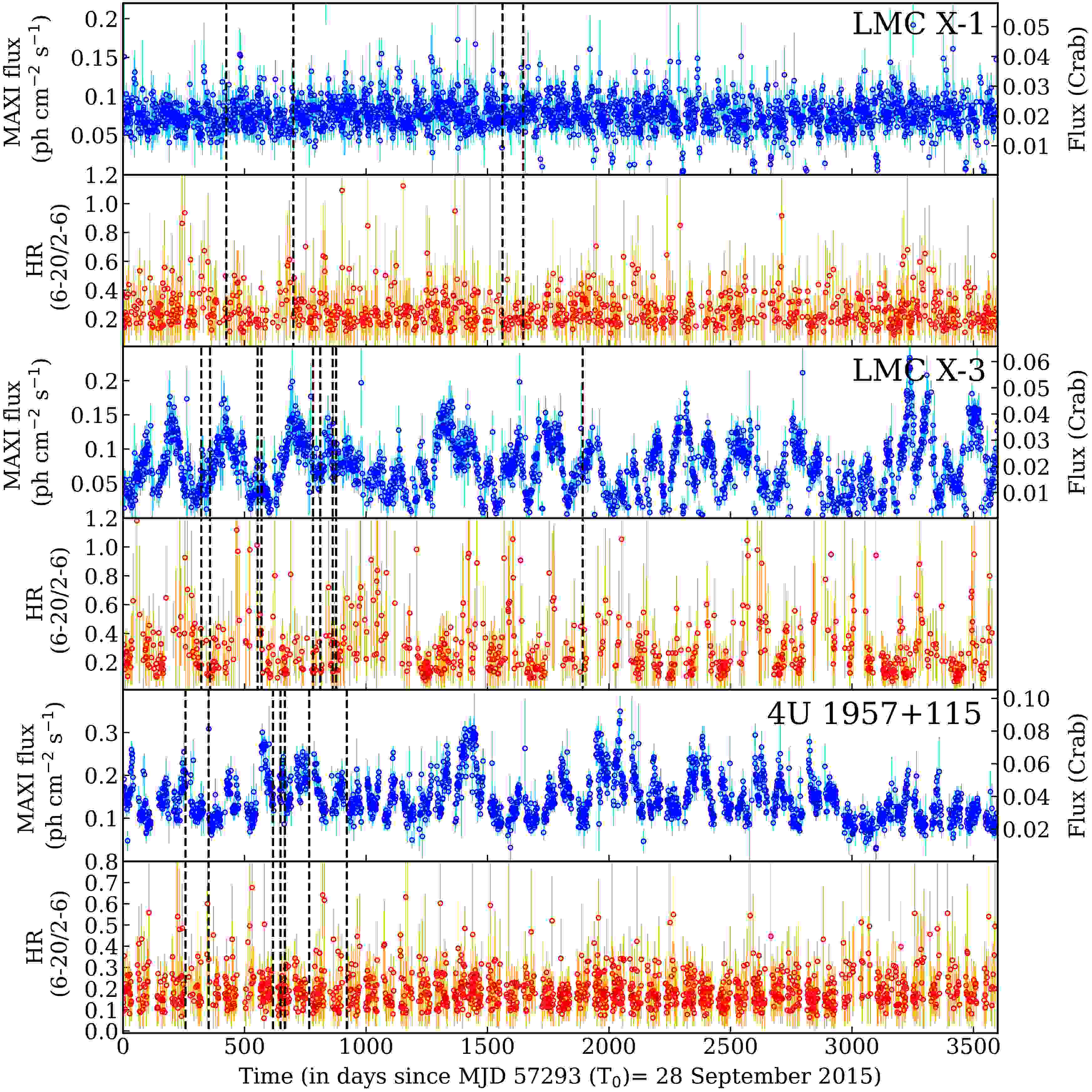}
    \end{center}
    \caption{{\it MAXI}/GSC monitoring of the persistent sources, LMC X$-1$, LMC X$-3$ and 4U 1957$+$115 in $2-20$ keV energy band during {\it AstroSat} observation. Variation of hardness ratio (HR) observed with {\it MAXI}/GSC is plotted for all sources. Vertical dashed lines indicate the epochs of {\it AstroSat} observations. The lightcurve and HR are obtained from $1$-day averaged {\it MAXI}/GSC data from the launch date of {\it AstroSat}. See text for details.
    }
    \label{fig:fig15}
\end{figure*}

We study the timing variability properties of five persistent BH–XRBs, namely LMC X$-1$, LMC X$-3$, 1E 1740.7$-$2942, GRS 1758$-$258, and 4U 1957$+$115, using LAXPC data obtained over several epochs. For LMC X$-$1 and LMC X$-$3, the fractional variability amplitude ($F_{\rm var}$) in the $2-20$ keV energy band during the $2016-2020$ observational epochs are found to be $6.5-14.9\%$ and $2.4-10.8\%$, respectively (see Table \ref{tab:Table-04}). In addition, we observe comparatively low variability in LMC X$-1$ and relatively higher variability in LMC X$-3$ in their long-term {\it MAXI}/GSC lightcurves, as illustrated in Fig. \ref{fig:fig15}. Note that none of the \textit{AstroSat} observations of these sources exhibit characteristic temporal features in the PDS except 4U 1957$+$115. Hence, PDS of these four sources are modelled using a simple \texttt{powerlaw} component to estimate the fractional rms variability amplitude. In the PDS, LMC X$-$1 shows moderate variability ranging from $7-16\%$ in the $3-20$ keV band, while LMC X$-$3 displays a total rms amplitude of $7-11\%$ in the same energy range \cite[see][for details]{Bhuvana-etal2021,Bhuvana-etal2022}. 

Note that PDS analysis is not performed for GRS 1758$-$258 and 1E 1740.7$-$2942 due to significant contamination in the \textit{LAXPC} data from nearby sources. However, $F_{\rm var}$ is estimated for both sources, yielding values of $6.1-13.8\%$ for GRS 1758$-$258 and $6.5-15.2\%$ for 1E 1740.7$-$2942. In case of 4U 1957$+$115, we report the detection of a LFQPO at $\sim 2.66$ Hz with rms$_{\rm QPO}\sim 2.1\%$ for the first time. The Q-factor is found to be $\geq 10$ with a significance of $\sim 4$. A soft-lag of $\sim 31 \pm 13$ ms is observed for this QPO. Note that the QPO feature is observed at low energies with no detection beyond $15$ keV. As shown in Fig. \ref{fig:fig15}, the \textit{MAXI} flux is observed to be $\sim 0.04$ Crab with HR $\sim 0.22$ indicating that the source is in softer state when the QPO is detected (always HR $\leq 1$ for this source). Following this, we indicate that the detected QPO is of type-A accompanied by a weak harmonic feature at $\nu_{\rm QPO} \sim 5.3$ Hz with an rms amplitude of $\sim 1.3\%$. Fig. \ref{fig:fig18} displays the model fitted PDS for two observations AS5 and AS3, i.e., with and without the detection of LFQPOs.

\subsubsection*{(b) Spectral Variability Properties}

\begin{figure}
    \includegraphics[width=0.9\columnwidth]{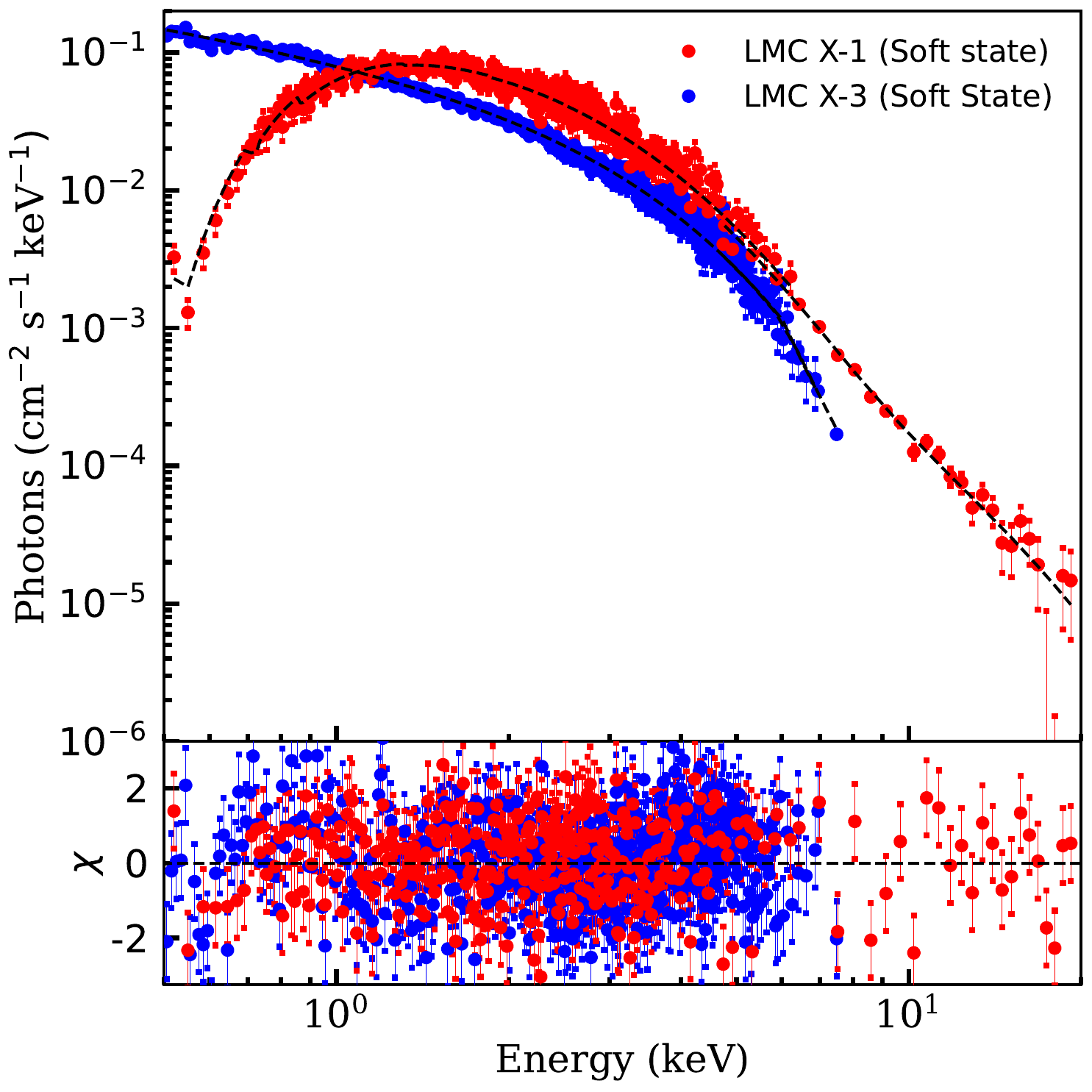}
    \caption{Best fit \textit{AstroSat} spectra of LMC X$-1$ ($0.5-20.0$ keV in red) and LMC X$-3$ ($0.5-7.0$ keV in blue) modelled using \texttt{tbabs$\times$(diskbb+nthcomp)}. The bottom panel shows the residual of the best fit. See text for details.
    }
    \label{fig:fig16}
\end{figure}

LMC X$-$1 is found to remain in the soft spectral state in all the observations considered in this study \cite[see][for details]{Bhuvana-etal2021,Bhuvana-etal2022}. The soft state spectrum is modelled using \texttt{tbabs$\times$(diskbb+nthcomp)}, as shown in Fig. \ref{fig:fig16}, and the corresponding best fit parameters are listed in Table \ref{tab:Table-11}. The hydrogen column density is found to be $n_{\rm H} \sim 1.22-1.48 \times 10^{22}$ atoms cm$^{-2}$. The spectral parameters remain largely consistent across observations, specifically, the disc temperature lies in the range $0.83-0.90$ keV, while $L_{\rm bol}$ remains in the range $\sim 8.01\%-11.43\%~{\rm L}_{\rm Edd}$. The $L_{\rm bol}$ is observed to decrease slightly with increase in disc temperature (see Table \ref{tab:Table-11}). 
Using {\tt kerrbb} model, the continuum fitting method yields a weighted mean spin value of $0.89 \pm 0.02$ for this source (see Table \ref{tab2} and \cite{Bhuvana-etal2021}). 

Although LMC X$-$3 is a persistent source, it was observed to undergo occasional transitions into hard and intermediate spectral states. However, during \textit{AstroSat} observations, the source consistently remains in the soft state. The best fit spectrum of LMC X$-$3 is shown in Fig. \ref{fig:fig16}, and the model parameters are listed in Table \ref{tab:Table-11}. The hydrogen column density is obtained as $n_{\rm H} \sim 0.04 \times 10^{22}$ atoms cm$^{-2}$. The disc component exhibits temperature slightly higher than LMC X$-$1, ranging from $0.99$ to $1.13$ keV, while the disc normalization remains largely consistent across observations. As expected, the flux remains approximately constant for persistent source. Meanwhile, the \texttt{kerrbb} model yields the weighted mean spin as $a_{\rm k} \sim 0.30 \pm 0.01$ (see Table \ref{tab2}).

\begin{figure*}
\centering
    \includegraphics[width=0.45\textwidth]{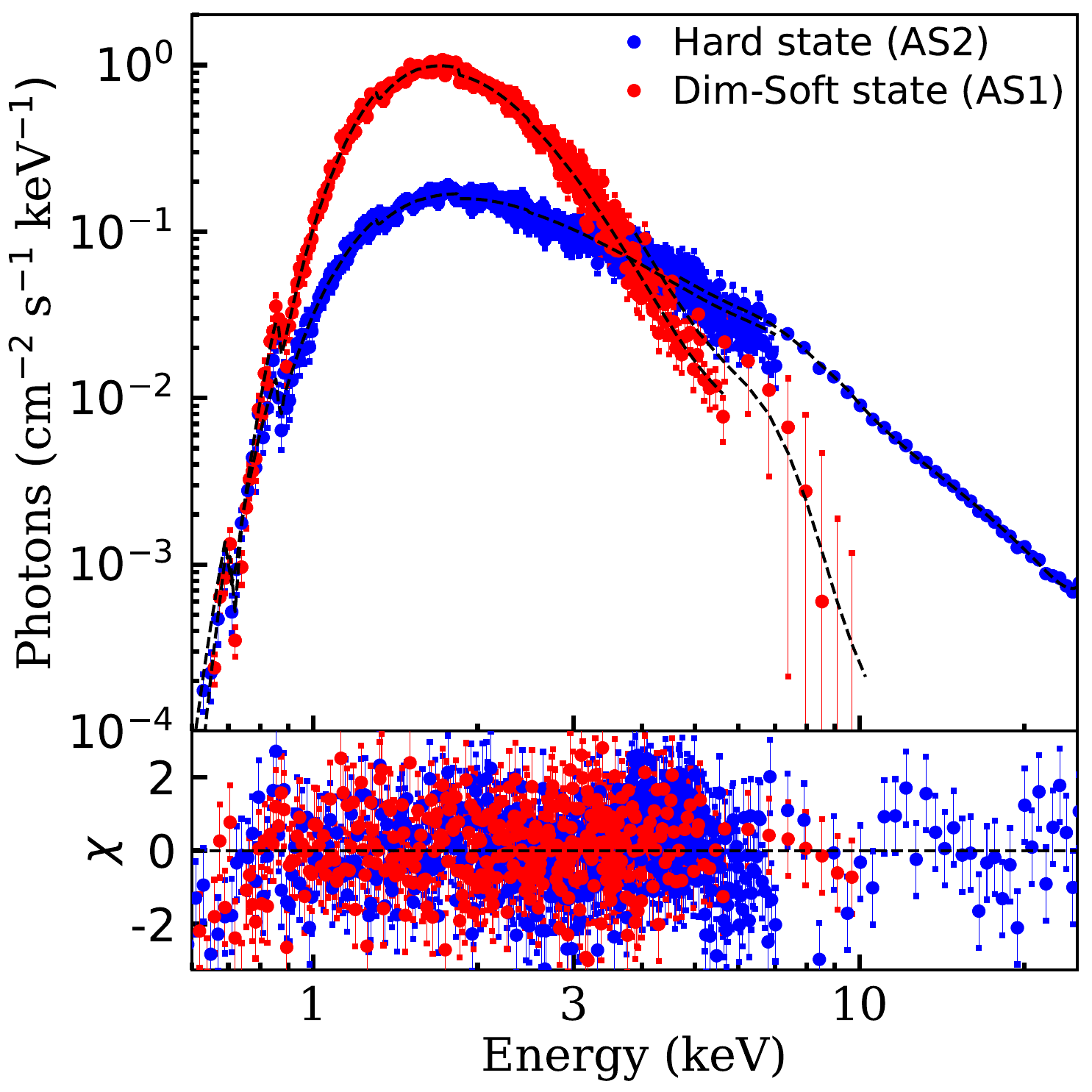}
    \includegraphics[width=0.45\textwidth]{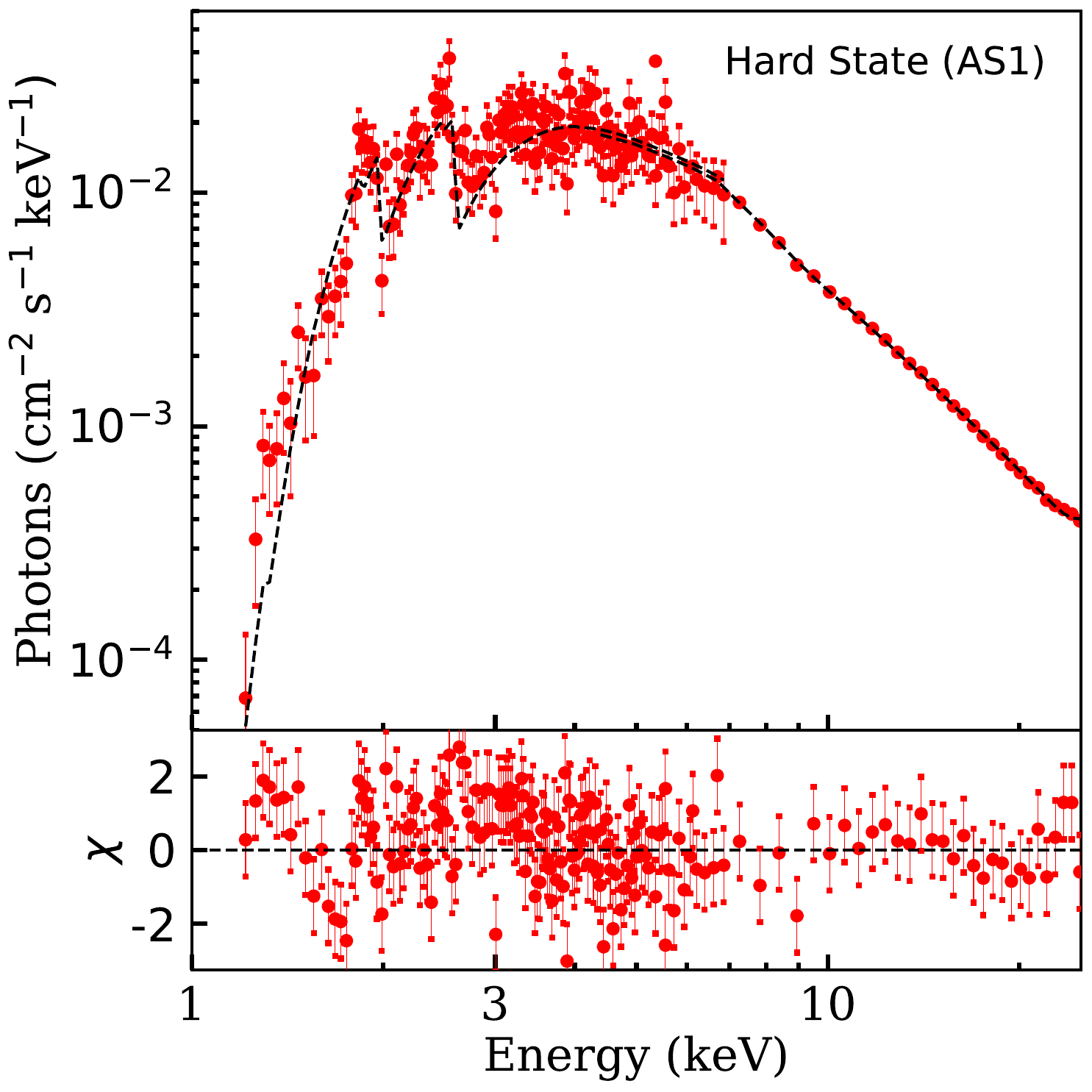}
    \caption{Best fit {\it AstroSat} spectra of GRS~1758$-$258 (left) and 1E~1740.7$-$2942 (right). The `dim-soft' and hard states of GRS~1758$-$258 are modelled with \texttt{tbabs$\times$(diskbb+nthcomp)} (red) and \texttt{nthcomp} (blue), respectively. The hard-state spectrum of 1E~1740.7$-$2942 is modelled with \texttt{tbabs$\times$nthcomp} (red). Residuals are shown in the lower panels. See text for details.
    } 
    \label{fig:fig17}
\end{figure*}

GRS 1758$-$258 was observed by \textit{AstroSat} in two spectral states, namely a hard state and the so called `dim–soft' state \cite[]{Bhuvana-etal2023}. The corresponding best fitted spectra are shown in Fig. \ref{fig:fig17}, and the model parameters are listed in Table \ref{tab:Table-11}. For this source, we obtain $n_{\rm H} \sim 2.4 \times 10^{22}$ atoms cm$^{-2}$. Thermal disc emission is detected only in the `dim–soft' state, where the spectrum is steep with photon index of $\sim 3.2$ (See Table \ref{tab:Table-11}) and disc temperature of $\sim 0.31$ keV. In contrast, the hard state is dominated by a Comptonized component accompanied by Fe emission features, and its spectral parameters remain largely unchanged across observations. Notably, the bolometric flux in the `dim–soft' state, generally expected to exceed that of the hard state, is instead found to be comparable, which is a defining property of this peculiar spectral state in GRS 1758$-$258. The presence of Fe lines enables reflection modelling using \texttt{relxill}, allowing constraints on the ionization parameter ($\log~\xi$), iron abundance ($A_{\rm Fe}$), inclination angle ($i$), and reflection fraction ($R_{f}$). The weighted mean values of these quantities are $\log \xi = 3.17 \pm 0.04$, $A_{\rm Fe} = 3.89 \pm 0.30$, $i = 61^\circ \pm 2^\circ$, and $R_{f} = 3.96 \pm 0.09$ \cite[see][for details]{Bhuvana-etal2023}. The weighted sum of the observed values ($\sum_i w_i x_i$) was computed by assigning weights $w_i = 1/\sigma_{\mathrm{sys},i}^2$, where $\sigma_{\mathrm{sys},i}$ denotes the systematic uncertainty of the $i$-th observation of each parameter. {\red During the \textit{AstroSat} observation, the source exhibited a bolometric luminosity of $\rm 1-6\%~L_{Edd}$. Considering the uncertainties in the mass ($\rm >8.6~M_{\odot}$) and distance ($8\pm1$ kpc), the luminosity is constrained to $\le 7.6~\%~{\rm L}_{\rm Edd}$.}

The spectral properties of 1E 1740.7$-$2942 closely resemble those of GRS 1758$-$258, except that the source remains consistently in the hard state and shows no evidence of state transitions within the \textit{AstroSat} observations. 1E 1740.7$-$2942 exhibits Fe line emission, and modelling with \texttt{relxill} yields constraints on the reflection parameters as $\log \xi = 3.17 \pm 0.06$, $i = 55.3^{\circ} \pm 1^{\circ}$, and $R_{f} = 3.21 \pm 0.08$. Unlike GRS 1758$-$258, no additional distant reflection component is detected. The best fitted reflection model parameters for both sources are listed in Table \ref{tab2}. Here, we obtain $n_{\rm H} \sim 10.8 \times 10^{22}$ atoms cm$^{-2}$. 

For 4U 1957$+$115, we perform a broad band ($0.7-20.0$ keV) spectral analysis as background dominates the spectrum above $\sim 20$ keV \cite[]{Mudambi-etal2022}. The fitted spectra yield $kT_{\rm in} \sim 1.30-1.60$ keV, $\Gamma_{\rm th}=2.0-2.9$ and $n_{\rm H} \sim 0.2 \times 10^{22}$ atoms cm$^{-2}$. The electron temperature cannot be constrained and is therefore fixed in the range $2-13$ keV. We obtain low covering fraction as $\sim 0.03-0.10$, which indicates that the emission is strongly disc dominated. {\red The bolometric luminosity during the \textit{AstroSat} observations is estimated to lie in the range $1.4-3.6~\%~{\rm L}_{\rm Edd}$. However, when the uncertainties in the mass ($\rm <6.2~M_{\odot}$) and distance ($20-40$ kpc) are taken into account, $L_{\rm bol}$ is constrained to $\rm >1.3\%~L_{Edd}$.} The best fitted spectral parameters are tabulated in Table \ref{tab:Table-11}.

\begin{figure*}
	\centering
	\includegraphics[width=0.9\textwidth]{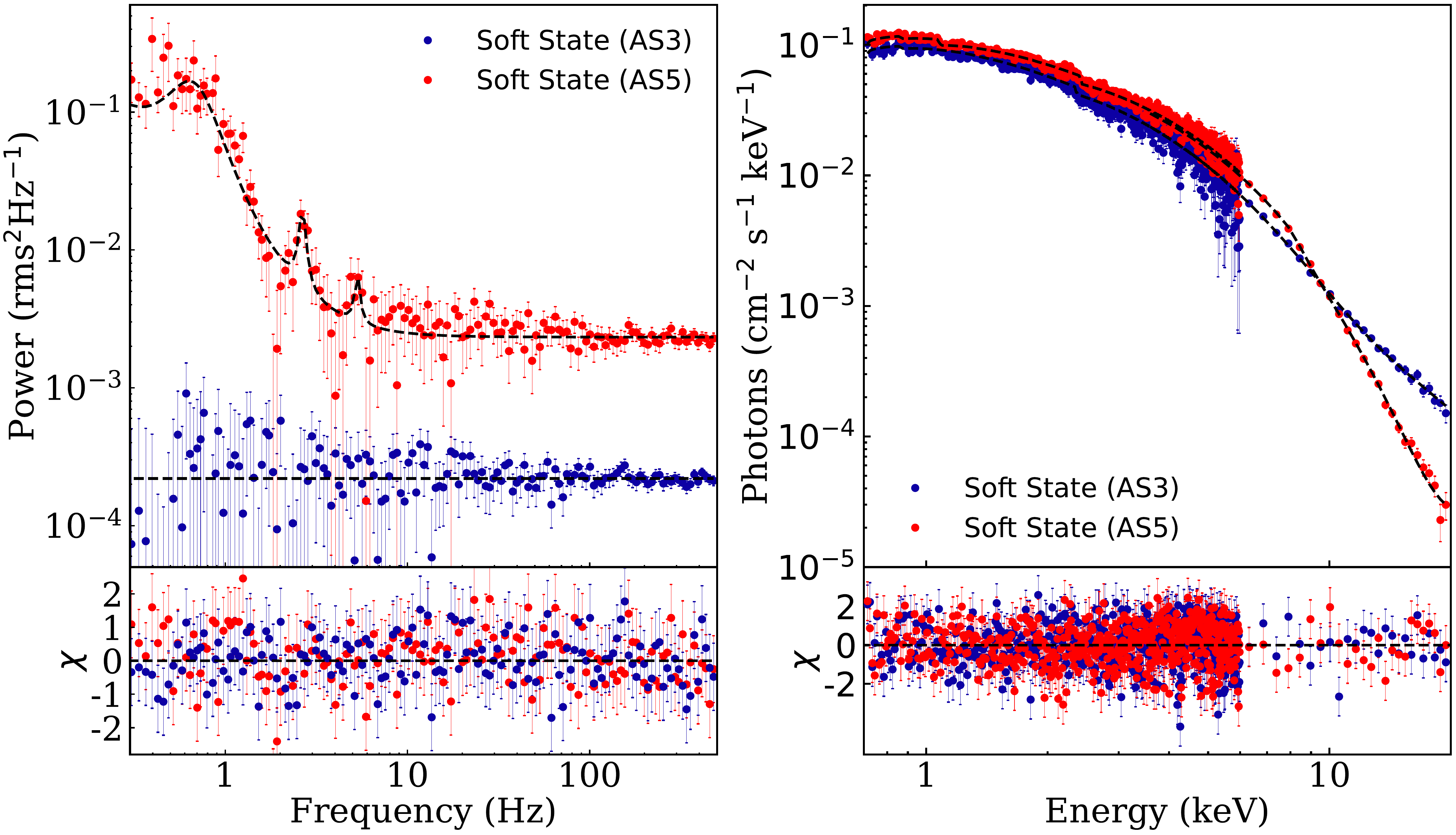}	
    \caption{\textit{Left:} Model fitted PDS of 4U 1957$+$115 using \textit{AstroSat} observations for AS3 (red) and AS5 (blue). \textit{Right:} Model fitted energy spectra of AS3 and AS5 using SXT and LAXPC data are plotted in red and blue. LAXPC spectra beyond $20$ keV is background dominated and hence, not considered. Residuals for respective model fits are shown in the bottom panel. See text for details.
    }
    \label{fig:fig18}
\end{figure*}

\begin{table*}
\caption{Wide-band spectral properties of persistent sources analysed using {\it AstroSat} data. The best fit spectral parameters are obtained using the model combination \texttt{tbabs$\times$(diskbb+nthcomp)} for LMC X$-1$, LMC X$-3$, GRS 1758$-$258 and 1E 1740.7$-$2942, while the model combination \texttt{tbabs$\times$(thcomp$\otimes$diskbb)} is employed for 4U 1957$+$115. The parameters $kT_{\rm in}$, $kT_{\rm e}$, $\Gamma_{\rm th}$, and cov$_{\rm frac}$ correspond to the inner disc temperature, electron temperature, photon index, and covering fraction, respectively. The flux components $F_{\rm disc}$ and $F_{\rm Comp}$ in $1-100$ keV represent the disc and Comptonized fluxes, and $L_{\rm bol}$ denotes the bolometric luminosity. Errors are calculated with $90\%$ confidence in model fitted parameters. See text for details.}
	
\renewcommand{\arraystretch}{1.4}
\resizebox{\textwidth}{!}{%
\begin{tabular}{l c c c c c c c c c c c c c c c} \hline

Source	& Epoch  & $kT_{\rm in}$ & $kT_{\rm e}$ & $\Gamma_{\rm th}$& cov$_{\rm frac}$ & $\chi^2\big/$dof & $F_{\rm disc}$ & $F_{\rm Comp}$ &  $L_{\rm bol}$ \\

& & (keV) & (keV) & &  &  & \multicolumn{2}{c}{($\tiny{10^{-10}}$ erg ${\rm cm}^{-2}$ ${\rm s}^{-1}$)}  & (${\%\rm L}_{\rm Edd}$) \\  \hline

LMC X$-$1 & AS1 & $0.83\pm0.01$ & $10.00^{*}$ & $4.08^{+0.11}_{-0.26}$ & $-$ & $700/529$  & $1.41$ & $1.31$ & $8.0$  &   \\
    & AS2 &$0.84\pm0.02$  & $40.00^{*}$ & $2.66\pm0.12$ &  $-$ & $606/417$   &$2.32$   &  $1.59$ & $11.4$ &   \\
	
    & AS3 & $0.90\pm0.01$ & $10.00^{*}$ & $4.98^{+0.84}_{-3.73}$ & $-$ &$548/480$  & $2.08$ & $0.44$  & $8.2$ &   \\
	
    & AS4 & $0.90\pm0.01$ & $4.00^{*}$ & $2.53^{+0.80}_{-0.39}$ & $-$ &$675/506$    & $2.20$ & $1.04$ & $9.4$ &   \\
    \hline
    
LMC X$-$3 & AS1 & $1.03\pm0.01$ & $-$ & $-$  & $-$ & $473/403$ & $2.30$  & $-$ & $7.9$  &   \\
	
    & AS2 & $1.03\pm0.01$ & $-$ & $-$ & $-$  & $456/387$  & $2.41$  & $-$  & $9.2$ &   \\
	 
    & AS3 & $0.94\pm0.01$ & $-$ & $-$ & $-$ & $446/417$ & $1.92$ & $-$ & $6.3$  &   \\
 	
    & AS4 & $0.98\pm0.01$ & $-$ & $-$  & $-$ & $338/318$   & $1.78$ & $-$ & $6.7$  &   \\
	
    & AS5 &$1.04\pm0.01$  & $-$ & $-$  & $-$ & $525/425$    & $5.34$  & $-$ & $9.2$ &   \\
    
    & AS6 & $1.07\pm0.01$ & $-$ & $-$ & $-$  & $450/452$   & $3.35$ & $-$ & $11.4$  &   \\
	
    & AS7 & $1.08\pm0.01$ & $-$  & $-$  & $-$  & $597/453$  & $3.83$ & $-$ & $12.1$ &   \\
	
    & AS8 & $1.04\pm0.01$ & $-$ & $-$  & $-$  & $533/447$ &  $3.72$ & $-$  & $12.4$ &   \\
	
    & AS9 & $0.98\pm0.01$ & $-$ & $-$  & $-$  & $670/480$   & $1.89$  & $-$  & $6.7$ &   \\
        
    \hline

GRS 1758$-$258$^{\dagger}$	& AS1 &$0.31\pm0.01$    & $1.00^{*}$      & $3.32^{+0.11}_{-0.14}$ &  $-$ &  $444/348$  &  $1.76$  & $0.79$ & 1.0&   \\
    
    & AS2 & $-$ &$4.70\pm0.10$   & $1.99\pm0.02$  &  $-$  &  $420/348$                 & $-$   & $38.1$ & $3.0$ &   \\
    
    & AS3 & $-$ &$5.40\pm0.10$  & $2.22\pm0.02$  & $-$   &  $431/345$                 &   $-$ & $46.0$ & $4.0$ &   \\
    
    & AS4 & $-$ & $3.50\pm0.10$  & $1.96\pm0.01$  &  $-$  &   $426/347$                &  $-$  &  $69.1$ & $6.0$ &     \\ 
    
    \hline
                    
1E 1740.7$-$2942 & AS1  & $-$ & $8.70\pm0.20$  & $2.32\pm0.01$           & $-$    &  $179/144 $ & $-$ & $2.26$ & $1.0$ &   \\
                      
    & AS2  & $-$ & $5.00\pm0.20$  &$1.81\pm0.20$           & $-$    &    $350/163$             & $-$  & $1.45$ &  $1.0$  &   \\
                      
    & AS3 & $-$  & $4.90\pm0.10$ &$1.97^{+0.02}_{-0.01}$  &   $-$      &    $267/168$             & $-$  & $1.48$  & $1.0$   &   \\
                      
    & AS4 & $-$  & $4.50\pm0.10$  &$1.99\pm0.01$ &  & $332/259$  &     $-$   & $1.49$          &  $1.0$  &   \\
                      
    & AS5 & $-$  & $50.00^{*}$      &$1.99^{+0.01}_{-0.05}$  &  $-$       &    $577/342$         & $-$   & $1.45$ &  $1.0$  &  \\
    
    \hline

4U 1957$+$115$^{\dagger}$ & AS1 &	 $1.63\pm0.03$&	 $5.00^*$&	$2.89\pm0.12$ &	$0.03^*$ &	$621/540$ &  $15.23$ &	$0.59$ & $2.4$\\
                     
    & AS2 &	$1.42\pm0.02$&	$2.00^*$&   $2.25\pm1.20$&  $0.05^*$	&$35/23$	   &   $5.83$& $0.17$&  $1.5$ \\
                     
    & AS3 &	$1.51\pm0.04$&	$6.05\pm2.30$&  $2.03\pm0.17$& $0.10^*$	&$630/549$&   $8.31$& $1.84$& $1.9$ \\
                     
    & AS4 &	$1.65\pm0.05$&	$10.00^*$& $2.34\pm0.23$& $0.10^*$	&$448/443$&    $11.40$& $1.42$& $1.5$ \\
                     
    & AS5 &	$1.37\pm0.02$&	$5.00^*$ &	$2.22\pm0.50$& $0.100\pm0.001$&	$589/545$& $6.82$ & $3.10$& $3.6$ \\
                    
    & AS7 &	$1.36\pm0.03$ &	$5.00^*$ & $2.52\pm0.22$ & $0.015\pm0.003$ & $374/331$ &  $8.16$ & $0.94$  & $1.4$ \\
 \hline
	\end{tabular}%
	}
\begin{tablenotes}
    \item $^{*}$Frozen values.
    \item $^{\dagger}${\red For the estimation of $L_{\rm bol}$, we consider $M_{\rm BH}=8.6~{\rm M}_{\odot}$ and $\rm D=8~kpc$ for GRS 1758$-$258, and $M_{\rm BH}=6~{\rm M}_{\odot}$ and $\rm D=10~kpc$ for 4U 1957$+$115.}
\end{tablenotes}

    \label{tab:Table-11}
\end{table*}

\subsubsection*{(c) UVIT study of LMC X$-$3}

Broad-band spectral energy distribution (SED) modelling of BH-XRBs using UVIT and SXT+LAXPC is challenging, particularly with imaging data, due to the limited number of UV data points. Continuous spectroscopic coverage enables more robust modelling across the UV and X-ray bands. Recently, \citet{Banerjee-etal2024} studied SED of MAXI J1820$+$070 using \textit{AstroSat} observations combining simultaneous or quasi-simultaneous multi-instrument data. The X-ray data constrained the inner accretion disc and corona, while the optical/UV data traced emission from the outer disk. The results show that X-ray irradiation dominates the optical/UV emission in both hard and soft states with a significantly higher reprocessing fraction in the hard state. Prominent UV emission lines ($e.g.$, Si IV, C IV, and He II), especially in the hard state, further support enhanced X-ray reprocessing.

Following \cite{Banerjee-etal2024}, we model the combined UVIT+SXT+LAXPC spectrum using the model \texttt{tbabs$\times$smedge$\times$(bbodyrad+diskbb+diskbb)}. Each component represents a physically distinct part of the spectrum. For example, \texttt{tbabs} accounts for interstellar absorption along the line of sight, \texttt{bbodyrad} is used to represent the emission from the B-type companion star (with a fixed temperature corresponding to $T \approx 15250 \pm 250$ K; \citealp{Orosz-etal2014}). The first \texttt{diskbb} component models the cooler outer accretion disc emission dominating in the UV band and the second \texttt{diskbb} describes the hotter inner accretion disc region responsible for the X-ray emission with temperature fixed at $kT_{\rm in}=1.03$ keV \cite[]{Bhuvana-etal2022}. Here, the normalization is allowed to vary to obtain best fit. The hydrogen column density ($n_{\rm H}$) is frozen at $0.02 \times 10^{-22}$ atoms cm$^{-2}$, and the blackbody temperature is also kept fixed. To account for the absorption like residuals at higher energies, we included a multiplicative \texttt{smedge} component, representing a smeared absorption edge due to reflection or partially ionized material in the accretion disc. The edge energy is allowed to vary, while the edge width is frozen to the best fitted value. The model fitted SED of LMC X$-$3 is shown in Fig. \ref{fig:fig20} along with model parameters presented in Table \ref{tab:Table-12}.

\begin{table}
\centering
\caption{Best fit multi-band spectral parameters of LMC X$-$3 of \textit{AstroSat} observations modelled with \texttt{tbabs$\times$(bbodyrad+diskbb+diskbb)$\times$smedge}. }
    
\begin{tabular}{lll} \hline
Component & Parameter & Value$^\dagger$ \\ \hline
\texttt{tbabs} 
    & $n_{\rm H}$ ($10^{22}$ atoms cm$^{-2}$) & 0.02 (frozen) \\ \hline
\texttt{bbodyrad} & $kT$ (keV) & $1.30\times10^{-3}$ (frozen) \\
    & norm & $(5.67 \pm 0.59)\times10^{10}$ \\ \hline
\texttt{diskbb (outer)}   & $T_{\rm in}$ (keV) & $(3.54 \pm 0.85)\times10^{-2}$ \\
    & norm & $(2.34 \pm 1.54)\times10^{6}$ \\ \hline
\texttt{diskbb (inner)} & $T_{\rm in}$ (keV) & 1.03 (frozen) \\
    & norm & $24.84 \pm 0.07$ \\ \hline
\texttt{smedge} & $E_{\rm edge}$ (keV) & $7.89 \pm 0.21$ \\
    & MaxTau & $7.88 \pm 2.75$ \\
    & width (keV) & 7.00 (frozen) \\ \hline
\texttt{gain} & slope & 1.00 (frozen) \\
    & offset (keV) & $0.028 \pm 0.002$ \\ \hline
\end{tabular}

\begin{tablenotes}
    \item $^\dagger$ Best fit yields higher $\chi^2_{\rm red} = 1.69$ for $446$ dof due to limited UV data.
\end{tablenotes}

\label{tab:Table-12}
\end{table}

\begin{figure}
    \begin{center}
    \includegraphics[width=0.9\columnwidth]{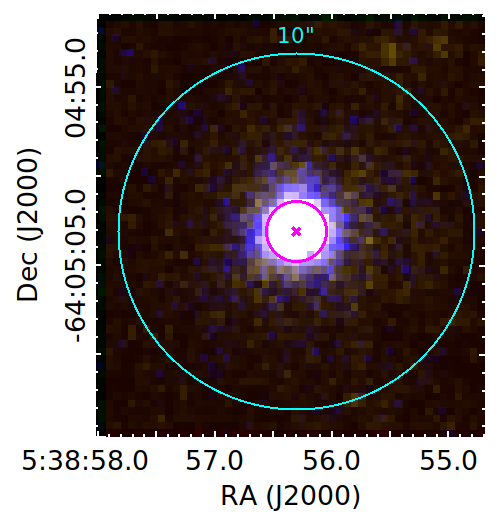}
    \end{center}
    \caption{Color composite image of LMC X$-$3 obtained using UVIT (red: NUV F5, green: NUV F5 and blue: FUV F1). A cyan circle with a radius of $10$ arcsec marks the region around the source, and the source centre is indicated by a magenta cross. At the centre (shown by the magenta circle of radius $1.7$ arcseconds), emission in both the NUV and FUV bands appears as a white composite region. Outside this region, distinct structures dominated by FUV (blue) and NUV (yellow) emission are observed around the source. See text for details.
    }
    \label{fig:fig19}
\end{figure}

\begin{figure}
    \begin{center}
    \includegraphics[width=0.9\columnwidth]{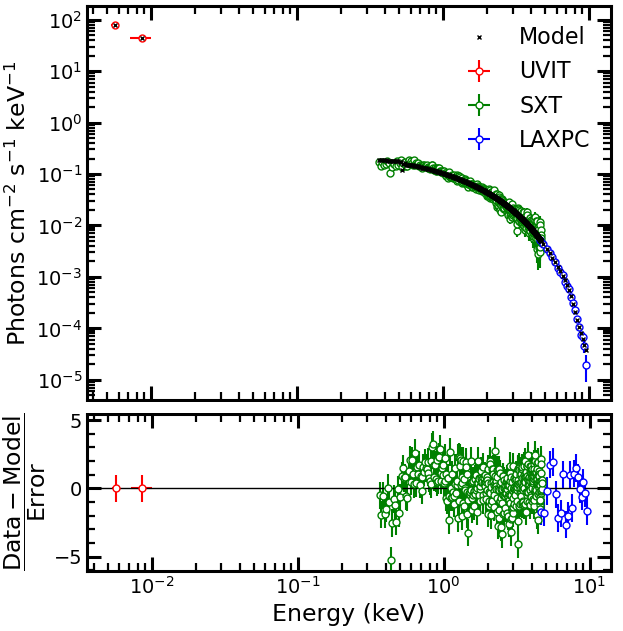}
    \end{center}
    \caption{Wide-band unfolded spectrum for LMC X$-$3. Red, green and blue circles represent UVIT, SXT and LAXPC data of \textit{AstroSat}. The best fit model fluxes are represented by the black crosses. See text for details.
    }
    \label{fig:fig20}
\end{figure}

\subsection{Transient Sources}\label{sec7:7.3_transi}

\begin{table*}
\caption{Temporal properties of transient sources observed with {\it AstroSat}. The notations and symbols have their usual meanings. See \S \ref{sec6:6.1_timing} and \S \ref{sec7:7.3_transi} for details.}
\renewcommand{\arraystretch}{1.4}
\resizebox{0.95\textwidth}{!}{%
\begin{tabular}{l c c c c c c c  c c c} \hline
Source & Epoch & $\nu_{\rm QPO}$ & FWHM & Significance & Q-factor  & rms$_{\rm QPO}\%$ & time-lag & rms$_{\rm total}\%$ & $\chi^2\big/$dof & QPO-Type \\
			
	& & (Hz) & (Hz) & ($\sigma$) &  & ($\%$) & (ms) &  &  & \\ \hline

IGR J17091$-$3624$^\ddagger$ & AS1 & $0.042_{-0.001}^{+0.001}$ & $0.012$ & $8.1$ & $3.5$ & $5.94\pm2.05$ & $-$ & $11.42\pm2.18$ & $583/472$ &  $-$ \\

&  & $0.082_{-0.003}^{+0.002}$ & $0.027$ & $8.3$ & $3.0$ & $3.24\pm1.12$ & $-$ & $-$ & $583/472$ &  $-$ \\

    & AS2 & $0.049_{-0.003}^{+0.003}$ & $0.016$  & $2.5$ & $3.1$ & $3.28\pm1.45$ & $-$ & $10.38\pm1.82$ & $443/533$ & $-$ \\ 

    &  & $0.084_{-0.002}^{+0.004}$ & $0.045$  & $5.3$ & $1.9$ & $3.77\pm1.15$ & $-$ & $-$ & $443/533$ & $-$ \\ 
    
    \hline
         
Swift J1727.8$-$1613$^{\dagger}$ & AS1 & $1.09_{-0.02}^{+0.02}$ & $0.05$ & 2.2 & 21.8 & $9.84\pm5.84$ & $-$ & $22.27\pm10.39$ & $44/44$ & C \\

    & AS2 & $1.43_{-0.02}^{+0.02}$ & $0.20$ & 5.5 & 7.2 & $10.95\pm2.14$ & $-$ & $21.88\pm1.71$ & $131/117$ & C \\

    & AS3  & $1.31_{-0.01}^{+0.01}$ & $0.25$ & 6.1 & 5.2 & $10.59\pm1.16$ & $-$ & $20.09\pm2.24$ & $145/115$ & C \\

    & AS4 & $1.45_{-0.01}^{+0.01}$ & $0.24$ & 24.6 & 6.1 & $9.55\pm0.97$ & $-$ & $19.44\pm1.15$ & $141/118$ & C \\
    
    & AS5 & $2.28_{-0.04}^{+0.03}$ & $0.33$ & 4.8 & 6.9 & $8.38\pm1.25$ & $-$ & $18.73\pm2.66$ & $154/115$ & C \\

    & AS6 & $2.48_{-0.03}^{+0.03}$ & $0.31$ & 5.3 & 8.0 & $8.11\pm1.09$ & $-$ & $17.85\pm1.13$ & $144/116$ & C \\

    & AS7 & $2.66_{-0.01}^{+0.01}$ & $0.53$ & 22.9 & 5.0 & $9.21\pm1.35$ & $-$ & $17.55\pm1.69$ & $130/115$ & C \\

    & AS8 & $1.89_{-0.01}^{+0.01}$ & $0.19$ & 9.5 & 9.9 & $5.57\pm0.84$ & $-$ & $17.71\pm3.58$ & $144/118$ & C \\ \hline
            
MAXI J0637$-$430 & AS1$-$AS4 & $-$ & $-$ & $-$  & $-$  & $-$  & $-$ & $4.40-8.30$ &  $-$ & $-$ \\ 

    \hline

MAXI J1535$-$571 & AS1 & 2.21 $\pm$  0.02 & 0.24 & 11.8  & 9.2  & 8.23 $\pm$ 0.49 & $2.9\pm 0.8$  &16.02 $\pm$1.14 &  116/118 & C   \\
			
	& AS2 & 2.65 $\pm$  0.01 & 0.31 & 13.8  & 8.5  & 8.51 $\pm$ 0.41 & $2.0\pm 0.6$ &15.82 $\pm$ 0.93  &  117/118 & C   \\				
	& AS3 & 2.54 $\pm$  0.01 & 0.29 & 9.2  & 8.8  & 8.42 $\pm$ 0.63 & $1.2\pm 0.9$  &15.51 $\pm$ 1.42  &  105/118 & C   \\	

	& AS4 & 2.56 $\pm$  0.01 & 0.27 & 9.9  & 9.5  & 8.46 $\pm$ 0.66 & $1.6\pm 0.8$  &15.41 $\pm$ 1.22  &  105/118 & C   \\			
    & AS5 & 2.08 $\pm$  0.01 & 0.18 & 9.9  & 11.6  & 8.42 $\pm$ 0.64 & $3.1\pm0.9$  &15.97 $\pm$ 1.10  &  129/118 & C   \\	
			
	& AS6 & 2.20 $\pm$  0.01 & 0.23 & 11.4  & 9.6  & 8.35 $\pm$ 0.42 & $3.7 \pm 0.8$  &15.74 $\pm$ 1.24  &  107/118 & C   \\	
			
    & AS7 & 2.17 $\pm$  0.01 & 0.23 & 13.7  & 9.4  & 8.44 $\pm$ 0.48 & $2.6\pm 0.6$  &15.98 $\pm$ 0.85  &  135/118 & C   \\							
	& AS8 & 2.31 $\pm$  0.01 & 0.24 & 8.6  & 9.6  & 8.73 $\pm$ 0.77 & $1.3\pm 1.0$  &15.70 $\pm$ 1.58  &  81/118 & C   \\				
	& AS9 & 2.59 $\pm$  0.01 & 0.30 & 15.5  & 8.6  & 8.86 $\pm$ 0.42 & $-1.2\pm 0.5$ &16.04 $\pm$ 1.04  &  198/119 & C   \\	
					
    & AS10 & 2.46 $\pm$  0.02 & 0.27 & 8.9  & 9.1  & 8.82 $\pm$ 0.69 & $-1.2\pm 1.0$ &15.40 $\pm$ 1.57  &  81/118 & C   \\	

    & AS11 & 2.11 $\pm$  0.02 & 0.22 & 10.0  & 9.6  & 8.89 $\pm$ 0.58 & 1.0 $\pm$ 0.8 &16.03 $\pm$ 1.61  &  159/118 & C   \\	
		
	\hline

Swift J1658.2$-$4242 & AS1 & $1.57\pm0.01$ & $0.20$ & $35.7$ & $7.9$ & $8.49\pm0.34$ & $-2.9\pm1.4$&$34.52\pm0.22$ & $152/227$& C \\
                    
    & AS2 & $1.78\pm0.01$ & $0.23$ & $16.4$ & $7.6$ & $9.13\pm0.69$ & $-8.0\pm3.5$&$35.22\pm0.51$ & $92/97$ & C \\
                    
    & AS3 & $6.58\pm0.04$ & $1.60$ & $18.4$ & $4.1$ & $5.12\pm0.26$ & $-3.5\pm2.7$& $10.62\pm0.25$& $214/157$& C \\
                    
    & AS4 & $7.00\pm0.01$ & $1.22$   & $12.2$ & $3.7$ & $3.64\pm0.23$ & $-6.0\pm2.8$&$8.87\pm0.71$ & $101/100$& A \\ 
                    
    & AS5 & $-$ & $-$ & $-$ & & $-$ & $-$ & $8.52\pm0.05$ & $101/100$ & $-$ \\ 
    
    & AS6 & $3.91\pm 0.04$ & $0.32$ & $19.0$ & $12.2$ & $2.15\pm0.18$ & $4.0\pm2.5$&$10.39\pm0.32$ &$142/101$ & B \\ 
    
    & AS7 & $4.17\pm0.01$ & $0.54$ & $20.1$ & $3.12$ & $3.12\pm0.34$ & $6.3\pm4.6$& $10.72\pm3.18$ & $146/99$ & B\\ 
                    
    & AS8 & $-$  & $-$  & $-$  & $-$  & $-$ & $-$ & $9.29\pm0.25$ &$101/100$ & $-$\\ 
    
    \hline

GRS 1716$-$249 &\red AS1 & $-$ & $-$ & $-$ & $-$ & $-$ & $-$&$34.54\pm0.33$ &$123/230$ & - \\

    & AS2 & $1.22\pm0.01$ & $0.33$ & $6.4$ & $3.7$ & $6.24\pm0.01$ & $5.4\pm3.5$&$33.43\pm0.22$ &$123/230$ & C \\
    
    & AS3  & $1.68\pm0.05$ & $0.33$ & $2.6$ & $4.7$ & $4.42\pm0.01$ & $10.4\pm3.3$&$35.10\pm0.34$ &$122/230$ & C \\
    
    \hline
    
MAXI J1348$-$630 & AS1 & $-$ & $-$ & $-$ & $-$ & $-$ & $-$ & $32.56\pm0.32$ & $138/233$ & $-$\\
    
    & AS2 & $-$ & $-$ & $-$ & $-$ & $-$ & $-$ & $25.97\pm0.95$ & $198/230$ & $-$ \\
    
    & AS3 & $0.94\pm0.01$ & $0.18$ & $5.2$ & $5.1$ & $4.73\pm0.01$ & $14.5\pm4.0$&$31.7\pm0.37$ & $122/230$& C \\ 
    
    \hline

MAXI J1803$-$298 & AS1 & $5.39\pm0.01$ & $0.55$ & $19.5$ & $9.8$ & $6.23\pm0.01$ & $-3.3\pm1.0$& $29.50\pm0.42$ &$227/230$ & C\\
    
    & AS2 & $5.97\pm0.02$ & $0.62$ & $20.6$ & $9.6$ & $5.61\pm0.01$ & $-4.1\pm1.1$& $\red 29.49\pm0.71$ &$199/233$ & C\\
    
    & AS3 & $5.54\pm0.02$ & $0.27$ & $2.9$ & $20.1$ & $2.72\pm0.01$ & $-4.9\pm1.0$& $\red 28.65\pm0.68$ &$189/235$ & C\\
    
    & AS4 & $5.46\pm0.02$ & $0.53$ & $7.5$ & $10.4$ & $4.06\pm0.01$ & $-4.9\pm1.0$& $30.00\pm0.70$ &$190/203$ & C\\
    
    & AS5 & $-$ & $-$ & $-$ & $-$ & $-$ & $-$ & $13.34\pm1.45$ &  $218/183$  & $-$ \\ 
    & AS6 & $0.68\pm0.03$ & $0.12$ & $2.7$ & $5.6$ & $2.38\pm0.02$ & $-$ & $31.34\pm1.45$ &  $131/236$  & A  \\ 
    
    \hline
4U 1630$-$472$^\ddagger$ & AS1 & $-$ & $-$ & $-$ & $-$ & $-$ & $-$ & $7.4\pm0.1$ & 110/236 & $-$ \\ 
    & AS2 & $-$ & $-$ & $-$ & $-$ & $-$ & $-$ & $2.5\pm0.1$ & 191/246 & $-$ \\
    & AS3$-$AS6 & $-$ & $-$ & $-$ & $-$ & $-$ & $-$ & $7.1\pm0.1$ & $1.10-1.32$ &$-$  \\
    & AS7$-$AS9 & $-$ & $-$ & $-$ & $-$ & $-$ & $-$ & $\le 4.0$ & $0.95-1.21$ & $-$ \\ 
    & AS10$-$AS13 & $-$ & $-$ & $-$ & $-$ & $-$ & $-$ & $7.1\pm0.1$ & $1.03-1.27$ &$-$ \\

  \hline
\end{tabular}%
}
\begin{tablenotes}
    \item $^{\dagger}$Time-lag estimation for Swift J1727.8$-$1613 is not performed due to the lack of low energy ($\lesssim 20$ keV) data from {\it AstroSat}. 
    \item $^\ddagger$rms estimates are expected to be affected by the contamination of nearby source in LAXPC FoV. rms$_{\rm QPO}$ is computed in the frequency range of $\nu_{\rm QPO}\pm {\rm FWHM}$ for IGR J17091$-$3624.
\end{tablenotes}

\label{tab:Table-13}
\end{table*}

\begin{table*}
\ContinuedFloat
\caption{Continued.}
\renewcommand{\arraystretch}{1.4}
\resizebox{\textwidth}{!}{%
\begin{tabular}{l c c c c c c c c c c} \hline

Source & Epoch & $\nu_{\rm QPO}$ & FWHM & Significance & Q-factor  & rms$_{\rm QPO}\%$ & time-lag & rms$_{\rm total}\%$ & $\chi^2\big/$dof & QPO-Type \\
			
	& & (Hz) & (Hz) & ($\sigma$) &  & ($\%$) & (ms) &  &  & \\ \hline
    4U 1630$-$472$^\ddagger$& AS14 & $-$ & $-$ & $-$ & $-$ & $-$ & $-$ & $4.15\pm0.22$  &  $70/54$ & $-$ \\ 
    & AS15 & $7.94\pm0.09$ & $1.48$ & $6.8$ & $5.2$ & $2.51\pm0.01$ & $-1.6\pm0.9$ & $22.16\pm0.22$  &  $201/161$ & B \\
    & AS16 & $4.79\pm0.09$ & $0.14$ & $3.8$ & $32.9$ & $0.96\pm0.01$ & $-0.4\pm0.3$ & $22.16\pm0.22$  &  $201/161$ & B \\
    & AS17 & $-$ & $-$ & $-$ & $-$ & $-$ & $-$ & $7.1\pm0.1$ & $151/145$ &$-$ \\  
  \hline   
4U 1543$-$47 & AS1$-$AS5 & $-$ & $-$ & $-$ & $-$ & $-$ & $-$ & $1.20-4.90$ & $1.12-1.35$ & $-$   \\            
    & AS6 & $2.45\pm0.08$ & $0.33$ & 2.6 & 7.4 & $0.43_{-0.01}^{+0.02}$  & $-$ & $1.90\pm0.60$ & $158/122$ & B  \\     \hline
    
MAXI J1820$+$070 & AS1 & $0.05_{-0.001}^{+0.002}$ & $0.009$ & 9.1 & 4.1 & $9.10\pm0.30$ & $-$ &  $28.77\pm0.10$& $110/100$ & C   \\ \hline

    \rowcolor{lightgray}           
GX 339$-$4 & AS1 & $-$ & $-$ &$-$  &$-$  & $-$ & $-$  &$2.82 \pm 0.02$  & $111/122$ & $-$  \\
    \rowcolor{lightgray}
    & AS2 & $-$ &$-$  & $-$ &$-$  & $-$ & $-$ & $9.93 \pm 0.01$ & $241/184$ &$-$    \\
    & AS3 & $-$ &$-$  & $-$ & $-$ & $-$ & $-$  &$14.82 \pm 0.01$  & $196/183$ & $-$   \\
    & AS4 & $-$  &$-$  & $-$ & $-$ & $-$ & $-$ & $15.220 \pm 0.002$ & $368/372$ &$-$    \\
    & AS5$^\dagger$ & $0.097 \pm 0.002$  & $0.02$ & $4.9$ &$5.7$  &$5.53 \pm 0.32$  & $-$ & $26.67 \pm 1.10$ &$519/499$  & C   \\
    & AS6$^\dagger$  &$0.18 \pm 0.01$  & $0.03$ &$5.7$  & $6.0$ &$4.66 \pm 0.23 $  & $-$ & $24.10 \pm 0.55$ &$582/451$  & C     \\
    & AS7 & $-$ &$-$  & $-$ & $-$ & $-$ & $-$ & $22.58 \pm 0.23$ &$443/372$  & $-$   \\
    & AS8 &  $-$  & $-$   &  $-$  & $-$   &$-$  & $-$ & $22.28 \pm 1.22$ & $179/185$ & $-$     \\
    & AS9 & $-$  &  $-$ &  $-$ &  $-$ & $-$  & $-$  & $22.61 \pm 0.60 $ &$359/307$  &  $-$   \\
    & AS10 &$5.37 \pm 0.02 $  & $1.21$ & $3.7$ &$4.4$  & $2.08 \pm 0.22$ & $8.0 \pm 0.7$  &$12.21 \pm 0.20$  &$205/174$  & B    \\
    & AS11 &$6.30^{+0.01}_{-0.02} $  &$0.06$  & $22.0$ & $9.0$ &$4.48 \pm 0.05 $  &$3.3 \pm 0.0$ &$5.03 \pm 0.55$  & $188/179$ & B   \\
	& AS12 & $-$ &$-$  & $-$ &$-$  & $-$ & $-$ &  $1.77 \pm 0.05$&$215/219$  & $-$   \\
	& AS13 & $-$ & $-$ &$-$  &$-$  & $-$ & $-$ & $4.53 \pm 0.12$ &$351/285$  &  $-$  \\
    & AS14  &  $-$ &  $-$ & $-$  & $-$  &  $-$ & $-$ &$4.09 \pm 0.34$  &$454/372$  &  $-$   \\
    & AS15 &$-$  & $-$ & $-$ & $-$ &$-$   & $-$ & $4.43 \pm 0.01 $ & $402/372$ &$-$    \\
	& AS16 & $-$ & $-$ &$-$  & $-$ &$-$  & $-$ &$2.26 \pm 0.01$  & $363/374$ &$-$    \\
	& AS17 & $-$ &$-$  &$-$  &$-$  &$-$  & $-$ & $0.75 \pm 0.01 $ &$389/374$  & $-$   \\
	& AS18 &$-$  & $-$ & $-$ & $-$ &$-$  & $-$ &$33.01 \pm 0.33$  & $132/140$ & $-$   \\ 
    & AS19 & $-$ & $-$ &$-$  & $-$ &$-$  & $-$ & $37.14 \pm 0.22 $ &$95/109$  & $-$   \\
	
    \rowcolor{lightgray}
	
    & AS20 &$-$  &  $-$ & $-$  &$-$   &  $-$ & $-$ & $0.180 \pm 0.002$  & $140/127$ & $-$   \\
	
    & AS21 & $ 4.43 \pm 0.31$ &$0.64$  &$6.1$  &$6.9$  & $9.97 \pm 0.40$ & $8.0 \pm 0.8$ & $47.10 \pm 1.31 $ &$200/146$  & B   \\
	
    & AS22 &  $-$  &  $-$  &   $-$ & $-$   &  $-$  & $-$  &$1.72 \pm 0.02$  &$101/127$  & $-$  \\
	
    & AS23 & $-$ & $-$ &$-$  & $-$ & $-$ & $-$ & $1.22 \pm 0.02$ &$170/160$  & $-$   \\
	
    \hline
					
H 1743$-$322 & AS1 & $0.600 \pm 0.002$ & $0.14$ & $17.9$ &  $4.2$ & $12.21 \pm 0.03$&$-4.3 \pm 3.4$ &  $23.63 \pm 0.35$ & $263/193$ & C   \\
	
    \rowcolor{lightgray}
	
    & AS2 & $-$  & $-$  & $-$  & $-$  &$-$  & - &$2.160 \pm 0.001 $  & $378/378$ &  $-$   \\
	
    \rowcolor{lightgray}
	
    & AS3 & $-$  &  $-$&$-$  &$-$  &$-$  & - &$1.41 \pm 0.23 $  & $321/379$ &  $-$  \\
	
    & AS4 & $0.434 \pm 0.002 $ &$0.05$  &$19.7$  & $8.5$ &$10.12^{+0.02}_{-0.03}$  &$-3.3 \pm 2.3$  &$21.27 \pm 0.04 $  & $224/185$ &  C  \\	\hline

{\red Swift J151857.0$-$572147}$^\boxdot$  & AS1 & $8.18\pm0.04$  & $2.59$  & $28.2$  & $3.2$  &$-$  & $-$ & $-$ & $146/113$ &  C   \\

& AS2 & $-$  & $-$  & $-$  & $-$  &$-$  & $-$ & $-$ & $116/113$ &  $-$   \\
\hline
                          
\end{tabular}%
}
\begin{tablenotes}
    \item Observations of GX 339$-$4 and H 1743$-$322 in QP are highlighted in light gray.
    \item $^\dagger$Time-lag is not reliably constrained due to poor statistics resulted from the limited number of large segments.
    \item $^\boxdot${\red QPO parameters are obtained in the $3-25$ keV energy band. Due to possible residual contamination from nearby Cir X$-$1 source, rms$_{\rm QPO}\%$, rms$_{\rm total}$ and time-lag values are not determined.}
\end{tablenotes}

\label{tab:Table-13}
\end{table*}

\subsubsection{Swift J1727.8$-$1613}

\begin{figure*}
    \begin{center}
    \includegraphics[width=\textwidth]{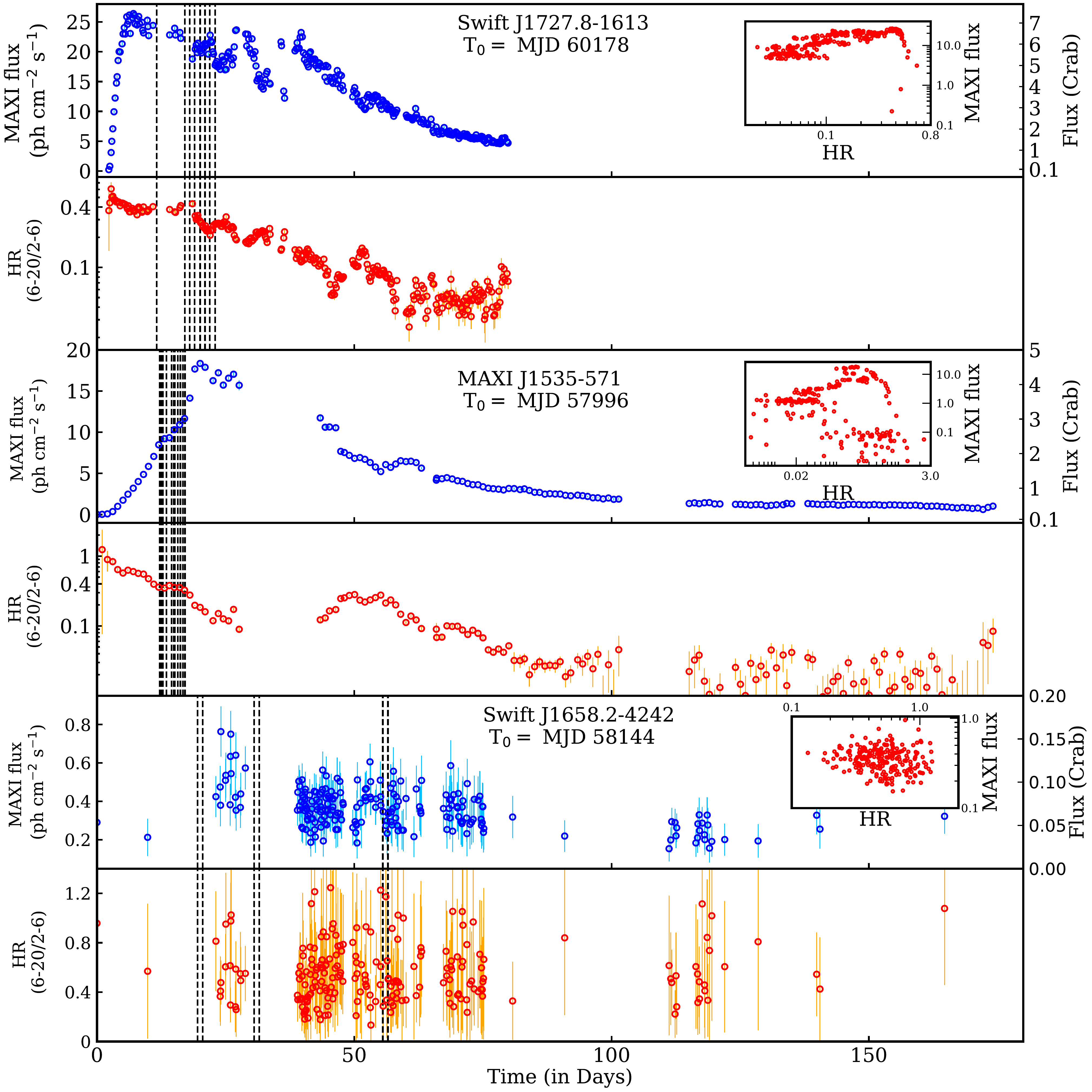}
    \end{center}
    \caption{{\it MAXI}/GSC monitoring of the transients Swift J1727.8$-$1613, MAXI J1535$-$571 and Swift J1658.2$-$4242 in $2-20$ keV energy band during their $2023$, $2017$ and $2018$ outbursts, respectively. The beginning of the outburst for each sources is mentioned as T$_0$. Variation of HR observed with {\it MAXI}/GSC is plotted for all sources. The vertical dashed lines indicate the epochs of {\it AstroSat} observations. The lightcurve and HR are obtained from 1-day averaged {\it MAXI}/GSC data from the launch date of {\it AstroSat}. HIDs of each sources are shown in the corresponding inset. See text for details.
    }
    \label{fig:fig21}
\end{figure*}

\subsubsection*{(a) Timing Variability Properties}

\begin{figure*}
    \begin{center}
    \includegraphics[width=0.45\textwidth]{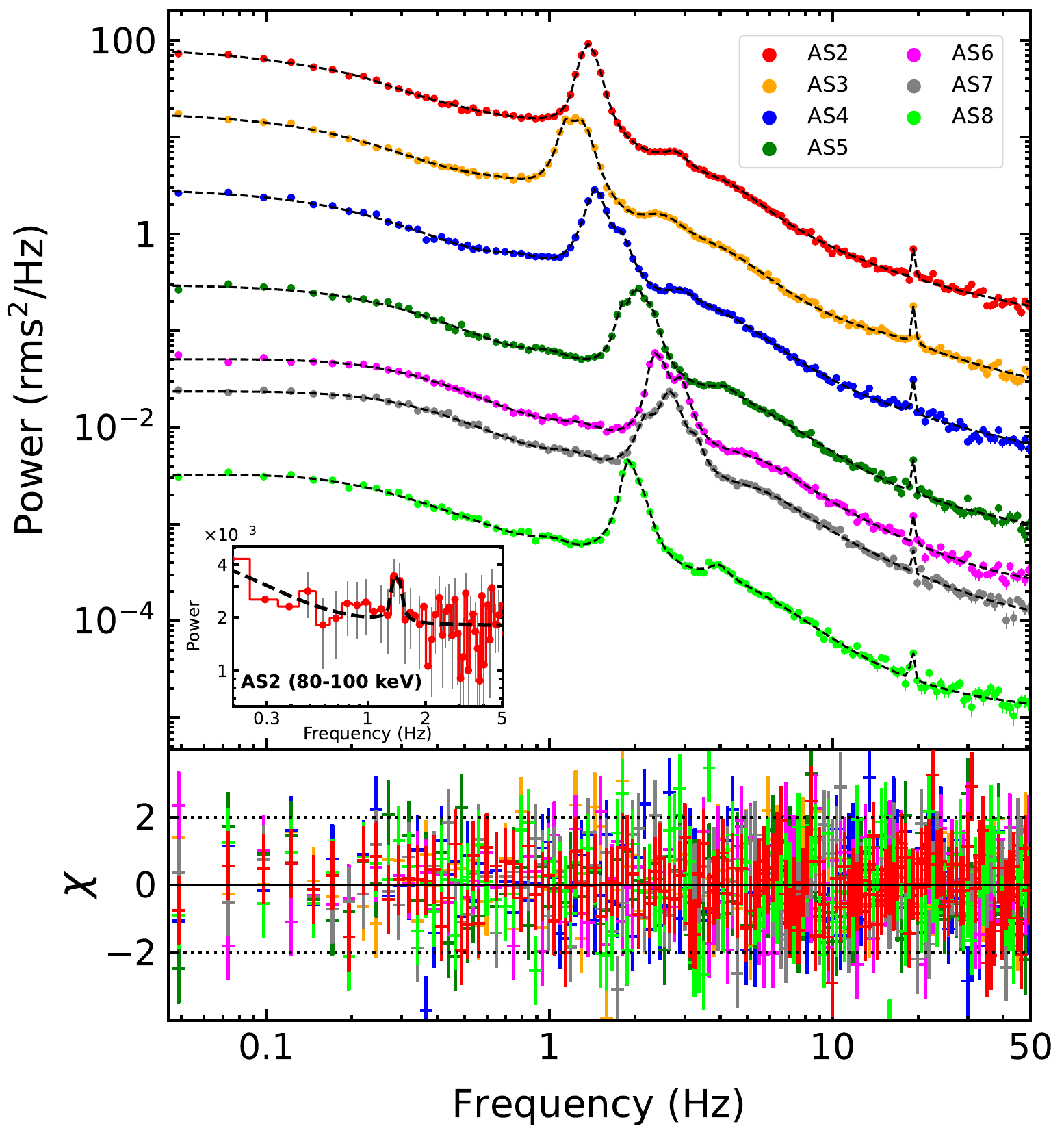}
    \includegraphics[width=0.45\textwidth]{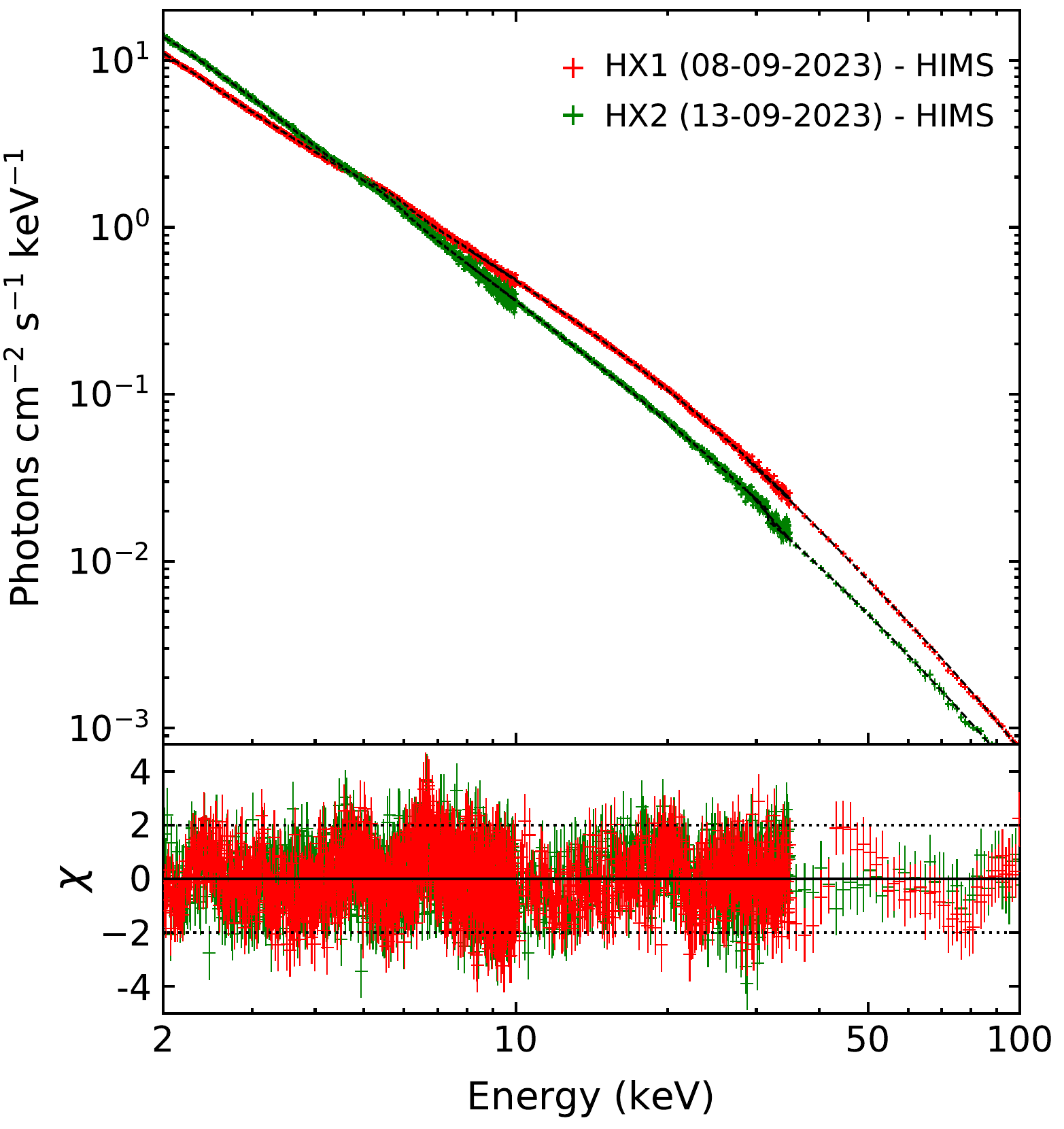}
    \end{center}
    \caption{{\it Left:} Best fit PDS of Swift J1727.8$-$1613 in $20-100$ keV energy range over different epochs, obtained from {\it AstroSat} observations. Each PDS is modelled with a combination of \texttt{powerlaw} and \texttt{Lorentzian} components. At the inset, PDS in $80-100$ keV energy band is shown for AS2 observation, displaying a clear detection of a QPO at $\sim 1.4$ Hz. {\it Right:} Best fit energy spectra of Swift J1727.8$-$1613 obtained from {\it HXMT} observations during epochs HX1 and HX2, modelled with \texttt{constant$\times$tbabs$\times$(thcomp$\otimes$diskbb+cutoffpl)}. See text for details.
    }
    \label{fig:fig22}
\end{figure*}

The outburst profile of the source observed with {\it MAXI} and the corresponding HR variation are shown in Fig. \ref{fig:fig21}. The \textit{AstroSat} and \textit{HXMT} observations marked in the figure correspond to HIMS during the outburst. The outburst profile of Swift J1727.8$-$1613 resembles the typical FRSD profile observed in BH-XRB transients and $F_{\rm var}$ varies as $7.2-10.2\%$ across spectral states.

We further investigate the power spectral properties using {\it AstroSat} data in $20-100$ keV band. The source exhibits an evolving type-C LFQPO, with the centroid frequency increasing from $1.09-2.66$ Hz, while the fractional rms ranges as $5.57-10.95\%$. The best fitted QPO parameters across all epochs are tabulated in Table \ref{tab:Table-13}. A strong energy dependence of these QPOs is reported by \cite{Nandi-etal2024}, which confirms the clear detection of QPOs in the hard X-ray band of $80-100$ keV (see inset of Fig. \ref{fig:fig22}). In addition, a QPO like feature at $\sim 19$ Hz is detected in the {\it AstroSat} observations during epochs AS2$-$AS8 \cite[see also][]{Nandi-etal2024}. The left panel of Fig. \ref{fig:fig22} presents the best fitted PDS of the source in the $20-100$ keV band for all epochs.

\subsubsection*{(b) Spectral Variability Properties}

We investigate the spectral properties of Swift J1727.8$-$1613 using \textit{HXMT} observations in $2-100$ keV band, incorporating data from the LE, ME, and HE instruments. We refrain analysing LAXPC10 spectral data as the source was observed in low gain mode \cite[]{Nandi-etal2024}. The details of the spectral modelling are provided in \S\ref{sec6:6.2_spec}. The best fitted results indicate a Comptonization dominated emission scenario ($F_{\rm Comp} > F_{\rm disc}$) with the covering fraction spanning $0.42-0.79$. The spectral photon index lies in the intermediate regime ($\Gamma_{\rm th} \sim 1.95-2.31$). The electron temperature varies between $6.33-9.63$ keV, while the inner disc temperature remains relatively low ($kT_{\rm in} \sim 0.38-0.52$ keV). {\red The bolometric luminosity of the source is estimated to lie in the range $\sim 17.8-20.1 \%~{\rm L}_{\rm Edd}$. Considering the uncertainty in the mass ($\rm >3.12~M_{\odot}$; see Table \ref{tab:Table-01}), however, $L_{\rm bol}$ is constrained to $\rm <24.8\%~L_{Edd}$, with no lower bound.} These properties are consistent with the canonical HIMS generally observed in transient BH-XRBs during outburst. The best fitted spectra for epochs HX1 and HX2 are shown in the right panel of Fig. \ref{fig:fig22}. 

\begin{table*}
\caption{Wide-band spectral properties of transient sources analysed using {\it AstroSat} data. The best fit spectral parameters are obtained using the model combination \texttt{tbabs$\times$(thcomp$\otimes$diskbb)}. The parameters $kT_{\rm in}$, $kT_{\rm e}$, $\Gamma_{\rm th}$, and cov$_{\rm frac}$ correspond to the inner disc temperature, electron temperature, photon index, and covering fraction, respectively. The flux components $F_{\rm disc}$ and $F_{\rm Comp}$ in $1-100$ keV represent the disc and Comptonized fluxes, and $L_{\rm bol}$ denotes the bolometric luminosity. Errors in model fitted parameters are calculated with $90\%$ confidence. See text for details.}
\renewcommand{\arraystretch}{1.4}
\resizebox{0.97\textwidth}{!}{%
\begin{tabular}{l c c c c c c c c c c c c c} \hline

Source & Epoch  & $kT_{\rm in}$ & $kT_{\rm e}$ & $\Gamma_{\rm th}$ & cov$_{\rm frac}$ & $\chi^2\big/$dof & $F_{\rm disc}$ & $F_{\rm Comp}$ & $L_{\rm bol}$ \\
& & (keV) & (keV) & & & & \multicolumn{2}{c}{(${10^{-8}}$ erg ${\rm cm}^{-2}$ ${\rm s}^{-1}$)} & ($\% {\rm L}_{\rm Edd}$) \\       \hline

IGR J17091$-$3624$^\dagger$ & AS1 & $1.97_{-0.10}^{+0.04}$ & $4.62_{-0.86}^{+0.43}$ & $2.51_{-0.22}^{+0.10}$ & $0.30^{*}$ & 425/378 & 1.41 & 0.16 & 18.0 \\
    
    & AS2  & $1.96_{-0.10}^{+0.08}$ & $7.00^{*}$ & $2.97_{-0.27}^{+0.48}$ & $0.31_{-0.11}^{+0.26}$ & 512/456 & 1.07 & 0.16 & 14.0 \\
        
	\hline
			
Swift J1727.8$-$1613 $^\dagger$ & HX1  & $0.38_{-0.01}^{+0.01}$ & $6.33_{-0.04}^{+0.08}$ & $1.95_{-0.02}^{+0.02}$ & $0.42_{-0.01}^{+0.01}$ & 1771/1455 & 2.92 & 9.75 & 20.1 \\
    
    & HX2 &  $0.38_{-0.01}^{+0.02}$ & $6.73_{-0.23}^{+0.18}$ & $1.96_{-0.05}^{+0.04}$ & $0.41_{-0.05}^{+0.05}$ & 1454/1455 & 2.66 & 8.56 & 18.7 \\
    
    & HX3 &  $0.42_{-0.02}^{+0.02}$ & $8.86_{-1.16}^{+1.24}$ & $2.17_{-0.04}^{+0.05}$ & $0.76_{-0.03}^{+0.02}$ & 1516/1455 & 5.24 & 18.27 & 17.8 \\
    
    & HX4 &  $0.52_{-0.02}^{+0.01}$ & $9.63_{-1.34}^{+1.44}$ & $2.31_{-0.05}^{+0.06}$ & $0.79_{-0.02}^{+0.03}$ & 1637/1455 & 7.27 & 15.82 & 18.1 \\
    
    \hline

MAXI J0637$-$430  & AS1 &  $0.63\pm0.06$ & $150.00^{*}$  & $2.32\pm0.08$ & $0.07\pm0.01$ & $486/291$ & 0.25 & 0.26 & 6.0  &   \\

    & AS2 & $0.58\pm0.05$  & $100.00^{*}$ & $2.33\pm0.09$  & $0.04\pm0.01$  & $508/397$ & 0.20 & 0.20 & 3.0 &   \\

    & AS3 &  $0.54\pm0.04$  & $10.00^{*}$ & $2.56\pm0.05$  & $0.28\pm0.02$  & $408/633$ & 0.11 & 0.09 & 3.0 &   \\

    & AS4 &  $0.53\pm0.01$  & $50.00^{*}$  & $2.58\pm0.09$  & $0.16\pm0.04$ & $478/393$ & 0.13 & 0.11 & 3.0 &   \\

    \hline 

MAXI J1535$-$571$^\dagger$ & AS1 & 0.25 $\pm$ 0.01  & 52 $\pm$ 10  & 2.28 $\pm$ 0.01  & 0.30 $\pm$ 0.04  & 1099/894   & 6.60 & 21.30 & 51.0 &   \\
			
    & AS2 &  0.24 $\pm$ 0.01  & $150^*$  & 2.36 $\pm$ 0.01  & 0.36 $\pm$ 0.04  & 1036/894 &7.07& 22.90 & 56.0 &   \\
	
    & AS3 &  0.27 $\pm$ 0.01  & $150^*$  & 2.31 $\pm$ 0.01  & 0.99 $\pm$ 0.40  & 1071/894  &2.13&16.90& 38.0 &   \\
			
    & AS4   &  0.24 $\pm$ 0.01  & $150^*$  & 2.34 $\pm$ 0.01  & 0.40 $\pm$ 0.05  & 1158/894  &5.37&22.30 & 57.0 &   \\
			
    & AS5& 0.24 $\pm$ 0.01  & 82 $\pm$ 48  & 2.27 $\pm$ 0.01  & 0.32 $\pm$ 0.05  & 981/894   &5.49&22.30& 58.0 &   \\
			
    & AS6  &  0.26 $\pm$ 0.01  & 48 $\pm$ 10  & 2.29 $\pm$ 0.01  & 0.35 $\pm$ 0.06  & 905/865   &5.75&21.08& 57.0 &   \\	
			
    & AS7  & 0.24 $\pm$ 0.01  & 27 $\pm$ 4  & 2.27 $\pm$ 0.01  & 0.30 $\pm$ 0.05  & 1046/896    &3.54&16.50& 43.0 &   \\
			
    & AS8  & 0.24 $\pm$ 0.01  & $150^*$  & 2.32 $\pm$ 0.01  & 0.39 $\pm$ 0.06  & 1073/894  &6.16 &24.50& 63.0 &   \\	
			
    & AS9 & 0.24 $\pm$ 0.01  & 34 $\pm$ 6  & 2.32 $\pm$ 0.01  & 0.38 $\pm$ 0.05  & 1053/894   &5.37&24.50& 63.0 &   \\
	
    & AS10  & 0.24 $\pm$ 0.01  & $150^*$  & 2.34 $\pm$ 0.01  & 0.34 $\pm$ 0.05  & 1153/894  &7.41&27.50& 71.0 &   \\	
    \hline
    
Swift J1658.2$-$4242 $^\dagger$ & AS1 & $1.21\pm0.19$ & $17.60\pm2.78$  &$1.91\pm0.02$ & $0.57\pm0.08$ &  $605/574$   & $0.21$& $0.28$& $1.3$ \\
                    
    & AS2 &  $1.51\pm0.17$ & $10.01\pm1.01$  &$1.75\pm0.07$ & $0.49\pm0.07$ &  $605/574$   & $0.22$& $0.31$& $1.4$ \\
                    
    & AS3 &  $0.90\pm0.17$  & $15.89\pm10.50$ &$2.17\pm0.13$ & $0.31\pm0.07$&	 $618/522$& $0.62$& $0.40$& $2.6$ \\
    
    & AS4 &  $1.11\pm0.11$  & $18.60\pm4.80$    &$2.12\pm0.01$ & $0.29\pm0.65$    &	 $618/522$& $0.70$&  $0.42$& $2.9$ \\
                    
    & AS5 &  $1.03\pm0.09$  & $12.29\pm6.55$  &$1.93\pm0.05$ & $0.15^*$     &	 $683/522$& $0.15$& $0.06$& $0.5$ \\
    
    & AS6 &  $1.21\pm0.09$  & $34.86\pm9.55$  &$1.98\pm0.03$ & $0.25^*$     &	 $605/574$& $0.54$& $0.35$& $2.3$ \\
    
    & AS7 &  $1.38\pm0.25$  & $25.00^*$ &$1.93\pm0.05$ & $0.25\pm0.02$     &	 $683/522$& $0.62$& $0.35$& $2.1$ \\
                    
    & AS8 & $0.90\pm0.11$  & $20.00^*$  &$2.02\pm0.04$ & $0.25\pm0.03$     &	 $618/522$& $0.51$& $0.36$& $2.2$ \\

    \hline

GRS 1716$-$249 & AS1 & $0.73\pm0.04$ & $12.80\pm3.80$ & $1.51\pm0.01$ & $0.50$ & $510/487$ &  $0.22$ & $\red 1.68$ & $11.7$\\

    & AS2 & $0.33\pm0.02$ &	$30.00^*$  & $1.68\pm0.01$ & $0.60^*$ & $567/542$&  $0.10$& $1.74$ & $12.2$\\
    
    & AS3 &  $0.36\pm0.04$ & $49.15\pm27.55$ & $1.71\pm0.01$ & $0.50^*$ & $541/473$ &  $0.06$ & $0.67$ & $5.0$\\
               
    \hline
    
MAXI J1348$-$630 & AS1 & $0.76\pm0.01$ & $30.00^*$ &$2.25\pm0.02$ &$0.10^*$&  $562/407$ &  $8.14$ & $0.99$& $7.7$\\
    & AS2 & 	$0.69\pm0.01$ & $10.00^*$ &$2.05\pm0.02$ &$\red 0.06$&  $562/407$ &  $8.72$ & $1.75$& $7.6$\\
    & AS3 & $1.14\pm0.04$  & $31.94\pm7.50$ &$1.53\pm0.01$ &$0.50^*$   &  $626/538$ &  $0.64$ & $2.58$& $2.3$\\
 
    \hline

MAXI J1803$-$298 $^\dagger$ & AS1  &  $0.76\pm0.01$  & $25.00^*$& $2.21\pm0.01$ &  $0.33\pm0.02$  &  $774/527$ &  $0.74$& $0.55$ & $17.7$   \\
                
    & AS2 & $0.91\pm0.02$    & $20.00^*$      & $2.05\pm0.02$  &  $0.18\pm0.02$ &  $887/482$ &  $0.83$& $0.42$ & $17.4$   \\
    & AS3 & $0.74\pm0.01$	 & $25.00\pm2.50$ & $2.26\pm0.01$  &  $0.50^*$ &  $704/477$ &  $0.52$& $0.54$ & $14.2$   \\       
    & AS4 & $0.80\pm0.02$	 & $18.45\pm5.10$ & $2.22\pm0.01$  &  $0.50^*$ &  $447/390$ &  $0.60$& $0.55$ & $15.6$	  \\
    & AS5 &	$0.67\pm0.01$	 & $10.00^*$      & $2.82\pm0.09$  &  $0.05^*$ &	 $576/490$ &  $0.12$& $0.01$ & $1.8$   \\
    & AS6 &	$0.63\pm0.01$	 & $10.00^*$      & $2.73\pm0.01$  &  $0.05^*$ &	 $576/491$ &  $0.21$& $0.01$ & $3.0$   \\
    \hline%
	\end{tabular}%
	}

\begin{tablenotes}
    \item $^{*}$Frozen values. 
    \item {\red $^{\dagger}$For the calculation of $L_{\rm bol}$, we employ $M_{\rm BH}=10.0~{\rm M}_{\odot}$ and $\rm D=11.0~kpc$ for IGR J17091$-$3624; $M_{\rm BH}=4.0~{\rm M}_{\odot}$ and $\rm D=2.7~kpc$ for Swift J1727.8$-$1613; $M_{\rm BH}=8.5~{\rm M}_{\odot}$ and $\rm D=4.1~kpc$ for MAXI J1535$-$571; $M_{\rm BH}=14.0~{\rm M}_{\odot}$ and $\rm D=6.3~kpc$ for Swift J1658.2$-$4242; and $M_{\rm BH}=6.5~{\rm M}_{\odot}$ and $\rm D=10.0~kpc$ for MAXI J1803$-$298.}
\end{tablenotes}

    \label{tab:Table-14}
\end{table*}

\begin{table*}
\ContinuedFloat
\caption{Continued.}
\renewcommand{\arraystretch}{1.2}
\resizebox{1.0\textwidth}{!}{%

\begin{tabular}{l c c c c c c c c c c} \hline

Source & Epoch  & $kT_{\rm in}$ & $kT_{\rm e}$ & $\Gamma_{\rm th}$ & cov$_{\rm frac}$ & $\chi^2\big/$dof & $F_{\rm disc}$ & $F_{\rm Comp}$ & $L_{\rm bol}$ \\
& & (keV) & (keV) & & & & \multicolumn{2}{c}{(${10^{-8}}$ erg ${\rm cm}^{-2}$ ${\rm s}^{-1}$)} & ($\% {\rm L}_{\rm Edd}$) & \\       \hline

4U 1630$-$472 $^\dagger$ &AS1 &$0.50^*$&	$1.41^{+0.88}_{-1.41}$  & $1.78\pm0.55$	& $0.90^*$& $46/66$ & $0.85$ & $1.82$  &$0.1$ \\
  &AS2	&$0.80\pm0.12$&	$10.00^*$    & $1.85\pm0.07$	& $0.37\pm0.07$& $383/406$ & $0.33$ & $2.45$  &$0.1$ \\ 
 &AS3	&$1.12\pm0.02$&	$3.00^*$          & $2.20\pm0.10$	& $0.10\pm0.01$& $716/528$ & $1.62$ & $0.06$  &$15.1$ \\
    &AS4  &$0.98\pm0.03$& $5.00^*$ & $2.38\pm0.09$   & $0.20^*$       & $794/632$ & $1.69$ & $0.26$  &$16.5$ \\
    &AS5  &$1.19\pm0.25$& $4.37\pm1.08$ & $3.01\pm0.15$   & $0.10^*$        & $781/531$ & $1.53$ & $0.08$  &$18.1$  \\
    &AS6  &$1.24\pm0.02$& $5.26\pm1.08$ & $2.54\pm0.11$   & $0.05^*$        & $660/519$ & $1.47$ & $0.06$  &$13.7$  \\
    &AS7  &$1.24\pm0.02$& $5.26\pm1.08$ & $2.54\pm0.11$   & $0.05^*$        & $660/519$ & $1.47$ & $0.06$  &$13.7$  \\
    &AS8  &$1.21\pm0.25$& $3.00^*$ & $2.86\pm0.22$   & $0.10^*$        & $607/505$ & $0.82$ & $0.03$  & $7.9$  \\
    &AS9  &$1.18\pm0.25$& $7.00^*$ & $2.28\pm0.22$   & $0.10^*$        & $529/456$ & $1.62$ & $0.001$  & $14.4$  \\
    &AS10  &$1.20\pm0.01$& $5.00^*$ & $2.85\pm0.22$   & $0.10^*$        & $24/20$ & $1.10$ & $0.07$  & $10.7$  \\
    &AS11  &$1.14\pm0.05$& $2.00^*$ & $2.69\pm0.32$	 & $0.20^*$       & $657/531$ & $1.47$ & $0.09$  &$13.0$ \\    
    &AS12  &$1.26\pm0.02$& $10.00^*$ & $2.50\pm0.03$	 & $0.10^*$       & $21/20$ & $1.33$ & $0.13$  &$13.5$ \\  
    &AS13  & $1.13\pm0.06$ & $3.00^*$ & $2.51 \pm 0.42$   & $0.10^*$	    & $738/535$ & $1.53$ &$0.24$   & $15.4$ \\
    &AS14   &$0.75\pm0.09$& $25.00^*$ & $2.52\pm0.01$   & $0.66 \pm 0.06$ & $734/578$ & $2.57$ & $2.28$ & $43.5$  \\
    &AS15  &$1.31\pm0.02$& $21.00\pm4.50$ & $2.21\pm0.11$   & $0.53\pm0.05$ & $746/565$ & $1.67$ & $1.35$  &$27.8$  \\
    &AS16  &$1.46\pm0.02$& $20.00^*$ & $2.39\pm0.11$   & $0.51\pm0.05$        & $716/565$ & $2.77$ & $1.52$  &$39.7$  \\
    &AS17  &$0.98\pm0.02$& $35.69\pm8.22$ & $2.48\pm0.11$   & $0.40^*$        & $817/562$ & $2.64$ & $1.18$  &$34.4$  \\
    \hline
    
4U 1543$-$47 & AS1 &  $0.99\pm0.01$ & $20.00^{*}$ &  $1.84^{+0.14}_{-0.12}$ & $0.39^{+0.07}_{-0.06}$  & $626/535$ & $11.41$ & $11.57$ & $103.0$ &   \\
    
    & AS2 & $0.97\pm0.02 $ & $20.00^{*}$ &  $2.30^{*}$  & $0.30\pm0.02$  &$ 674/531$  & $16.55$ & $16.59$ & $75.0$ &   \\
        
    & AS3 & $0.98\pm0.01$ & $20.00^{*}$ &  $2.26^{+0.06}_{-0.05}$ & $0.20^{*}$ & $671/520$ & $11.08$ & $11.13$ & $49.0$ &   \\
        
    & AS4 & $0.96\pm0.01$ & $20.00^{*}$ &  $1.92^{+0.30}_{-0.25}$ & $0.11^{+0.05}_{-0.03}$  & $678/530$ & $7.29$ & $7.23$ & $33.0$ &   \\
    
    & AS5 & $0.91\pm+0.01$ & $20.00^{*}$ &  $1.77\pm0.03$ & $0.12\pm0.05$  & $539/490$ & $7.43$ & $7.09$ & $35.0$ &   \\
        
    & AS6 & $0.91\pm0.01$ & $20.00^{*}$ &  $1.94\pm0.02$ & $0.40^*$  & $668/510$ & $2.82$ & $2.86$ & $42.0$ &   \\
        
    \hline

MAXI J1820$+$070 & AS1 & $0.49\pm0.01$ & 20.00$^{*}$ & $1.65\pm0.01$ & $0.28\pm0.01$ & $806/510$ & $3.70$ & $25.95$ & $12.9$ &   \\
        
    & AS2 & $0.67\pm0.01$ & $20.00^*$ & $2.06_{-0.02}^{+0.02}$ & $0.03\pm0.01$ & $870/515$ & $8.28$ & $9.31$ & $8.6$ &   \\
    & AS3 & $0.58 \pm 0.01 $ & $6.49\pm0.75$ & $2.75_{-0.05}^{+0.01}$ & $0.05^*$ & $747/501$ & $4.88$ & $3.09$ & $4.8$ &   \\
        
    \hline
\rowcolor{lightgray}

GX 339$-$4$^{\dagger}$ & AS1 & $0.42 \pm 0.12$ & $10.00^*$  & $1.88 \pm 0.10$  & $0.14 \pm 0.01$ & $62/55$ &$0.002$  & $0.003$ & $0.04$  \\
	\rowcolor{lightgray}
	
    & AS2&$0.32 \pm 0.06$ & $10.00^*$ &$1.92^{+0.16}_{-0.17}$&$0.14_{-0.4}^{+0.05}$ & $145/113$ & $0.003$& $0.001$ &$0.03$\\
		 
    & AS3 & $0.20 \pm 0.11$ & $26.83 \pm 15.13$ & $1.61^{+0.01}_{-0.03}$ & $0.74_{-0.03}^{+0.24}$ & $93/108$ & $0.001$ & $0.060$ & $0.4$ \\
		
    & AS4&$0.17^{+0.02}_{-0.05}$&$18.96 \pm 3.65$&$1.55 \pm 0.01$ &$0.46_{-0.17}^{+0.23}$ & $511/478$&$0.001$ &$0.275$&$2.2$\\
		
    & AS5& $0.21 \pm 0.02$&$14.79_{-0.56}^{+0.72}$&$1.65 \pm 0.01$&$0.46^{+0.09}_{-0.16}$&$717/570$&$0.02$& $0.86$&$6.3$\\
	
    & AS6&$0.26 \pm 0.01$& $56.01_{-18.03}^{\boxdot} $&$1.65 \pm 0.01$&$0.56_{-0.06}^{+0.14}$&$834/595$&$0.04$&$1.21$ &$8.7$ \\
	
    & AS7&$0.36 \pm 0.04$&$147.17_{-57.04}^{\boxdot}$&$1.64 \pm 0.01$&$0.57 \pm 0.11$&$423/413$ &$0.06$&$0.77$&$6.4$\\
	
    & AS8& $0.32 \pm 0.05$ &$40.23_{-13.40}^{\boxdot}$&$1.66 \pm 0.01$& $0.52_{-0.13}^{+0.21}$ &$440/340$ &$0.05$ &$0.77$&$6.6$\\
	
    & AS9&$0.29 \pm 0.03$ &$144.17_{-24.20}^{\boxdot}$ &$1.67 \pm 0.01$& $0.51^{+0.16}_{-0.08}$&$339/372$ &$0.05$&$0.80$&$6.8$\\
	
    & AS10 & $0.80_{-0.01}^{+0.02}$ & $84.45_{-45.08}^{\boxdot}$ & $2.44_{-0.06}^{+0.10}$ & $0.36_{-0.04}^{+0.06}$ & $886/605$ & $0.85$ & $0.36$ & $8.6$ \\
	
    & AS11& $0.82 \pm 0.03$& $62.45_{-32.17}^{\boxdot}$&$2.40 \pm 0.05$ & $0.21_{-0.05}^{+0.06}$ & $981/583$ & $1.29$ & $0.34$ & $11.3$\\
	
    & AS12&$0.81 \pm 0.01$ &$74.0_{-39.24}^{\boxdot}$&$2.49_{-0.06}^{+0.07}$&$0.18 \pm 0.04$&$810/594$&$1.11$&$0.24$&$9.6$ \\
	
    & AS13&$0.83 \pm 0.01$&$56.94_{-14.97}^{+8.97}$&$2.27_{-0.10}^{+0.23}$&$0.12 \pm 0.03$&$707/594$&$0.93$&$0.18$&$8.0$ \\
	
    & AS14&$0.83 \pm 0.01$&$33.32 \pm 13.69$&$2.32_{-0.14}^{+0.13} $&$0.09 \pm 0.01$&$676/594$&$1.30$&$0.19$&$10.7$ \\
	
    & AS15&$0.84 \pm 0.01$&$24.66_{-3.83}^{\boxdot}$&$2.02_{-0.16}^{+0.24}$&$0.04 \pm 0.01$ &$674/594$&$1.33$&$0.15$&$10.7$ \\
	
    & AS16&$0.86 \pm 0.01$ & $10.00^{*}$ & $2.53 \pm 0.03$ & $0.02^{*}$ & $705/527$ & $1.46$ & $0.04$ & $11.0$\\
	
    & AS17&$0.83 \pm 0.01$ &$10.00^{*}$ &$2.28_{-0.19}^{+0.21}$&$0.02^*$& $545/482$ & $1.39$ & $0.05$ & $10.6$\\
	
    & AS18&$0.45^{+0.12}_{-0.06}$& $13.31^{+1.26}_{-0.92}$&$1.56 \pm 0.01$&$1.00 \pm 0.04$&$551/527$&$0.03$ &$0.18$ &$1.5$  \\
	
    & AS19&$0.56 \pm 0.30$&$19.61 \pm 2.26$ &$1.56 \pm 0.20$&$1.00 \pm 0.02$ &$631/526$&$0.02$&$0.26$&$1.8$ \\

	\hline        
	\end{tabular}%
	}
\begin{tablenotes}
    \item $^{*}$Frozen values. 
    \item $^{\boxdot}$Pegged at hard limit. Observations of GX 339$-$4 and H 1743$-$322 in QP are highlighted in light gray color. 
    \item {\red $^{\dagger}$For the calculation of $L_{\rm bol}$, we consider $M_{\rm BH}=10.0\,{\rm M}_{\odot}$ and $D=11.5$\,kpc for 4U 1630$-$472, and $M_{\rm BH}=8.3\,{\rm M}_{\odot}$ and $D=8.4$\,kpc for GX 339$-$4.}
\end{tablenotes}

    \label{tab:Table-14}
\end{table*}

\begin{table*}
\ContinuedFloat
\caption{Continued.}
\renewcommand{\arraystretch}{1.2}
\resizebox{1.0\textwidth}{!}{%

\begin{tabular}{l c c c c c c c c c c} \hline

Source & Epoch  & $kT_{\rm in}$ & $kT_{\rm e}$ & $\Gamma_{\rm th}$ & cov$_{\rm frac}$ & $\chi^2\big/$dof & $F_{\rm disc}$ & $F_{\rm Comp}$ & $L_{\rm bol}$ \\
& & (keV) & (keV) & & & & \multicolumn{2}{c}{(${10^{-8}}$ erg ${\rm cm}^{-2}$ ${\rm s}^{-1}$)} & ($\% {\rm L}_{\rm Edd}$) & \\       \hline

    \rowcolor{lightgray}
	
GX 339$-$4    & AS20 &$0.66 \pm 0.04$ &$10.00^{*}$&$1.31 \pm 0.03$&$0.02 \pm 0.01$&$176/127$&$0.001$&$0.001$&$0.01$\\
	
    & AS21&$0.68_{-0.02}^{+0.01}$& $9.34_{-2.34}^{+0.94}$&$2.42_{-0.03}^{+0.04}$ & $0.51_{-0.03}^{+0.02}$& $829/669$& $0.96$&$0.80$&$12.1$ \\
	
    & AS22 &$0.65 \pm 0.01$&$31.77^{*}$&$1.92 \pm 0.19$&$0.02 \pm 0.01$&$458/469$&$0.54$&$0.05$&$4.3$ \\
	
    & AS23&$0.61 \pm 0.01$&$138.94 \pm 81.70$&$2.09_{-0.06}^{+0.05}$ & $0.06 \pm 0.01$& $447/554$&$0.38$&$0.06$&$3.2$\\
\hline
H 1743$-$322 & AS1 & $1.03_{-0.28}^{+0.26}$ & $\red 16.39\pm2.52$ & $1.66 \pm 0.01$ & $0.87 \pm 0.12$ & $536/500$ & $0.05$ & $0.39$ & $2.7$   \\
	
    \rowcolor{lightgray}
	
    & AS2 & $0.59 \pm 0.09$ & $20.83^{*}$ & $2.07 \pm 0.18$ & $0.15 \pm 0.08$ & $234/179$ & $0.006$ & $0.007$ & $0.1$ \\
	
    \rowcolor{lightgray}
	
    & AS3 & $1.18 \pm 0.31$ & $3.59_{-0.47}^{+0.42}$ & $1.42 \pm 0.18$ & $0.46 \pm 0.28$ & $64/43$ & $0.007$ & $0.014$ & $0.1$ \\
			
    & AS4 & $0.41 \pm 0.20$ & $23.49 \pm 7.79$ & $1.64 \pm 0.03$ & $1.00 \pm 0.41$ & $663/560$ & $0.02$ & $0.43$ & $2.7$ \\  \hline

{\red GRS 1739$-$278$^{\dagger}$} & AS1 & $0.11_{-0.04}^{+0.02}$ & $20.00^*$ & $1.92\pm0.03$ & $0.10^*$ & $133/97$ & $0.003$ & $0.01$ & $0.1$   \\ \hline

{\red XTE J2012$+$381$^{\dagger}$} & AS1 & $0.58\pm0.03$ & $9.27_{-2.31}^{+8.96}$ & $1.97_{-0.14}^{+0.12}$ & $ 0.02\pm0.01 $ & $101/91$ & $0.30$ & $0.03$ & $0.8$   \\  \hline
        
	\end{tabular}%
	}
\begin{tablenotes}
    \item $^{*}$Frozen values. 
    \item $^{\boxdot}$Pegged at hard limit. Observations of GX 339$-$4 and H 1743$-$322 in QP are highlighted in light gray color.
    \item {\red $^{\dagger}$For the calculation of $L_{\rm bol}$, we consider $M_{\rm BH}=9.5\,{\rm M}_{\odot}$ and $D=8.5$\,kpc for GRS 1739$-$278, and $M_{\rm BH}=11.8\,{\rm M}_{\odot}$ and $D=5.4$\,kpc for XTE J2012$+$381.}
\end{tablenotes}

    \label{tab:Table-14}
\end{table*}

\subsubsection{MAXI J1535$-$571}

\subsubsection*{(a) Timing Variability Properties}

\begin{figure*}
\centering
	\includegraphics[width=0.49\textwidth]{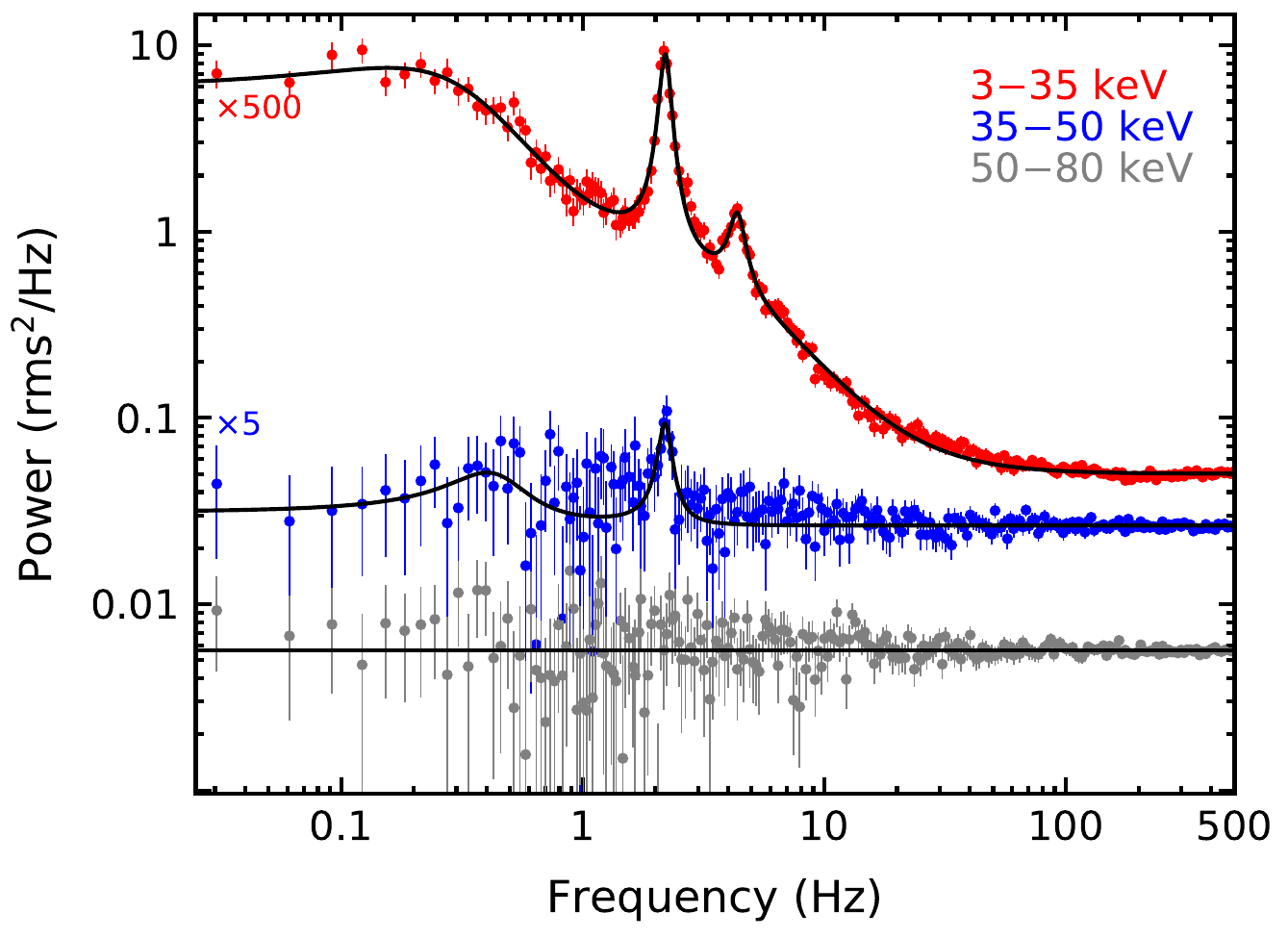}
    \includegraphics[width=0.49\textwidth]{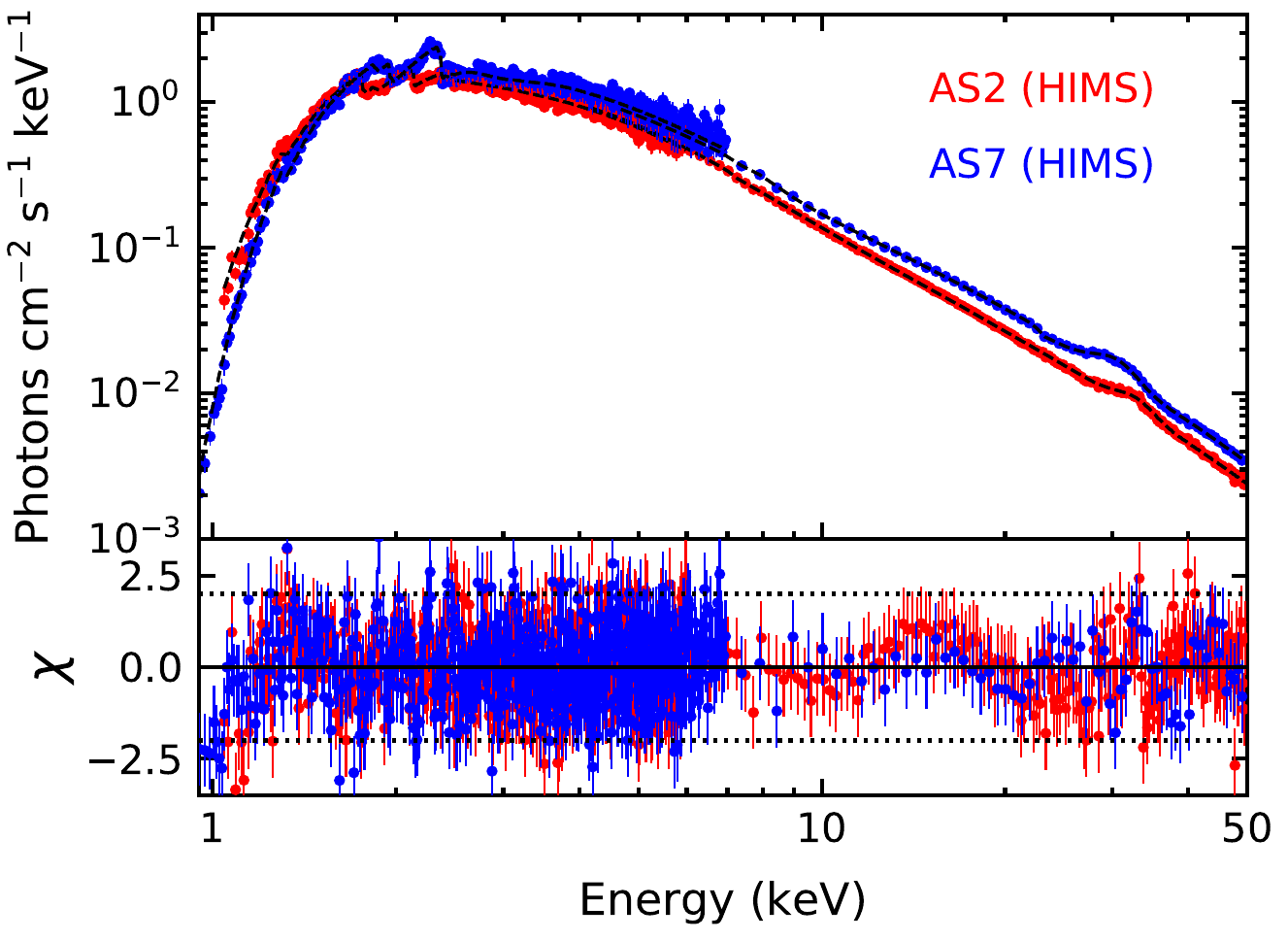}
	\caption{{\it Left:} Energy dependent PDS of MAXI J1535$-$571 with low frequency QPO detected below $50$ keV and its harmonic is seen below $35$ keV. {\it Right:} Wide-band energy spectra of MAXI J1535$-$571 observed on AS2 and AS7 epochs. See text for details.
    }
    \label{fig:fig23}
\end{figure*}

We examine the temporal variability of MAXI J1535$-$571 using \textit{AstroSat} observations during its $2017$ outburst (see Fig. \ref{fig:fig21}). Our temporal analysis confirms the presence of LFQPOs in the frequency range of $\nu_{\rm QPO} \sim 2.08-2.65$ Hz along with harmonic features around $\sim 4$ Hz. Furthermore, we perform energy dependent power spectral analysis in $3-35$ keV, $35-50$ keV and $50-80$ keV energy bands, respectively. We find that the fundamental QPO continues to present up to $\sim 50$ keV, whereas the harmonic feature appears only up to $\sim 35$ keV, as shown in the left panel of Fig. \ref{fig:fig23}. All the best fitted and estimated PDS parameters are tabulated in Table \ref{tab:Table-13} \cite[see also][]{Sreehari-etal2019b}. It is worth mentioning that the QPO properties of the source exhibit marginal variation across the \textit{AstroSat} observational epochs. We find that in most of the observations, this source exhibits hard-lag at $\nu_{\rm QPO}$ varying in the range of $1.0-3.7$ ms. However, in epochs AS9 and AS10, we observe soft-lag ($-1.2$ ms), which is associated with a relatively higher QPO centroid frequency of $2.46-2.59$ Hz. Considering the origin of QPOs to be associated with the modulation of the Comptonizing corona, hard-lag is possibly resulted due to the Compton up-scattering of soft photons at lower QPO frequencies. Note that both hard- and soft-lags of relatively small magnitudes are also reported for MAXI J1535$-$571 using \textit{AstroSat}, \textit{NICER}, and \textit{HXMT} observations \cite[]{Garg-etal2022, Bollemeijer-etal2025}.

\begin{figure*}
	\centering
	\includegraphics[width=0.9\textwidth]{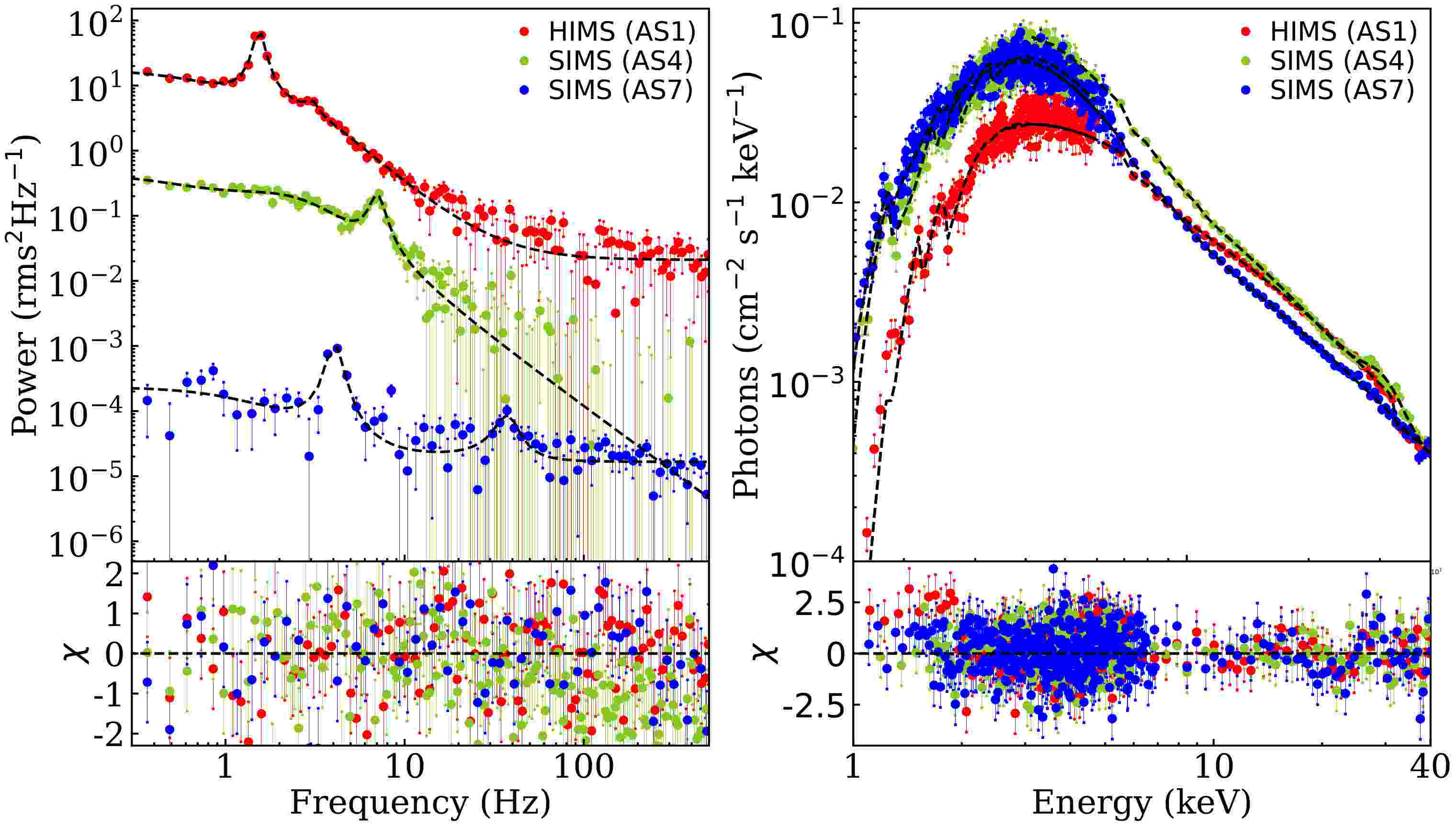}\\	
    \caption{\textit{Left:} Best fit PDS of Swift J1658.2$-$4242 during HIMS and SIMS. Red, green and blue denote PDS including type-C, type-A and type-B QPOs. \textit{Right:} Wide-band energy spectra (SXT and LAXPC) of Swift J1658.2$-$4242, where red, green and blue markers denote spectra for AS1, AS4 and AS7 epochs, respectively.
    }
    \label{fig:fig24}
\end{figure*}

\subsubsection*{(b) Spectral Variability Properties}

Wide-band energy spectra of MAXI J1535$-$571 are modelled using the basic model prescription discussed in \S \ref{sec6:6.2_spec} along with a \texttt{Gaussian} component at $\sim 6.4$ keV accounting for the Fe K$\alpha$ emission. All the best fitted spectral parameters corresponding to each observation are tabulated in Table \ref{tab:Table-14}. We find that the photon index ($\Gamma_{\rm th}$) and the inner disc temperature ($kT_{\rm in}$) exhibit marginal variations around $\sim 2.3$ and $\sim 0.24$ keV, respectively, over different {\it AstroSat} epochs. Furthermore, we observe that $kT_{\rm e}$ is best constrained within $27-82$ keV. However in some cases, $kT_{\rm e}$ could not be constrained and the best fit yielded a value of $150$ keV. On the other hand, the covering fraction remains around $\sim 0.3$ for most cases. Interestingly, the spectral flux associated with the Comptonizing component is found to be $3-4$ times higher than the disc flux contribution. At the same time, we find the presence of hard-lags ($\sim 0.8-3.7$ ms) at $\nu_{\rm QPO} \sim 2.08-2.65$ Hz in most of the {\it AstroSat} observations (see Table \ref{tab:Table-13}). This predominantly suggests that photons in the higher energy band ($6-20$ keV) undergo multiple Compton up-scatterings in the corona before reaching the observer, producing the observed time-lag (see \S\ref{sec8:8.2_spec_tempo_correlation} and \S\ref{sec10p5:QPO-Lag}). Moreover, the source is found to accrete in the sub-Eddington regime during the {\it AstroSat} observation with a bolometric luminosity in the range of $38-71\%~{\rm L}_{\rm Edd}$ estimated using the revised source distance of $4.1$ kpc (see \S\ref{sec3:transi}). {\red Accounting for the uncertainty in the black hole mass ($5.1-7.8\,\rm M_{\odot}$), the inferred bolometric luminosity lies in the range of $\sim 24-116\%$ of $\rm L_{\rm Edd}$.} Notably, \citet{Sreehari-etal2019b} reported a near Eddington accretion state for MAXI J1535$-$571 during HIMS. In Fig. \ref{fig:fig23}, we present two representative spectra of the source, illustrating spectral variability with $\Gamma_{\rm th}$ varying from $2.36$ (AS2) to $2.27$ (AS7).

\begin{figure*}
    \begin{center}
        \includegraphics[width=\textwidth]{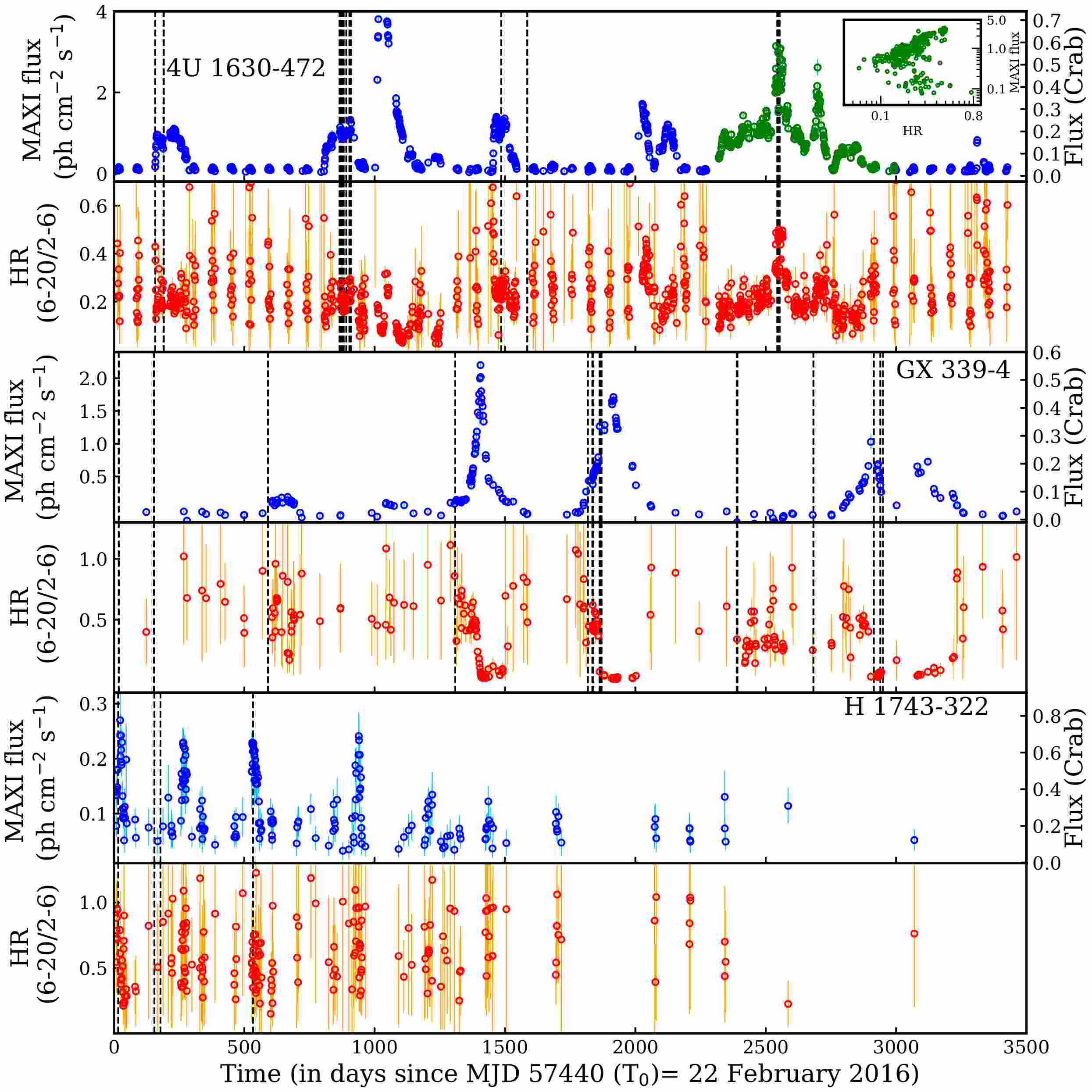}
    \end{center}
    \caption{{\it MAXI}/GSC monitoring of the transients 4U 1630$-$47, GX 339$-$4 and H1743$-$322 in $2-20$ keV energy band during their multiple outbursts. The evolution of HR, derived from \textit{MAXI}/GSC observations, is shown for all three sources. Vertical dashed lines mark the epochs of the \textit{AstroSat} observations. The light curves and HR profiles were generated using $1$-day averaged \textit{MAXI}/GSC data spanning the \textit{AstroSat} mission era. The 2023 super-outburst of 4U 1630$-$47 is highlighted in green in the top panel with the corresponding HID shown in the inset. See the text for details.
    }
    \label{fig:fig25}
\end{figure*}

\subsubsection{Swift J1658.2$-$4242}

\subsubsection*{(a) Timing Variability Properties}

The outburst profile of Swift J1658.2$-$4242, generated from \textit{MAXI} observations along with the corresponding HR, is shown in Fig.~\ref{fig:fig21}. During the \textit{AstroSat} observations, the source is found to be in the HIMS with fluxes of $\sim 25-50$ mCrab and HR of $\sim 0.7-1.2$. In the SIMS, the source exhibits rapid flux variability ranging from $\sim 50$ to $160$ mCrab (see Fig.~\ref{fig:fig21}). This variability produces an overall top-hat shaped profile, also referred to as a flip–flop behaviour \cite[]{Bogensberger-etal2020}. In this state (${\rm HR} \sim 0.1-1.0$), the flip–flop persists for $\sim 10$ days and display the variation with a characteristic timescale of $\sim 1-2$ days.

The epochs of the \textit{AstroSat} observations are marked with vertical dashed lines. Using LAXPC data in the $3-60$ keV band, we examine the timing properties and identify a sequence of LFQPOs. Type-C QPOs are detected in the range $1.5-6.6$ Hz with rms $5-9\%$ during HIMS. Near the outburst peak, during the HIMS–SIMS transition, a type-A QPO emerges at $\nu_{\rm QPO} \sim 7$ Hz with rms $\sim 3.6$\% and Q-factor $\sim 3.7$. The subsequent observation shows no detectable QPO, after which a type-B QPO appears at $\nu_{\rm QPO} \sim 4$ Hz with rms $\sim 2-3$\% and Q-factor $\sim 3-12$. The type-C and type-A QPOs exhibit soft time-lags of $\red \sim 2.9-8.0$ ms and $\sim 6.0$ ms, respectively, whereas a positive lag of $\red \sim 4.0-6.3$ ms is observed for the type-B QPO. The left panel of Fig.~\ref{fig:fig24} displays representative examples of the type-C, type-B, and type-A QPOs from observations AS1, AS4, and AS7, respectively. 

\subsubsection*{(b) Spectral Variability Properties}

We examine the wide-band ($0.7-50.0$ keV) spectral properties of the source using \textit{AstroSat} observations. During HIMS, the photon index is found to be $\Gamma_{\rm th} \sim 1.7-2.2$, with an inner disc temperature of $kT_{\rm in} \sim 0.9-1.5$ keV and electron temperature $kT_{\rm e} \sim 10-17$ keV. In SIMS, the disc temperature increases to $kT_{\rm in} \sim 1.0-2.1$ keV, while $kT_{\rm e}$ lies in the range $12-37$ keV and $\Gamma_{\rm th}$ remains within $1.6-1.9$. The luminosity in the HIMS ranges from $1.2$ to $2.6\%~{\rm L}_{\rm Edd}$, while in the SIMS it spans $0.54-2.91\%~{\rm L}_{\rm Edd}$. Within the SIMS, the minimum luminosity corresponds to observations showing without QPOs feature, whereas the maximum luminosity ($\sim2.9\%~{\rm L}_{\rm Edd}$) is associated with the presence of a type-A QPO. {\red However, due to the substantial uncertainty in the source distance ($>3$\,kpc), the inferred $L_{\rm bol}$ is constrained only to be $>0.5\%$ of $\rm L_{\rm Edd}$, with no upper bound.}
The representative energy spectra corresponding to observations showing type-C, type-B, and type-A QPOs AS1, AS4 and AS7, respectively are presented in the right panel of Fig. \ref{fig:fig24}.

\subsubsection{Multiple Outbursting BH-XRBs}

In this section, we present the results of timing and spectral analyses for three BH-XRBs, namely GX 339$-$4, H 1743$-$322 and 4U 1630$-$472, that exhibited multiple outbursts during the \textit{AstroSat} era. In Fig. \ref{fig:fig25}, we show the MAXI lightcurves and HR variations of these sources, where \textit{AstroSat} observations are marked using dashed vertical lines.

\subsubsection*{(a) Timing Variability Properties}

GX 339$-$4 exhibits four outbursts during \textit{AstroSat} era with peak flux reaching nearly $\sim 0.6$ Crab, while $F_{\rm var}$ remained relatively low in the range $\sim1.3\%-10.0\%$.  It exhibits total rms variability of $0.18-9.93\%$, $14.82-37.14\%$, $1.22-47.10\%$, and $0.75\%$ in the QP, LHS, IMS and HSS, respectively. During four epochs (AS5, AS6, AS10 and AS11) of the 2021 outburst, the source shows prominent QPO features, where the fundamental QPO frequency ($\nu_{\rm QPO}$) varies between $0.10$ Hz and $5.37$ Hz with $\sigma \sim 3.66-21.96$ and rms$_{\rm QPO}\sim 2.08-5.53\%$. We find that $\nu_{\rm QPO}$ increases monotonically with time as indicated in the Table \ref{tab:Table-13}. A rapid evolution occurs between epochs AS10 and AS11 (separated by a day), where $\nu_{\rm QPO}$ changes from $5.37$ Hz to $6.30$ Hz. During AS10, GX 339$-$4 exhibits a sub-harmonic at $\sim 2.86$ Hz and a second harmonic near $9.83$ Hz. However, in AS11, the sub-harmonic disappears while the second harmonic persists at $\sim 9.78$ Hz. Similar behaviour was also reported in previous studies \cite[]{Mondal-etal2023, Peirano-etal2023}. Recently, during the $2023$-$2024$ outburst, QPOs are detected in epoch AS21 with $\nu_{\rm QPO}=4.43$ Hz, $\sigma=6.1$, and rms$_{\rm QPO}=9.97\%$, along with a second harmonic at $\sim 9.32$ Hz. We classify QPOs in AS5 and AS6 as type-C, and those in AS10, AS11, and AS21 as type-B (see Table \ref{tab:Table-13} and also Table \ref{tab:Table-06} for details). These QPOs are associated with either the LHS or IMS. Note that GX 339$-$4 exhibits only hard-lags during $2016-2024$. All model fitted and estimated timing parameters are summarized in Table \ref{tab:Table-13}.

\begin{figure*}
	\centering
	\includegraphics[width=0.45\textwidth]{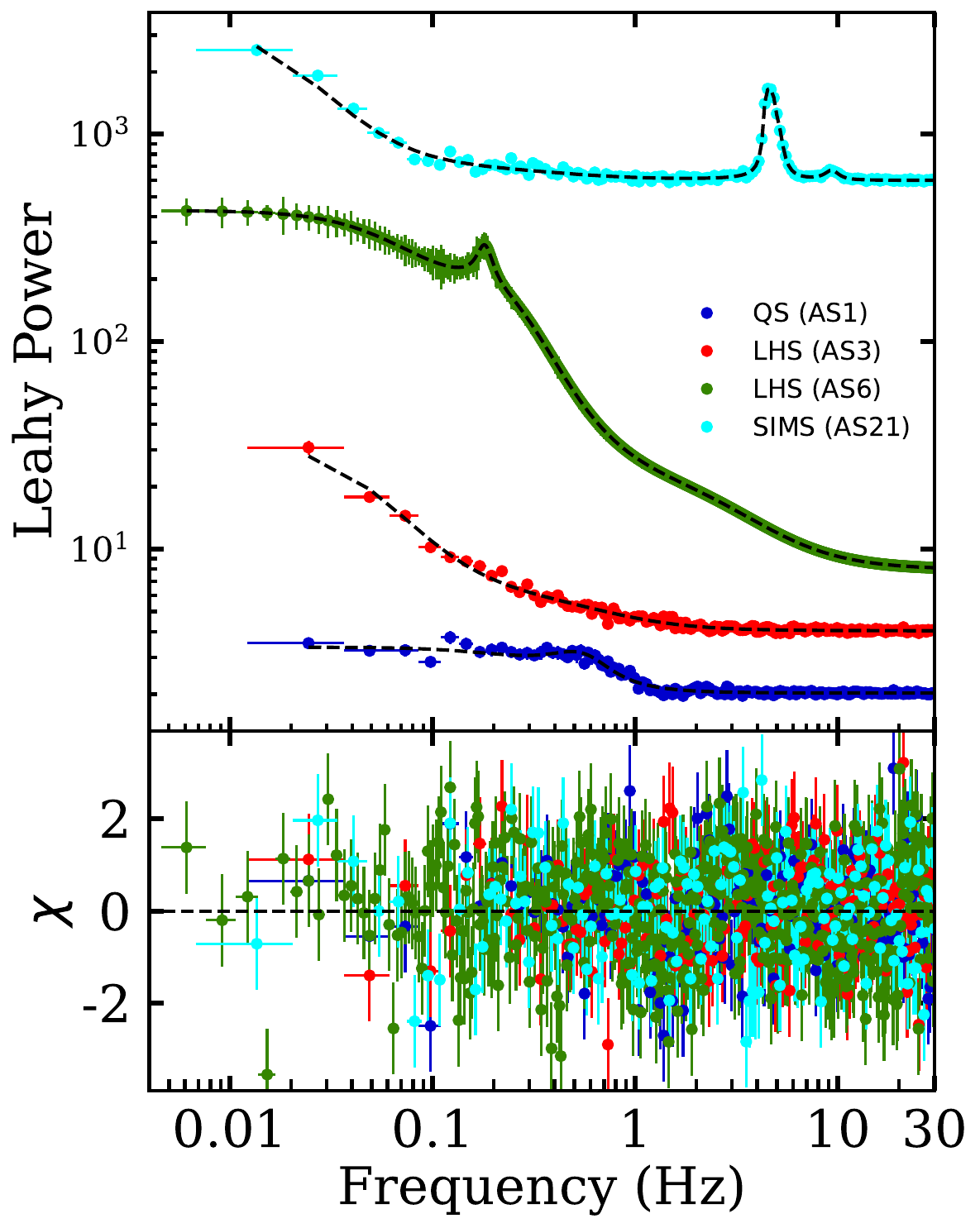}
	\includegraphics[width=0.45\textwidth]{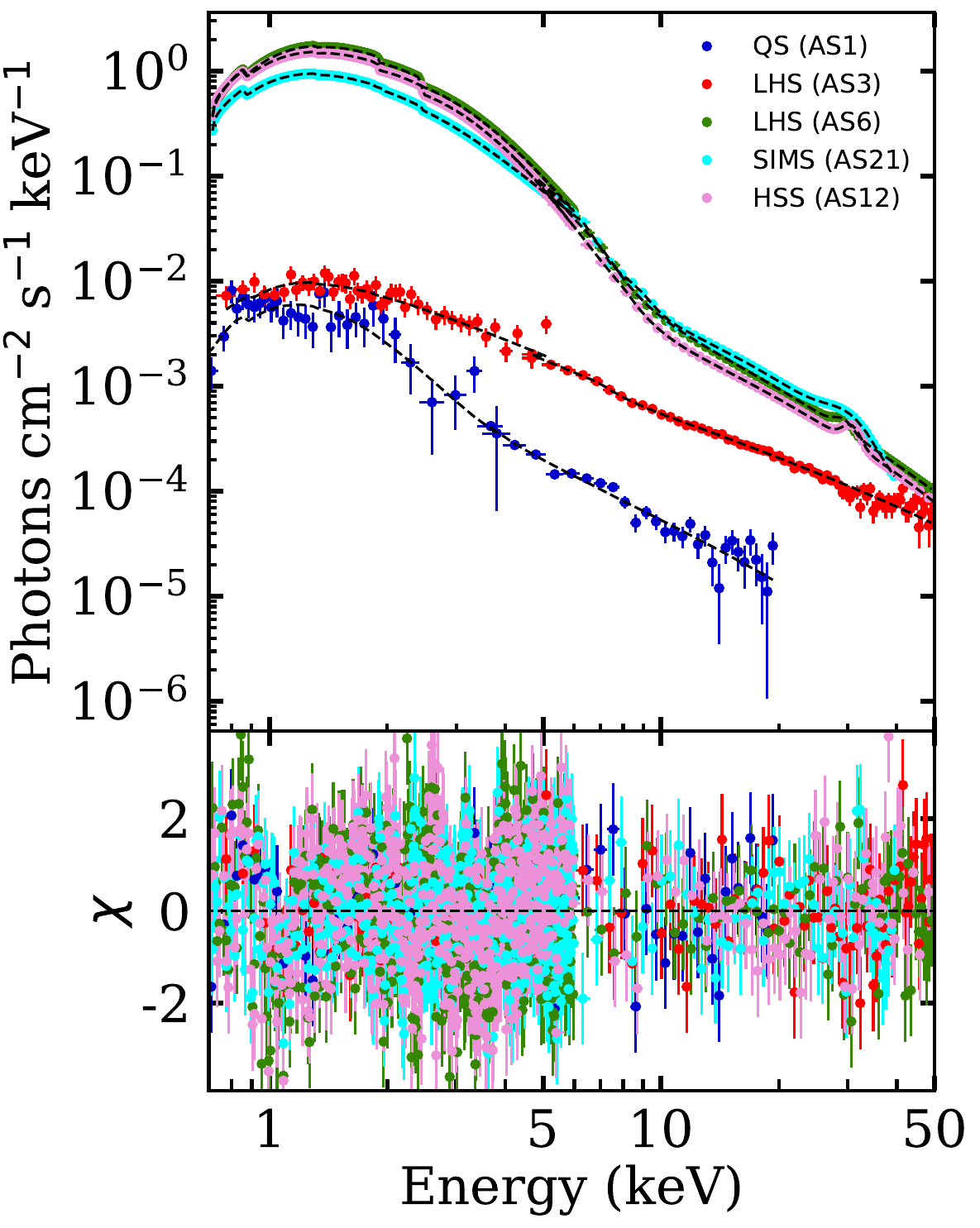}
	\caption{{\it Left}: Best fit PDS of GX 339$-$4 during QS, LHS and SIMS in the frequency range of $0.001-30$ Hz. {\it Right}: Wide-band energy spectra (SXT and LAXPC) of GX 339$-$4 over various epochs during multiple outbursts. See text for details.
    }
    \label{fig:fig26}
\end{figure*}

\begin{figure*}
	\centering
	\includegraphics[width=0.45\textwidth]{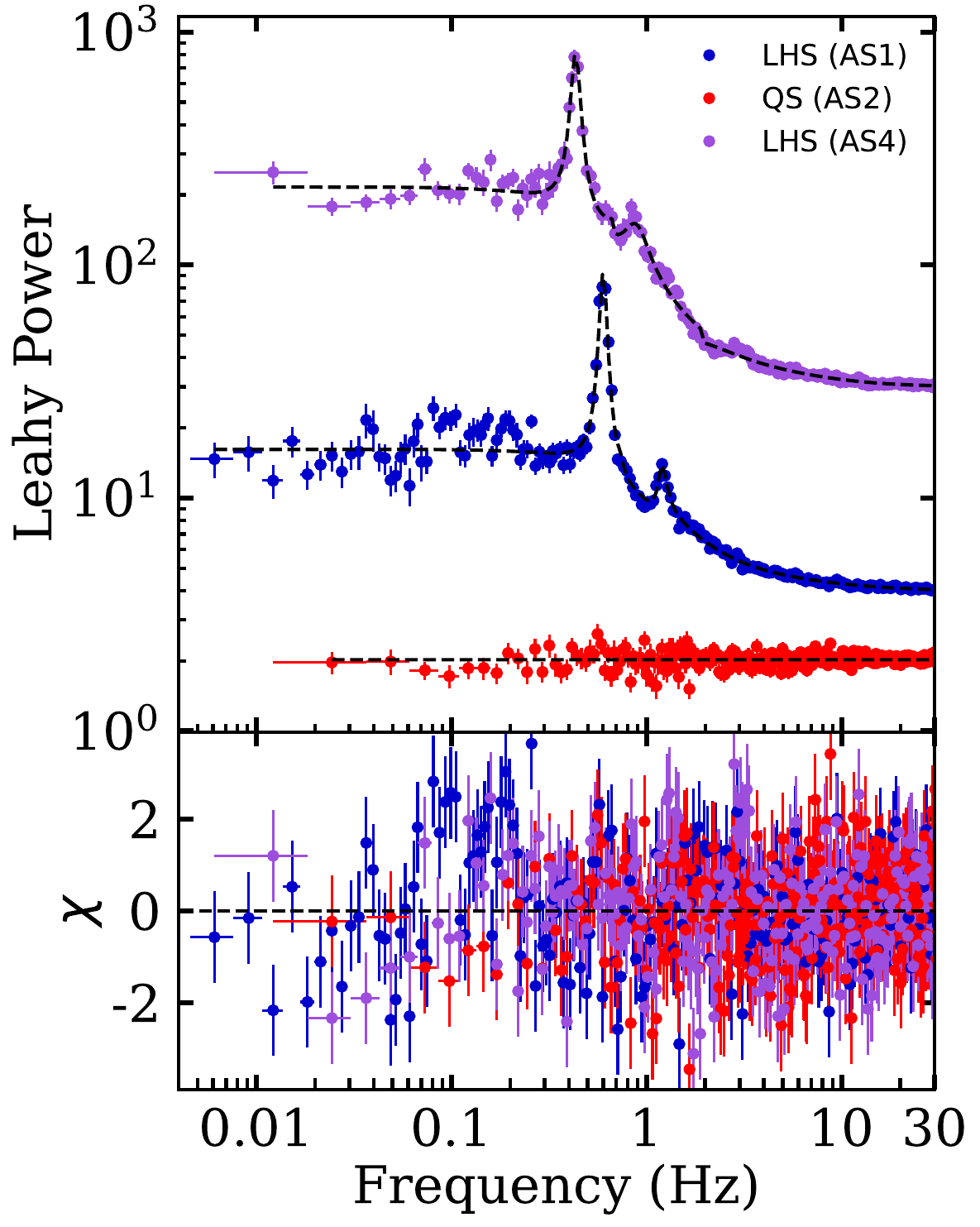}
	\includegraphics[width=0.45\textwidth]{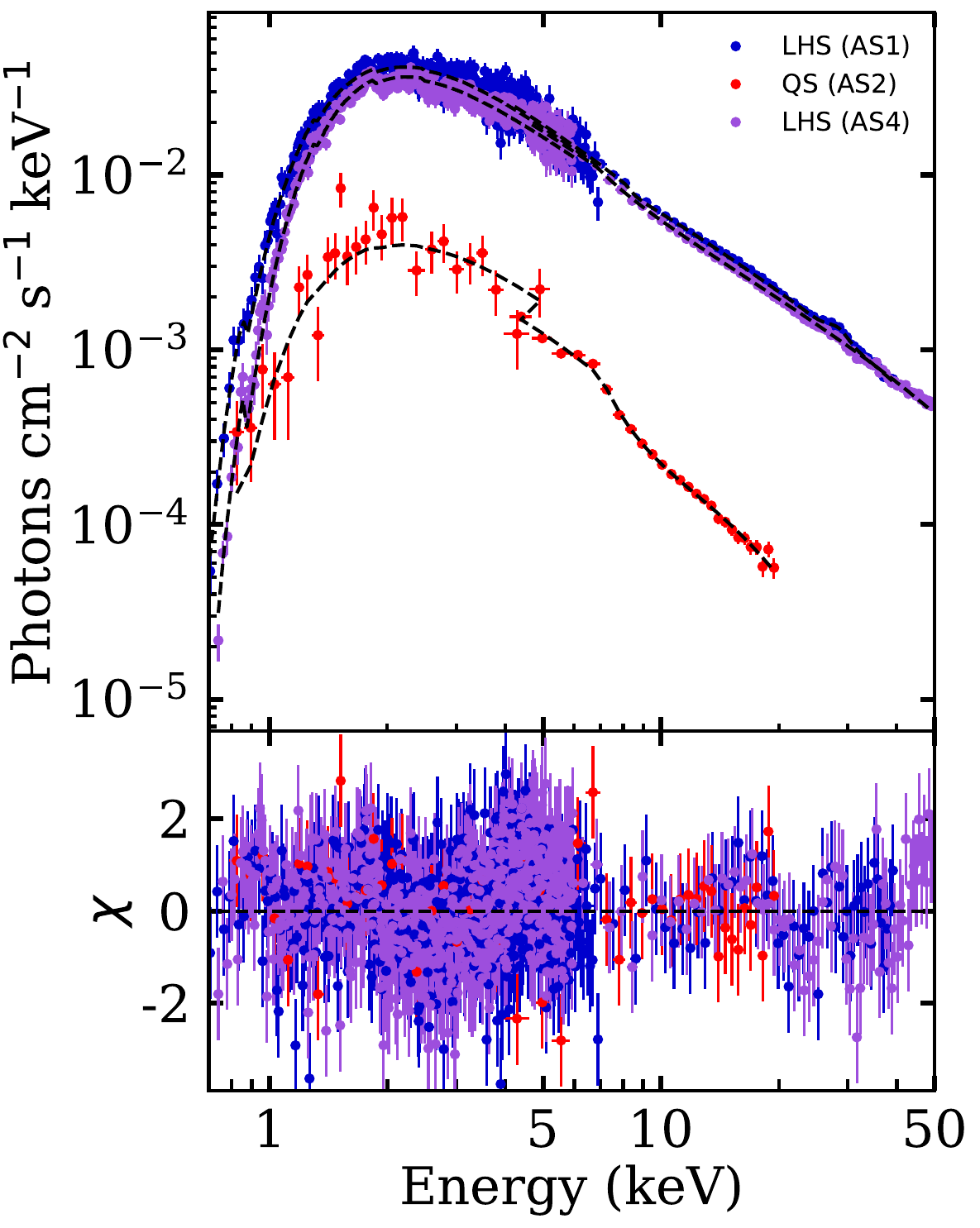}
	\caption{{\it Left}: Best fit PDS of H 1743$-$322 during different observations (QS and LHS) in the frequency range of $0.001-30$ Hz. {\it Right}: Wide-band energy spectra (SXT and LAXPC) of H 1743$-$322 over different \textit{AstroSat} observational epochs during multiple outbursts. See text for details.    
    }
    \label{fig:fig27}
\end{figure*}

H 1743$-$322 exhibits multiple outbursts during \textit{AstroSat} era with $F_{\rm var}$ in the range $3.7-9.3\%$. Fig. \ref{fig:fig27} (left panel) shows the PDS for selected epochs. QPO features are detected during epochs AS1 and AS4 with $\nu_{\rm QPO}$ in the range $0.43-0.60$ Hz, significance $\sigma \sim 17.87-19.74$, and rms$_{\rm QPO} \sim 10.12-12.21\%$. We observe the presence of harmonic at $1.20$ Hz and $0.87$ Hz in epochs AS1 and AS4 along with the $\nu_{\rm QPO}$. The source exhibits total variability of rms$_{\rm total} \sim 1.41-2.16\%$ in QP and rms$_{\rm total} \sim 21.28-23.63\%$ in LHS. Furthermore, H 1743$-$322 shows only soft-lags associated with type-C QPOs during \textit{AstroSat} observations as listed in Table \ref{tab:Table-13}.

4U 1630$-$47 exhibits recurrent outbursts after a quiescent interval of approximately $500-600$ days. This source was observed by {\it AstroSat} in 2016, 2018, 2020 and 2023. During its 2016 and 2018 outbursts, the source remained in active phase for $\sim 100$ and $\sim 150$ days, respectively. The outburst profile showed a similar pattern during these two outbursts, where the flux increased from $\sim 0.10$ Crab to $\sim 0.35$ Crab. However, source exhibited a super-outburst during 2023, where the flux  reached upto $> 0.7$ Crab, and the outburst lasted for $\sim 400$ days. During 2016 and 2018 outbursts, the LAXPC observations are found to be contaminated by a nearby source IGR J16320$-$4751 \cite[]{Baby-etal2020}. In order to account for this contamination, we examine {\it Swift}/BAT flux quasi-simultaneous with the {\it AstroSat} observations for 2023 outburst. Note that the observed flux of 4U 1630$-$472 is $14$ times compared to the observed flux of IGR J16320$-$4751 (see Fig. \ref{fig:fig39}), that indicates marginal contamination by the IGR J16320$-$4751 source as it was not in active phase. \\

The variability timescale of IGR J16320$-$4751 during previous outbursts was found to be $\sim 3-5$ ks, whereas 4U 1630$-$47 showed no such signature in previous outbursts in {\it AstroSat} observations. However, in {\it RXTE} observations of 1998 outburst, it exhibited variability pattern at timescale $\sim 10$ s \cite[]{Zhao-etal2023}. During the 2023 outburst, 4U 1630$-$47 exhibits pronounced variability on timescales of $40-500$ s (see Fig. \ref{fig:fig39}) in the LAXPC data, which is similar to IGR J17091$-$3624 and GRS 1915$+$105 \cite[]{Athulya-etal2022,Katoch-etal2021,Majumder-etal2022}. Simultaneous observations by SXT and {\it NICER} had shown variability of the timescale $\sim 50$ s and $\sim 200$ s, respectively, which is in close agreement with observations reported using LAXPC data. It is worth mentioning that SXT and \textit{NICER} observations are not affected by the contamination from IGR J16320$-$4751 due to smaller FoV. During the 2016 and 2018 outbursts, $F_{\rm var}=1.9-5.6\%$, except for AS1 ($F_{\rm var} \sim 18.2\%$) due to severe contamination, whereas during the 2023 outburst $F_{\rm var} \sim 15.0 -47.0\%$ (see Table \ref{tab:Table-05}).

The PDS of the 2016 and 2018 outbursts of 4U 1630$-$47 do not show any QPO features with rms$_{\rm total } \sim 2.5-7.4\%$. In contrast, \textit{AstroSat} detected type-B QPOs for the first time using {\it AstroSat} observations during the 2023 outburst. Type-C and Type-B QPO have been reported earlier using {\it RXTE} observations \cite[]{Motta-etal-2015, Choudhury-etal-2015, vanDoesburgh-etal2020}. During 2023 outburst, type-B QPO is observed at $\nu_{\rm QPO} \sim 4.8$ and $\sim 7.7\,\mathrm{Hz}$ with corresponding rms$_{\rm QPO} \sim 0.83\%$ and $\sim 1.96\%$. A soft-lag of $\sim 0.4-1.5\,\mathrm{ms}$ is associated with these QPOs. The timing parameters are listed in Table~\ref{tab:Table-13}. 

\begin{figure}
	\centering
    \includegraphics[width=0.9\columnwidth]{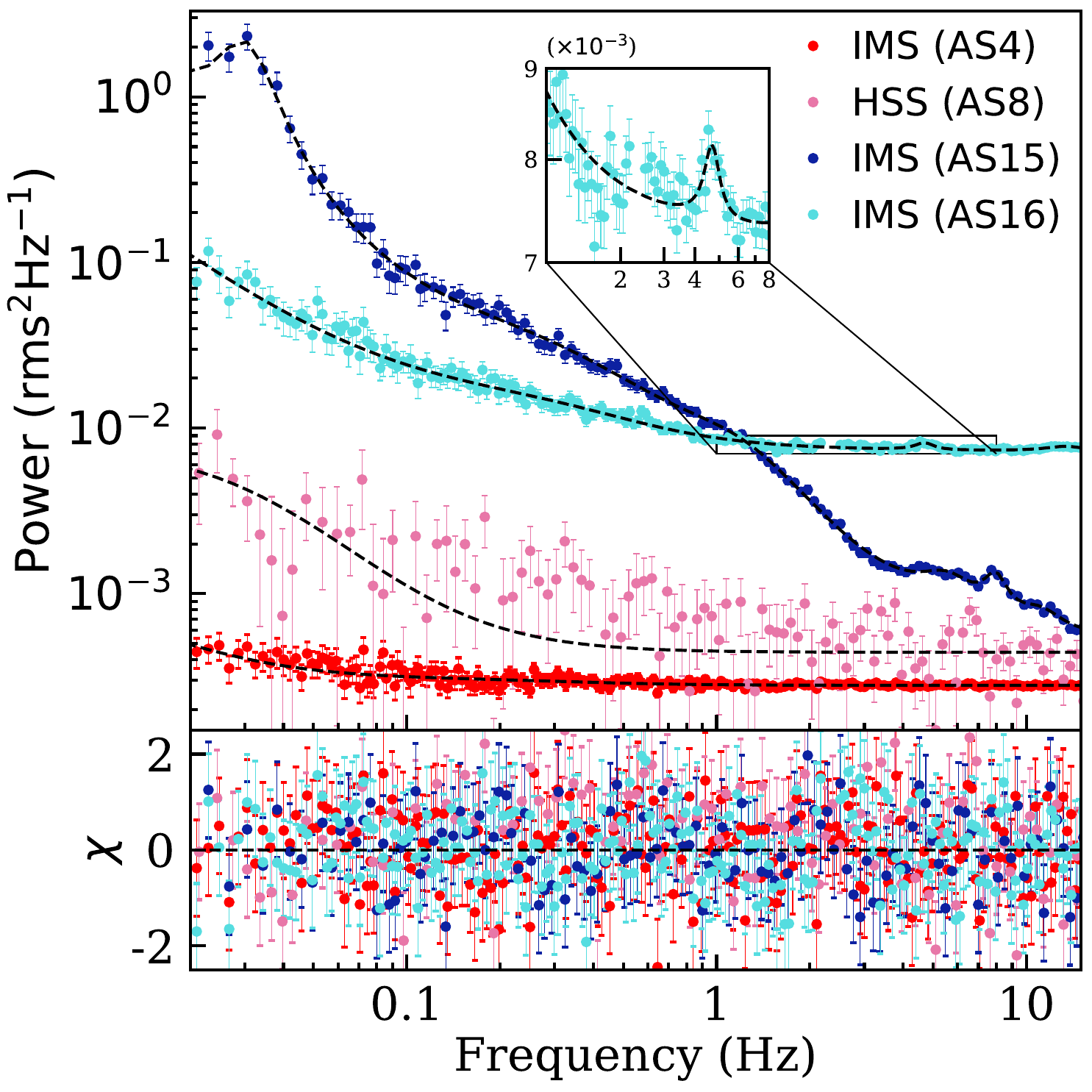}
    \includegraphics[width=0.9\columnwidth]{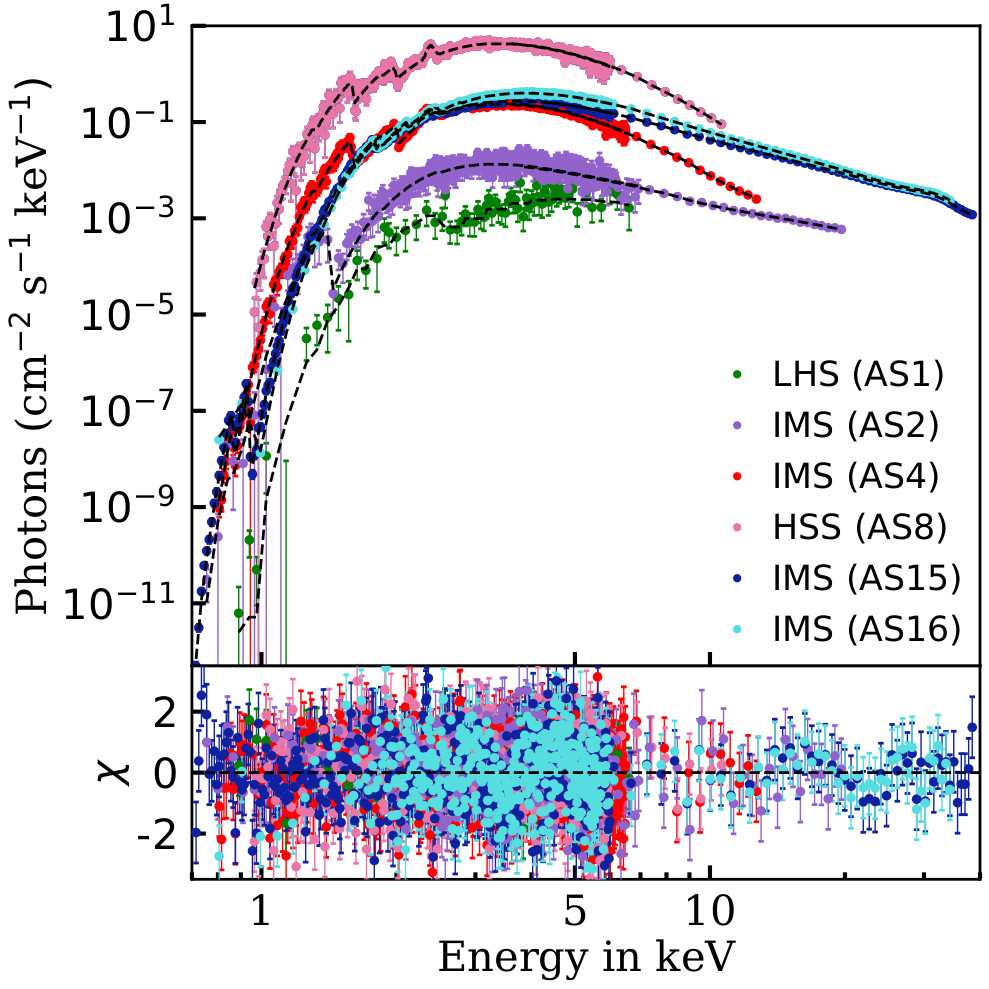}
    \caption{\textit{Top:} PDS in IMS (AS4) and in HSS (AS8) without QPO detection. Type-B QPOs are observed during epoch AS15 and AS16 in IMS (2023 outburst). PDS in epochs AS4, AS8, AS15 and AS16 are scaled for clarity purpose by multiplying $0.025$, $3$, $2$ and $10$, respectively. \textit{Bottom:} Model fitted energy spectra of 4U 1630$-$472 for \texttt{AstroSat} observations in LHS (AS1), IMS (AS2, AS4, AS15, AS16) and HSS (AS8). See text for details.
    }
    \label{fig:fig28}
\end{figure}

\subsubsection*{(b) Spectral Variability Properties}

During the \textit{AstroSat} observations, GX 339$-$4 underwent spectral transitions between quiescent and outburst phases. We note that during the outburst, the source was observed in LHS and IMS. In Fig.~\ref{fig:fig26} (right panel), we show the energy spectra of GX 339$-$4 in different spectral states. Below, we discuss the spectral properties of the source in each state.

In QP (AS1, AS2, AS20), the photon index $\Gamma_{\rm th}$ varies in the range $1.31-1.92$, indicating that the spectrum tends towards a harder state. The inner disc temperature is obtained as $kT_{\rm in} \sim 0.32-0.66$ keV with the covering fraction in the range $0.02-0.14$, while the electron temperature is kept fixed at $kT_{\rm e}=10$ keV. The bolometric luminosity of the source $L_{\rm bol}$ is found to be low at $0.01-0.04 \% \,{\rm L}_{\rm Edd}$. In LHS, GX 339$-$4 was observed during nine epochs (see Table \ref{tab:Table-14}). Spectral analysis indicates that $\Gamma_{\rm th}$ varies as $\sim 1.55-1.67$. The inner disc temperature is found to be $kT_{\rm in} \sim 0.17-0.56$ keV, while the electron temperature spans $kT_{\rm e} \sim 13.31-147.17$ keV. The covering fraction lies in the range $0.46-1.00$. The bolometric luminosity is estimated as $L_{\rm bol} \sim 0.4-8.7 \% \,{\rm L}_{\rm Edd}$, which is significantly higher than in the quiescent state. \textit{AstroSat} observed GX 339$-$4 in the IMS on eleven occasions. The inner disc temperature is estimated as $kT_{\rm in} \sim 0.61-0.87$~keV indicating a hotter disc, while $\Gamma_{\rm th}$ varies as $\sim 1.92-2.44$. In addition, the electron temperature spans $kT_{\rm e} \sim 9.34-138.94$~keV, and the covering fraction lies in the range $0.006-0.51$. The bolometric luminosity during IMS is found to vary between $L_{\rm bol} \sim 3.2-12.1\% \, {\rm L}_{\rm Edd}$. The best fit spectral parameters are listed in Table \ref{tab:Table-14}. {\red During the \textit{AstroSat} observations, GX 339$-$4 spanned a luminosity range of $\sim 0.01-12.1\%$ of $\rm L_{\rm Edd}$. Considering the uncertainty in the black hole mass ($8.3-11.9\,\rm M_{\odot}$), the inferred luminosity may range from $\sim 0.01\%$ to $17.5\%$ of $\rm L_{\rm Edd}$.}

H 1743$-$322 remained predominantly in quiescence during the \textit{AstroSat} era undergoing occasional outbursts (see Fig. \ref{fig:fig25}). \textit{AstroSat} observed the source twice in the QP. The spectra exhibit a hard tail with $\Gamma_{\rm th}$ varying between $1.42$ and $2.07$. The inner disc temperature lies in the range $kT_{\rm in} \sim 0.59-1.18$ keV, while the electron temperature spans $kT_{\rm e} \sim 3.59-23.49$ keV. The covering fraction is found to be $0.15-0.46$. The source is observed in a low-luminosity state with bolometric luminosity $L_{\rm bol} \sim (0.09-0.11)\%\,{\rm L}_{\rm Edd}$. Two more \textit{AstroSat} observations were carried out during LHS, where spectral fits yield $\Gamma_{\rm th} \sim 1.64-1.66$. The inner disc temperature is $kT_{\rm in} \sim 0.41-1.03$ keV and the electron temperature is $kT_{\rm e} \sim 16.4-23.5$ keV. The source luminosity during LHS is significantly higher than in QP with $L_{\rm bol} \sim (2.66-2.71)\% \,{\rm L}_{\rm Edd}$. The right panel of Fig. \ref{fig:fig27} illustrates the energy spectra corresponding to these states, and a summary of the fitted and derived spectral parameters is provided in Table \ref{tab:Table-14}.

We also investigate the spectral properties of 4U 1630$-$472 using \textit{AstroSat} observations in the $0.7-10$ keV energy range and the source is observed in various spectral states. Note that the LAXPC spectrum is affected by contamination (see \S \ref{sec5:5.6_remove_contamination} for details) during its 2016 and 2018 outbursts, and therefore we restrict the LAXPC data to $10$ keV, beyond which the spectrum is dominated by background. During 2016 outburst, the source was in LHS at the beginning of the outburst and made a transition to disc dominated state within few hours \cite[]{Baby-etal2020}. The LAXPC spectra corresponding to these observations are highly contaminated, therefore we only utilize SXT spectra. We obtain $\Gamma_{\rm th} \sim 1.7$ and $L_{\rm bol}\sim 1.03\%~{\rm L}_{\rm Edd}$. In the subsequent observations, the best fit parameters indicate that the source was in a disc dominated state with $F_{\rm Comp} < F_{\rm disc}$. The disc temperature is constrained to $kT_{\rm in} \sim 1.2$ keV and $\Gamma\sim 2.86$ during HSS. The electron temperature is fixed at $kT_{\rm e} \sim 3$ keV. During IMS, we obtained $kT_{\rm in}\sim 0.9-1.1$ keV. The photon index is constrained as $\Gamma\sim 2.2-2.5$, with $kT_{\rm e}$ fixed at $\sim 2-10$ keV. The bolometric luminosity is $\sim 16\%-39\%~{\rm L}_{\rm Edd}$ in IMS. During its 2023 outburst, the flux of 4U 1630$-$47 was very high compared to IGR J16320$-$4751 and the contaminating source background was not subtracted. The source is observed in IMS with $\Gamma_{\rm th}\sim 2.21-2.52$, $kT_{\rm e}\sim 20-35$ keV, $\rm cov_{frac}\sim 0.4-0.6$, $kT_{\rm in}\sim 0.75-1.46$ keV during when the bolometric luminosity is observed to be $\sim 27\%-43\%~L_{\rm Edd}$. {\red Throughout the observational period, the inferred $L_{\rm bol}$ spans $\sim 0.06-43.51\%$ of $\rm L_{\rm Edd}$. Considering the uncertainties in the black hole mass ($3-9\,\rm M_{\odot}$) and distance ($11.5\pm0.3$\,kpc), the corresponding luminosity range broadens to $\sim 0.04-91.62\%$ of $\rm L_{\rm Edd}$.} The bottom panel of Fig. \ref{fig:fig28} depicts the energy spectra, and the spectral parameters are listed in Table \ref{tab:Table-14}. 

\subsubsection{Single Outbursting BH-XRBs}

\begin{figure*}
    \begin{center}
        \includegraphics[width=0.8\textwidth]{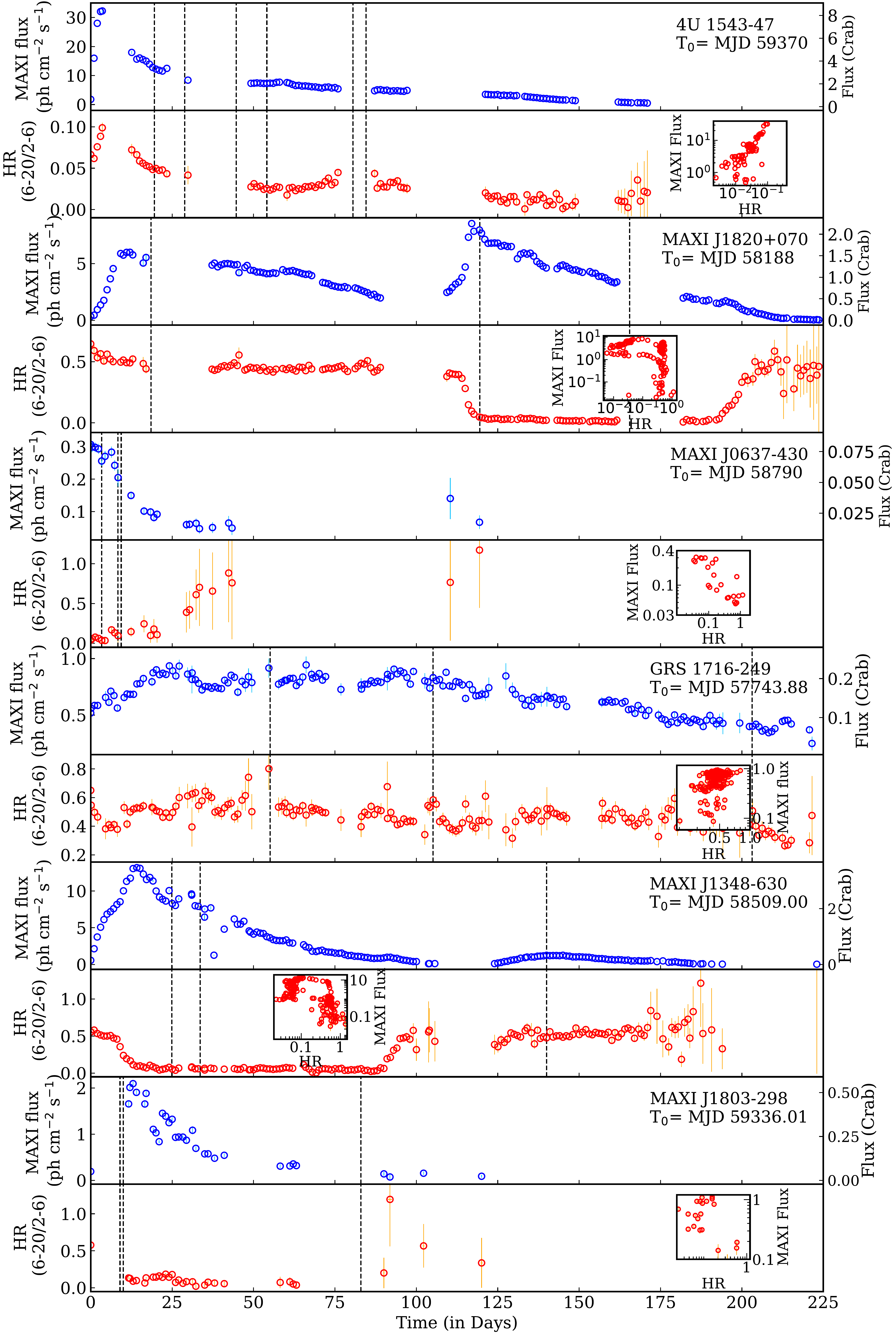}
    \end{center}    
    \caption{{\it MAXI}/GSC monitoring of the transients 4U 1543$-$47, MAXI J1820+070, MAXI J0637$-$430, GRS 1716$-$249, MAXI J1348$-$630 and MAXI J1803$-$298 in $2-20$ keV energy band. The beginning of the outburst for each sources is mentioned as T$_0$. Variation of HR observed with \textit{MAXI}/GSC is plotted for all sources. The vertical dashed lines indicate the epochs of \textit{AstroSat} observations. The lightcurve and HR are obtained from $1$-day averaged {\it MAXI}/GSC data from the launch date of \textit{AstroSat}. HIDs of each sources are shown in the corresponding inset. See text for details. 
    }
    \label{fig:fig29}
\end{figure*}

In this section, we examine seven BH-XRBs that experienced a single outburst during the \textit{AstroSat} era. Fig. \ref{fig:fig29} shows the outburst profiles and HR variations of these sources (except IGR J17091$-$3624) with \textit{AstroSat} observations indicated by dashed vertical lines. Below, we summarize the timing and spectral properties of these sources.

\subsubsection*{(a) Timing Variability Properties}

\begin{figure}[h]
	\centering 
    \includegraphics[width=0.8\columnwidth]{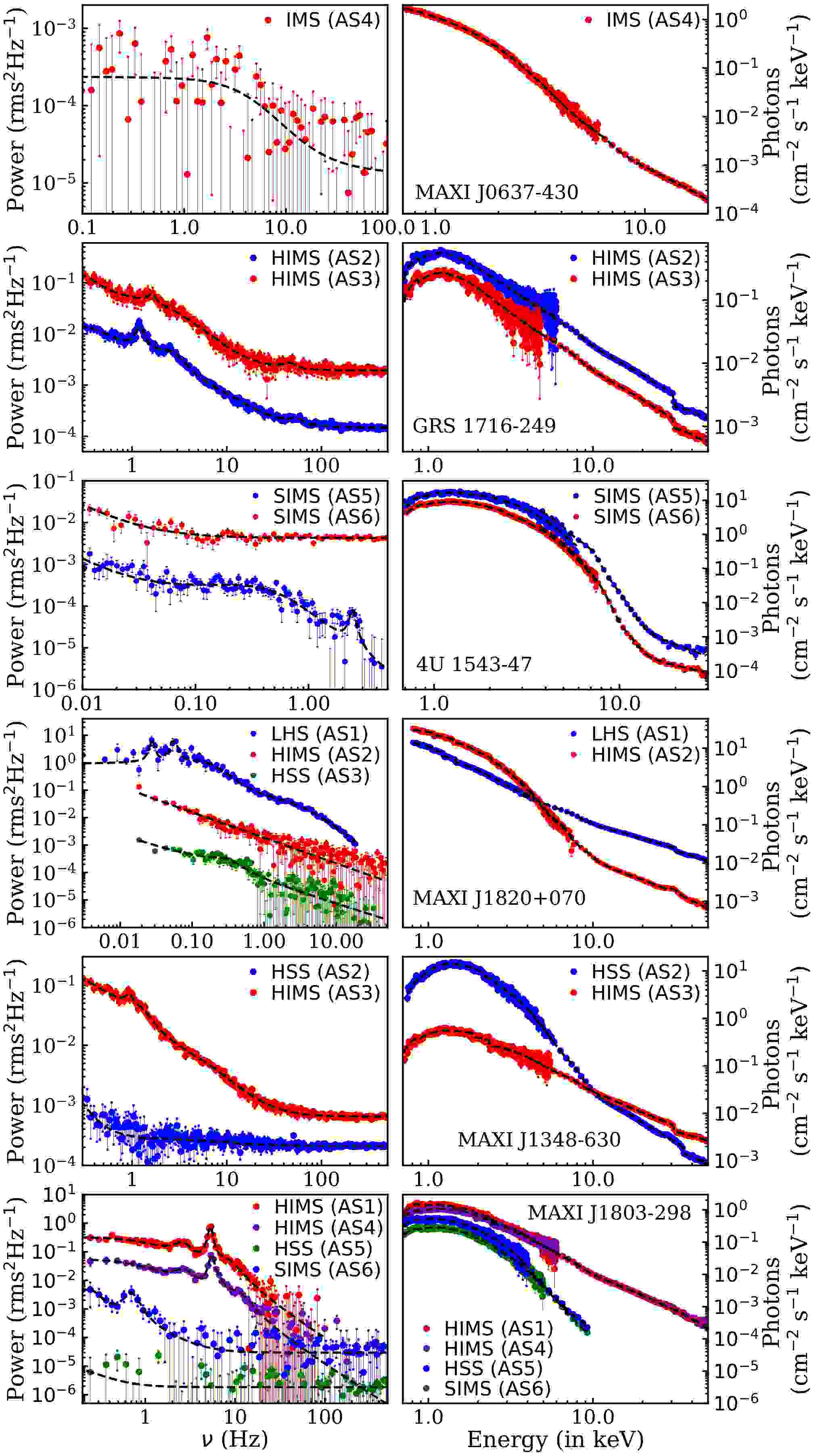}
	\caption{\textit{Left:} Best fit PDS of MAXI J0637$-$430 during IMS (AS4), GRS 1716$-$249 during HIMS (AS1 and AS2), 4U 1543$-$47 during SIMS (AS5 and AS6), MAXI J1820$+$070 during LHS (AS1), HIMS (AS2) and HSS (AS3), MAXI J1348$-$630 during HSS (AS2) and HIMS (AS3), and MAXI J1803$-$298 during HIMS (AS1 and AS4), HSS (AS5) and SIMS (AS6), respectively. {\it Right}: Wide-band energy spectra (SXT and LAXPC) of these sources over different epochs during their outbursts. See text for details.
    }
    \label{fig:fig30}
\end{figure}

\textit{AstroSat} observed 4U 1543$-$47 on six occasions during its $2021$ outburst. The \textit{MAXI}/GSC lightcurve (Fig.~\ref{fig:fig29}) shows a sharp rise followed by a slow decay with a peak luminosity of $\sim 8$ Crab. Power spectra is modelled using a power law and a broad Lorentzian when required in the $0.001-5$ Hz range. The PDS is featureless across all epochs except AS6, where a type-B QPO at $2.45\pm0.08$ Hz is detected (see left panel of Fig.~\ref{fig:fig30}) with a Q-factor of $7.4$ and a significance of $2.6$, without any significant time-lag. 

The $2018$ outburst of MAXI J1820$+$070 comprised two phases, namely a ‘failed’ outburst (Outburst-I) which did not reach the HSS and a complete outburst (Outburst-II). The failed one is comparatively harder, while the complete outburst was extremely bright and soft \cite[]{Prabhakar-etal2022}. In Fig.~\ref{fig:fig29}, we show the MAXI lightcurve and HR variations, where HID is displayed in the inset. Power spectral analyses of AS1 confirms the type-C QPO at $0.05$ Hz with Q-factor $\sim 4.1$, $\sigma \sim 9.1$, and rms$_{\rm QPO} \sim 9.10\%$, as shown in the left panel of Fig. \ref{fig:fig30} and tabulated in Table \ref{tab:Table-13}.

\textit{AstroSat} observed MAXI J0637$-$430 four times during its peak and decay phase with low HR indicating the source remains in IMS (see Fig.~\ref{fig:fig29}). As the source makes transition from HSS to LHS via IMS during decay phase, HR gradually increases and reaches maximum $HR \sim 1$. The HID displays a distinctive ‘C’-shaped track rather than the usual ‘q’-shape \cite[see][for details]{Baby-etal2021}. PDS does not reveal any QPO like features, instead a broad Lorentzian near $\sim 2$ Hz is present with rms$_{\rm total}$ below $5\%$ in the $3-12$ keV band. The left panel of Fig. \ref{fig:fig30} shows the model fitted PDS where no QPO is observed during the IMS.

We examined the timing variability of GRS 1716$-$249 using three \textit{AstroSat} observations during $2017$ outburst. Fig.~\ref{fig:fig29} shows the outburst profile and HR evolution with the HID displayed in the inset. The source is observed in HIMS exhibiting type-C QPOs with $\nu_{\rm QPO} \sim 1.2-1.6$ Hz and rms$_{\rm QPO} \sim 4-6\%$. These QPOs exhibit hard time-lag of $\sim 5-8$ ms. The power spectral distribution is shown in Fig. \ref{fig:fig30}.

\textit{AstroSat} observed MAXI J1348$-$630 twice during the decay phase of the outburst and once during the secondary outburst. The outburst profile of the source is displayed in Fig.~\ref{fig:fig29}, where HR variation and HID are also shown. Power spectral analysis confirms the type-C QPO (epoch AS3) at $\nu_{\rm QPO}\sim 0.94$ Hz with rms$_{\rm QPO} \sim 4.7\%$ (see Table \ref{tab:Table-13} and left panel of Fig. \ref{fig:fig30}) and a hard-lag of $\sim 14$ ms. However, PDS is featureless during AS1 and AS2 epochs.

The source MAXI J1803$-$298 was observed by \textit{AstroSat} three times. The source outburst profile, HR variation and HID is shown in Fig. \ref{fig:fig29}. The power spectral analysis exhibits type-C QPOs at $\nu_{\rm QPO} \sim 5.4-6.0$ Hz with rms$_{\rm QPO} \sim 2.7-6.2 \%$ (see Table \ref{tab:Table-13} and left panel of Fig. \ref{fig:fig30}). The source exhibits soft-lag of $\sim 3.3-4.9$ ms. During the decay phase of the outburst, a type-A QPO is detected at $\nu_{\rm QPO} \sim 0.68$ Hz with rms$_{\rm QPO}\sim 2.38\%$. The Q-factor and $\sigma$ is obtained as $\sim 5.6$ and $2.6$. The QPO feature is present in $6-25$ keV without any detection in low energy band ($3-6$ keV).

IGR J17091$-$3624 was observed with {\it AstroSat} during several outbursts of the source spanning $2016-2022$. Owing to strong contamination from nearby sources within the FoV of LAXPC, we consider two representative observations of the source and perform temporal analysis. We find that the source exhibits $\rm F_{var}$ in the range of $12.8-15.5\%$, with average LAXPC count rates of 2248 (AS1) and 1472 (AS2). Power spectral analysis reveals a total rms amplitude of $10.38-11.42\%$, indicating marginal variability, generally observed during the softer state of black hole transients. In addition, we find the presence of multi-peaked `heartbeat' QPO features with frequencies varying in the range $0.042-0.084$ Hz and rms amplitudes of $3.24-5.94\%$ (see Table \ref{tab:Table-13}). It is worth noting that similar QPOs are also reported by \citet[]{Katoch-etal2021} for IGR J17091$-$3624 using {\it AstroSat} observations.

{\red 
It is noteworthy that the newly discovered black hole candidate Swift J151857.0$-$572147 was observed with \textit{AstroSat} between 16 and 18 March 2024 (MJD $60385-60387$), when the source reached a high flux of $\sim967$ mCrab. However, the analysis of this source is complicated due to contamination from the nearby bright X-ray binary Cir X$-$1, located only $\sim0.2^\circ$ away \citep{Peng-etal2024b,Chatterjee-etal2024}. We extracted the light curves and PDS of Swift J151857.0$-$572147 in the $3-25$ keV band for two observational epochs, AS1 and AS2 (see Table~\ref{tab:Table-05}). The source count rate was significantly higher during AS1 than during AS2. The PDS of AS1 reveals a QPO at a centroid frequency of $8.18$ Hz, with a Q-factor of $3.2$ and a significance of $28.2$ (Table~\ref{tab:Table-13}). In contrast, no QPO was detected during AS2. A similar result was reported to investigate the timing and spectral properties of the source using \textit{AstroSat} data without applying contamination correction \cite[]{Ram-etal2027}. On the other hand, \citet{Mandal-etal2026b} performed spectral analysis in $5-25$ keV energy band for possible removal of contamination. Since the \textit{MAXI} lightcurve of Swift J151857.0$-$572147 is severely affected by contamination from Cir X$-$1, we present the \textit{Swift}/BAT outburst profile in the left panel of Fig.~\ref{fig:fig40}, while the corresponding lightcurves and PDS for the two epochs are shown in the right panel. Owing to the possible residual contamination from Cir X$-$1, no spectral analysis was performed for this source.
}

\subsubsection*{(b) Spectral Variability Properties}

We study the spectral properties of 4U 1543$-$47 in the $0.5-30$ keV energy range using the model prescription described in \S \ref{sec6:6.2_spec} and present the best fit spectral parameters in Table \ref{tab:Table-14}. A broad absorption feature at approximately $8-11$ keV is detected in the spectra across all epochs. This feature is confirmed using \textit{NuSTAR} data for the same period, where it appears even stronger \cite[]{Prabhakar-etal2023}, and it is well described with a Gaussian absorption component (\texttt{gabs}). However, the spectral modelling could not constrain the electron temperature ($kT_{\rm e}$). Similarly, the widths ($\sigma$) of both \texttt{Gaussian} emission line and absorption feature remain unconstrained and are kept fixed at $20$ keV and $0.6$ keV, respectively, during the fitting. The inner disc temperature ($kT_{\rm in}$) shows a gradual decline as the outburst progresses, and the bolometric luminosity follows the similar trend. The high value of $\Gamma_{\rm th}$ persists throughout the outburst, while only a small fraction (cov$_{\rm frac}$) of the soft disc photons are Comptonized in the corona. The strength of the absorption feature appears to be variable. The source is found to be super-Eddington during Epoch AS1 and $L_{\rm bol}$ gradually decreases thereafter (see Table \ref{tab:Table-14}). The model fitted energy spectra is shown in the right panel of Fig. \ref{fig:fig30}. Furthermore, we test for reflection features in the spectra of 4U 1543$-$47 and find that the \texttt{relxilllp} model combined with the \texttt{diskbb} component provides an acceptable fit (see Table \ref{tab:4u-reflec-param}). 

The $2018$ outburst of MAXI J1820$+$070 exhibited an initial hard‑state phase followed by an extremely bright soft‑state phase. During Outburst‑I, the source remained in the LHS, whereas in Outburst‑II, it was observed to be in HIMS by {\it AstroSat}. In AS1, the source is in the LHS ($\Gamma_{\rm th} \sim 1.65$, ${\rm cov}_{\rm frac} \sim 0.28$, and $kT_{\rm in} \sim 0.49$ keV), showing strong Comptonized emission. However, the low level of variability suggests a relatively stable coronal geometry in the early stages of the outburst. AS2 and AS3 correspond to the HIMS ($\Gamma_{\rm th} \sim 2.06-2.75$, ${\rm cov}_{\rm frac} \sim 0.03-0.05$, and $kT_{\rm in} \sim 0.58-0.67$ keV), where the Comptonized emission reduces and moderate variability observed from disk-corona interactions. The electron temperature is fixed at $kT_{\rm e} \sim 20$ keV in AS1 and AS2, while it is observed to be $\sim 6.49$ keV during AS3. The bolometric luminosities in LHS are estimated as $\sim 12.0\% \,{\rm L}_{\rm Edd}$, and the same in HIMS are found to be $\sim 8.0\% \,{\rm L}_{\rm Edd}$ and $\sim 4.8\% \,{\rm L}_{\rm Edd}$, respectively. The model fitted parameters are listed in Table \ref{tab:Table-14} and bolometric luminosity in the unit of $\rm L_{Edd}$ is estimated using source distance of $2.96$ kpc \cite[]{Atri-etal2020} and mass of $8.5~\rm M_{\odot}$ \cite[]{Torres-etal2020}. The best fit energy spectra are depicted in the right panel of Fig. \ref{fig:fig30}. MAXI J1820$+$070 also show strong reflection features which are modelled using various reflection models revealing key properties such as the accretion disc structure, black hole spin, and system inclination (see Table \ref{tab:maxi-simult-refl}). Outburst-I (epoch AS1) shows high ionization and iron abundance, while Outburst-II (Epochs AS2 and AS3) exhibits increased inner disc temperature, steep photon index ($\Gamma_{\rm th}$), high reflection fraction, and overabundance of iron ($A_{\rm Fe}$) content. A dynamic corona with varying height ($h$) is observed \cite[see][for details]{Prabhakar-etal2022}. 

Four \textit{AstroSat} observations of MAXI J0637$-$430 during the early $2019$ outburst captured the source in an intermediate state, possibly transiting from the IMS to LHS. The disc temperature is decreased from $\sim 0.63$ keV to $\sim 0.53$ keV along with the reduction of unabsorbed flux (see Table \ref{tab:Table-14}). The model fitted spectra is shown in the right panel of Fig. \ref{fig:fig30} and best fit spectral parameters are listed in Table \ref{tab:Table-14}. Furthermore, positive residuals in $5-7$ keV range is detected in the spectrum (AS4; 21 November 2019). This feature indicates the presence of disc reflection, which we model using {\tt relxillCP} to constrain spin that yields $a_{\rm k}=0.96^{+0.02}_{-0.73}$ via MCMC (chain length $= 10000$; burn $= 5000$; walkers $= 100$). The best fit provides $\log\xi = 4.03 \pm 0.06$~erg~cm~s$^{-1}$ and reflection fraction $5.35^{+0.48}_{-8.56}$ (see Table \ref{tab2}).

We study the wide-band spectral properties of GRS 1716$-$249 in $0.7-50.0$ keV range. During LHS, the photon index $\Gamma_{\rm th}$ is obtained as $\sim 1.50$, $kT_{\rm e} \sim 30.00$ keV, cov$_{\rm frac} \sim 0.50$, and $kT_{\rm in} \sim 0.30$ keV. During the HIMS, where type-C QPO is observed, the photon index is constrained as $\Gamma_{\rm th} \sim 1.71$, $kT_{\rm e} \sim 12-49$ keV, cov$_{\rm frac} \sim 0.60$, and $kT_{\rm in} \sim 0.53$ keV. The source exhibits a bolometric luminosity $L_{\rm bol}\le11.0\%~{\rm L}_{\rm Edd}$ during LHS. During HIMS, the luminosity is obtained as $5-12\%~{\rm L}_{\rm Edd}$. The right panel of Fig. \ref{fig:fig30} shows the energy spectrum of GRS 1716$-$249.

MAXI J1348$-$630 is observed with {\it AstroSat} in both HIMS and HSS. During HIMS, when type-C QPO is observed, the spectral parameters are obtained as $\Gamma_{\rm th} \sim 1.53$, $kT_{\rm e} \sim 30.00$ keV, cov$_{\rm frac} \sim 0.05$, and $kT_{\rm in} \sim 1.10$ keV. In the HSS, the corresponding parameters are $\Gamma_{\rm th} \sim 2.05-2.25$, $kT_{\rm e} \sim 10-30$ keV, cov$_{\rm frac} \sim 0.06-0.10$, and $kT_{\rm in} \sim 0.69-0.76$ keV. In the HIMS, the source exhibits a bolometric luminosity of approximately $2.3\%~{\rm L}_{\rm Edd}$. In contrast, during the HSS, the bolometric luminosity increases to $7.7\%~{\rm L}_{\rm Edd}$. 

MAXI J1803$-$298 is observed with \textit{AstroSat} during HIMS at the beginning of the outbursting phase. During the decay phase, it is observed in HSS and SIMS. In the HIMS, we find $\Gamma_{\rm th} \sim 2.05-2.23$, while the other parameters are constrained as $kT_{\rm e} \sim 18-25$ keV, cov$_{\rm frac} \sim 0.20-0.50$, and $kT_{\rm in} \sim 0.74-0.91$ keV, respectively. The bolometric luminosity is estimated as $\sim 14-17\%~{\rm L}_{\rm Edd}$. At the decay phase during HSS, we obtain $\Gamma_{\rm th}\sim 2.82$, $kT_{\rm e}\sim 10.00$ keV, cov$_{\rm frac} \sim 0.05$ and $kT_{\rm in}\sim 0.67$. The bolometric luminosity is found to be $\sim 1.8\%~{\rm L}_{\rm Edd}$. In SIMS, the spectral parameters are obtained as $\Gamma_{\rm th} \sim 2.73$, $kT_{\rm e} \sim 10.00$ keV, cov$_{\rm frac} \le0.05$, and $kT_{\rm in} \sim 0.63$ keV with a bolometric luminosity of approximately $\sim 3.0\%~{\rm L}_{\rm Edd}$. {\red Assuming a source distance of $8.0$\,kpc, the inferred $L_{\rm bol}$ during the \textit{AstroSat} observations spans $\sim 1.8-17.7\%$ of $\rm L_{\rm Edd}$. Incorporating the uncertainty of black hole mass of $3-10\,M_{\odot}$, the luminosity range is obtained as $\sim 1.2-38.4\%$ of $\rm L_{\rm Edd}$.}

Using \textit{AstroSat} observations from the 2016 and 2022 outbursts, we explore the spectral variability of IGR J17091$-$3624. Owing to the contamination due to the nearby source in FoV of LAXPC, spectral modelling is restricted to energies within $\sim 30$ keV. The best fit spectra from epochs AS1 and AS2 indicate a steep photon index ($\Gamma_{\rm th} = 2.50-3.00$) accompanied by weak Comptonized emission (cov$_{\rm frac} \sim 0.30$). The inner disc temperature peaks at $\sim 2.00$ keV, and the disc flux exceeds the Comptonized flux by a factor of $\sim 9$. These results collectively point to a disc dominated accretion scenario for IGR J17091$-$3624 during \textit{AstroSat} monitoring with luminosity range of $\rm \sim14- 18\%~L_{Edd}$ estimated using source distance of $11$ kpc. {\red However, accounting for the uncertainties in the black hole mass ($8.7-15.6\,\rm M_{\odot}$) and source distance ($11-17$\,kpc), the bolometric luminosity is estimated to lie in the range of $\sim 7-37\%$ of $\rm L_{\rm Edd}$.}

\subsubsection{Sources with Limited \textit{AstroSat} Observations}

Among the $34$ BH-XRBs, limited observations are available for $\red 10$ sources due to the absence of noticeable X-ray activity during the operational period of \textit{AstroSat}, as marked in red in Table \ref{tab:Table-01}. The \textit{MAXI}/GSC flux of these sources in the $2-20$ keV energy range is found to lie between $3$ and $7$ mCrab. For our analysis, we consider archival observations of sources with fluxes $>10$ mCrab, since below this threshold the LAXPC background is known to dominate. Considering this, spectro-temporal analyses of two such sources are discussed below.

\subsubsection*{(a) Spectro-temporal Properties}

XTE J2012$+$381 was observed with \textit{AstroSat} on $27$ March $2023$ (MJD $60030$) during the decay phase of its $2023$ outburst. The source flux in the $2-20$ keV band is measured to be $\sim 32$ mCrab. The background corrected mean count rate is found to be $79$ cts s$^{-1}$. To investigate variability in the lightcurve, a PDS was extracted, which is entirely dominated by Poisson noise. Wide-band spectral analysis ($0.8-30.0$ keV) is performed using the model $\texttt{tbabs}\times (\texttt{thcomp}\otimes\texttt{diskbb})$ that yields $\chi^{2}_{\rm red} = 1.10$. The hydrogen column density is obtained as $1.74 \times 10^{22}$ atoms cm$^{-2}$ with $kT_{\rm in} \sim 0.58 \pm 0.03$ keV, $\Gamma \sim 1.97^{+0.12}_{-0.14}$ and $kT_{\rm e} \sim 9.27^{+8.96}_{-2.31}$ keV, respectively. The cov$_{\rm frac}$ is obtained to be $0.02\pm0.01$. We refrain from attempting for the spectral evolution study of the source due to single day observation by \textit{AstroSat}. 

GRS 1739$-$278 was observed on $26$ July $2021$ (MJD 59421) with \textit{AstroSat}. The long‑term MAXI lightcurve indicates that the source flux in the $2$–$20$ keV band was $\sim 11$ mCrab at the time of the \textit{AstroSat} observation. The background corrected mean count rate is found to be $24$ cts s$^{-1}$. The PDS is completely dominated by Poissonian noise with no detectable features up to $100$ Hz. Spectral analysis was carried out in the $0.8-25$ keV energy range, yielding $n_{\rm H} \sim 1.36 \times 10^{22}$ atoms cm$^{-2}$, $\Gamma \sim 1.92 \pm 0.03$, $kT_{\rm in} \sim 0.11^{+0.02}_{-0.04}$ keV, $kT_{\rm e} \sim 20$ keV and cov$_{\rm frac} \sim 0.1$, respectively.

{\red 
It is worth noting that uncertainties in black hole mass and distance introduce systematic uncertainties in the estimated Eddington scaled luminosities presented in this work. Since ${\rm L}_{\rm Edd}$ depends directly on the adopted black hole mass \cite[]{Frank-etal2002}, inaccuracies in mass measurements propagate into the Eddington luminosity estimate. Similarly, the source distance is required to convert the observed flux into intrinsic bolometric luminosity affecting the derived Eddington ratio, $L_{\rm bol}/{\rm L}_{\rm Edd}$. Consequently, sources with poorly constrained masses and/or distances may have less reliable Eddington scaled luminosities, and the corresponding values should be interpreted with due caution. The adopted mass and distance estimates, along with their associated uncertainties, are summarized in Table~\ref{tab:Table-01} and discussed in \S\ref{sec3:BH_overview}.}

\section{Correlation Study of BH-XRBs}\label{sec8:correlation}

In this section, we investigate the spectro-temporal correlations of the BH-XRBs using {\it AstroSat} observations. We begin by examining the correlations among the spectral parameters and then explore their relationship with the timing properties in the following subsections.

\subsection{Correlation Among Spectral Parameters}\label{sec8:8.1_spec_correlation}

\subsubsection{Spectral Correlation in GRS 1915$+$105}

\begin{figure*}
    \begin{center}
    \includegraphics[width=0.495\textwidth]{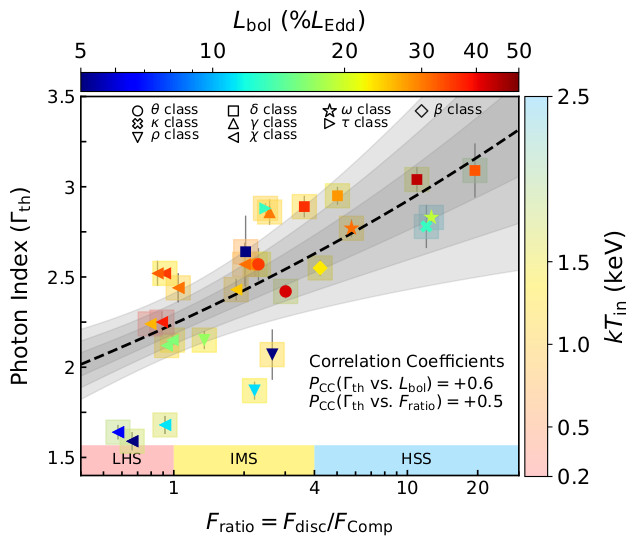}
    \includegraphics[width=0.495\textwidth]{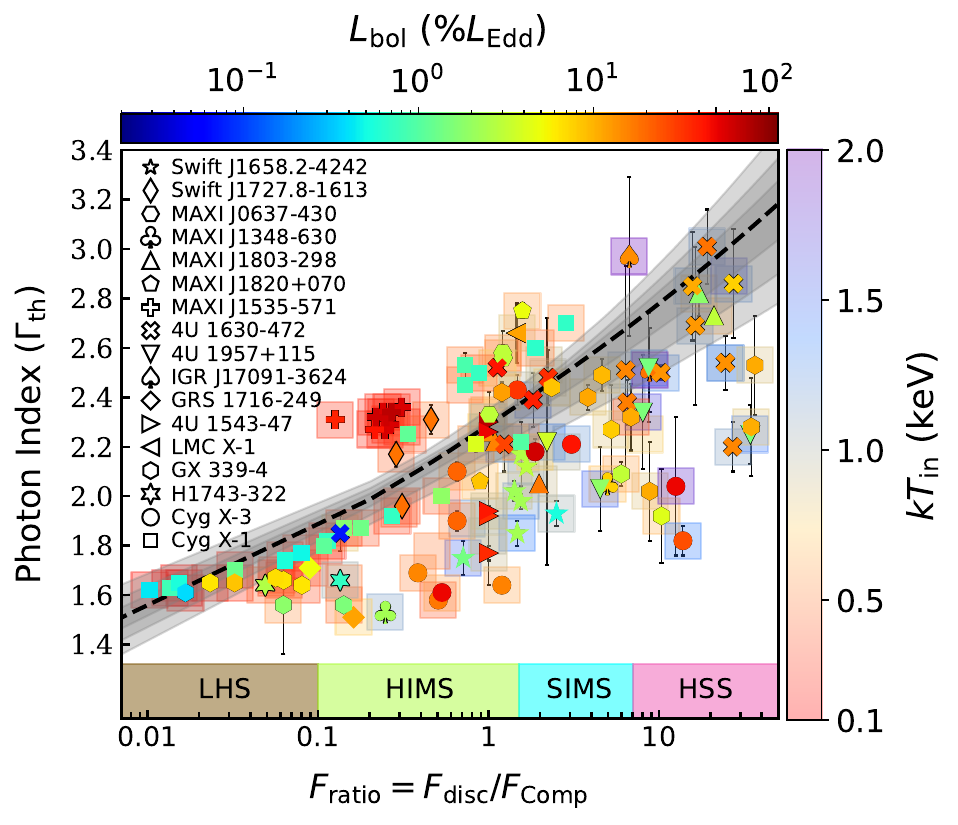}
    \end{center}
    \caption{{\it Left:} Variation of photon index ($\Gamma_{\rm th}$) with the disc to Comptonized flux ratio ($F_{\rm ratio}$) in different variability classes of GRS $1915+105$ observed with {\it AstroSat}. The colorbar represents the variation of bolometric luminosity ($L_{\rm bol}$) in the respective observations. {\it Right:} Variation of photon index ($\Gamma_{\rm th}$) with $F_{\rm ratio}$ for different sources as shown using different markers. The horizontal colorbar denotes bolometric luminosity $L_{\rm bol}$, whereas the vertical colorbar is used to depict the inner disc temperature ($kT_{\rm in}$). Different spectral states are marked based on the variation of $\Gamma_{\rm th}$ and $F_{\rm ratio}$. The dashed curves in both panels are used to represent the best fit, while the grey shaded regions with increasing transparency indicate $1 \sigma$, $2 \sigma$ and $3 \sigma$ confidence intervals. Here, the sources Cyg X$-$3, MAXI J1535$-$571 (AS2), GX 339$-$4 (AS22, AS23) and MAXI J1803$-$298 (observation in HSS) are excluded from fitting. In addition, LMC X$-$3, GRS 1758$-$258 and 1E 1740.7$-$2942 are not considered due to the absence of disc component. See text for details.
    }
    \label{fig:fig31}
\end{figure*}

We examine the correlation between spectral parameters of GRS 1915$+$105 across its different variability classes. In Fig. \ref{fig:fig31} (left panel), we depict the variation of photon index ($\Gamma_{\rm th}$) obtained from the \texttt{thcomp} model as a function of the ratio of disc flux to Comptonized flux ($F_{\rm ratio}=F_{\rm disc}/F_{\rm Comp}$; see Table \ref{tab:Table-08}). The color represents the bolometric luminosity ($L_{\rm bol}$) with the color bar at the top denoting its range. The Pearson correlation coefficient ($P_{\rm CC}$) between $\Gamma_{\rm th}$ and $F_{\rm ratio}$ is found to be $0.5$ indicating a moderate positive correlation. This trend reflects the spectral softening of the source as the relative disc contribution increases in the wide-band energy spectral distribution. A similar positive correlation is obtained between $\Gamma_{\rm th}$ and $L_{\rm bol}$ with $P_{\rm CC} = 0.6$, suggesting that high luminous states generally correspond to softer spectra for most of the softer variability classes of GRS 1915$+$105 \cite[]{Majumder-etal2022,Majumder-etal2025a}. 

Interestingly, we notice that the photon index is coarsely distinguished in two domains with $\Gamma_{\rm th} < 2.6$ for $\theta$, $\chi$, $\beta$ and $\rho$, whereas a steeper spectral index ($\Gamma_{\rm th} > 2.6$) for the remaining softer variability classes ($\delta$, $\gamma$, $\omega$, $\kappa$, $\tau$). Moreover, the variation of $F_{\rm ratio}$ further allows one to define three spectral states for GRS 1915$+$105 analogous to those observed in transient BH-XRBs. Specifically, we find that the observations of $\chi$ class variants mostly remain within $F_{\rm ratio} \lesssim 1$ with few exceptions exhibiting Comptonization dominated spectral characteristics similar to those of LHS (see Fig. \ref{fig:fig31}). On the other hand, $\theta$, $\rho$, $\tau$, $\beta$, and $\gamma$ classes lie within $1 \lesssim F_{\rm ratio} \lesssim 4$ showing IMS (as HIMS and SIMS remain indistinguishable) like characteristic. Finally, the $\delta$, $\kappa$, and $\omega$ classes exhibit canonical HSS behaviour with $F_{\rm ratio} \gtrsim 4$. It is worth noting that the softer variability classes in GRS 1915$+$105 are primarily characterized by state-A and state-B, or a combination of them. In general, state-B represents a high flux phase where thermal emission from inner accretion disc is dominated, whereas state-A corresponds to a low intensity soft state \cite[]{Belloni-etal2000}. The distinct variability signatures observed in these variability classes arise from the rapid transition of the source between these two soft states or undergoing brief excursions into the harder state-C. On the other hand, the presence of a strong state-C is generally observed for the former classes through the elongated branch towards higher HR2 values in the CCD (see Fig. \ref{fig:fig9} and \citet[]{Belloni-etal2000}). Note that, state-C resembles the quiescent state (equivalently hard state) observed in BH-XRBs, and therefore, these classes often exhibit intermediate to harder spectral characteristics. Consistently, LFQPOs and power spectral breaks are commonly observed in the $\theta$, $\chi$, $\beta$, and $\rho$ classes of GRS 1915$+$105 (see Fig. \ref{fig:fig9};~\cite{Majumder-etal2022};~\cite{Majumder-etal2026b}). 

Further, we examine the correlation between $\Gamma_{\rm th}$ and $F_{\rm ratio}$, we fit an empirical power-law function as $\Gamma_{\rm th}=a\times F_{\rm ratio}^{b}$. We perform $10^4$ bootstrap realizations by resampling the data within its measurement errors assuming a Gaussian distribution \cite[see][for details]{Curran-2014} to determine the parameter uncertainties. In the left panel of Fig. \ref{fig:fig31}, dashed curve represents the best fit with $\Gamma_{\rm th}=(2.23\pm 0.08)\times F_{\rm ratio}^{(0.12 \pm 0.02)}$ and grey shaded regions with increasing transparency correspond to $1 \sigma$, $2\sigma$ and $3\sigma$ confidence intervals. In addition, the variation of $kT_{\rm in}$ is depicted using colour with the vertical colour bar showing its range. During LHS, $kT_{\rm in}$ is in the range $\sim 0.20-0.50$ keV, whereas in IMS it is $\ge 0.50$ keV. In HSS, $kT_{\rm in}$ lies in the range $\sim 1.50-2.50$ keV.

\subsubsection{Spectral Correlation in a Sample of BH-XRBs}

We investigate correlations among spectral parameters for BH-XRBs under considerations except GRS 1915$+$105. The right panel of Fig. \ref{fig:fig31} shows the variation of $\Gamma_{\rm th}$ with $F_{\rm ratio}$, where colorbars at the top and right denote $L_{\rm bol}$ and $kT_{\rm in}$, respectively.

Several BH-XRBs, namely H1743$-$322, GX 339$-$4, Cyg X$-$1, Cyg X$-$3 and MAXI J1820$+$070, were observed by \textit{AstroSat} in the LHS, which exhibit photon indices as $\Gamma_{\rm th} \sim 1.50-1.70$. We find the flux ratios in the range $F_{\rm ratio} \sim 0.01-0.10$ and luminosities $\leq 5 \% \,{\rm L}_{\rm Edd}$. In contrast, sources such as MAXI J1803$-$298, Swift J1727.8$–$1613, MAXI J1535$-$571, Cyg X$-$1, MAXI J1348$-$630, GX 339$-$4 and Swift J1658.2$-$4242 were observed in the HIMS, where $\Gamma_{\rm th}$ spans $1.50-2.40$, luminosity is typically $\sim 10 \% \,{\rm L}_{\rm Edd}$, and $0.10 \lesssim F_{\rm ratio} \lesssim 1.50$. However, MAXI J1535$-$571 shows a lower luminosity of $\sim0.1-1.0 \% \, {\rm L}_{\rm Edd}$, whereas Cyg X$-$3 shows a much higher luminosity of $\sim 50 \% \, {\rm L}_{\rm Edd}$ with $F_{\rm ratio}\sim0.50$. During both LHS and HIMS, these sources show type-C QPOs with $F_{\rm ratio} \lesssim 1.5$ indicating the prevailing Comptonized emission in these accretion states.

As $F_{\rm ratio}$ exceeds unity, the disc emission starts to dominate resulting in spectral softening and sources transits towards SIMS with $F_{\rm ratio} > 1.50$. During SIMS, the photon index typically lies in the range $\Gamma_{\rm th} \sim 1.90-2.70$ with luminosities of $\sim 2-30 \% \, {\rm L}_{\rm Edd}$ and $1.5 \lesssim F_{\rm ratio} \lesssim 7$. Sources such as Swift J1658.2$-$4242, 4U 1543$-$47, Cyg X$-$1, Cyg X$-$3 and GX 339$-$4 are observed in this state with the detection of type-B QPOs in some of them. In the subsequent HSS, the spectrum becomes even softer with $\Gamma_{\rm th} \geq 2.20$, with luminosities $\ge 50 \% \, {\rm L}_{\rm Edd}$ (occasionally emitting at super-Eddington luminosity), and $F_{\rm ratio} \geq 7.00$. Note that, Cyg X$-$3, 4U 1957$+$115, 4U 1543$-$47, 4U 1630$-$47, GX 339$-$4, MAXI J1820$+$070, MAXI J1348$-$630, and MAXI J1803$-$298 are observed in the HSS. In addition, the persistent source 4U 1957$+$115 shows comparatively low luminosity ($\sim 1 \% \, {\rm L}_{\rm Edd}$) during the HSS.

Overall, we find that $\Gamma_{\rm th}$ remains nearly constant for $F_{\rm ratio} < 1.00$, whereas $L_{\rm bol}$ increases marginally. As the disc flux becomes dominant ($F_{\rm ratio} > 1.00$), we observe spectral softening ($\Gamma_{\rm th} > 2.00$) along with the increase of source brightness. For $F_{\rm ratio} > 7$, the source predominantly resides in the thermal disc dominated HSS with $\Gamma_{\rm th} > 2.20$. This comprehensive spectral state classification based on $\Gamma_{\rm th}$, $F_{\rm ratio}$, and $L_{\rm bol}$ provides useful framework, with exceptions only in few observations of Cyg X$-$3, GX 339$-$4 and MAXI J1803$-$298.

As before, we examine the correlation between $\Gamma_{\rm th}$ and $F_{\rm ratio}$ by fitting an empirical power-law function as $\Gamma_{\rm th}=a\times F_{\rm ratio}^{b}$. Here, we have excluded observations of Cyg X$-$3, MAXI J1535$-$571, GX 339$-$4 (AS22, AS23) and MAXI J1803$-$298 (observation in HSS) due to anomalous characteristics which we discuss below. The best fit parameters are obtained as $a=2.29\pm0.04$ and $b=0.09\pm0.01$. In Fig. \ref{fig:fig31} (right), the dashed line denotes the best fit power-law function and grey shaded regions with increasing transparency represent $1 \sigma$, $2 \sigma$ and $3 \sigma$ confidence intervals. In addition, we estimate the inner disc temperature $kT_{\rm in}$ for each observation and represent it using a color scale with the corresponding range of $kT_{\rm in}$ displayed in the vertical color bar. During HIMS, $kT_{\rm in}$ lies in the range $\sim 0.50-1.50$ keV, while in the SIMS it is typically $\sim 0.70-1.50$ keV. In the HSS, $kT_{\rm in}$ is found in the range $\sim 0.70-1.98$ keV. Overall, $kT_{\rm in}$ increases systematically with both $\Gamma_{\rm th}$ and $F_{\rm ratio}$ reflecting the combined influence of spectral softening and disc dominance on the inner disc temperature.

Cyg X$-$3 exhibits unusually high luminosity in all spectral states reaching super‑Eddington values in the SIMS and HSS. Its high luminous hard-state has been attributed either to state transitions occurring at intrinsically high accretion rates or to a comparatively massive compact object \cite[]{Hjalmarsdotter-etal2008}. \citet{Lasota-King2023} classify Cyg X$-$3 as a ULX proposing that disc‑wind beaming enhances its apparent luminosity, while \citet{Veledina-etal2024} argue that geometrical beaming from a narrow outflow with a half opening angle of $\sim 15^{\circ}$ and an orbital inclination of $29.5^{\circ} \pm 1.2^{\circ}$ can reproduce the observed luminosity. This geometry places the line of sight close to the outflow axis supporting beaming as the primary cause of the high observed flux. The comparatively low photon index ($\Gamma_{\rm th}$) observed in the HSS of Cyg X$-$3 (with lower $kT_{\rm in}$) may arise from inefficient Compton cooling of the corona due to (a) wind mediated obscuration of seed photons by the dense Wolf–Rayet wind \cite[]{Hjalmarsdotter-etal2008}, and (b) persistent coronal activity and beaming at high accretion rates, which may prevent the source from attaining a canonical HSS \cite[]{Veledina-etal2024}.

MAXI J1535$-$571 exhibits high luminosity $\ge20\%~L_{\rm Edd}$ during HIMS. However, despite showing such high luminosity, the $kT_{\rm in}$ is observed to be low $\le 0.50$ keV, which may be attributed to high accretion rate \cite[]{Sreehari-etal2019b}.

For GX 339$-$4, two observations (AS22 and AS23) from the $2024$ outburst show relatively low $\Gamma_{\rm th} \sim 2.00$ and a bolometric luminosity of $\sim 4 \%\,{\rm L}_{\rm Edd}$, while the flux ratio remains high ($F_{\rm ratio} \sim 10$). Although such a high flux ratio might suggest a high‑soft state, the spectral parameters are more consistent with the SIMS. The comparatively low inner disc temperature ($kT_{\rm in} \sim 0.65$ keV) relative to other HSS observations naturally explains both reduced $L_{\rm bol}$ and harder $\Gamma_{\rm th}$. This possibly happens due to the limited soft‑photon supply that limits Compton cooling and maintains a hotter corona. At these luminosities, the disc is likely not fully extended to the ISCO allowing partial truncation and supporting a moderately strong coronal component, which in turn produces the harder Comptonized spectrum. The lower overall flux observed during the $2024$ outburst (see Fig. \ref{fig:fig25}) compared to previous outbursts further suggests a reduced mass‑accretion rate and a correspondingly less extended disc. Although the disc still contributes substantial seed photons, the relatively large corona enhances hard X‑ray emission yielding a high $F_{\rm ratio}$ together with a low $\Gamma_{\rm th}$. Altogether, the spectral properties indicate that the source is more likely in a SIMS rather than HSS.

MAXI J1803$-$298 exhibits relatively low luminosity $\sim 1.8\%-2.9\%~L_{\rm Edd}$ during the decay phase in HSS/SIMS compared to other spectral states. The observed inner disc temperature is low ($kT_{\rm in} \sim 0.63-0.67$ keV) with weak Comptonized emission indicating a reduced supply of seed photons. Indeed, the reduced mass accretion rate during these observations naturally lead to the observed low luminosity during SIMS and HSS.

\subsection{Spectro-temporal Correlation}\label{sec8:8.2_spec_tempo_correlation}

\begin{figure*}[h]
    \begin{center}
    \includegraphics[width=\textwidth]{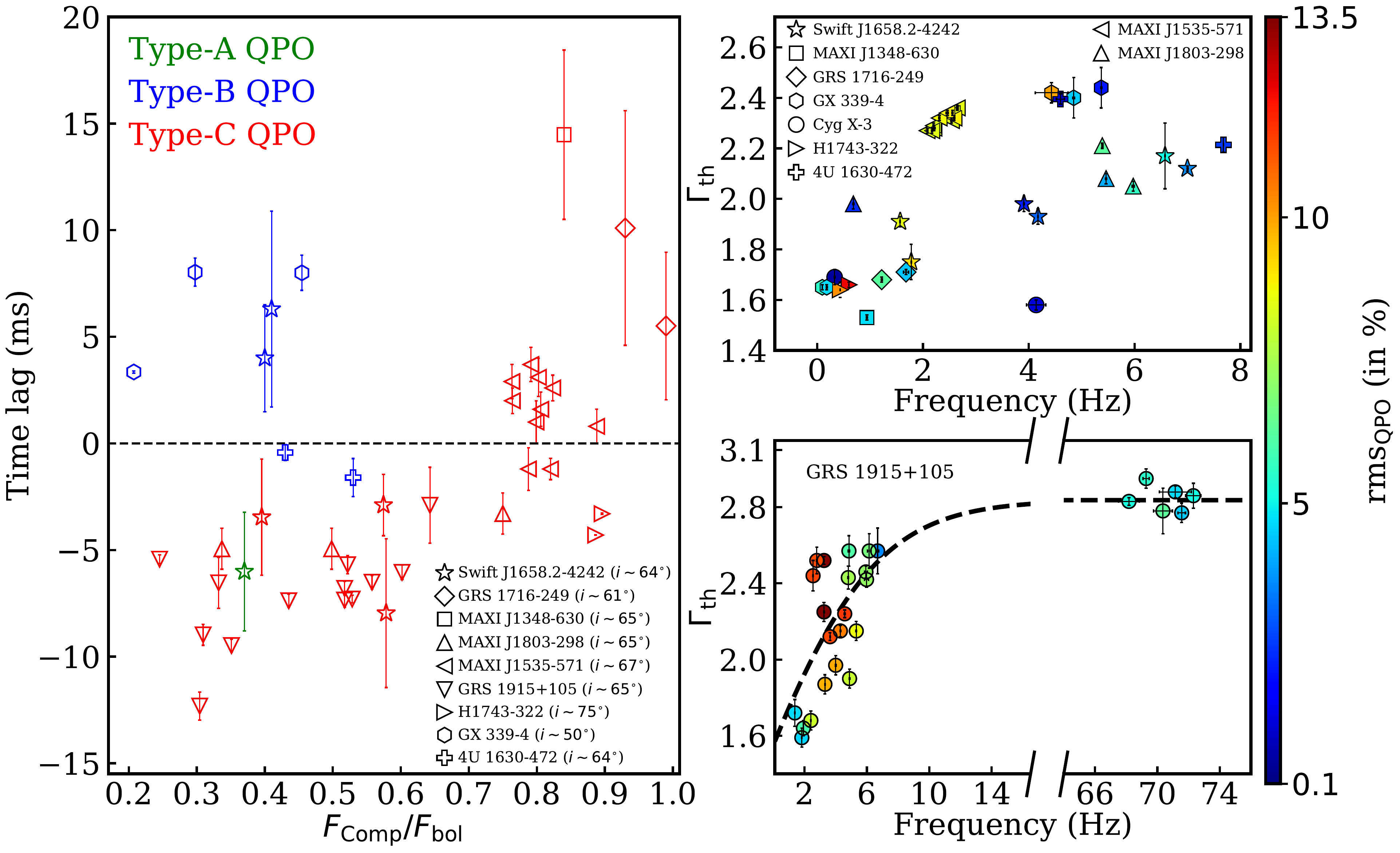}
    \end{center}
    \caption{{\it Left:} Variation of time-lag (in ms) with normalised Comptonized flux ($F_{\rm Comp}/F_{\rm bol}$) for Swift J1658.2$-$4242, GRS 1716$-$249, MAXI J1348$-$630, MAXI J1803$-$298, MAXI J1535$-$571, GRS 1915$+$105, H1743$-$322, GX 339$-$4 and 4U 1630$-$472. Green, blue and red denote type-C, type-B and type-A QPOs. {\it Right-Top:} Variation of $\Gamma_{\rm th}$ with LFQPO frequency ($\nu_{\rm QPO}$) for different BH-XRBs. Colorbar represents rms$_{\rm QPO}\%$. {\it Right-Bottom:} Variation of $\Gamma_{\rm th}$ as function of $\nu_{\rm QPO}$ for GRS 1915$+$105, where dashed curve represents the empirical fitted function. See text for details.
    }
    \label{fig:fig32}
\end{figure*}

We perform a correlation study between timing parameters, such as time-lag ($t_d$), $\nu_{\rm QPO}$ and rms$_{\rm QPO}$, and spectral parameters including the normalised Comptonised flux ($F_{\rm Comp}/F_{\rm bol}$) and the photon index ($\Gamma_{\rm th}$). The left panel of Fig. \ref{fig:fig32} illustrates the correlation between the time-lag and $F_{\rm Comp}/F_{\rm bol}$, where different symbols denote different sources, and green, blue, and red colors represent type-A, type-B, and type-C QPOs, respectively (see Table \ref{tab:Table-06} for details). In this work, BH-XRBs under consideration are highly inclined systems ($i > 60^\circ$), except GX 339$-$4 with $i \sim 50^{\circ}$. Note that GRS 1716$-$249, MAXI J1348$-$630, MAXI J1535$-$571, GRS 1915$+$105, and H1743$-$322 exhibit type-C QPOs, while Swift J1658.2$-$4242 shows all three types of QPOs. Interestingly, GX 339$-$4 exhibits both type-B and type-C QPOs, where as 4U 1630$-$472 shows only type-B QPOs. We observe that type-C QPOs are generally associated with soft-lags for $0.2 < F_{\rm Comp}/F_{\rm bol} < 0.9$, however MAXI J1535$-$571 shows both hard- and soft-lags at higher $F_{\rm Comp}/F_{\rm bol}$. Moreover, GRS 1716$-$249 and MAXI J1348$-$630 show only hard-lags with large error bars. Type-B QPOs show hard-lags at $F_{\rm Comp}/F_{\rm bol} \lesssim 0.5$, whereas type-A QPOs exhibit soft-lag. These findings are in good agreement with the previous study \cite[][for details]{Choudhury-etal2025}.

The right-top panel of Fig. \ref{fig:fig32} illustrates the correlation among $\nu_{\rm QPO}$, $\Gamma_{\rm th}$ and rms$_{\rm QPO}$ for the same sources considered in left panel of Fig. \ref{fig:fig32}. The color denotes rms$_{\rm QPO}$ and colorbar at the right refers to its range. We find a modest correlation between $\nu_{\rm QPO}$ and $\Gamma_{\rm th}$ for Swift J1658.2$-$4242, GRS 1716$-$249, MAXI J1348$-$630, MAXI J1803$-$298, Cyg X$-$3, and MAXI J1535$-$571 indicating transitions from harder to softer spectral states with increasing $\nu_{\rm QPO}$. However, no definite correlation is observed between $\nu_{\rm QPO}$ and rms$_{\rm QPO}$.

The right-bottom panel of Fig. \ref{fig:fig32} shows the correlation results for GRS 1915$+$105. We observe that $\Gamma_{\rm th}$ increases from $\sim 1.70$ to $2.60$ as $\nu_{\rm QPO}$ rises from $\sim 1.40$ to $70$ Hz, while for HFQPOs, $\Gamma_{\rm th}$ tends to saturate at $\sim 2.80$ with increasing frequency. This behaviour indicates an overall increase in $\Gamma_{\rm th}$ with $\nu_{\rm QPO}$ for type-C QPOs followed by saturation beyond a characteristic frequency. In addition, we observe a systematic decrease of rms$_{\rm QPO}$ with increasing $\nu_{\rm QPO}$ and $\Gamma_{\rm th}$. Furthermore, we empirically use a mathematical relation \cite[]{Titarchuk-Fiorito-2004} to fit $\nu_{\rm QPO}-\Gamma_{\rm th}$ correlation, which is given by,
\begin{equation}
\Gamma_{\rm th}=A-DB\ln\left[\exp\left(\frac{\nu_{\rm tr}-\nu}{D}\right)+1\right],
\end{equation}
where $A$ is the saturation photon index, $B$ is the slope in the low-frequency regime, $\nu_{\rm tr}$ is the transition frequency at which the $\Gamma_{\rm th}$–$\nu_{\rm QPO}$ relation tends to saturate, and $D$ controls the sharpness of the transition. It is worth noting that this correlation was earlier studied for LFQPOs \cite[]{Shaposhnikov-etal-2007a} using $D=1.00$ Hz to estimate the mass of GRS 1915$+$105 as $\sim 15\,\mathrm{M}_{\odot}$ via the scaling method. We extend this study for HFQPOs as well, where empirical good fit yields $D=3.30$. With this consideration, the best fit model parameters are estimated as $A=2.84\pm0.01$, $B=0.26 \pm 0.04 ~ \mathrm{Hz}^{-1}$, and $\nu_{\rm tr} = 4.11 \pm 0.04$ Hz, which indicates that the saturation occurs during the harder to softer state transition. 

Following \cite{Shaposhnikov-etal-2007}, we estimate source mass as $M_{\rm GRS} = (B_{\rm GRS}/B_{\rm GRO}) \times M_{\rm GRO}$. We consider GRO J1655$-$40 as a reference source adopting $M_{\rm GRO} \sim 6.3 \,{\rm M}_\odot$ and $B_{\rm GRO} \sim 0.122 \pm 0.008$ Hz$^{-1}$. Using the value $B_{\rm GRS} = 0.26 \pm 0.04 ~ \mathrm{Hz}^{-1}$ for GRS 1915$+$105, we compute its mass as $\sim 12.30 \pm 2.25 \,{\rm M}_{\odot}$, which is consistent with dynamically measured mass estimate.

\section{Accretion disc dynamics: Model prescription}\label{sec9:accretion_model}

Motivated by the observational results presented in the previous sections for BH–XRBs observed with \textit{AstroSat}, we investigate the accretion dynamics by adopting two complementary approaches. First, we carry out time-dependent study of the accretion flow to investigate the origin of the QPOs observed in the BH-XRBs. Second, we perform wide-band spectral modelling using a hybrid two-component accretion flow framework relevant for BH–XRBs. The details of these approaches are described in the subsequent subsections.

\subsection{Simulation of Accretion Flow}\label{sec9:9.1_simulation}

In black hole accretion, the infalling matter initially starts its journey subsonically from the outer edge of disc with negligible radial velocity. As it spirals inward under the influence of strong gravity, the radial velocity increases, and the flow becomes supersonic after passing through a critical point \cite[][and references therein]{Bondi1952,Ichimaru1977,Fukue1987,Chakrabarti1989,Chakrabarti-Das2004}. Depending on the angular momentum, the flow experiences centrifugal barrier in the vicinity of the horizon, causing the flow to decelerate and pile up. This eventually trigger shock transition producing a hot, dense post-shock region (equivalently post-shock corona, hereafter PSC), which acts as an effective Comptonizing boundary layer. Numerous analytical studies have investigated such shocks and their physical implications \cite[]{Fukue1987, Chakrabarti1989, Lu-etal1999, Becker-Kazanas2001, Das-etal2001a,Fukumura-Tsuruta2004, Chakrabarti-Das2004, Das2007,Aktar-etal2015, Aktar-etal2017, Sarkar-etal2018, Dihingia-etal2018a,Mitra-Das2024,Kumar-etal2025} while studying accretion flow properties around black holes. On the numerical front, time dependent simulations provide a more realistic description of accreting flows that harbour shock waves. Several numerical investigations explored shock oscillation in black hole accretion \cite[][and references therein]{Molteni-etal1996,Chakrabarti-etal2004c, Giri-etal2010, Lee-etal2011, Das-etal2014,Okuda-Das2015, Lee-etal2016, Okuda-etal2019, Okuda-etal2022, Okuda-etal2023, Sukova-etal2017,SDebnath-etal2024,Mao-etal2025,Dihingia-etal2025}. It is worth noting that most of these models primarily focused on LFQPOs. Recently, \cite{Aktar-etal2026} examine both LFQPOs ($\nu \lesssim 30$ Hz) and HFQPOs ($\nu \gtrsim 40$ Hz) using a hydrodynamic simulation framework around a rotating black hole, demonstrating that such temporal variabilities strongly depend on the flow parameters (e.g., specific energy and angular momentum) as well as the black hole spin. Motivated by these developments, we investigate the possible origin of both LFQPO and HFQPO features observed in GRS 1915$+$105 with \textit{AstroSat}.

\subsubsection{Simulation Setup and Governing Equations}

We conduct a time-dependent, axisymmetric, inviscid, and non-radiative simulation of accretion flow around a rotating black hole. The present analysis utilizes the hydrodynamics (HD) module of the publicly available numerical simulation code, PLUTO\footnote{\url{http://plutocode.ph.unito.it}} \cite[]{Migone-etal2007}. We adopt a unit system in which $ G = M_{\rm BH} = c = 1 $, where $G$ is the gravitational constant, $ M_{\rm BH} $ is the mass of the black hole, and $c$ is the speed of light. Consequently, we define the measurements of distance, velocity, and time as follows: $ r_g = G M_{\rm BH}/c^2$, $c$, and $t_g = G M_{\rm BH}/c^3$, respectively.

The governing equations for hydrodynamics of accretion flow around black hole are described as \cite[]{Migone-etal2007},
\begin{align}
& \label{governing_eq_1} \frac{\partial \rho}{\partial t} + \nabla \cdot (\rho \bm{v}) =0,\\
& \label{governing_eq_2} \frac{\partial (\rho \bm{v})}{\partial t} + \nabla \cdot (\rho \bm{v} \bm{v}) + \nabla P = -\rho \nabla \Phi^{\rm eff}, \\
& \frac{\partial E }{\partial t} + \nabla \cdot (E+P)\bm{v}  = -\rho \bm{v}.\nabla \Phi^{\rm eff},
\end{align}
where $\rho$, $\bm{v}$, and $P$ are the mass density, fluid velocity, and gas pressure, respectively. $E$ is the total energy density which is given by,
\begin{align}
E = \frac{P}{\gamma -1} + \frac{1}{2} \rho v^2 + \Phi^{\rm eff},
\end{align}
where constant $\gamma$ is the adiabatic index. Here, $\Phi^{\rm eff}$ refers to the effective potential of the rotating black hole \cite[]{Dihingia-etal2018b}.

Following \cite{Aktar-etal24a,Aktar-etal24b}, we implement $\Phi^{\rm eff}$ in the PLUTO code considering cylindrical coordinates ($r$, $\phi$, $z$). The details of the effective potential are given by \cite[]{Dihingia-etal2018b},
\begin{align} \label{DDMC18_pot}
\Phi^{\rm eff}\left(r,z,a_{\rm k}, \lambda\right)= \frac{1}{2} \ln \left(\frac{A (2 R - \Sigma) r^2 - 4 a_{\rm k}^2 r ^4}{\Sigma \lambda \left(\Sigma \lambda  R^2 + 4 a_{\rm k} r ^2 R - 2 \lambda  R^3\right)- A \Sigma r ^2 }\right),
\end{align}
where $R~(=\sqrt{r^2+z^2})$ is the spherical radial distance, $\Delta=a_{\rm k}^2+R^2-2 R$, $\Sigma = R^2 + a_{\rm k}^2 z^2/R^2$ and $A=\left(a_{\rm k}^2+R^2\right)^2 - a_{\rm k}^2 r^2 \Delta/R^2$. Here, $\lambda$ and $a_{\rm k}$ are specific flow angular momentum and black hole spin, respectively. The dimensionless black hole spin is defined as $a_{\rm k} = J/M_{\text{BH}}$, where $J$ represents the angular momentum of the black hole. The Keplerian angular momentum can be obtained from equation (\ref{DDMC18_pot}) in the equatorial plane $(z \rightarrow 0)$ as $\lambda^2_K = r^3 \left(\partial \Phi^{\rm eff}/\partial r\right)_{\lambda \rightarrow 0}$. The angular frequency is obtained as $\Omega = \lambda/r^2$. 

It is worth mentioning that \cite{Dihingia-etal2018b} performed detailed analytical comparisons between general relativity (GR) and the effective Kerr potential showing that the latter reproduces GR transonic properties well within the semi-relativistic regime around Kerr black holes. This validates the use of the effective Kerr potential, which captures the essential spacetime features while enabling significantly higher spatial resolution than GRMHD simulations for the same computational resources \cite[]{Aktar-etal24b, Aktar-etal25,Mao-etal2025,Dihingia-etal2025}.

\begin{table*}
\centering
\begin{threeparttable}
\caption{Details of the simulation models. In column 1, we mention the model, while model parameters are listed in columns $2-7$. In last column, we tabulate the estimated $\nu_{\rm QPO}$ values. See text for details.} \label{tab:Table-15}
\begin{tabular}{c c c c c c c c}
\hline
Model & $a_{\rm k}$ & $E$  & $\lambda$ &  $v_{\rm out}$  & $T_{\rm out}$ & $h_{\rm out}/r_{\rm out}$ & $\nu_{\rm QPO}$ \\
      &  & ($c^2$) & ($r_g c$) &  ($c$) & ($K$) &  &   (Hz)  \\ \hline 
NOQPO  & 0.98 & 0.001  & 2.00  & $-0.101348$ & $7.6859 \times 10^9$      & 0.372981 & --- \\ 

LFQPO  & 0.98 & 0.002  & 2.42  & $-0.088559$ & $1.0730 \times 10^{10}$      & 0.440700 & $9.95  \pm 0.28 $, $20.95 \pm 0.38$ \\ 

HFQPO  & 0.98 & 0.0005 & 2.25 & $-0.111679$ & $5.5620 \times 10^9$      & 0.317336 & $71.98 \pm 0.48$, $143.45 \pm 0.22$, $214.65 \pm 0.28$ \\ \hline
\end{tabular}
\end{threeparttable}
\end{table*}

\subsubsection{Initial and Boundary Conditions}\label{initial_cond}

We simulate two-dimensional hydrodynamic flows in cylindrical coordinates $(r, \phi, z)$ using the PLUTO code. In doing so, we employ the HLL Riemann solver along with second-order linear interpolation in space and the second-order Runge-Kutta algorithm in time \cite[]{Migone-etal2007}. The computational domain is defined as $0.1 r_g \leq r \leq 100r_g$ in the radial direction and $-100 r_g \leq z \leq 100r_g$ in the vertical direction. {\red The large extent of the outer boundary effectively minimizes any spurious influence of boundary conditions on the shock dynamics.} We set the grid resolution to $(n_r, n_z) = (400, 800)$, maintaining uniform grid spacing in both radial and vertical directions. {\red Here, we note that increasing the grid resolution further, the key results remain qualitatively unaltered \cite[and references therein]{Aktar-etal2026}.} The effective inner boundary is located at $R_{\rm in} = (R_{\rm H} + 0.2) r_g$, which represents the location of the horizon \cite[]{Aktar-etal24a}. Here, $R_{\rm H}~\left[= 1 + \sqrt{1 - a_{\rm k}^2}\right]$ is the event horizon's location. At this effective inner boundary, where $r \leq R_{\text{in}}$, the radial velocity is set to the free-fall value. The other variables are determined through interpolation from the outer region near the boundary. All other boundaries are set to outflow boundary conditions \cite[]{Aktar-etal24a,Aktar-etal24b,Aktar-etal2026}. This setup ensures that the inner boundary minimally influences the shock oscillations, as inflowing matter transits to a supersonic state at this boundary. {\red For simplicity, the present study neglects explicit dissipation and radiative effects within the accretion flow. Furthermore, the use of two-dimensional simulations is warranted because the flow maintains approximate symmetry with respect to the disc equatorial plane \cite[]{Aktar-etal25,Dihingia-etal2025}.}

To begin with, we specify the initial conditions at the outer boundary ($r_{\rm out}$), including the flow angular momentum ($\lambda$), radial speed ($v_{\rm out}$), temperature ($T_{\rm out}$), and the spin ($a_{\rm k}$) of the black hole. The input flow parameters for the simulation are derived from $1.5-$dimensional steady state model calculations under the geometrically thin flow approximation \cite[]{Aktar-etal2015,Aktar-etal2017}, utilizing the effective Kerr potential as outlined in equation \eqref{DDMC18_pot}. Moreover, we calculate the local disc height at $r_{\rm out}$ as $h_{\rm out} \simeq c_s \sqrt{ \frac{r_{\rm out}}{\gamma}} \left( \frac{d\Phi_{\rm eff}}{dr}\right)^{-1/2}_{z < r} $, where $c_s$ is the local sound speed and $\gamma$ denotes the adiabatic index considered to be constant $(\gamma=4/3)$. Furthermore, in this work, we estimate the internal energy loss per unit volume (entire computational domain) due to free-free emission (Bremsstrahlung), which does not affect the flow configuration. The corresponding amplitude of the volume integrated luminosity is calculated as $L_{\rm ff} = \int{q_{\rm ff} dV}$, where $q_{\rm ff}$ represents the free-free emissivity \cite[]{Okuda-etal2019, Aktar-etal24a}.

\begin{figure*}[h]
	\begin{center}
        \includegraphics[width=0.33\textwidth]{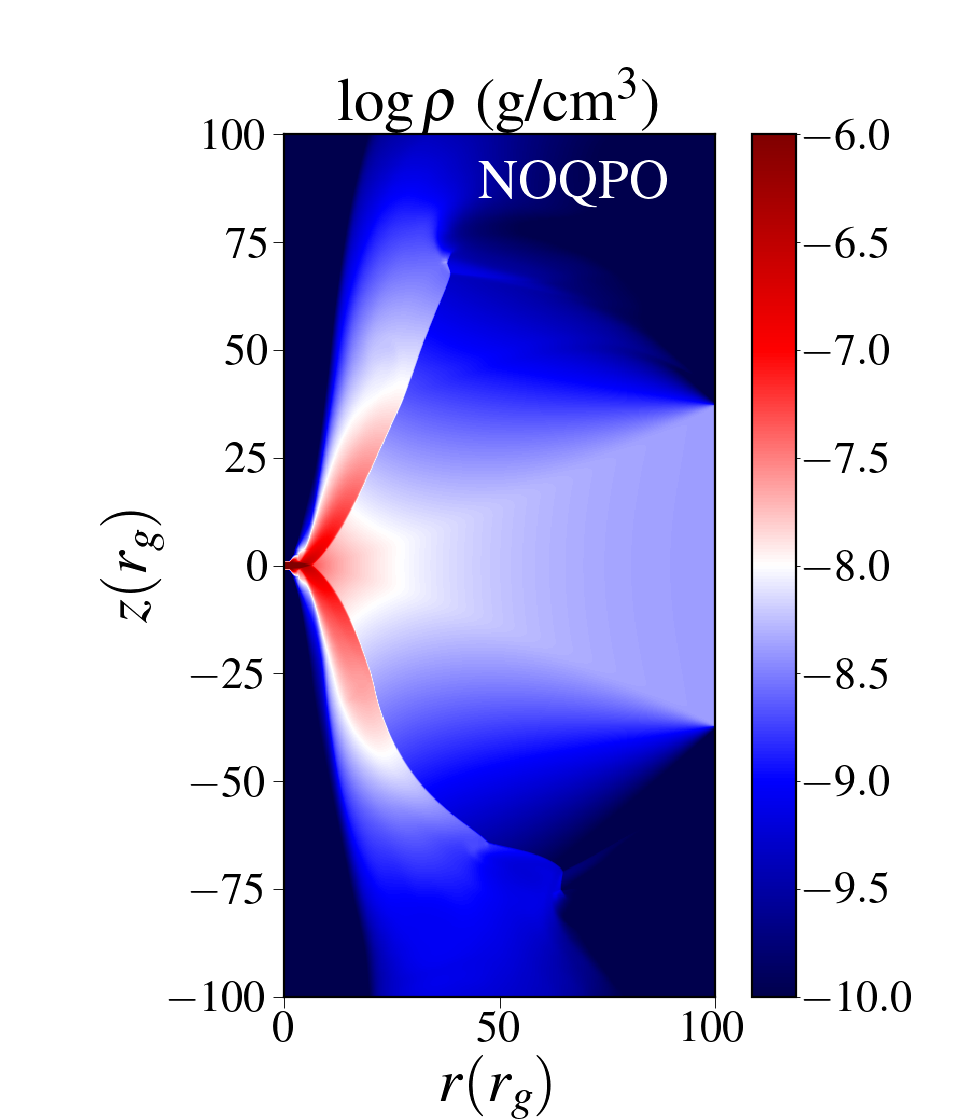} 
        \includegraphics[width=0.33\textwidth]{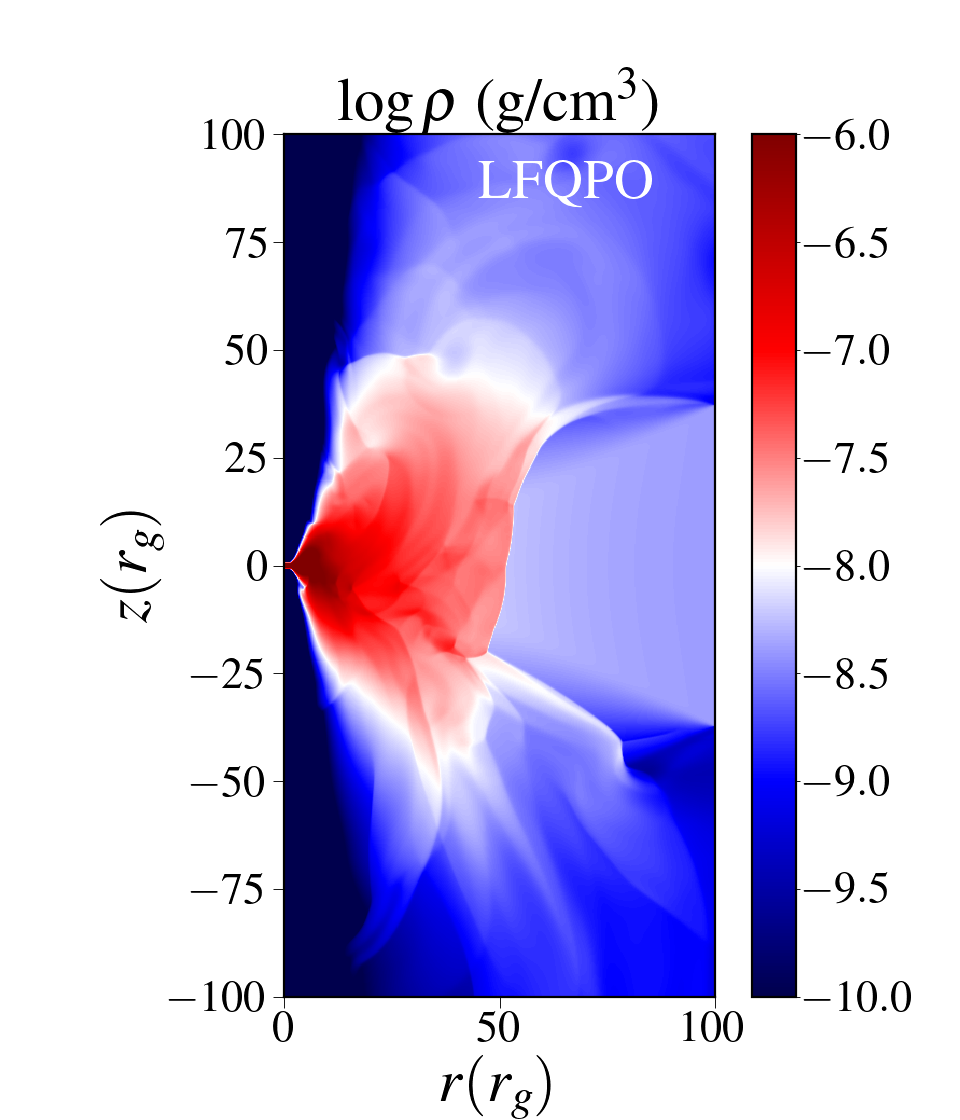} 
        \includegraphics[width=0.33\textwidth]{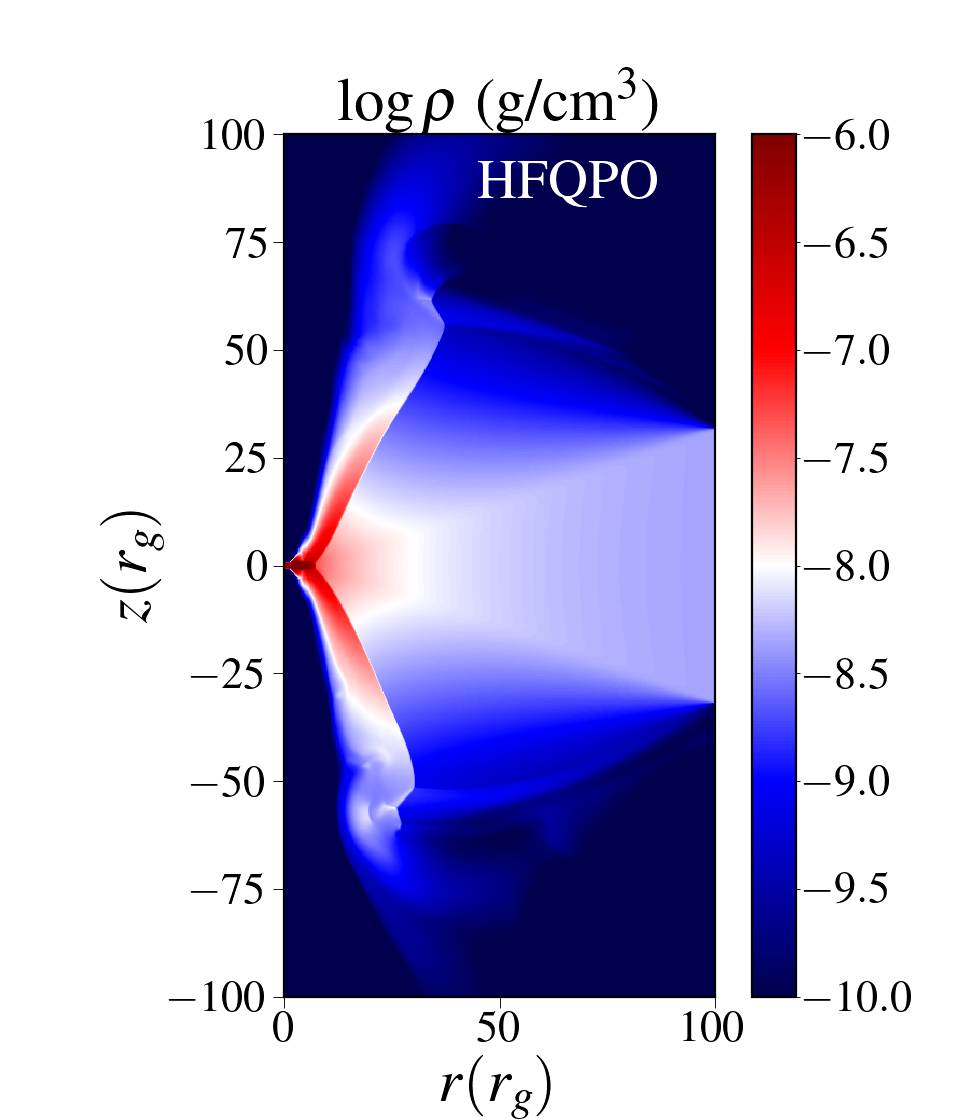}
	\end{center}
	\caption{Two-dimensional spatial distributions of mass density ($\rho$) in the meridional ($r$–$z$) plane are presented for three simulation models, such as NOQPO, LFQPO, and HFQPO at time $t = 5$ seconds. The region with high density contrast near horizon is referred as PSC. See text for details.
    }
    \label{fig:fig33}
\end{figure*}

\subsubsection{Simulation Results: Application to GRS 1915$+$105}

\begin{figure*}[h]
	\begin{center}
        \includegraphics[width=0.9\textwidth]{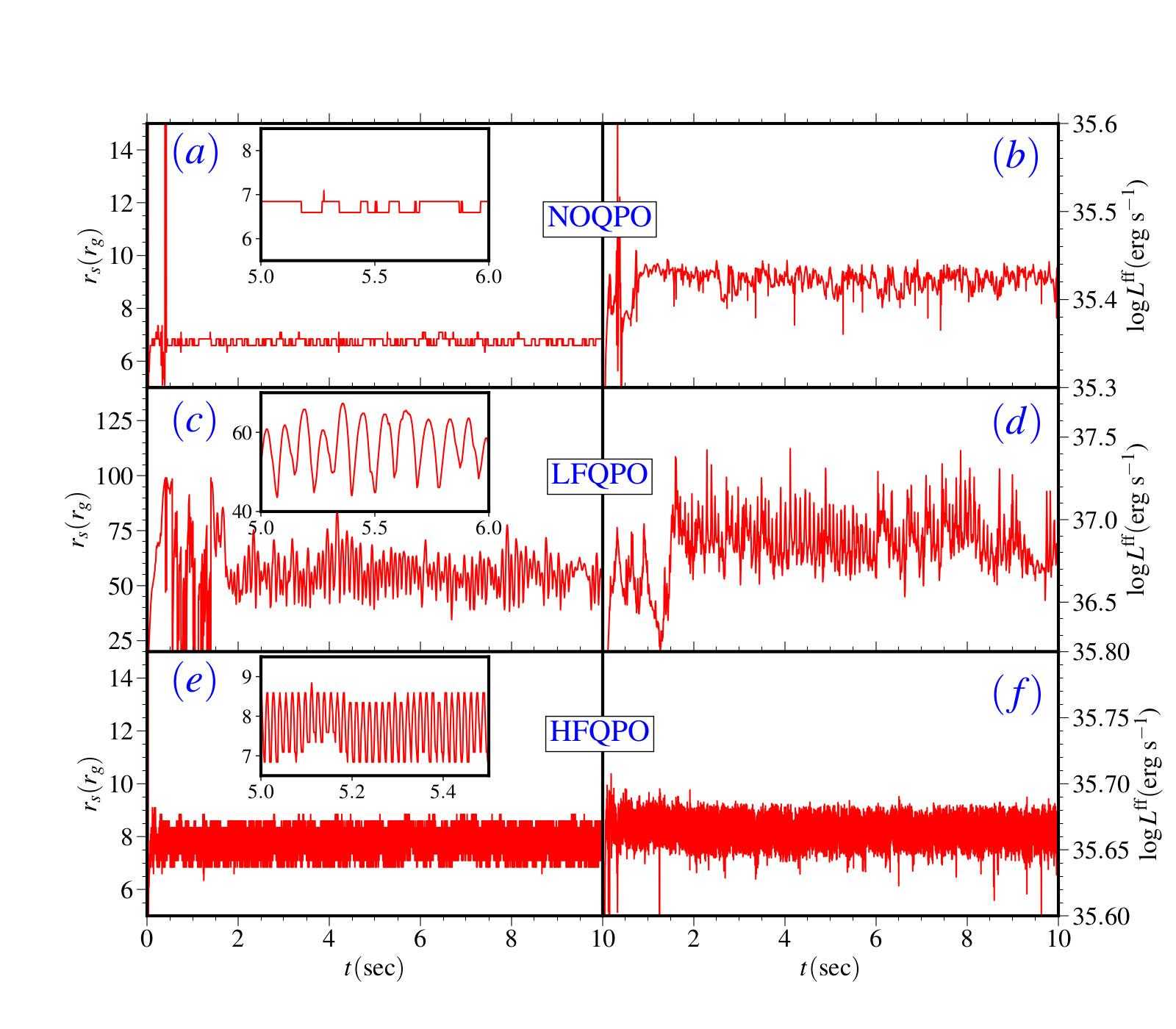} 
	\end{center}
    \vskip -1 cm
	\caption{Time variations of shock location (left panel) and luminosity (right panel) for NOQPO model (top), LFQPO model (middle), and HFQPO model (bottom). The inset shows a magnified view of the variation of shock location. See text for details.
    }
    \label{fig:fig34}
\end{figure*}

In this work, we focus on simulating the QPOs commonly observed from GRS 1915$+$105 together with the corresponding dynamical configuration that does not exhibit any QPO signature (hereafter NOQPO). This investigation is based on the hydrodynamical shock oscillation framework \cite[]{Aktar-etal2026}. Here, we consider $M_{\rm BH} = 12\,{\rm M}_{\odot}$ and $a_k = 0.98$ for GRS 1915$+$105. Three simulation models are outlined to investigate the HFQPO, LFQPO, and NOQPO scenarios in GRS 1915$+$105 and the corresponding input flow parameters are summarised in Table \ref{tab:Table-15}. The spatial distribution of density ($\rho$) in the ($r$$-$$z$) plane for these models is illustrated in Fig. \ref{fig:fig33}. In each case, the high density region near the equatorial plane identifies the PSC consistent with previous studies \cite[]{Das-etal2014, Aktar-etal2015, Okuda-etal2019,SDebnath-etal2024,Mao-etal2025}. The modulation characteristics of the PSC differ among the models resulting in the distinct QPO signatures.
In Fig. \ref{fig:fig34}, we examine the temporal evolution of the shock location ($r_s$) and the associated free–free luminosity ($L_{\rm ff}$) for three simulation models. The upper, middle, and lower rows correspond to the NOQPO, LFQPO, and HFQPO model, respectively. For the NOQPO model, the variations of $r_s$ and $L_{\rm ff}$ are presented in Fig. \ref{fig:fig34}a and \ref{fig:fig34}b. The inset in Fig. \ref{fig:fig34}a shows negligible fluctuations in $r_s$, which remains consistent with a steady shock configuration \cite[]{Chakrabarti-1990,Das-etal2001a,Das2007,Aktar-etal2015,Dihingia-etal2018a,Mitra-Das2024,Kumar-etal2025}. Figs. \ref{fig:fig34}c and \ref{fig:fig34}d illustrate the temporal variation of $r_s$ and $L_{\rm ff}$ for the LFQPO model. We find moderate variability in the shock front at a frequency of $\sim 10$ cycles per second with its location varying between $\sim 45r_g$ and $\sim 65r_g$, as seen in the inset of Fig. \ref{fig:fig34}c. The corresponding luminosity variations display a similar modulation pattern (Fig. \ref{fig:fig34}d). For the HFQPO model, the evolution of the shock location and luminosity is shown in Figs. \ref{fig:fig34}e and \ref{fig:fig34}f. In this case, the shock undergoes rapid oscillations between $\sim 6r_g$ and $9r_g$, as highlighted in the inset of Fig. \ref{fig:fig34}e. The PSC becomes significantly smaller and is located very close to the event horizon \cite[see][]{Aktar-etal2026}, which can also be visualized in the 2D density distributions in Fig. \ref{fig:fig33}. Consequently, the amplitude of the luminosity variations appears to be comparatively low (Fig. \ref{fig:fig34}f) with respect to LFQPO scenario.

\begin{figure}[h]
	\begin{center}
        \includegraphics[width=\columnwidth]{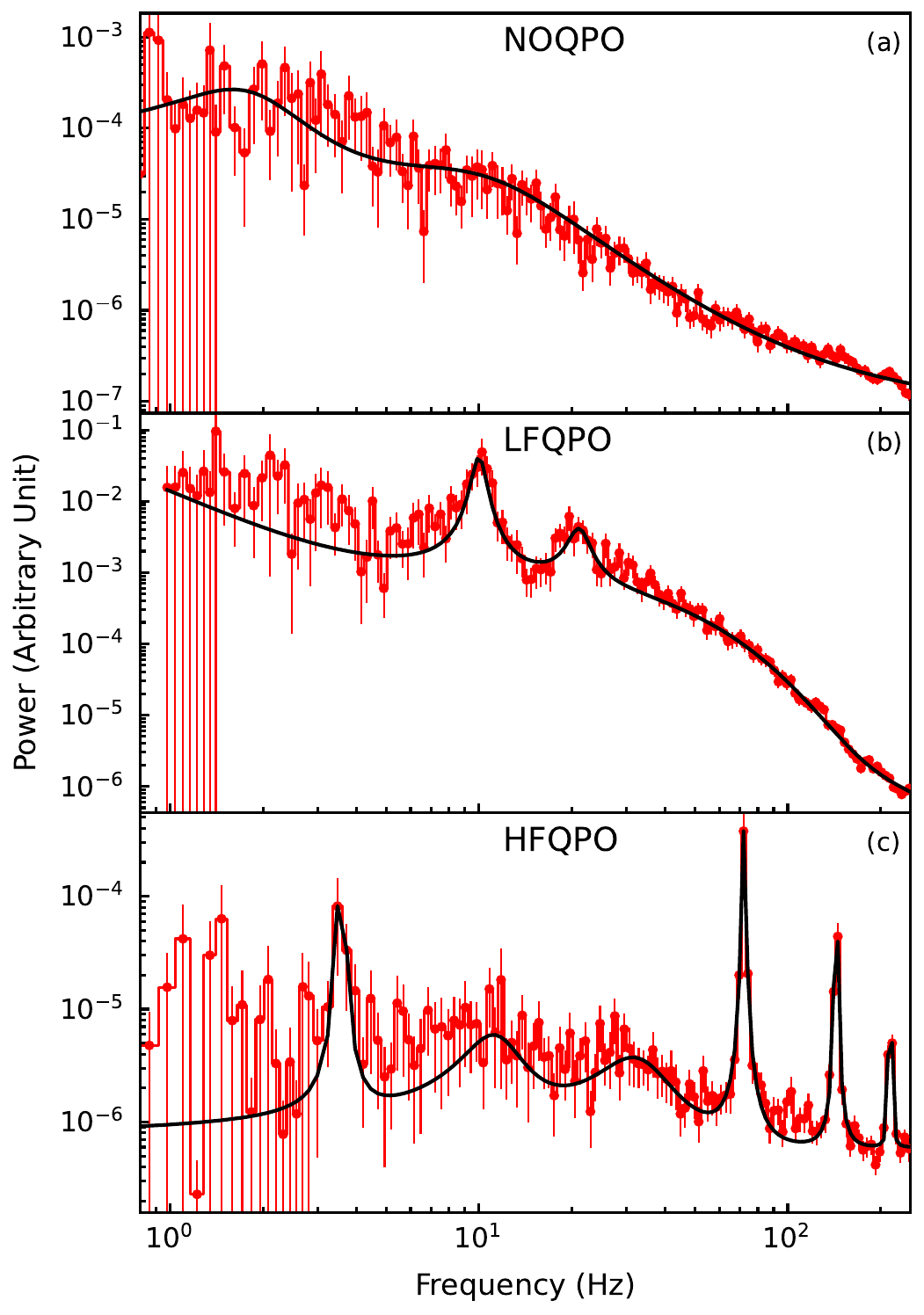}  
	\end{center}
	\caption{Power density spectra generated from the luminosity variations for NOQPO model (top), LFQPO model (middle), and HFQPO model (bottom). See text for details.
    }
    \label{fig:fig35}
\end{figure}

To explore the temporal variabilities in the luminosity, we compute PDS from the luminosity variations for all models, as shown in Fig. \ref{fig:fig35}. Each PDS is fitted using multiple Lorentzian components and a power-law. As expected, the NOQPO model does not exhibit any distinct QPO signature (Fig. \ref{fig:fig35}a). The LFQPO model displays a dominant peak at $\nu_{\rm QPO} \sim 10$ Hz accompanied by a harmonic at $\nu_{\rm QPO} \sim 20$ Hz (Fig. \ref{fig:fig35}b). The HFQPO model reveals a high frequency peak at $\nu_{\rm QPO} \sim 72$ Hz, along with two harmonics at $\nu_{\rm QPO} \sim 144$ Hz and $\nu_{\rm QPO} \sim 215$ Hz (Fig. \ref{fig:fig35}c). In addition, a LFQPO like feature at $\sim 3.5$ Hz is also present. The simultaneous appearance of LFQPO and HFQPO signatures has been reported in a few BH–XRBs \cite[]{Cui-etal2000, Strohmayer2001b, Majumder-etal2025b}. The model fitted QPO frequencies for all three simulations are listed in the last column of Table \ref{tab:Table-15}.

Overall, we examine three representative models that reproduce NOQPO, LFQPO ($\nu_{\rm QPO} \sim 10$ Hz), and HFQPO ($\nu_{\rm QPO} \sim 72$ Hz) scenarios, each of which satisfactorily accounts for the observed QPO properties of GRS 1915$+$105. In agreement with the shock oscillation framework, PSC resides much closer to the event horizon and is considerably more compact for HFQPO model than the LFQPO model (Fig. \ref{fig:fig33}-\ref{fig:fig34}). Dissipation processes, magnetic fields, and radiation processes \cite[]{Das-etal2014, Okuda-etal2019,Okuda-etal2022,Okuda-etal2023}, which are known to influence QPO dynamics, are not included in the present simulations and will be presented elsewhere.

\subsection{Spectral Modelling of Accretion Flow}\label{sec9:9.2_HTCM}

The spectral properties of BH-XRBs serve as powerful tools for understanding accretion physics inferring black hole spin, and elucidating the geometry of matter in strong gravitational fields. Following the simulation results discussed in the previous section, we now examine the observational characteristics of GRS 1915$+$105 through wide-band X-ray spectral modelling. This source is particularly well suited for bridging theoretical predictions with measurable radiative signatures as energy spectra encapsulate the coupled behaviour of the Keplerian disk, the `hot' Comptonizing region, and the dynamical structure predicted by simulations. In BH–XRBs, the primary source of soft X-ray radiation arises from the optically thick accretion disk, which emits a multicolor blackbody radiation \cite[]{Shakura-Sunyaev1973,Novikov-etal1973}. These soft photons are subsequently inverse-Comptonized \cite[]{Sunyaev-Titarchuk1980,Sunyaev-Titarchuk1985,Titarchuk1994} by the `hot' corona \cite[]{Haardt-Maraschi1991,Chakrabarti-Titarchuk1995,Narayan-Yi1995,Mandal-Chakrabarti2005,Chakrabarti-Mandal2006,Iyer-etal2015,Gonzalez-etal2017} resided in the innermost regions of the accretion flow.

To connect these theoretical insights with observations, we analyse selected \textit{AstroSat} observations of GRS 1915$+$105 in the $\chi$, $\delta$, and $\omega$ variability classes (see \S \ref{sec7:7.1_pcv}). The combined wide-band ($0.7-50.0$ keV) spectral modelling enables us to constrain key physical parameters, such as mass accretion rates and the size of the Comptonizing region, which can be directly compared with the flow properties predicted by our hydrodynamic simulations.

\subsubsection {Hybrid Two-Component Accretion Flow Model}

The two-component accretion flow model \cite[]{Chakrabarti-Titarchuk1995} assumes a Keplerian disc on the equatorial plane and a sub-Keplerian flow sandwiched between the equatorial disc from top and bottom. The sub-Keplerian flow may create a centrifugal barrier leading to the formation of a shock, and the post-shock flow acts as the `hot' corona (equivalently PSC; see \S \ref{sec9:9.1_simulation} for details). In the presence of a shock, the Keplerian disc is truncated at the shock radius, otherwise, it extends till the innermost stable circular orbit (ISCO). Soft photons from the Keplerian disc are intercepted at the PSC and upscattered to produce high energy photons via inverse-Compton scattering. Considering this, \cite{Mandal-Chakrabarti2005} extended the seminal work of \cite{Chakrabarti-Titarchuk1995} by incorporating synchrotron processes and particle acceleration (primarily electrons) across the shock front leading to a hybrid (a mixture of thermal and non-thermal) electron distribution in the PSC. The general characteristics of the accretion process are then described within the framework of a hybrid two-component flow model (hereafter \texttt{HTCM}). \cite{Chakrabarti-Mandal2006} applied this upgraded model to explain LHS and HSS of Cyg X$-$1, although their approach primarily involved custom modelling without quantitative assessment of goodness-of-fit parameters. A simplified table model was subsequently developed \cite[]{Iyer-etal2015} within the \texttt{Xspec} framework \cite[see also][for an alternative implementation]{Debnath-etal2014}. Indeed, synchrotron emission and particle acceleration were not taken into account in this work due to computational constraints in generating template spectra across the entire parameter range of such a complex model, while estimating the mass of BH-XRB source IGR J17091$-$3624.

In the recent past, \cite{Baby-etal2020} followed similar methodologies to determine the mass of 4U 1630$-$47. In this model, the input parameters are considered as black hole mass ($M_{\rm BH}$ in ${\rm M}_\odot$), Keplerian mass accretion rate (${\dot m}_d$ in Eddington rates), sub-Keplerian mass accretion rate (${\dot m}_h$ in Eddington rates), shock radius ($r_s$ in $r_g$) and normalization. The table model implemented in \texttt{Xspec} \cite[]{Iyer-etal2015} admits limiting accretion rate ${\dot m}_d \sim 10$ due to computational limitations. The spectral modelling of GRS 1915$+$105, in particular $\delta$ and $\omega$ classes, generally requires high accretion rate, which is outside the range of the implemented table model in \texttt{Xspec}. Therefore, we manually fit the data running the source code (excluding synchrotron emission and particle acceleration processes) for every iteration within the \texttt{Xspec} environment.

\begin{figure}
\vskip -2.5 cm
    \begin{center}   
    \includegraphics[width=\columnwidth]{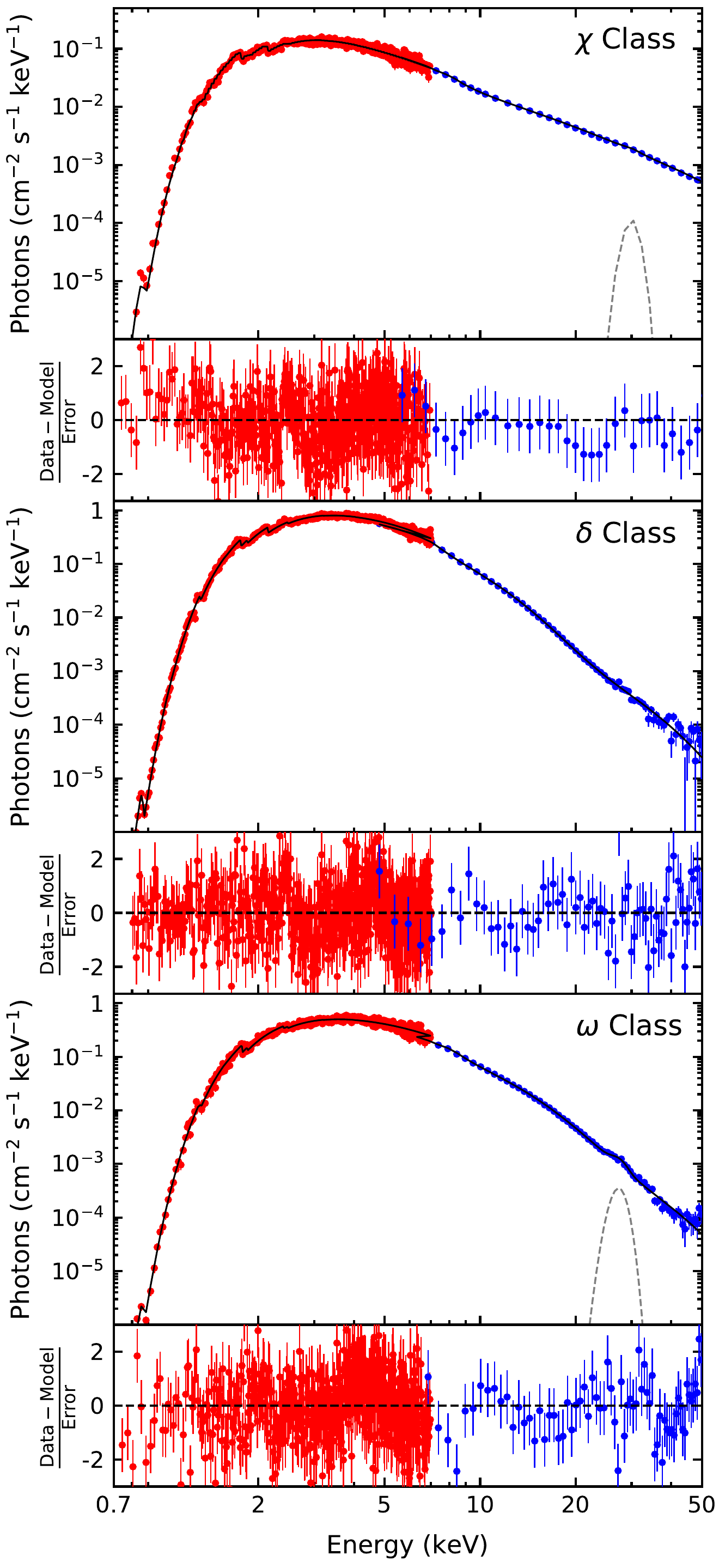}
    \end{center}
    \vskip -0.75 cm
    \caption{Simultaneous spectral modelling of AstroSat data during class-$\chi$ (top), class-$\delta$ (middle) and class-$\omega$ (bottom) of GRS 1915$+$105. Here, red and blue represent the SXT and LAXPC data, respectively, while the model spectra are represented by black lines and the \texttt{Gaussian} profile at 30 keV with a grey dashed line indicates the Xe K fluorescence in the LAXPC data.
    }
    \label{fig:fig36}
\end{figure}

\subsubsection{Methods of HTCM Spectral Modelling and Assumptions}

We carry out the spectral modelling of three variability classes $\chi$, $\delta$, and $\omega$ of GRS 1915$+$105. In doing so, we adopt a black hole mass of $12.5\,{\rm M}_\odot$, a spin parameter of $a_{\rm k}=0.98$, a source distance of $8.5$ kpc, and an inclination of $65^\circ$. The line-of-sight column density is fixed at $6 \times 10^{22}\, {\rm  atoms~cm}^{-2}$, and interstellar absorption is described using \texttt{tbabs} with the \texttt{wilm} abundance \cite[]{Wilms-etal2000}. We consider edge components at $1.75$ and $2.14$ keV for absorption features present in SXT spectra. To correct the gain shift, we apply the \texttt{gain fit} command with an \texttt{offset} of $\sim 0.06$. The LAXPC spectra show a blurred absorption feature between $7-12$ keV, which is likely to be due to broadened Fe K absorption edge. We use a \texttt{smedge} component to address this. Further, LAXPC data show a hump like feature near $30$ keV due to Xe K fluorescence and a \texttt{Gaussian} component is added to the spectral model whenever required. The composite spectral model is therefore considered as \texttt{constant$\times$tbabs$\times$edge$\times$edge$\times$smedge(HTCM+Gaussian)}, where a \texttt{constant} component accounts for calibration offset between instruments. All fits are performed using \texttt{XSPEC version 12.15.0}, and $\chi^2$ statistics are adopted to assess the goodness of fit. The model fitted parameter uncertainties are quoted at the $90\%$ confidence level.

{\red It is wroth noting that in the present analysis, the connection between the HD simulations and spectral modelling lies in their common shock-induced accretion geometry. We emphasize that the simulation results are not directly used as inputs to the spectral modelling, rather similar PSC geometries (associated with HFQPO and LFQPO) obtained in the HD simulations, are adopted to describe the radiative properties of the hot Comptonizing regions. In this framework, the spectral characteristics are primarily governed by physical quantities such as the shock location, compression ratio, accretion rate, and flow geometry \citep{Chakrabarti-Titarchuk1995,Mandal-Chakrabarti2005,Chakrabarti-Mandal2006,Iyer-etal2015,Baby-etal2020}.}

\begin{table}
    \caption{Summary of the best fit spectral parameters obtained from HTCM.}
    \label{tab:Table-16}
    \centering
    \begin{tabular}{|c|c|c|c|c|}
        \hline
        {Class} & \multicolumn{3}{c|}{Model: HTCM} & {$\chi^2/\text{dof}$} \\ 
        \cline{2-4} 
         & \multicolumn{1}{c|}{$\dot m_d$ ($\dot{M}_{\rm Edd}$)} & \multicolumn{1}{c|}{$\dot m_h$ ($\dot{M}_{\rm Edd}$)} & \multicolumn{1}{c|}{$r_s$ ($r_g$)} & \\
        \hline   
        $\chi$ & $1.5\pm0.05$ & $0.1\pm0.03$ & $93.0\pm9.0$ & $586/548$ \\
        \hline
        $\delta$ & $17.8\pm0.6$ & $0.5\pm0.01$ & $7.9\pm0.2$ & $765/677$ \\
        \hline        
        $\omega$ & $18.6\pm1.0$ & $0.7\pm0.10$ & $7.0\pm2.1$ & $759/610$ \\
        \hline
    \end{tabular}
\end{table}

\subsubsection{Results from HTCM Spectral Modelling}\label{sec9:9.2.3_htcm_results}

The model fitted spectra of three variability classes ($\chi$, $\delta$ and $\omega$) along with the corresponding residuals are presented in Fig. \ref{fig:fig36}. The SXT and LAXPC data are plotted in red and blue, respectively, and the best fitted model spectra are shown using solid lines. The \texttt{Gaussian} component at $30$ keV, marked with a grey dashed line, represents the Xe K fluorescence present in the LAXPC data. Overall, the adopted model provides a good fit to all variability classes, and the \texttt{HTCM} model fitted parameters are listed in Table \ref{tab:Table-16}. The $\chi$ class (top panel of Fig. \ref{fig:fig36}) represents a `canonical' hard state with a low ${\dot m}_d$ and a larger truncation radius ($r_s$). The resulting corona is more extended and hotter due to the reduced supply of soft photons from the disc, yielding a harder spectral slope. In contrast, $\delta$ class spectrum (middle panel of Fig. \ref{fig:fig36}) corresponds to the `canonical' soft state characterized by a very high ${\dot m}_d$ relative to ${\dot m}_h$ and a small $r_s$. Consequently, the emission is dominated by the Keplerian accretion disc with the inner edge extending close to the central compact object. This geometry naturally produces a steeper photon index and a higher disc temperature. The inferred shock radius in the $\chi$ class and the $\delta$ class are consistent with the same obtained from hydrodynamic simulations (see Fig. \ref{fig:fig34}). The $\omega$ class spectrum (bottom panel of Fig. \ref{fig:fig36}) reveals a similar Keplerian disc accretion rate (${\dot m}_d$) as in the $\delta$ class (see Table ~\ref{tab:Table-16}). However, a relatively high sub-Keplerian rate (${\dot m}_h$) produces a slightly harder photon index than the $\delta$ class. Therefore, a radially extended corona (a truncated accretion disc at $r_s$ as assumed in \texttt{HTCM} model) plausibly explains both spectral and timing behaviours. However, the spectral analysis indicates that both $\delta$ and $\omega$ classes require unusually high ${\dot m}_d$. A possible reason is that in the present model, the Keplerian disc does not extend to the ISCO even in a `canonical' soft state. Therefore, a higher accretion rate is required to produce the same inner disc temperature at the truncated radius. An alternative physical picture may involve a two-phase fluid scenario, where the Keplerian disc extends to the ISCO (i.e., the Keplerian disc is not truncated at $r_s$) in the soft state, and the sub-Keplerian disc coexists with a shock located at $r_s$ above the Keplerian disk. This scenario is expected to reduce the Keplerian accretion rate, and the timing properties will be governed by the post-shock sub-Keplerian flow (PSC), which sandwiches the Keplerian disc on the equatorial plane. A detailed investigation of this scenario is beyond the scope of the present work and will be explored in future studies.

{\red Overall, in this effort, we attempt to explain the spectral characteristics of arguably the most complex BH-XRB source, GRS 1915$+$105, as a representative case. We further emphasize that the spectral model employed here does not use the full MHD solution directly as an input to the radiative transfer calculation. Instead, the magnetic field is treated parametrically while modelling the synchrotron emission and Comptonization of the synchrotron photons in shocked advective flows \citep{Mandal-Chakrabarti2005,Chakrabarti-Mandal2006}. A fully self-consistent treatment would require coupling the density, temperature, velocity, and magnetic field distributions obtained from MHD simulations with the relevant radiative processes, either through simultaneous radiation-MHD calculations or via detailed post-processing of the simulation outputs. Such a fully coupled radiation-MHD framework is computationally demanding and lies beyond the scope of the present work. The development of a full-featured table model based on \texttt{HTCM} \citep{Chakrabarti-Mandal2006} is currently in progress, and a comprehensive spectral modelling of \textit{AstroSat} observations of BH-XRBs using this framework will be presented elsewhere \citep{Mandal-etal2026}.}

\section{Discussion and Conclusion}\label{sec10:discussion}

\begin{figure*}
	\centering 
		\includegraphics[width=0.98\textwidth]{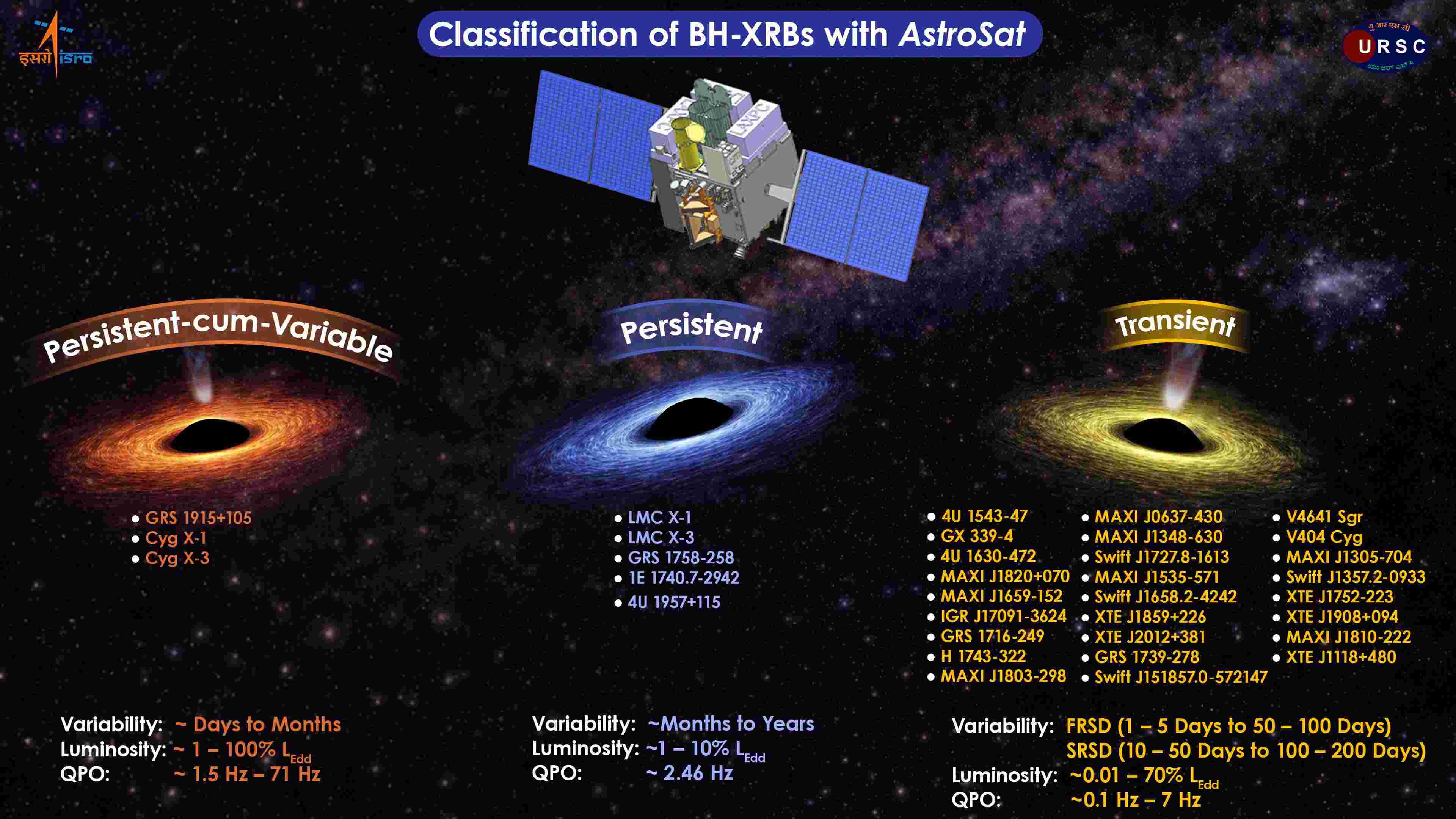}
	\caption{Infographic illustration of the classification of $34$ BH‑XRBs observed with \textit{AstroSat} into persistent‑cum‑variable, persistent, and transient sources. The classification is based on key observed properties, namely the QPO frequency, bolometric luminosity, and characteristic long-term variability timescale, as highlighted in the figure. The Milky Way and Large Magellanic Cloud (LMC) are shown in the background to indicate the host galaxy of these sources. See text for details.
    }
    \label{fig:fig37}
\end{figure*}

In this section, we present and discuss the key results obtained from decade long study of BH-XRBs with \textit{AstroSat}. We examine the observed spectral and timing properties of $3$ persistent-cum-variable, $5$ persistent, and $26$ transient BH-XRBs using combined \textit{AstroSat} and \textit{MAXI} data (see Table \ref{tab:Table-01} and Fig. \ref{fig:fig37}). We explore the spectro-temporal correlations obtained from these observations and plan to interpret the findings in the context of the underlying accretion physics. Further, we intend to put efforts to address the possible physical origin of LFQPOs as well as HFQPOs and investigate the associated time-lag behaviour to gain insight into the geometry and dynamics of the accretion flow.

\subsection{Persistent-cum-Variable Sources}

GRS 1915$+$105 has been monitored with \textit{AstroSat} for nearly a decade ($2016-2025$), yielding $47$ pointed observations. Initially, the source showed strong flux variability of $\sim 0.1-2.5$ Crab (January $2016$ to April $2018$), during which HR ranges between $0.2$ and $0.9$. Being persistently bright and variable for $\sim 25$ years, the source finally went to an obscured state in $2019$ with fluxes dropping below $0.05$ Crab and average HR values $\sim 1$ (see Fig.~\ref{fig:fig6}). Intermittent re‑brightening episodes occurred during this obscured phase \cite[]{Athulya-etal2023}, accompanied by lower HR values of $\sim 0.3$. Throughout the \textit{AstroSat} monitoring, the luminosity is observed as $\sim 2-44\%\,{\rm L}_{\rm Edd}$. The source is also known to exhibit substantial flux variability on timescales of minutes to hours or even days.

In this work, we carry out an in-depth spectro-temporal study of the source considering $25$ pointed observations over nine variability classes  \cite[$\beta$, $\theta$, $\chi$, $\omega$, $\delta$, $\rho$, $\kappa$, $\gamma$, and $\tau$;][]{Athulya-etal2022,Majumder-etal2022,Majumder-etal2025b} including five intra-class variants of the $\chi$ class \cite[$\chi^{}_1$, $\chi^{}_2$, $\chi_2^{\prime}$, $\chi^{}_3$, $\chi^{}_4$;][]{Majumder-etal2026b}. The first \textit{AstroSat} observations of the source led to the detection of LFQPOs in the $2-8$ Hz range during a radio-quiet $\chi$ class, showing rapid evolution of the QPO frequency (a change of $\sim 4$ Hz within $\sim 5$ hours) correlated with flux variations \cite[]{Yadav-etal2016}. Using a $\sim 90$ ks continuous {\it AstroSat} exposure, \cite{Rawat-etal2019} reported a transition from a non-variable $\chi$-like state to a periodic heartbeat state via an intermediate state characterized by nearly periodic $100-150$ s oscillations. It was noted that although the millihertz variability evolved substantially, the $3-5$ Hz LFQPO remained comparatively stable implying a coupled inner disc–corona accretion flow configuration largely unaffected by the onset of large amplitude flares. Furthermore, \cite{Misra-etal2020} reported a strong correlation between the LFQPO frequency, the inner disc radius, and the accretion rate. This finding appears to be consistent with the dynamical frequency model, in which the frequency is inversely proportional to the sound crossing time, as suggested by relativistic thin disc theory \cite[][]{Novikov-etal1973} for rapidly spinning black holes. Using {\it AstroSat} observations, \cite{Garg-etal2020} proposed that LFQPO ($\sim 3-4$ Hz) and its harmonics likely arise from coupled fluctuations in the local accretion rate and truncated inner disc radius driven by variations in the coronal optical depth and heating rate.

Meanwhile, energy dependent HFQPOs ($\sim 70$ Hz) have been detected in five variability classes ($\delta$, $\gamma$, $\kappa$, $\omega$, and $\tau$) using the large effective area of \textit{AstroSat}/LAXPC surpassing the capabilities of \textit{RXTE}/PCA \cite[]{Belloni-etal2019, Sreehari-etal2020, Majumder-etal2022, Majumder-etal2025b,Dhaka-etal2025}. Wide-band spectro-temporal variability associated with these HFQPOs, observed up to photon energies of $\sim 25$ keV, provides compelling evidence for an oscillating inner compact corona as their possible origin \cite[]{Sreehari-etal2020, Majumder-etal2022, Majumder-etal2025a,Harikesh-etal2025}.

Cyg X$-$1 has been extensively monitored with \textit{AstroSat} over nearly a decade. Among more than $50$ pointed observations, we analyse $21$ representative datasets to investigate its spectro‑temporal behaviour. Unlike GRS 1915$+$105, Cyg X$-$1 has remained persistently active for over five decades and continued to show variability throughout the \textit{AstroSat} monitoring period. During these observations, the \textit{MAXI}/GSC flux varied between $0.25$ and $1.25$ Crab, while HR remained in the range of $0.20-0.75$ (see Fig.~\ref{fig:fig6}). The luminosity was observed in the range of $0.4-4.0\%\,{\rm L}_{\rm Edd}$. Cyg X$-$1 generally exhibits flux variability on timescales from a few days to several months.

Depending on the relative contributions of the disc and coronal components, the source traces between LHS, IMS, and HSS. In LHS, the Comptonized flux is strongest accompanied by a PDS exhibiting three broad \texttt{Lorentzian} components. The IMS shows reduced Comptonization and two broad timing features, while in the HSS the accretion disc dominates with the maximum $kT_{\rm in}$, and the PDS becomes featureless. These transitions reaffirm the long term dynamical nature of the accretion flow in Cyg X$-$1, which has remained active for more than five decades \cite[]{Ling-etal1983,Jiang2024}. Notably, \textit{AstroSat} captured the `softest' spectral state of this source to date \cite[]{Kushwaha-etal2021} during which the spectral index reached its maximum value. On MJD $58050$, Cyg X$-$1 exhibited an extremely soft spectrum, which was quantified using the prescription of \citet{Wilms-etal2006} and \citet{Grinberg-etal2014} based on the low energy photon index $\Gamma_{1}$ of a broken power law. \cite{Kushwaha-etal2021} obtained $\Gamma_{1} = 4.13^{+0.06}_{-0.08}$ during the \textit{AstroSat} observation, which is steeper than in previously reported softest states \cite[]{Zdziarski-etal2004b, Kawano-etal2017}. The \textit{MAXI} hardness ratio on MJD $58050$ drops to $\sim 0.229$, lower than the value ($\sim 0.238$) during the softest \textit{Suzaku} epoch reported by \citet{Kawano-etal2017}. Such a steep photon index at low energy and the minimum HR indicate a disc dominated spectrum providing ideal conditions for continuum fitting based black hole spin estimation. Indeed, \citet{Kushwaha-etal2021} used this state to infer an almost maximally rotating black hole of spin $a_{\rm k} > 0.9981$, which is consistent with earlier measurements \cite[]{Gou-etal2014}.

Cyg X$-$3 is observed with \textit{AstroSat} between $2016$ and $2019$, and we carry out a detailed spectro‑temporal analysis of $14$ pointed observations. During this period, the \textit{MAXI}/GSC flux varies between $\sim 0.1$ and $0.5$ Crab, while the HR lies in the range $\sim 0.25-1.3$ (see Fig. \ref{fig:fig6}). The luminosity is found to vary as $\red \sim 14-61\%\,{\rm L}_{\rm Edd}$ (see Table \ref{tab:Table-10}). Cyg X$-$3 generally shows flux variability on month long timescales. Based on the \textit{MAXI}/GSC flux and HR, the source is found to be in LHS, IMS, and HSS during these epochs. We detect LFQPOs at $\sim 0.3$ Hz and $\sim 4.1$ Hz in two observations, both characterised by relatively low rms$_{\rm QPO}$. However, owing to limited photon statistics, the corresponding time-lags could not be constrained reliably. The energy spectra are well described by a combination of a thermal disc component and a non‑thermal Comptonized component. The spectral states are further classified into LHS, HIMS, SIMS, and HSS based on $\Gamma_{\rm th}$ and flux ratios ($F_{\rm ratio}$). The spectral modelling also reveals emission features consistent with Fe K$\alpha$ and Si XIII \cite[]{Kallman-etal2019, Chaurasia-etal2025} likely originating from the accretion disc and the dense Wolf–Rayet stellar wind, respectively. In addition, Cyg X$-$3 exhibits an unusually high luminosity, which may be attributed to geometric beaming associated with the strong outflow \cite[]{Veledina-etal2024}.

\subsection{Persistent Sources}

The persistent source LMC X$-$1 exhibits marginal variability in \textit{MAXI}/GSC flux (see Fig. \ref{fig:fig15}) of $\sim 0.02$ Crab, HR of $\sim 0.2$ with $F_{\rm var}\sim 5-14\%$. Its bolometric luminosity remains relatively stable in the range $8.0-11.4\%\,{\rm L_{Edd}}$. In contrast, LMC X$-$3 shows noticeable variability with flux varying between $\sim 0.01$ and $0.06$ Crab (see Fig. \ref{fig:fig15}), HR covering the range $0.2-1.0$ with $F_{\rm var}\sim 2-10\%$. Compared to LMC X$-$1, its luminosity varies more widely, lying between $6.3$ and $12.4\%\,{\rm L_{Edd}}$. While LMC X$-$3 displays flux variations on timescales of a few days to several months, LMC X$-$1 remains effectively steady with no appreciable long term variability.

LMC X$-$1 and LMC X$-$3 are typically observed in the soft spectral state, characterised by dominant thermal emission from a standard accretion disc and only a weak coronal contribution, consistent with the canonical disc-corona configuration. The dominance of thermal emission also explains the low MAXI HR as shown in Fig. \ref{fig:fig15}. Although \textit{AstroSat} has not captured either source in other spectral states, previous studies indicate that LMC X$-$3 does undergo transitions into intermediate and hard states \cite[][]{Bhuvana-etal2022}. During the intermediate state, the coronal emission increases while the overall luminosity decreases, and in the hard state, the thermal disc weakens further with the spectrum becoming dominated by the Comptonized coronal component. This behaviour is broadly consistent with the standard accretion paradigm in which high mass accretion rates allow the inner accretion disc to extend close to the black hole resulting in a soft state. At lower luminosities, the disc temperature decreases and the disc normalisation increases, implying a receding inner disc radius. Such trends support the truncated disc scenario \cite[][]{Done-etal2007}, where the inner disc is replaced by a `hot', advective flow during the hard state. Interestingly, for LMC X$-$1, \textit{XMM–Newton} observations at a luminosity of $\sim 0.3\,{\rm L}_{\rm Edd}$, higher than that typically associated with the soft state, show a pronounced spectral broadening that can be explained by a `slim disc' model \cite[]{Bhuvana-Nandi2025}. Similar broadened spectra have also been reported by \citet{Sutton-etal2017} for LMC X$-$3 and GX 339$-$4 suggesting that such deviations from the standard disc model may be common in scenario with high luminosity. \textit{AstroSat} has enabled precise measurements of the soft state spectral properties of both LMC X$-$1 and LMC X$-$3 allowing us to estimate their black hole spins using the continuum fitting method. For LMC X $-$1, we obtain a moderately high spin of $a_{\rm k} \sim 0.89$, whereas LMC X$-$3 consistently indicates a low spin of $a_{\rm k} \sim 0.3$, reinforcing the view that it hosts a slowly rotating black hole. 

Long term monitoring of GRS 1758$-$258 does not show any variability in \textit{MAXI}/GSC lightcurve, whereas 1E 1740.7$-$2942 exhibits a long term variability in the time scale of several months to years \cite[]{Main-etal1999,Smith-etal2002b} since its discovery, and follows similar variability time-scale during \textit{AstroSat} era. The \textit{AstroSat} observations of these sources exhibit variation of bolometric luminosities of $\sim 1-6\%$ and $\sim 1 \%\,{\rm L_{Edd}}$ (see Table \ref{tab:Table-11}), respectively. Both sources generally remain in the hard spectral state, characterised by a dominant Comptonized component, weak disc emission, low disc temperature, and a large inferred inner disc radius. However, the soft state observed in GRS 1758$-$258 is unusual. Despite a steep spectrum and reduced hard X-ray contribution characteristic of a soft state, the luminosity does not increase significantly compared to the hard state, and the disc temperature remains low ($\sim 0.5$ keV). This deviation suggests a non-standard soft state, likely due to a persistent but obscured disc \cite[]{Soria-etal2011}. In these sources, the disc may remain present always with low accretion rate and an advective component appears above and below the disc obscuring its emission. When the accretion rate increases, the corona responds more rapidly due to its shorter viscous timescale, while the thin disc takes longer timescale ($\sim$ weeks) to adjust. During this delay, the source enters a transitional phase showing a cool, `dim disc' with steep spectral shape leading to a `dim-soft' state, which is a distinctive spectral configuration arising from delayed disc response during state transitions.

Leveraging the large effective area of \textit{AstroSat}/LAXPC at high energies, we have obtained high quality wide-band spectra for both GRS 1758$-$258 and 1E 1740.7$-$2942 enabling self‑consistent spectral modelling with state‑of‑the‑art relativistic reflection models such as \texttt{relxill}. This approach allows us to constraint the accretion disc parameters of GRS 1758$-$258 that include iron abundance ($\sim 4$ times solar abundance) and disc inclination ($\sim 61^\circ$) providing valuable insights of the system’s accretion geometry. However, the super-solar Fe abundance may reflect degeneracy between the Fe abundance and ionization parameter in \texttt{relxill} model.

Similarly, for 1E 1740.7$-$2942, the high energy sensitivity of LAXPC yielded hard state spectra in which both gravitationally broadened Fe line and the associated `Compton reflection hump' are detected. Modelling with \texttt{relxill} constrained the disc inclination to $\sim 55^\circ$, consistent with \textit{NuSTAR} measurements, demonstrating \textit{AstroSat}’s capability to deliver robust constraints on reflection parameters in hard state of BH-XRBs.

4U 1957$+$115 was observed by \textit{AstroSat} between $2016$ and $2018$. During this period, the source exhibits \textit{MAXI}/GSC flux of $0.02-0.08$ Crab with HR in the range $\sim 0.1-0.6$ with flux variability occurring on timescales of several days to a few months (see Fig. \ref{fig:fig15}). Based on the \textit{MAXI}/GSC flux and HR values, the source is consistently found in the HSS. Note that we observe LFQPO with frequency $\sim 2.66$ Hz in the soft state for the first time, which is classified as type‑A (see Table \ref{tab:Table-06}). The model fitted spectra comprise of thermal and Comptonization components, indicating that the emission is strongly dominated by the thermal disc. Despite being in the HSS, the source exhibits a relatively low bolometric luminosity $\sim 1.4\%-3.6\%~L_{\rm Edd}$.

\subsection{Transient Sources}

\textit{Swift J1727.8$-$1613:} The $2023-2024$ outburst profile of this source was observed as FRSD in nature with maximum \textit{MAXI}/GSC luminosity reaching upto $\sim 7$ Crab during the rising phase of $\sim 6$ days. Subsequently, the flux is observed to decay for more than $100$ days. The HR is observed to decay from 0.4 to 0.04 during the whole outburst. The bolometric luminosity of this source during \textit{Insight-HXMT} observations are found to lie in range $17.8-20.1\%\,{\rm L_{Edd}}$.

In this work, we investigate eight epochs of {\it AstroSat} observations of Swift J1727.8$-$1613 during its HIMS, confirming the presence of a LFQPO with an evolving centroid frequency from $1.09$ to $2.66$ Hz detected up to $\sim 100$ keV. The initial discovery of such evolving LFQPOs in this source was reported by \cite{Nandi-etal2024}. Using quasi-simultaneous {\it Insight-HXMT} observations, we further examine the spectral energy distribution and find that the emission is dominated by a strong Comptonized component with a substantial flux contribution, while the thermal disc remains comparatively weak. The combined spectral–timing results support an accretion geometry in which the disc is truncated at a large radius and a radially extended Comptonizing corona occupies the inner region, consistent with polarimetric constraints for the source in the HIMS \cite[]{Ingram-etal2024, Majumder-etal2025b}. Within this framework, radial modulation of the extended corona can naturally account for both hard X-ray origin of the LFQPO and its evolving frequency (see also \citet{Nandi-etal2024}). In addition to the multi-temperature disc and thermal Comptonization components, {\it Insight-HXMT} observations reveal an additional high energy spectral feature that has been interpreted as a `Compton hump' potentially arising from a strong jet \cite[]{Yang-etal2024, Peng-etal2024a}. This has led to an alternative accretion scenario involving strong coupling between the corona and the jet in Swift J1727.8$-$1613. Furthermore, the type-C QPO rms amplitude is observed to remain significantly higher above $40$ keV compared to below $20$ keV, suggesting that the additional hard spectral component may originate from jet emission, with the excess variability plausibly explained by precession of the jet base \cite[]{Yang-etal2024}.

\textit{MAXI J1535$-$571:} During $2017$, MAXI J1535$-$571 exhibits SRSD-type outburst profile of $\sim 175$ days,  where maximum \textit{MAXI}/GSC luminosity reached up to $\sim 4.5$ Crab within the first $\sim 20$ days of the outburst and then decayed for more than $\sim 150$ days. The HR is observed to decay from $\sim 1$ to $0.02$ within $\sim 150$ days. The bolometric luminosity was observed in the range $38-71\%\, {\rm L_{Edd}}$. {\it AstroSat} observed MAXI J1535$-$571 in its intermediate spectral state during the outburst. LAXPC detected LFQPOs at $\sim 2$ Hz while the source remained bright ($\sim 71\%~{\rm L}_{\rm Edd}$) with an almost constant photon index of $\sim 2.3$ and a covering fraction exceeding $0.3$. Using {\it AstroSat} data, \cite{Sreehari-etal2019b} constrained the black hole mass to $\sim 6.4\,{\rm M}_{\odot}$ and inferred a halo accretion rate of $\sim 0.2\,\dot{m}_{\rm Edd}$ together with a disc accretion rate close to $\dot{m}_{\rm Edd}$, indicating a disc dominated emission geometry with a compact corona corresponding to inner disk radius of $\sim 16\,r_{\rm g}$. The QPO properties were cross validated using dead‑time corrected LAXPC data and pile‑up corrected {\it Swift} observations, and their energy dependence was explored. More recently, \cite{Garg-etal2025} proposed a radiative feedback between the disc and corona to explain these QPOs. Following $2017$ main outburst, the source also exhibited four shorter yet similar mini‑outbursts \cite[]{Cuneo2020}.

\textit{Swift J1658.2$-$4242:} During $2018$ outburst, the source exhibits flip-flop behaviour in X-ray flux in the time scale of days and the bolometric luminosity ranges as $0.5-2.9\%\,{\rm L_{Edd}}$. The source flux was observed to increase initially for $\sim 20$ days, followed by two flip-flop intervals, whereas source flux subsequently decays for more than $\sim 100$ days \cite[see][for details]{Bogensberger-etal2020}. During the HIMS, LAXPC detected type‑C QPOs in the PDS with centroid frequencies in the range $\nu_{\rm QPO}\sim 1.5-6.6$ Hz, while the X‑ray intensity remained between $\sim 25$ mCrab and $50$ mCrab. In the subsequent SIMS, where the source exhibited a flip–flop pattern \cite[]{Bogensberger-etal2020}, type‑A and type‑B QPOs were observed. The type‑C and type‑A QPOs showed soft-lags, whereas the type‑B QPOs exhibited hard-lags. 

\textit{GX 339$-$4:} During a decade of \textit{AstroSat} operation, GX 339$-$4 exhibited four outbursts, one of which was a mini-outburst. During the $2017-2018$ mini-outburst of $\sim 120$ days, the source reached a maximum flux of $\sim 0.1$ Crab within $\sim 42$ days with an HR of $\sim 0.25-1.00$. In the $2019-2020$ outburst of duration of $\sim 300$ days, the \textit{MAXI}/GSC flux peaked at $\sim 0.6$ Crab in $\sim 115$ days, with HR values between $0.05$ and $1.00$. The $2021$ outburst attained a maximum flux of $\sim 0.5$ Crab in $\sim 150$ days and an HR range of $0.01-1.00$. During the $2023-2024$ outburst, the peak flux rose to $\sim 0.3$ Crab in $\sim 120$ days with HR spanning $0.10-1.00$ \cite[see][for details]{Aneesha-etal2024}. In quiescence, the HR remains high, averaging $\sim 0.75$. The bolometric luminosity varies in a wide range from $0.01$ to $12.07\%\,{\rm L_{\rm Edd}}$. \textit{AstroSat} observed the source in quiescence between $2016$ and $2023$, as well as during its outbursts in $2017$, $2018$, $2021$, $2022$, and $2024$. Across these outbursts, the source was found in the LHS, HIMS, and SIMS. Type‑C QPOs were detected in the LHS and HIMS, while a type‑B QPO was observed during the SIMS. The type‑B QPO exhibited a hard-lag, whereas the time‑lag associated with the type‑C QPOs could not be constrained. The observed time‑lag behaviour is consistent with the findings of the low‑inclination systems during episodes close to jet ejections \cite[]{Choudhury-etal2025}.

\textit{H 1743$-$322:} This source has undergone multiple outbursts of typical SRSD nature and was observed by \textit{AstroSat} in $2016$ and $2017$, both during its outbursting phases and in quiescence. During the outburst period of $\sim 100$ days, the maximum \textit{MAXI}/GSC flux reached to $\sim 0.8$ Crab within $\sim 24$ days, while average HR remains in the range $\sim 0.25-1.00$. The bolometric flux was observed to lie in range $0.1-2.7\%\,{\rm L_{\rm Edd}}$. During the LHS and HIMS, the source exhibited type‑C QPOs with centroid frequencies in the range $\nu_{\rm QPO} \sim 0.43 - 0.60$ Hz. These QPOs showed soft time‑lags for observations with $F_{\rm th} \leq 0.9$, which is consistent with highly inclined system during the HIMS \cite[]{Eijnden-etal2017, Choudhury-etal2025}.

\textit{4U 1630$-$472:} Over the past decade, 4U 1630$-$472 exhibited six outbursts in $2016$, $2018$, $2019$, $2020$, $2021$–$2022$, and $2022$–$2023$. Among these, four were SRSD-type mini‑outbursts, one was a FRED‑type outburst, and one displayed a complex evolution like super-outburst \cite[see Fig. \ref{fig:fig25} and][]{Baby-etal2020}. The durations of the mini‑outbursts observed in 2016, 2018, 2020, and $2021$–$2022$ were each $\sim150$ days, reaching peak fluxes of up to $\sim0.35$ Crab within $\sim40$ days. The successive mini‑outbursts were separated by intervals of $\sim500$–$600$ days. Following the 2018 mini‑outburst, the source underwent a FRED-type outburst in 2019, during which the flux increased rapidly to $\sim1.7$ Crab within $\sim7$ days and subsequently decayed over $\sim120$ days. During the super-outburst of $2022$–$2023$, the source flux gradually increased to $\sim 0.7$ Crab over $\sim 220$ days, followed by a decline lasting $\sim100$ days, and again reached a secondary peak of $\sim0.6$ Crab within the next $\sim 60$ days. During $2016$, $2018$ and $2020$ outbursts observed by \textit{AstroSat}, the source is captured in LHS, IMS, and HSS without any QPO features in the PDS. On the other hand, type-B QPOs are detected during $2023$ outburst. The bolometric flux was found to be in range $0.1-18.0\%~{\rm L_{\rm Edd}}$, while during 2023 outburst, the source was only observed in IMS with $L_{\rm bol}\sim 29.0-43.0\%~{\rm L}_{\rm Edd}$.

\textit{MAXI J0637$-$430:} This source was observed by \textit{AstroSat} during a single outburst in which the \textit{MAXI}/GSC intensity decayed from approximately $75$ mCrab to $25$ mCrab (see Fig. \ref{fig:fig29}). Over the same period, the HR increased from about $0.10$ to $1.00$. The bolometric luminosity of the source was found to vary within the range $3-6\%~ {\rm L_{Edd}}$. The source was observed in IMS during its $2019$ outburst without any detection of QPO feature. The inner disc temperature remains relatively low ($kT_{\rm in} \sim 0.50-0.60$ keV), while the photon index remains above $\Gamma_{\rm th} \ge 2.30$. The thermal disc and Comptonized components contribute nearly equally during the IMS with $F_{\rm ratio} \sim 1$. Across all \textit{AstroSat} observations, MAXI J0637$-$430 exhibits spectral behaviour broadly consistent with standard accretion scenario. However, \textit{NuSTAR} soft state spectra reveal an additional excess component beyond the thermal disc and Comptonized emissions. This blackbody like excess has been interpreted as emission from the plunging region inside the ISCO \cite[]{Lazar-etal2021, Mummery-etal2024}, particularly during high luminosity epochs. Similar high luminosity excess emission is also observed in LMC X$-$1, as discussed above. Such features are also been reported in other transient systems such as MAXI J1820$+$070 \cite[]{Fabian-etal2020}.

\textit{GRS 1716$-$249:} During its $2017$ activity, GRS 1716$-$249 underwent a failed outburst and remained in LHS and HIMS (see Fig. \ref{fig:fig29}). The \textit{MAXI}/GSC intensity of the source increased from $\sim 0.15$ to $\sim 0.25$ Crab, followed by a gradual decrease to below $0.1$ Crab. Throughout this period, HR remained nearly constant, with an average value of $\sim 0.5$. The luminosity of the source was estimated to lie in the range $\red 5.0-12.2\%\,{\rm L_{Edd}}$. The PDS shows type‑C QPOs at $\sim 1.20-1.60$ Hz, while spectral analysis yields $\Gamma_{\rm th} \sim 1.50$ in the LHS and $\sim 1.71$ in the HIMS. The disc temperature is moderately low ($\sim 0.33-0.53$ keV) in both states. The disc appears truncated at smaller inner radius of $r_{\rm in}\sim 20\,{\rm r}_{\rm g}$ \cite[]{Jiang-etal2020}.

\textit{4U 1543$-$47:} The 2021 activity of this source exhibits a FRED‑type outburst profile, during which the \textit{MAXI}/GSC flux increased to $\sim8$ Crab within $\sim5$ days and subsequently decayed to $\sim150$ mCrab over a period of $\sim170$ days (see Fig. \ref{fig:fig29}). HR was also observed to decline from $0.10$ to $0.01$. During the \textit{AstroSat} observations, the bolometric luminosity of the source decreased from $103\%$ to $35\%\, {\rm L}_{\rm Edd}$. The 2021 outburst of 4U 1543$-$47 is particularly unique. The spectra exhibit a strong, evolving absorption feature in the $8-11$ keV range, whose strength initially increases and later declines over the course of the outburst. This feature is attributed to a relativistic accretion disc wind \cite[see][]{Prabhakar-etal2023}. The source reaches a peak Eddington luminosity, which steadily decreases thereafter. Strong reflection signatures are present throughout the outburst, and the accretion disc appears to be highly ionized with an enhanced iron abundance. The source remains in the HSS for the entire duration of the outburst. Consequently, the PDS are featureless and exhibit low rms variability.

\textit{MAXI J1820$+$070:} The $2018$ outburst of MAXI J1820$+$070 comprises a hard state phase (Outburst‑I) of duration $\sim 90$ days with peak flux $\sim 1.5$ Crab, followed by an extremely bright soft state phase  (Outburst‑II) of $\sim 100$ days with peak flux $\sim 2$ Crab (see Fig. \ref{fig:fig29}). During the Outburst-I, the HR was observed to be $\sim 0.50$ whereas during Outburst-II, the HR reduces to $\sim 0.05$. Interestingly, during both outburst phases, the HR is observed to be almost constant. The bolometric flux of this source was observed to lie in the range $4.8-12.9\%~{\rm L_{Edd}}$.  During Outburst-I, the source exhibited a moderately hot accretion disk ($kT_{\rm in} \sim 0.50$ keV), while the spectrum was primarily dominated by Comptonized emission. A type-C QPO accompanied by strong X-ray variability was observed during this phase. As the outburst progresses, the source makes transition through a short HIMS into a very soft state. The X-ray variability dropped noticeably during the soft state transition and recovered gradually during the decay phase \cite[]{Prabhakar-etal2022}. Strong reflection features are present throughout both outburst phases. At the onset of the outburst, the corona height is significantly large, resulting in a low reflection fraction. Thereafter, the coronal height decreases leading to stronger reflection. 

\textit{MAXI J1348$-$630:} The source underwent a complete outburst in $2019$. The source exhibited a rapid rise in \textit{MAXI}/GSC flux, reaching a peak of $\sim3$ Crab within $\sim16$ days, followed by a decay to $\sim30$ mCrab over $\sim90$ days (see Fig. \ref{fig:fig29}). This main outburst was subsequently followed by a mini‑outburst with a peak flux of $\sim0.3$ Crab. During the primary outburst phase, the HR was observed to be $\sim0.02$, whereas during the mini‑outburst phase it increased to $\sim0.50$. The bolometric luminosity of the source during this activity was estimated to lie in the range $2.3-7.7\%\,\mathrm{L_{Edd}}$. \textit{AstroSat} observed the source during the HIMS and HSS. The PDS is featureless in the HSS, whereas a type‑C QPO is detected during the HIMS. The HIMS observed by \textit{AstroSat} corresponds to the decay phase of the outburst. The type‑C QPO exhibits a hard time-lag, consistent with a high Comptonized fraction ($F_{\rm th} \ge 0.8$), likely arising from an extended corona.

\textit{MAXI J1803$-$298:} The source exhibited a FRED type outburst profile during its $2021$ outburst, where \textit{MAXI}/GSC flux peaked at $\sim0.45$ Crab within $\sim 15$ days and subsequently decayed to $\sim 60$ mCrab over $\sim 90$ days (see Fig. \ref{fig:fig29}). During the outburst, HR was observed to be low at $\sim 0.01$, and gradually increased to nearly $\sim 1$ toward the end of the outburst. The bolometric luminosity of the source during this period was estimated to vary from $1.8$ to $17.7\%\,\mathrm{L_{Edd}}$. This source is observed by {\it AstroSat} during HIMS at the beginning of the outburst and at HSS and SIMS during the decay phase. The source exhibits type-C QPO during HIMS, whereas type-A QPO is observed at the decay phase in SIMS. The observation of soft-lag corresponding to type-C QPO is consistent with \cite{Choudhury-etal2025} as $F_{\rm Comp}/F_{\rm bol}$ is $\sim 0.5$ indicating the origin of type-C QPO from the corona of a smaller radius, as the source approaches towards the peak of the outburst. The energy spectrum steepens as the source moves from HIMS to HSS and SIMS, which is indicated by the increase in photon index. However, during HSS and SIMS, the inner disc temperature is observed to be lower than that in HIMS. This is attributed to lower bolometric luminosity during HSS and SIMS as the source is in decay phase.

\textit{IGR J17091$-$3624:} The black hole candidate IGR J17091$-$3624 has shown recurrent outburst activity during 1994, 1996, 2001, 2003, 2007, and 2011 \cite[]{Revnivstev-etal2003,Int-Zand-etal2003,Capitanio-etal2009,Krimm-etal2011b,Sreehari-etal2018}. During its 2011 outburst, the source underwent spectral state transitions from the hard to intermediate states within the first $\sim 12$ days, followed by a later phase after $\sim 45$ days in which it exhibited variability classes similar to those observed in GRS 1915$+$105, including the detection of HFQPOs \cite[]{Altamirano-Belloni2012}. IGR J17091$-$3624 was subsequently observed by \textit{AstroSat} during its outbursts between $2016$ and $2022$. However, strong contamination from the nearby source GX 349$+$2 affects the LAXPC data. In this work, we investigate the wide-band spectro–temporal variability of the source using two \textit{AstroSat} observations from $2016$ and $2022$. Power spectral analysis reveals broad QPO features in the $42-84$ mHz range, which are consistent with the reported results of \cite{Katoch-etal2021}. During both observations, the source is found to be in the softer state, characterized by an inner disc temperature of $\sim 1.97$ keV and a photon index between $2.5$ and $2.9$, with only marginal Comptonization. Our findings suggest that the source remained in the sub-Eddington accretion regime with luminosity reaching up to $\sim 18\%$ of ${\rm L}_{\rm Edd}$ throughout the observation period. Note that `heartbeat' like variability in IGR J17091$-$3624 has previously been studied using \textit{AstroSat} data \cite[]{Katoch-etal2021}, further emphasizing its remarkable similarities to GRS 1915$+$105.

\subsection{Spectro-Temporal Correlation: Physical Insights of Accretion Dynamics}

In \S\ref{sec8:correlation}, we investigate possible correlations among the spectro-temporal properties of BH-XRBs across different spectral states. In particular, we find a positive correlation between $\Gamma_{\rm th}$ and $F_{\rm ratio}$ (see Fig. \ref{fig:fig31}). This trend indicates that BH-XRB spectra soften as disc emission becomes dominant with $\Gamma_{\rm th}>2.40$ for $F_{\rm ratio}>1$, and harden to $\Gamma_{\rm th} \lesssim 2.40$ when Comptonized emission dominates ($F_{\rm ratio} \lesssim 1$). These results are consistent with the view that the spectral states of BH-XRBs are governed by the underlying accretion dynamics. In the following, we discuss plausible disc–corona configurations within the framework of the hybrid two-component accretion flow model (HTCM; see \S \ref{sec9:9.2_HTCM}) that accounts for the distinct spectral states and temporal variabilities observed in $34$ BH-XRBs using {\it AstroSat}.

\begin{figure*}[h]
	\centering 
		\includegraphics[width=0.87\textwidth]{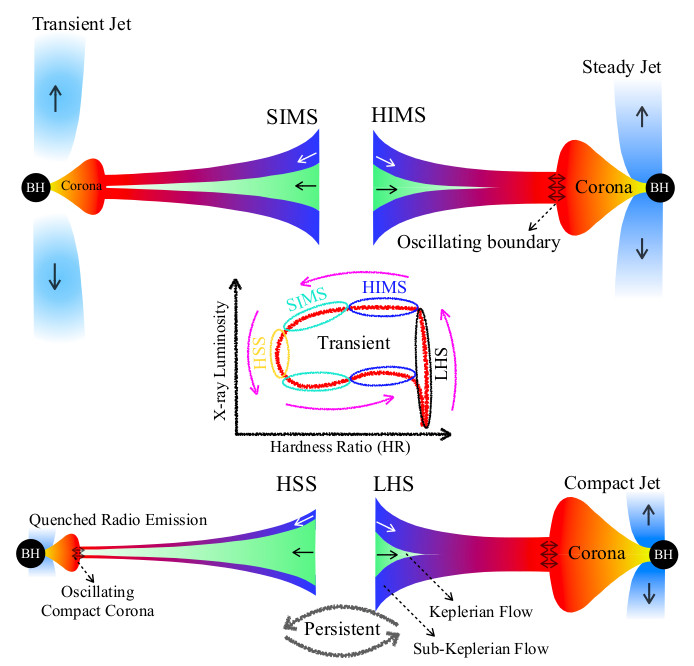} 
	\caption{Schematic representations of the disc-corona-jet geometries in different spectral states of BH-XRBs are shown, based on the framework of two-component advective accretion flow model around black holes. The Keplerian and sub-Keplerian flow components, as shown in the representation for LHS, are also present in rest of the representations for respective spectral states. The concept of the schematic is adapted from \citet{Chakrabarti-Titarchuk1995}. See text for details. 
    }
    \label{fig:fig38}
\end{figure*}

In HTCM, the accretion flow around a black hole comprises of a high viscous Keplerian disc flanked by the low angular momentum sub-Keplerian components. The sub-Keplerian flow typically gives rise to the post-shock corona (equivalently Comptonizing corona, $i.e.$, PSC) located in the inner regions of the disc \cite[]{Chakrabarti-Titarchuk1995, Molteni-etal1996, Ebisawa-etal1996, Chakrabarti-etal2008, Nandi-etal2012, Das-etal2014,Iyer-etal2015}. In this model, the inner edge of the Keplerian disc smoothly joins with the outer boundary of the PSC. Soft photons emitted from the Keplerian disc get inverse-Comptonized by the `hot' electrons in the corona and produce the hard spectral tail commonly observed in BH-XRBs at higher energies. In order to describe distinct spectral states of BH-XRBs along with their appearance in a typical HID of transient systems, we illustrate the physically motivated disc-corona-jet schematics in Fig. \ref{fig:fig38}. Generally, in the LHS, the Keplerian disc is truncated far away from the black hole and the inner region comprises of a radially extended Comptonizing corona (PSC), which serves as the base of a compact radio jet \cite[see Fig. \ref{fig:fig38};][]{Chakrabarti-1999,Das-etal-2001,Chattopadhyay-Das07,Das-Chattopadhyay08,Kumar-Chattopadhyay13,Aktar-etal2015,Aktar-etal2017,Das-etal-2022,Joshi-etal-2022}. Accordingly, energy spectra remain mostly dominated by the strong Comptonized emission along with marginal contribution from thermal disc (see \S\ref{sec9:9.2_HTCM}). This disc-corona configuration applies to LHS of all BH-XRBs along with harder variability classes ($\chi$ variants) of GRS 1915$+$105 observed with {\it AstroSat}.

Despite the presence of a relatively prominent Keplerian disc component ($i.e.$, thermal emission), Comptonized emission dominates in the HIMS of BH-XRBs (see Fig. \ref{fig:fig31}). Steady and strong radio jets are also found to be associated with HIMS \cite[see Fig. \ref{fig:fig38} and][for details]{Choudhury-etal2025}. We detect strong type-C LFQPO features in several BH-XRBs within the frequency range of $0.10-6.60$ Hz (see Fig. \ref{fig:fig32} and Table \ref{tab:Table-13}) during HIMS. In addition, GRS 1915$+$105 exhibits LFQPOs in the range of $1.38-6.69$ Hz in its $\theta$ and $\chi$ variability classes. These coherent LFQPOs are generally explained by the aperiodic modulation of the PSC in BH-XRBs during HIMS \cite[see Fig. \ref{fig:fig35}b;][]{Molteni-etal1996, Chakrabarti-etal2008,Das-etal2014,SDebnath-etal2024}.

Time-lag analysis at type-C QPO frequencies reveals both soft- and hard-lags, which show a positive correlation with the fractional Compton flux ($F_{\rm Comp}/F_{\rm bol}$). Notably, time-lag can result from multiple physical processes near a black hole. Among these, Comptonization plays a major role in explaining both soft- and hard-lags. A higher fractional Compton flux indicates a larger radial extent of the corona with higher electron temperature, which efficiently up-scatters seed photons from the accretion disc (see Fig. \ref{fig:fig32} and Fig. \ref{fig:fig38}). Since these multiple up-scattered high energy photons take relatively longer time to reach the observer as compared to the direct low energy thermal photons, this results in a hard-lag. Conversely, a lower fractional Compton flux suggests a corona with smaller radial extent and relatively cooler electron temperature. The cooler corona can initially up-scatter seed photons, some of which escape as high energy photons. However, photons that fail to escape may lose energy through reprocessing in the disc and eventually be emitted as soft photons, resulting in a soft-lag. Interestingly, during HIMS, the photon index remains in the intermediate regime ($\Gamma_{\rm th} \sim 1.50-2.40$) with relatively higher $F_{\rm ratio}$ compared to the LHS (see Fig. \ref{fig:fig31}).

In contrast, the SIMS of BH-XRBs are typically characterized by enhanced disc emission and the presence of transient, strong jets that are decoupled from the inner Compton corona and often move at relativistic speeds. During this state, the Keplerian disc extends close to the black hole, while the corona contracts significantly compared to the LHS and HIMS (see Fig. \ref{fig:fig38}). The spectral shape of BH-XRBs in SIMS becomes much steeper with $\Gamma_{\rm th} \gtrsim 2$ and $2 \lesssim F_{\rm ratio} \lesssim 7$ (see Fig. \ref{fig:fig31}). We also detect type-B LFQPOs in the SIMS of Swift J1658.2$-$4242 and GX 339$-$4 (see Table \ref{tab:Table-13}) that exhibit hard-lags. These QPOs are generally observed during HIMS to SIMS transition, which also seems to be associated with quasi-simultaneous radio flaring activity. This finding suggests a strong coupling between the corona and jet formation. During this transition, the corona likely becomes vertically elongated before being ejected as a jet (see Fig. \ref{fig:fig38}). Such vertical expansion of corona reduces the probability of up-scattered photons returning to the accretion disc, and therefore allows hard photons to escape directly to the observer. This geometry is potentially viable to explain the observed hard-lags associated with type-B QPOs \cite[see][for details]{Choudhury-etal2025}. Recently, \cite{Ram-etal2026} carried out the power–colour analysis using \textit{AstroSat} to generate a state dependent characterization of variability in BH–XRBs. 

Finally, in the HSS of BH-XRBs with enhanced luminosity, the Keplerian disc extends very close to the black hole resulting in a highly compact corona located at the inner accretion disc (see Fig.~\ref{fig:fig38}). This configuration explains the disc dominated spectro-temporal characteristics observed in several transient and persistent sources during their HSS (see Fig. \ref{fig:fig31}). For example, in the ‘softer’ variability classes ($\delta$, $\kappa$, $\omega$, $\gamma$, and $\tau$) of GRS 1915$+$105, thermally dominated spectral features are observed in the presence of a compact corona. Furthermore, the radial oscillation of this compact corona near the black hole is inferred to be responsible for generating HFQPOs (see Fig.~\ref{fig:fig38}) frequently observed in GRS 1915$+$105 \cite[]{Sreehari-etal2020, Majumder-etal2022, Majumder-etal2024, Majumder-etal2025a, Harikesh-etal2025}.

Overall, Fig. \ref{fig:fig31} indicates that the Keplerian component correlates with the bolometric luminosity ($L_{\rm bol}$), and evolves on viscous time scales. Beyond this, variations in the sub‑Keplerian flow and changes in the Comptonization geometry play a key role in governing the observed variability and the diversity of spectro‑temporal properties primarily through changes in the disc flux ($F_{\rm disc}$) and the Comptonized flux ($F_{\rm Comp}$). These trends collectively point to the evolving sub‑Keplerian flow together with the size of the Comptonizing region (equivalently PSC), as the primary regulator of variability in BH‑XRBs apart from the Keplerian disc.

\subsection{On the Origin of LFQPOs \& HFQPOs and Lag Properties}\label{sec10p5:QPO-Lag}

Despite extensive observational and theoretical efforts \cite[]{Morgan-etal1997, Remillard-etal1999, Bachetti-etal2015, Ingram-etal2019, Zhang-etal-2020, Ma-etal2021}, the physical origin of QPOs remains inconclusive. Proposed models invoke geometric effects, accretion flow instabilities, or propagating mass accretion rate fluctuations \cite[]{Lyubarskii-etal1997, Kotov-etal2001, Ingram-van-2013, Misra-etal2020, Liu-etal2021, Rawat2023}. Among these, the Lense–Thirring (LT) precession model attributes LFQPOs to the precession of the inner accretion flow, where relativistic Doppler boosting modulates the X-ray flux \cite[]{Stella-etal1999, Ingram-etal-2009, Ingram-Motta2014}. In contrast, transition layer and radiation hydrodynamic models associate LFQPOs with instabilities in the Comptonizing region \cite[]{Titarchuk-Osherovich1999, Titarchuk-Fiorito-2004, Kumar-etal2014, Karpouzas-etal-2020}. However, in the widely considered LT precession model \cite[]{Ingram-Motta2014}, sustaining an extended, luminous precessing region within the inner accretion flow \cite[]{Bakala-etal2014} over timescales of days to weeks remains challenging in a highly turbulent accretion environment. Moreover, the model does not naturally explain the observed QPO evolution including time-lag reversal. The systematic evolution of type-C QPOs during the rising and declining phases of outbursts is one of the most significant discoveries of the RXTE era \cite[]{Vignarca-etal2003,Chakrabarti-etal2008, Chakrabarti-etal2009, Titarchuk-Seifina2009, Shaposhnikov-Titarchuk2009,Debnath-etal2010, Nandi-etal2012, Debnath-etal2013, Stiele-etal2013, Debnath-etal2014,Radhika-Nandi2014,Iyer-etal2015, Dutta-etal2018, Ingram-etal2019, Chatterjee-etal2020,Baby-Ramadevi2022}, and has been shown to trace geometric variations in the accretion flow \cite[]{Chakrabarti-etal2009, Dutta-Chakrabarti2010, Dutta-etal2018}.

Within the two component accretion flow framework \cite[]{Chakrabarti-Titarchuk1995}, extensive observational and numerical studies support the coexistence of a Keplerian disc and a sub-Keplerian flow \cite[]{Smith-etal2001b, Wu-etal2002, Smith-etal2007}, which sets the size of the Comptonizing region. Interestingly, the resonance between heating and cooling timescales causes the modulation of the Comptonizing region, producing the type-C QPOs \cite[]{Molteni-etal1996, Garain-etal2014}. In the Propagating Oscillatory Shock (POS) model, QPO frequency ($\nu_{\rm QPO}$) corresponds to the shock location, and inward shock motion during the rising phase increases density and optical depth ($\tau_{\rm e}$), thereby driving spectral state transitions \cite[]{Chakrabarti-etal2009, Dutta-Chakrabarti2010,Nandi-etal2012,Iyer-etal2015,Debnath-etal2026}. These oscillations modulate Comptonized photons, linking QPO frequency to coronal size and optical depth \cite[]{Vadawale-etal2001b, Rao-etal2000}. While some studies favor geometric origins \cite[]{Schnittman-etal2006, Homan-etal-2006, Ingram-Motta2014, Motta-etal2015}, others associate LFQPOs with luminosity changes driven by accretion rate variations \cite[]{Tagger-Pellat1999, Cabanac-etal2010}. 

HFQPOs, though rare and transient, provide a powerful probe of accretion dynamics in the strong gravity regime. The first HFQPO detection at $\sim67$ Hz was reported from GRS 1915$+$105 \cite[]{Morgan-etal1997}, with subsequent detections in a handful of BH-XRBs at frequencies of a few hundred Hz, although their origin is still not well understood. Early interpretations linked HFQPOs to dynamical frequencies at the innermost regions of the accretion flow, including the Keplerian frequency at the innermost stable circular orbit (ISCO) or trapped disc oscillation modes \cite[]{Nowak-etal1997}. Other models associate HFQPOs with variability in a compact Comptonizing corona during soft or intermediate states \cite[]{Remillard-etal2002, Mendez-etal2013, Aktar-etal2017, Sreehari-etal2020, Majumder-etal2022}. Resonance models propose that non-linear coupling between radial and vertical epicyclic motions in Kerr spacetime can generate commensurate frequency ratios, most notably the frequently observed $3\!:\!2$ ratio \cite[]{Abramowicz-etal2001}. Within the relativistic precession model, HFQPOs are attributed to fundamental geodesic frequencies of matter orbiting close to the black hole \cite[]{Stella-Vietri98, Stella-etal1999}, although the occurrence of stable frequency pairs is not a generic prediction. Alternative approaches based on oscillations of sub-Keplerian tori also reproduce harmonic ratios with weak dependence on the disc geometry \cite[]{Rezzolla-etal2003}. Despite these advances, a unified physical framework capable of explaining the sporadic nature, weak coherence, and state dependence of HFQPOs remains elusive.

In the \textit{AstroSat} era, coordinated wide-band spectro-temporal studies have significantly advanced our understanding of accretion dynamics and the physical origin of QPOs in BH-XRBs. LFQPOs of GRS 1915$+$105 with frequencies $\nu_{\rm QPO} \sim 1.38-7.38$ Hz are detected across different classes, with rapid class transitions occurring on timescales of a few hours \cite[]{Athulya-etal2022, Majumder-etal2022, Athulya-etal2023,Majumder-etal2026b}. These transitions are accompanied by correlated changes in X-ray flux, QPO frequency, and spectral parameters, indicating rapid modifications in the accretion geometry on timescales much shorter than the viscous timescale. Such behaviour supports the presence of a dynamically significant sub-Keplerian flow component that intermittently plunges into the black hole on a free-fall timescale \cite[]{Majumder-etal2026b}. Energy dependent time-lag studies further show that type-C LFQPOs trace characteristic accretion timescales, with enhanced Comptonized flux corresponding to higher QPO rms and increased hard-lags, while both $\nu_{\rm QPO}$ and the photon index ($\Gamma_{\rm th}$) decrease \cite[]{Belloni-etal2024, Majumder-etal2026b}. The hard-lag transition reported during the \textit{RXTE} era is not observed with \textit{AstroSat}. Moreover, QPO lag properties associated with type-C QPO depend on source inclination \cite[]{Ram-etal2026} with high-inclination systems predominantly showing soft-lags, while type-A and type-B QPOs exhibit distinct inclination dependent lag behaviour, indicating a geometric evolution of the Comptonizing region \cite[]{Choudhury-etal2025}.

Recent studies have revealed diverse QPO and time-lag phenomenology across BH-XRBs, reflecting complex disc--corona dynamics. In MAXI~J1803$-$298, an evolving QPO in the $5.31-7.61$~Hz range was reported, with swing between hard and soft time-lags interpreted as a shift in the dominant variability origin between the corona and the accretion disc \cite[]{Jana-etal2022a, Pradhan-etal2025}. The observed increase in soft-lag with energy in this source has been attributed to inclination-dependent effects \cite[]{58_chand_2022_maxij1803}, while in H1743$-$322 similar soft-lags were linked to enhanced reflection as the truncated disc moved closer to the black hole \cite[]{Sahu-etal2024}. Variability properties including time-lags, QPO frequency, and rms amplitude have also been explored within the fluctuation propagation framework in several sources, such as MAXI~J1348$-$630, Swift~J1658$-$4242, MAXI~J1820$+$070, Cyg~X$-$1, H1743$-$322, and MAXI~J1535$-$571 \cite[]{Misra-etal2017, Chand-etal2021, Bhargava2019}. In contrast, the non-evolving sharp type-C QPOs in MAXI~J1535$-$571 have been attributed to non-dissipative shock oscillations \cite[]{Chatterjee2021}. Observations of GX~339$-$4 during its 2021 outburst support disc--corona geometry models through viscous evolution and recession of the truncated disc \cite[]{Husain-etal2022, Sahu-etal2024}. At higher energies, evolving type-C QPOs have been detected up to $80$--$100$~keV in Swift~J1727.8$-$1613 and other systems \cite[]{Nandi-etal2024, Aneesha-etal2024}, while a double peaked QPO in IGR~J17091$-$3624 during heartbeat variability inferring a maximum accretion rate ratio of $\sim 25:1$ relative to GRS 1915$+$105 \cite[]{Katoch-etal2021}.

Furthermore, several detections of HFQPOs ($\sim 67.96-70.62$ Hz) from GRS 1915$+$105, observed during radio quiet and `softer' variability classes ($\delta$, $\kappa$, $\omega$, and $\gamma$), have been extensively investigated using AstroSat observations \cite[]{Sreehari-etal2020, Majumder-etal2022, Dhaka-etal2023, Majumder-etal2024, Majumder-etal2025a}. Wide-band ($0.7-50.0$ keV) spectro-temporal analyses further indicate that modulated Comptonized photons emerging from a compact corona of size $\sim 2.8-16.0\,r_{\rm g}$ are possibly responsible for the HFQPOs, which implies that HFQPOs arise due to oscillations of this compact coronal region \cite[]{Majumder-etal2025a}. The associated soft time-lags exhibit a gradual increase with photon energy and are interpreted as resulting from the reflection of hard photons reflected off the cooler accretion disc. Recently, \cite{Harikesh-etal2025} performed a detailed investigation of HFQPOs during the `dip' and `non-dip' segments of the $\kappa$ and $\omega$ classes and found a clear correlation between the spectral properties and the presence or absence of HFQPOs. More specifically, during the `non‑dip' intervals of the light curve, HFQPOs are detected and the spectra become relatively harder compared to the `dip' intervals. At the same time, the accretion disc moves closer to the black hole resulting in a compact corona in the innermost region. 

The QPO features can be interpreted as arising from the modulation of Comptonized photons produced by oscillations of the corona. The interplay between the sub‑Keplerian and Keplerian accretion components regulates the thermodynamic and radiative properties of the corona, thereby shaping the observed spectral characteristics. When the Keplerian component dominates, the corona weakens and collapses into a smaller, more compact region, while the jet is largely quenched, producing the softer HSS with a higher photon index (see Fig. \ref{fig:fig31}). In this state, rapid oscillations of the compact corona close to the event horizon (see Fig. \ref{fig:fig38}) can give rise to HFQPOs, in contrast to the more extended and slowly oscillating PSC associated with LFQPOs in the LHS and HIMS (see Figs. \ref{fig:fig34}–\ref{fig:fig35} and Fig. \ref{fig:fig38}). Conversely, scenarios of non-detection of QPO correspond to configurations in which the corona exhibits negligible or stable oscillations (see Fig. \ref{fig:fig35}). Overall, LFQPOs originate from oscillations of an extended corona in the LHS and HIMS, whereas HFQPOs arise from rapid oscillations of a compact corona in the HSS. These geometrical changes as shown in Fig. \ref{fig:fig38}, coupled with variations in the accretion rate, naturally drive the spectral state transitions discussed in \S\ref{sec9:9.2.3_htcm_results}.

{\red
Indeed, an alternative interpretations of LFQPOs have also been proposed in the literature. As already indicated, the Lense-Thirring (LT) precession model provides a viable geometrical framework for explaining several observed properties of type-C LFQPOs, including their inclination dependent amplitudes and phase-lag behaviour \citep{Schnittman-etal2006,Ingram-etal2009,Motta-etal-2015,van-den-Eijnden-etal2016,Vincentelli-etal2026}. The LT scenario has also been invoked to explain the fundamental QPO, harmonics and relative variability amplitude \citet{Stella-etal1999,Veledina-etal2013}, demonstrating that the angular dependence of radiation from a precessing Comptonizing flow. However, observational studies indicate that the evolution of QPO frequencies cannot always be attributed solely to changes in the geometry of the precessing inner flow (i.e. the disc truncation radius) inferred within the LT framework \citep{Buisson-etal2019,Nathan-etal2022,Prabhakar-etal2022}. 
In this context, it is noteworthy that the presence of harmonic features may not be unique to the LT precession model. Harmonics can also arise naturally from nonlinear variability associated with the dynamics of the accretion flow. For example, oscillations of the post-shock Comptonizing region are capable of producing QPOs together with harmonic frequency components through nonlinear and non-sinusoidal variations of the flow variables and the emergent radiations \citep{Molteni-etal1996,Ryu-etal1997,Chakrabarti-etal2004c,Okuda-etal07,Torok-etal06,Das-etal2014,Debnath-etal2025,Dihingia-etal2025b}. Similarly, models involving disc-corona coupling and radiative feedback have also been proposed to explain the origin of QPOs, their harmonics, and the associated broad-band variability \citep{Garcia-etal-2021,Garg-etal2026}. Therefore, we argue that the existence of harmonics and related timing properties does not uniquely favour a geometrical precession origin, and a comprehensive description of the observed spectro-temporal evolution may require consideration of the underlying dynamical evolution of the Comptonized flow \cite[]{Chakrabarti-Manickam00, Chakrabarti-etal2008, Nandi-etal2012,Das-etal2014,Iyer-etal2015,Nandi-etal2018,Debnath-etal2026}.
}

Fig. \ref{fig:fig32} (left) shows that, for type‑C QPOs, the soft lag amplitude decreases systematically with increasing normalised Comptonized flux ($F_{\rm Comp}/F_{\rm bol}$). Hard-lags appear only when $F_{\rm Comp} \gtrsim 0.8$, observed in GRS 1716$-$249, MAXI J1535$-$571 and MAXI J1348$-$630, while all observations with lower flux exhibit exclusively soft-lags. Notably, type-B QPOs are observed for $F_{\rm Comp} \lesssim 0.5$, which is possibly coupled with the outflow/jet emission \cite[]{Choudhury-etal2025}. The right panel of Fig. \ref{fig:fig32} further demonstrates that higher QPO frequencies, generally correspond to the reduced corona size, are associated with higher photon indices indicating spectral softening, as also evident from Fig.~\ref{fig:fig31}. These findings are consistent with earlier \textit{AstroSat} results \citep{Majumder-etal2025a, Majumder-etal2026b}. A decrease in $\nu_{\rm QPO}$ implies an outward expansion of the coronal boundary and a reduction in optical depth ($\tau_{\rm e}$), leading to enhanced Compton up‑scattering and the production of hard-lags, whereas an increase in $\tau_{\rm e}$ favours down‑scattering and soft-lags \citep{Reig-etal2000}. In GRS 1915$+$105, \textit{AstroSat} detected only soft-lags, consistent with its lower Comptonized flux compared to the \textit{RXTE} era, where a soft‑to‑hard transition near $\nu_{\rm QPO} \sim 2$ Hz was observed \citep{Majumder-etal2026b}. Several alternative mechanisms have been proposed to explain the frequency dependence of the time-lag in MAXI J1535$-$571. These include a transition from disc origin QPOs at $\nu_{\rm QPO}<2.20$ Hz to corona origin QPOs at higher frequencies \citep{Garg-etal2022}, two-component Comptonization model \citep{nobili2000}, and frequency dependent disc-corona feedback \citep{Karpouzas-etal2021}. Overall, the observed lags likely result from a combination of physical processes, such as Comptonization \citep{Miyamoto-etal1988}, outflows/jets \citep{Patra-etal2019, Chatterjee-etal2020, Mendez-etal2022}, reflection and propagation delays \citep{poutanen_fabian1999, Ross-Fabian2007, Chatterjee-etal2020}, coronal geometry, and inclination effects \citep{Dutta-etal2016, van-den-Eijnden-etal2016, Chatterjee-etal2020, Choudhury-etal2025}. While this framework explains the prevalence of soft-lags at higher frequencies and hard-lags at low frequencies, the recurrence and inclination dependent nature of the $\sim 2.20$ Hz lag transition, particularly in GRS 1915$+$105, remains an open problem.

{\red Further, in GRS~1915$+$105, the structured variability classes (for example $\rho$ and $\kappa$) have been interpreted within the framework of radiation pressure driven thermal-viscous instabilities in the accretion disc \cite[]{Belloni-etal1997,Belloni-etal-2000,Janiuk-etal2001,Neilsen-etal2011,Neilsen-etal2012}. Specifically, the $\rho$ class may result from a combination of thermal-viscous and radiation-pressure instabilities, the local Eddington limit, and episodic radiation-pressure driven mass ejections from the inner disc \cite[]{Neilsen-etal2012}. However, the observed complexity of the spectral and timing properties, together with the presence of variability classes over a wide luminosity range \cite[]{Athulya-etal2023}, suggests that disc instabilities alone may not provide a complete explanation. Alternative scenarios invoking local fluctuations in the accretion flow coupled with jets and outflows have also been proposed \cite[]{Chakrabarti-Nandi2000,Nandi-etal2001b,Nandi-etal2001a}. Within this broader picture, it is reasonable to consider that variability classes, LFQPOs, and HFQPOs need not originate from a single physical mechanism. Instead, changes in the disc geometry and physical conditions of the inner accretion flow, driven by the combined effects of disc instabilities, accretion rate fluctuations, and disc-corona-jet interactions, possibly provide a unified framework for understanding the diverse timing and spectral properties observed in BH-XRBs including GRS~1915$+$105.
}

HFQPOs are detected when the disc flux is decreased and Comptonization becomes stronger \citep{Majumder-etal2022, Majumder-etal2024, Majumder-etal2025a} at lower $\tau_{\rm e}$ during radio quiet and softer variability classes ($\delta$, $\kappa$, $\omega$, and $\gamma$). As $\tau_{\rm e}$ increases, the time‑lag amplitude declines, implying weaker Comptonization and shorter delays for high energy photons. This reduces the hard-lag magnitude, which {\red may} eventually lead to a transition to a soft-lag. Thus, low optical depth favours efficient up‑scattering of disc photons resulting in hard-lags, whereas higher $\tau_{\rm e}$ allows repeated down‑scattering, causing high energy photons to lose energy and yield soft-lags \citep{Reig-etal2000}. HFQPOs disappear abruptly when the inner disc radius moves inward, increasing the optical depth and leading to a more compact corona with reduced Comptonized flux, thus highlighting their strong dependence on the Comptonizing region. When $\tau_{\rm e}$ exceeds a threshold ($\tau_{\rm e} \sim 8.8$), the electron temperature drops to $\sim 2.0-2.6$ keV, insufficient for effective Comptonization \citep{Majumder-etal2025a}. The covering fraction then decreases, and the HFQPOs disappear. Variations in the coronal optical depth, likely driven by changes in accretion rate, jet or outflow activity, and associated alteration of the coronal geometry, can therefore regulate the appearance of HFQPOs. Oscillations of a compact corona thus provide a plausible origin for the HFQPOs, consistent with \cite{Sreehari-etal2020,Majumder-etal2022,Majumder-etal2025a,Harikesh-etal2025}. 

{\red While these observational results favour the origin associated with a compact inner Comptonizing region, alternative interpretations have also been suggested. In particular, the relativistic precession model has been invoked to explain the $\sim67$ Hz feature in GRS~1915$+$105 as the nodal precession frequency of matter orbiting close to the ISCO \cite[]{Motta-Belloni2024}. With this, the observed frequency implies a black hole spin of $a_{\rm k}\sim0.71$. However, the inferred spin is substantially lower than the high values ($a_{\rm k}\gtrsim0.90$) obtained from continuum fitting method \cite[]{McClintock-etal2006,Sreehari-etal2020,Wang-etal2023b}, highlighting the model dependence of the inferred source parameters. Moreover, the $\sim67$ Hz oscillation is identified as a type-C QPO \cite[]{Motta-Belloni2024}, yet it is observed during relatively softer spectral states. This behaviour differs from that of conventional type-C QPOs, which are generally associated with harder states and its origin possibly connected with the modulation of the radially extended Comptonizing region \cite[]{Nandi-etal2012,Nandi-etal2018,Sreehari-etal2020,Majumder-etal2022,Nandi-etal2024,Harikesh-etal2025,Majumder-etal2025b,Majumder-etal2026b}. These observational differences suggest that the physical conditions giving rise to the $\sim67$ Hz oscillation may not be fully captured by a straightforward extension of the Lense--Thirring precession framework developed for low-frequency type-C QPOs. Consequently, the origin of the $\sim67$ Hz feature remains inconclusive, and alternative scenarios involving oscillations of a compact inner disc-corona configuration warrant further investigation.
}
Within the present formalism, we have highlighted several observational features that can be naturally interpreted within the framework of the Hybrid Two-Component Model (HTCM), emphasizing the role of the sub-Keplerian flow in explaining fast time variability and spectral properties of BH-XRBs. Moreover, the HTCM offers a more comprehensive description of the accretion process to interpret the spectro-temporal behaviour of BH-XRBs, potentially providing a unified framework that offers advantages over existing models.

\section{Limitations, Challenges and Future Prospects}\label{sec11:limitations}

\textit{AstroSat} has significantly advanced the study of BH-XRBs by enabling simultaneous timing and spectral observations over a broad energy range. Nevertheless, a number of instrument specific and operational constraints limit the achievable sensitivity and complicate the interpretation of the resulting data products. These effects reduce the detectability of weak timing signatures, hinder the characterisation of subtle spectral features, and introduce additional uncertainties in joint modelling across multiple instruments.

\subsection{Limitations, Challenges and Way Forward}\label{sec11:subsec1}

In what follows, we outline the instrumental, modelling and observational limitations relevant to the present analysis.

\begin{itemize}
    \item Impact of Deadtime on LAXPC Timing Analysis: LAXPC provides excellent timing capabilities with a time resolution of $10~\mu{\rm s}$, which is ideal for searching for the very fast variability features across a broad frequency range. The key challenge for timing analysis is the detector deadtime of approximately $42~\mu{\rm s}$. Deadtime introduces a frequency dependent modulation of the Poisson noise level in the power density spectrum, suppressing high frequency power and complicating the recovery of intrinsic source variability. Therefore, the deadtime correction is mandatory, and in practice, the correction is sensitive to count rate, energy, and detector properties. Following \cite{Zhang-etal1995}, \cite{Yadav-etal2016} addressed deadtime correction in LAXPC adopting an empirical systematic uncertainty of $0.5\%$ in quadrature to PDS and adequately estimated the deadtime corrected Poisson noise power level. However, the impact of the adopted $0.5\%$ systematic uncertainty on the physical quantities derived from timing analysis, such as time-lag and coherence, has not been investigated in detail \cite[]{Rout-etal2025}. Moreover, different modes of \textit{RXTE/PCA} data exhibited different deadtime \cite[]{Jahoda-etal2006}, whereas average deadtime of LAXPC data is estimated using all three proportional counters where the contribution from different events are not considered. Nevertheless, several avenues remain for improving the deadtime modelling of LAXPC detectors including the incorporation of variable deadtime models \citep{Tuo-etal2024} and the computation of cross spectra from identical detectors \cite[see][for details]{Bachetti-etal2015,Rout-etal2025}.
    
    \item LAXPC Background and Spectral Response: As a gas filled proportional counter, LAXPC experiences substantial instrumental background that becomes dominant for relatively faint sources and toward higher energies. However, owing to \textit{AstroSat}’s near equatorial orbit, LAXPC exhibits relatively smaller background variations compared to RXTE/PCA, which operated in a high inclination orbit \cite[]{Jahoda-etal2006,Antia-etal2021}. Indeed, the background estimation of LAXPC is generally performed either from blank sky observations closest to the observation date or through the empirical background model \cite[]{Misra-etal2021} for faint sources, which scales the high energy count rate to predict the background at lower energies. While \cite{Antia-etal2022} improved the LAXPC background model by incorporating the quasi-diurnal variation of the satellite, the background estimation still suffers from a systematic uncertainty of approximately $3\%$. The systematic can be reduced further by calibrating the detector gain through long term monitoring of the Xe K-edge features at $\sim 34$ keV \cite[]{Antia-etal2022}. Apart from that, \textit{LAXPC} spectra generally exhibits feature of Xe L-edge at $\sim 4.8-5.5$ keV and K$_\alpha$ at $\sim 29$ keV, which are often mitigated by incorporating the model components \texttt{smedge} and \texttt{edge}. This eventually improves the fit quality but at the cost of reduced accuracy, particularly in constraining reflection features and high-energy cut-offs. Further refinement of the response can be achieved using multiple Crab observations and Crab scaling techniques, analogous to \textit{pcacorr} approach developed for \textit{RXTE}/PCA \cite[]{Garcia-etal2014b}. In addition, the impact of induced radioactivity may be incorporated into the background modelling, following the methodology adopted for \textit{RXTE}/PCA \cite[]{Jahoda-etal2006}. Spectral analysis is further constrained by the moderate energy resolution of LAXPC ($\sim 20\%$ at $6.4$ keV). This resolution blends intrinsically narrow spectral features including fine structure in the Fe K$\alpha$ lines from neutral to highly ionized species into broader apparent lines. Consequently, key diagnostics such as ionization balance, relativistic broadening, and weak absorption or emission components cannot be well separated. The moderate spectral resolution affects the application of sophisticated reflection models (\texttt{relxill} family of models; see \citet[][and references therein]{Garcia-etal2014,Dauser-etal2016}) commonly used to constrain black hole spin, disc inclination, and coronal geometry through line profiles and the `Compton reflection hump'. 
  
    \item Hardware Degradation and Detector Reliance: Early mission issues reduced both effective collecting area and the redundancy of the three LAXPC units. In particular, a gas leak in LAXPC30 and gain instabilities in LAXPC10 often made it difficult to operate all three detectors simultaneously, leaving LAXPC20 as the most consistently reliable unit. Generally, the leakage in gas filled proportional counters is quite common as experienced in the outermost veto layers of \textit{RXTE}/PCA (PCU0 and PCU1) in 2000 and 2006, respectively\footnote{\url{https://heasarc.gsfc.nasa.gov/docs/xte/whatsnew/big.html}}. In practice, a reduction of the effective area in LAXPC by a factor of $3$ lowers the sensitivity by approximately a factor of $1.7$. As a result, the full scientific capability of the complete LAXPC configuration can be recovered only for sources that are about $1.7$ times brighter. Despite this limitation, LAXPC remains a powerful instrument for studying bright black hole transients. The reduced throughput, however, decreases photon statistics, affecting sensitivity to low amplitude timing features and limiting the characterization of spectral distribution at higher energies.

    \item SXT Pile-up, Timing, and Low Energy Response: SXT complements LAXPC by providing simultaneous low energy coverage essential for wide-band spectral modelling. For bright sources, however, SXT data are susceptible to photon pile‑up, which can distort the spectral shape unless mitigated through annular extraction of the point-spread function at the core \cite[]{Banerjee-etal2024}. While effective, this approach reduces photon statistics and can introduce flux uncertainties if the encircled energy fraction is not accurately accounted for. However, it is noteworthy that the SXT experiences less pile up ($\sim 40$ cts/s) as compared to XMM-Newon/EPIC ($\sim 0.5$ cts/s)\footnote{\url{https://xmm-tools.cosmos.esa.int/external/xmm_user_support/documentation/uhb/XMM_UHB.pdf}}. The intrinsic timing resolution of SXT ($2.4$ s) further limits sensitivity to study rapid variability and put constrain for combined time resolved spectroscopy. At low energies, the calibration uncertainties, particularly near the Si and Au edges, can introduce artificial features in the $0.5-2.5$ keV band. Further, low energy response can be improved via quantum efficiency correction through cross-calibration with simultaneous \textit{NICER}, SXT observations of supernova remnants and bright AGNs \cite[]{Plucinsky-etal2012}.          
    
    \item Cross-calibration and Operational Complexities: One of the major limitations of \textit{AstroSat} for wide-band black hole spectroscopy arises from cross-calibration uncertainties and operational complexities between its instruments. However, for cross instrument calibration, the prerequisite is that each individual instrument must itself be calibrated to the highest possible accuracy. In this context, the large systematic uncertainties associated with LAXPC represent a major hurdle. With the currently prescribed $\sim 3 \%$ systematic errors, the LAXPC spectral response and background modelling are not sufficiently accurate for rigorous cross-instrument studies. Simultaneously, SXT possesses residual calibration uncertainties at low energies, resulting the wide-band ($0.7-25.0$ keV) spectral fitting that significantly underutilizes \textit{AstroSat}’s capability. On the other hand, CZTI data with its current calibration allow analysis over a broad energy range of $20.0-200.0$ keV. Beyond $\sim 200.0$ keV, CZTI data becomes increasingly background dominated requiring non-simultaneous blank sky background subtraction. Consequently, joint spectral fitting among SXT, LAXPC, and CZTI may reduce degeneracies in continuum modelling and improve the robustness of subtle spectral features, such as electron temperature and reflection signatures. \textit{AstroSat} further encounters certain operational constraints when observing X-ray bright sources located in the Galactic plane. This is primarily due to limitations as well as simultaneous requirements imposed by the UVIT instrument. These constraints often lead to restrict observing windows of the X-ray observations.
        
    \item Mitigation through Pipelines, Calibration, and Software: Several challenges encountered in \textit{AstroSat} data analysis can be mitigated through continued improvements in data processing pipelines, calibration, and background modelling. Continued refinement of instrument response matrices together with time dependent calibration updates would enhance the reliability of both spectral and timing measurements. Advanced background estimation techniques including model based and data driven approaches could improve sensitivity and model fit for faint sources at high energies. From the analysis perspective, unified and well documented software capable of handling multi‑instrument data in a seamless manner would reduce systematic inconsistencies. In the present work, additional limitations arise from the exclusion of CZTI data due to the asymmetry in the count statistics, where the count in low energy is significantly higher than count in higher energy. Hence, even with the LAXPC-CZTI joint spectral analysis, the spectral parameters are expected to be influenced by the LAXPC statistics which is dominated by $3\%$ systematics. Further calibration of CZTI and joint analyses of LAXPC-CZTI observations could provide constraints on spectral properties, especially for reflection and hard X‑ray components. 
    
    \item Orbit‑driven Observing Efficiency and Mitigation Strategies: The five \textit{AstroSat} payloads are well suited for detailed observations of X-ray binaries and transient events, although long term monitoring is constrained by orbital and environmental factors. The mission’s $6^{\circ}$ inclination equatorial orbit at an altitude of approximately $650$ km offers more opportunities for repeated observations of many sources relative to a polar orbit, but it still imposes significant scheduling and data quality limitations. With roughly $14-15$ orbits per day, instrument operations must account for SAA passages, Earth occultations that depend on source declination, and avoidance of bright regions such as the Earth limb. These factors generate bad time intervals and reduce observation efficiency across all payloads. One remedy would be to design future missions with highly elliptical orbits, as used by \textit{Chandra}, \textit{XMM-Newton}, and \textit{INTEGRAL}, which avoid prolonged exposure to enhanced charged particle regions in the Van Allen belts and provide long, uninterrupted observation windows. These long observation windows will enable to observe sources during hour scale variations as well as their class transitions similar to GRS 1915$+$105 like BH-XRBs. Moreover, observing very low frequency variation may also be possible for long uninterrupted lightcurves through the implementation of such elliptical orbits. Such orbits would also require careful planning for ground station contact to ensure reliable commanding and telemetry without downtime. In parallel, improved planning tools for limb and SAA avoidance would further increase the scientific yield under current orbital constraints.

    \item Methodological Challenges in Background Modelling: A pronounced $\sim 23.5$ hours background modulation, not previously accounted for prior to $2022$, introduces spurious spectral features at $\lesssim 5 \%$ level of the background, mimicking intrinsic source variability if left uncorrected. The effective $\sim 3\%$ systematic uncertainty floor for bright sources is therefore dominated by background subtraction residuals and constitutes the primary limitation for exposures $\gtrsim 5$ ks. \textit{AstroSat} lacks a continuous, energy resolving particle monitor, preventing direct tracking of sub‑orbital particle flux variations that drive LAXPC background changes. The CPM onboard \textit{AstroSat} functions largely as a binary SAA flag and is insensitive to these effects. Consequently, the current background model relies on a statistical Fourier representation in latitude and longitude rather than physically motivated predictors such as Earth elevation or local particle flux.
    
\end{itemize}

Despite these limitations, \textit{AstroSat} observations have facilitated robust estimates of black hole mass and spin in several BH‑XRBs \citep{Sreehari-etal2019b, Sreehari-etal2020, Baby-etal2020, Bhuvana-etal2021}. Looking ahead, continued improvements in calibration, background characterisation, and multi‑instrument cross-calibration will be essential for maximising the scientific return of wide-band X‑ray observations. Advances in spectral resolution, readout capabilities, and unified analysis frameworks are likely to reduce systematic uncertainties and strengthen constraints on accretion geometry, coronal physics, reflection features, and black hole spin in the \textit{AstroSat} era and beyond.

\subsection{Overcoming the Legacy of \textit{RXTE}: Future Science Prospects with \textit{AstroSat}} 

The era of \textit{RXTE} ($1996-2012$) marked a major turning point in the study of accreting compact objects, as it enabled high sensitivity, wide‑band spectro‑temporal investigations for the first time. Its primary instruments, the PCA with a large collection area and microsecond timing capability and HEXTE provided access to higher energy photons, together opened an unprecedented observational window to examine rapid X‑ray variability and wide-band spectral characteristics of BH-XRBs. \textit{RXTE} provided continuous sky monitoring through the ASM enabling long term tracking of outbursts, HID evolution and state transitions. Its unprecedented timing capability led to the discovery and detailed characterization of rapid variability phenomena including LFQPOs, HFQPOs, inclination dependent time-lags, wide-band noise components, and empirical correlations between spectral and timing properties. Several landmark detections, namely the stable $\sim 70$ Hz HFQPO in GRS 1915$+$105, the twin peak HFQPOs at $\sim 300$ and $450$ Hz in GRO J1655$-$40 and several detection of HFQPOs in other BH-XRBs, were possible only because of PCA’s large area and high time resolution. However, \textit{RXTE} also suffered from limitations: (a) the PCA’s coarse energy resolution restricted the study of Fe line features and introduced degeneracies in reflection based spin measurements, (b) the rapid decline of effective area at higher energies hindered detailed investigation of high energy variability, and (c) HEXTE’s increasing background uncertainties in later years forced reliance on PCA spectra. Furthermore, high time resolution modes restricted flexible energy band selection, while the lack of coverage below $\sim 2$ keV limited the ability to constrain the properties of the thermal disk. These instrumental constraints complicated energy dependent timing study and reduced sensitivity to faint features at high energies.

With the launch of \textit{AstroSat} in September 2015, spectro‑timing studies entered a new phase shifting from the discovery oriented focus of \textit{RXTE} to more physically grounded interpretations of accretion flows around black holes. LAXPC with its large effective area above $>20$ keV and $\sim 10~\mu$s timing resolution enables detection of rapid variability such as HFQPOs, time-lag energy spectra, and wide‑band timing behaviour. The synergy of LAXPC with SXT provides simultaneous coverage from $\sim 0.3-80.0$ keV that strengthens constraints on the disc and coronal emission. Although LAXPC requires better background modelling and SXT offers limited low energy timing, their combined capability, supplemented by \textit{AstroSat}'s UV and soft X‑ray instruments, extends the spectro–temporal diagnostics pioneered by \textit{RXTE}. Indeed, this allows better modelling of QPO evolution, time-lags, inclination dependent variability, and accretion geometry. As a result, \textit{AstroSat}, builds on the phenomenological framework established by \textit{RXTE}, provides improved understand of more physically grounded insights into disk–corona coupling in BH-XRBs.

The transition from \textit{RXTE} to \textit{AstroSat} outlines how differences in mission orbits, instrument design, and pointed as well as coordinated long term operational strategy enhance the quality and continuity of high energy astrophysical data. Although both operate in LEO and stop operating during SAA passages, \textit{AstroSat} benefits from its near equatorial orbit ($\sim 6^{\circ}$), reducing radiation exposure relative to \textit{RXTE} ($\sim 23^{\circ}$) and improving duty cycle \citep{Jahoda-etal2006}. However, this advantage is offset by challenges like instrument design with higher dead-time of LAXPC ($\sim 42\,\mu\text{s}$) and its gain instability, in contrast to the more stable gain, low dead‑time ($\sim 10\,\mu\text{s}$) of PCA \citep{Antia-etal2021}. The multi‑instrument, multi‑wavelength nature of \textit{AstroSat}, especially for the UVIT instrument, complicates the coordination and pointing observations of BH-XRBs. In comparison, \textit{RXTE} excelled in ample monitoring, building a vast and continuous archive that remains unmatched for spectral studies during long outbursts. Instrumentally, PCA’s five proportional counters provided redundancy and high photon throughput, whereas the gas loss in LAXPC30 and gain instability of LAXPC10 reduced the net effective area of LAXPC which has been discussed earlier in \S \ref{sec11:subsec1}. Despite these constraints, \textit{AstroSat} remains a powerful wide-band observatory whose capabilities complement the stable and highly sensitive legacy of \textit{RXTE}.

\textit{AstroSat} was designed as a multi-wavelength observatory for simultaneous broadband X-ray observations and has made substantial contributions to the spectro-temporal investigation of accreting compact objects \citep{Agrawal2006, Paul2013}. This achievement builds upon the legacy of the Indian X-ray Astronomy Experiment (\textit{IXAE}), the first Indian X-ray astronomy payload flown onboard the IRS-P3 mission nearly three decades ago \citep{Agrawal-etal1996}. In particular, LAXPC has played a pivotal role in detecting HFQPOs and strong LFQPOs in BH‑XRBs including extensive timing studies of GRS~1915$+$105 \citep{Sreehari-etal2020, Majumder-etal2022, Majumder-etal2025a}, and has enabled QPO measurements up to $\sim 100$ keV \citep{Nandi-etal2024}. Its large effective area at high energies allows energy resolved timing analyses that were beyond the reach of \textit{RXTE}/PCA. In contrast, SXT effectively demonstrated soft X‑ray spectroscopy for bright BH-XRBs enabling improved disc component modelling, although pile up persists for a few brightest sources. A major scientific highlight of CZTI is its ability to measure hard X‑ray polarization in the $100-380$ keV range. The detection of polarization fraction of $\sim 23 \pm 4\%$ in Cyg~X$-1$ \cite[]{Chattopadhyay-etal2024} provides unique and valuable information about the geometry of the corona and the jet.

To further enhance the scientific capabilities of \textit{AstroSat}, several calibration issues need to be addressed. For the existing observational data, a well defined cross‑calibration strategy, similar to those used for \textit{XMM‑Newton} and \textit{Chandra} (IACHEC guidelines\footnote{\url{https://iachec.org/}}), would improve the consistency of SXT and LAXPC responses. A practical approach involves modelling instrumental backgrounds and then determining the SXT$+$LAXPC cross‑normalisation in the $3-7$ keV overlap using BH‑XRB data in hard state. This can be followed by validating energy dependent effective areas through comparisons of spectral indices, plasma temperatures, and Fe line shapes. The calibration can be anchored by tying SXT to IACHEC low energy standards and using CZTI to define the high energy domain of the LAXPC response. In a way, the combined SXT+LAXPC performance alignes with the global IACHEC reference ensuring consistency based on physical modelling rather than arbitrary systematics.

A similar multi‑step approach can improve LAXPC$+$CZTI cross‑calibration. First, a time series of the Crab cross‑normalisation in the $20$–$30$ keV range across all available CZTI Crab observations can be constructed to track epoch dependent variation. Next, spectral slope agreement ($\Delta\Gamma$) across different sources and spectral states can be evaluated to quantify consistency between the two instruments. From this, a recommended cross‑normalisation constant with $1\sigma$ uncertainties can be provided for users performing joint LAXPC$+$CZTI spectroscopy. The procedure can also identify which of the existing CZTI observations are directly compatible with LAXPC for reliable $3-200$ keV joint analysis and which require spectral state dependent corrections that help ensure robust hard X‑ray spectroscopy with \textit{AstroSat}.

With improved calibration, background modelling, and cross‑instrument consistency, \textit{AstroSat} has the potential for simultaneous wide-band spectroscopy from soft to hard X‑rays extending into the soft gamma‑ray regime with CZTI. Enhanced calibration would expand the effective LAXPC band ($\sim 3-40$ keV), while joint SXT$+$LAXPC$+$CZTI observations would provide robust energy coverage ($\sim 0.3-200.0$ keV) with systematic uncertainties reduced from $\sim 3\%$ to below $1\%$. With full cross‑calibration, \textit{AstroSat} can deliver $\sim 0.3-380.0$ keV spectroscopy together with hard X‑ray polarization measurements in the $100-380$ keV range, a capability unmatched by any mission in operation. These measurements are expected to allow a unified investigation of accretion processes from the disk through the corona to the jet. CZTI polarimetry, previously demonstrated for only a few sources such as Cyg~X$-$1, could then be extended to bright BH‑XRBs including GRS~1915$+$105 and GX~339$-$4. When combined with IXPE polarization in the $2-8$ keV band and wide-band spectroscopy across the intermediate $\sim 8-100$ keV range, \textit{AstroSat} can resolve long standing questions about coronal geometry, emission processes, and disk–jet coupling by constraining the coronal electron temperature and its contributions in the hard X‑ray band. Furthermore, improved calibration would reveal high energy cut-offs and reflection features in the extended $25-40$ keV band of BH‑XRB spectra and enable spectro‑polarimetric studies of $\sim 10$ bright sources. The high polarization fraction ($\sim 23\%$) measured for Cyg~X$-$1 illustrates the potential of fully calibrated CZTI data, where the energy dependence of polarization can disentangle jet and coronal components offering new insight into accretion–ejection physics. Indeed, a fully calibrated \textit{AstroSat} offers a major opportunity to refine the classification of spectral states in BH-XRBs beyond traditional framework in terms of `q'-diagram \cite[]{Belloni-etal2005}, which compress intrinsically multi‑dimensional accretion physics into a single disk‑to‑corona flux ratio. Because these methods rely on decomposing the $2-10$ keV (or $2-20$ keV for \textit{RXTE}) spectrum into disk and power‑law components, the inferred results remain ambiguous without independent hard X‑ray constraints. High energy band diagnostics, such as the $\Gamma$-Flux plane in the $\sim 22-100$ keV range demonstrated for Cyg~X$-$1 \cite[]{Lubinski-etal2020}, instead provide a clearer mapping of accretion regimes and direct access to the physical parameters of Comptonization ($kT_{\rm e}$, $\tau_{\rm e}$).

Moreover, \textit{AstroSat} may enable simultaneous measurements of multiple accretion diagnostics on timescales of $\sim 100$ s or less. The SXT constrains the disc temperature and inner disc radius, while the LAXPC measures the Comptonization slope ($\Gamma_{\rm th}$) and the high energy cut-off in the $3-40$ keV band. The CZTI provides complementary measurements of the hard X-ray flux and spectral slope in the $20-100$ keV range. Furthermore, the $\sim 10~\mu$s timing resolution of the LAXPC facilitates detailed investigations of PDS features. No earlier mission has offered such simultaneous wide-band and high time resolution capability. This unified parameter space combining spectral, timing, and polarimetric \cite[hereafter STP;][]{Nandi2018,Majumder-etal2026} information enables a physically grounded multi‑dimensional classification scheme surpassing both 1-dimensional HID and 2-dimensional hard X‑ray planes used previously. It allows direct tests of state transition hysteresis and examine the proposed six distinct accretion geometries in hard X-ray flux \cite[]{Lubinski-etal2020}, while probing the physical origin of different QPO types, exploring possible links between HFQPOs and ISCO related frequencies, and identifying $kT_{\rm e}$ as a key tracer of coronal evolution. Crucially, this framework supports real time classification on $\sim 100$ s timescales and is ideally suited for complex systems such as GRS~1915$+$105. In addition, \textit{AstroSat}’s growing archive holds significant potential for phase resolved QPO spectroscopy, energy dependent lag studies, and frequency resolved spectroscopy for the BH-XRBs.

\subsection{Future Prospects}

Future space-based X‑ray observatories are increasingly geared toward integrating spectroscopy, timing, and polarimetry (STP) to study BH-XRBs. Although spectro‑temporal analyses have made significant progress, key uncertainties persist which are driven in part by degeneracies among competing accretion and ejection geometries. Wide-band timing and spectro‑polarimetric observations offer a promising pathways to breaking these degeneracies and constraining the disc–corona–jet configuration in the vicinity of the black hole. Some future prospects are outlined below.

\begin{itemize}
    \item High Time Resolution and Large Effective Area: Future X‑ray missions would benefit significantly from combining ultra‑fast timing capability with a large collecting area. Time resolution of $\sim \mu$s would enable robust studies of rapid variability phenomena such as kilo‑Hertz QPOs and high frequency Fourier diagnostics including time‑lags and coherence. At the same time, large effective areas ($\gtrsim 10^{4}$ cm$^{2}$) would improve photon statistics, enhance sensitivity to weak or transient timing signatures, and mitigate deadtime limitations that increasingly dominate at high count rates \cite[see][]{Bachetti-etal2015}. Together, these capabilities would allow more stringent tests of QPO models, especially those invoking relativistic dynamics in the innermost accretion flow close to the black hole.
    
    \item Wide-band Spectral Capability with Large Effective Area: Wide-band spectral coverage, combined with large effective area and improved energy resolution, will be essential for advancing studies of accretion geometry and radiative processes in BH-XRBs. Energy resolution of a few eV would significantly improve constraints on Fe line profiles and reflection components which are central to black hole spin measurements. Micro‑calorimeter detectors, as demonstrated by missions such as \textit{XRISM/Resolve} demonstrate the potential of achieving $\sim 4.5$ eV spectral resolution together with microsecond scale timing. Large collecting areas further enable studies of faint sources and weak spectral features, while well characterized, overlapping energy bands remain critical for robust wide-band spectral modelling.

    \item Polarization Measurements and Scientific Prospects: X‑ray polarimetry offers a powerful and largely unexplored probe of emission geometry and scattering environments in accreting black hole systems. Polarization measurements can directly constrain the geometry and location of the corona, complementing spectral estimates of coronal temperature and optical depth, since the degree and angle of polarization are intrinsically linked to scattering processes in the disc–corona system. Polarimetry also provides an independent handle on black hole spin by enabling accurate determination of disc inclination, reducing uncertainties inherent in continuum fitting approaches. Although \textit{IXPE} has already provided insightful polarimetric results \cite[][and references therein]{Dovvciak-etal2024,Majumder-etal2026} at low energies ($2-8$ keV), extending such measurements to broader energy ranges would be crucial for fully characterising coronal emission, reflection, and disc winds. When combined with wide-band spectroscopy and fast timing, dedicated polarimetric capability would enable a more complete and self‑consistent picture of accretion physics in black hole X‑ray binaries.

    \item Multi-wavelength Observations: A dedicated multi-wavelength UV/X-ray mission, overcoming the limitations of \textit{AstroSat}, would enable a comprehensive view of accretion systems across bands that probe distinct regions: UV/optical emission traces the outer accretion disc, X-rays originate from the hot inner flow, and radio emission is associated with relativistic jets. {\red Wide-band spectral energy distributions (SEDs) enable the simultaneous characterization of the disc, corona, and jet, while coordinated timing studies constrain disc-corona coupling and the propagation of fluctuations within the accretion flow. What is more is that simultaneous UV and X-ray lag measurements provide an especially powerful probe of these processes. Although such studies are currently not feasible for BH-XRBs because of limitations in temporal resolution, similar lag analyses have been carried out for AGNs using SXT and UVIT onboard \textit{AstroSat} \cite[and references therein]{Kumari-etal2023,Kumari-etal2024,Reshma-etal2026}.}
    
\end{itemize}

Overall, the scientific legacy of \textit{AstroSat} demonstrates the importance of coordinated multi‑band observations of X‑ray binaries from a single platform enabling robust wide-band and time resolved studies despite operational constraints. Building on this experience, future missions would benefit from extended spectral coverage, wide field monitoring for rapid transient detection, and fast slew capabilities to capture early outburst phases. Large effective areas with high time resolution ($\sim \mu$s) will be essential for probing rapid variability and faint sources. Incorporating intelligent on‑board processing and accommodating wide dynamic range across instruments would further maximise observing efficiency and data quality. Collectively, these considerations reflect how the lessons learned from \textit{AstroSat} can guide the design of next generation missions \cite[][]{Nandi2018} targeting accreting compact objects.

\section{Acknowledgment}

{\red Authors thank the anonymous reviewer for the constructive comments and valuable suggestions that helped to improve the quality of the manuscript.} Authors acknowledge the stimulating scientific interactions and fruitful discussions carried out through the ongoing virtual platform \textit{Exploring Saga of Compact Objects (ESCO)}, which has been active since 2020. Authors also acknowledge the insightful discussions during the 6th national conference \textit{REcent Trends in the Study of Compact Objects: Theory and Observation (RETCO-VI)} held at IIT Indore ($10-12$ March 2025) and the international conference \textit{A Decade of AstroSat Observations: Science Outcomes and Future Prospects} held in Bengaluru (30 January–1 February 2026). These interactions contributed significantly to the development and refinement of the ideas presented in this work. Authors are especially grateful to Prof. P. C. Agrawal, the founding Principal Investigator of \textit{AstroSat}, for his pioneering contributions to the conceptualization, development, and realization of the mission. AN and SD further thank H. M. Antia for fruitful discussions and valuable suggestions that improved the quality of this study. AN, AK, SS, RBT, VKA thank Group Head (Space Astronomy Group – SAG), Deputy Director (Payload, Data Management \& Space Astronomy Area – PDMSA), and the Director (U R Rao Satellite Centre – URSC) for encouragement and continuous support to carry out this research. MCR thanks Program Director and the Director, URSC. SDC, SM and SD acknowledge the Department of Physics, IIT Guwahati, for providing the necessary facilities to complete this work. SM acknowledges support from JSPS Postdoctoral Fellowship {\red (Fellowship ID: P25337 and KAKENHI Grants-in-Aid number: 25KF0225)}, during which part of this work was carried out at the Cosmic Ray Laboratory, Department of Physics, Kyoto University. SDC acknowledges financial support from Prime Minister Research Fellowship (PMRF) program. BGR thanks Gifu University, Japan for support. {\red RA acknowledges the access to high-performance computing facilities {\it Kawas} cluster (Theory Cluster and Storage) provided by ASIAA.} PM, BGD and AN acknowledge the support from ISRO sponsored project (DS-2B-1313(2)/6/2020-Sec.2). {\red PM and BGD acknowledge IUCAA, Pune for the Visiting Associateship Program}. KPS thanks the Indian National Science Academy for support under the INSA Senior Scientist Programme. This publication uses the data from the \textit{AstroSat} mission of the Indian Space Research Organisation (ISRO), archived at the Indian Space Science Data Centre (ISSDC). This work has used the data from the Soft X-ray Telescope (SXT) developed at TIFR, Mumbai, and the SXT-POC at TIFR is thanked for verifying and releasing the data and providing the necessary software tools. This work has also used the data from the LAXPC Instruments developed at TIFR, Mumbai, and the LAXPC-POC at TIFR is thanked for verifying and releasing the data. We also thank the \textit{AstroSat} Science Support Cell hosted by IUCAA and TIFR for providing the LAXPCSOFT software which we used for LAXPC data analysis. The authors thank \textit{RXTE}, \textit{HXMT} and \textit{MAXI} instrument teams for processing the data and providing the necessary software for analysis. {\red The Indian Space Research Organisation (ISRO) funded, managed and facilitated the overall {\it AstroSat} project.}

\section{Data Availability}

\textit{AstroSat} data used for this publication are available at the Astrobrowse (\textit{AstroSat} archive) website {\url{https://astrobrowse.issdc.gov.in/astro_archive/archive}} of the Indian Space Science Data Center (ISSDC). \textit{RXTE} and \textit{HXMT} data used in this work are archived at the HEASARC data center {\url{https://heasarc.gsfc.nasa.gov/cgi-bin/W3Browse/w3browse.pl}} and \url{https://archive.hxmt.cn/proposal}, respectively. The \textit{MAXI}/GSC on-demand data are available at {\url{http://maxi.riken.jp/mxondem}}.

\bibliographystyle{elsarticle-harv} 
\bibliography{ms_bh_astrosat}


\appendix

\section{Summary of supplementary information}

\begin{table*}[h]
\begin{center}
\caption{Summary of UVIT Observations for BH-XRBs.} \label{uvit_summary}
\begin{tabular}{llll}
\hline																		
Source Name	& Observation Date & Filters & Remarks	\\	\hline
LMC X$-$1 &	2020-01-07	& FUV F2 & Imaging \\
          & 2020-03-31  & FUV F2 & Imaging \\	\hline 		
LMC X$-$3 &	2016-08-14	& NUV F2; FUV F1 & Imaging \\
		   & 2016-09-20	 & NUV F2; FUV F1 &	Imaging	\\		
          & 2017-04-03  &	NUV	F5;	FUV	F1 & Imaging \\
          & 2017-04-19  & NUV F5; FUV F1 & Imaging \\
          & 2017-11-17  & NUV F5; FUV F1 & Imaging \\
          & 2017-12-18  & NUV F5; FUV F1 & Imaging \\
          & 2018-02-06  & FUV F1 & Imaging \\
          & 2020-01-13  & FUV	F1 & Imaging \\ \hline 
          
GX 339$-$4 & 2021-03-05 & FUV F5 & Non-detection \\						   
           & 2021-03-31	& FUV F2, F4(G), F6(G) & Non-detection \\			   
           & 2022-09-11	& FUV F1, F2, F3, F5   & Non-detection \\
           & 2024-03-11 & FUV F1, F2, F3, F5	&Non-detection	\\	\hline 
       
MAXI J1820$+$070 & 2018-03-31 & FUV F4(G) &	Spectroscopy	\\
                & 2018-08-25 &	FUV	F4(G), F6(G) & Spectroscopy	\\ \hline 																		
IGR J17091$-$3624 &	2022-04-02 & FUV F5	& Non-detection	\\	\hline 																		
MAXI J0637$-$430 & 2019-11-20 &	FUV	F5 & Non-detection	\\
                & 2019-11-21 &	FUV	F5 & Non-detection	\\
                & 2019-11-23 &	FUV	F5 & Non-detection	\\ \hline 																		
GRS 1758$-$258	& 2017-07-28 & NUV F2, F6; FUV F5 &	Non-detection	\\	\hline  
\end{tabular}
\end{center}
\end{table*}

\begin{table*}[h]
\begin{center}
\caption{Parameters associated with physical modelling of the soft state sources (LMC X$-$1 and LMC X$-$3) with \texttt{kerrbb} and hard/intermediate state sources (GRS 1758$-$258, 1E 1740.7$-$2942, and MAXI J0637$-$430) with \texttt{relxillCp}. Uncertainty of model parameters are presented with $90\%$ confidence intervals, while frozen parameters are listed without uncertainties. 
}
\label{tab2}
\begin{tabular}{|l|lllll|}
\hline
Parameters  & LMC X$-$1 & LMC X$-$3 & GRS 1758$-$258 & 1E 1740.7$-$2942 & MAXI J0637$-$430  \\ \hline
& \texttt{kerrbb}  & \texttt{kerrbb}   & \texttt{relxillCp}    & \texttt{relxillCp}  & \texttt{relxillCp}      \\ \hline
Spin $a_{\rm k}$ & $0.890 \pm 0.020$ & $0.300 \pm 0.010$ & $0.998$ & $0.998$ & $0.960^{+0.020}_{-0.730}$     \\
Inclination angle $i^{\circ}$ &   $36$ (a)      &   $69$ (b)     &    $61\pm2$      &    $55\pm1$  &   $46$ (c)     \\
Black hole mass $M_{\rm BH}$ (M$_{\odot}$)       &    $8.82$ (d)     &  $6.48$ (d)    &      -        &     - &     -      \\
Mass accretion rate$^{\dagger}$ $\dot{M}$ (g s$^{-1}$)   &   $1.46\pm0.05$    &   $3.44\pm0.02$     &      -        &      -         & -   \\
Distance $d$ (kpc)   &    $48$ (a)    &    $48$ (a)    &       -      &      -       &     -   \\  
Ionization log $\xi$  (erg cm s$^{-1}$)      &     -    &   -      &     $3.17\pm0.04$   &   $3.17\pm0.06$  &  4.03$\pm$0.06    \\
Iron abundance $A_{\rm Fe}$ ($A_{\rm Fe, \odot}$)       &    -     &   -      &     $3.89\pm0.30$  &      $2.00$ (e)      &  $1.00$   \\
Spectral index $\Gamma^{\dagger}$               &    -     &      -   &      $1.74\pm0.01$   & $1.76\pm0.01$ & $2.01^{+0.21}_{-0.16}$     \\
Electron temperature $kT_{\rm e}$ (keV)         &     -    &     -    &  $3.03^{\dagger}\pm0.03$      &   $10.00-50.00$      & $60.00$     \\
Reflection fraction $R_{\rm f}$         &   -      &     -    &      $3.96\pm0.09$     &    $3.21\pm0.08$   &  $5.35^{+0.48}_{-8.56}$     \\ \hline
\end{tabular}
\begin{tablenotes}
	\item[] $^{\dagger}$Averaged over considered observations. (a) \citet{Orosz-etal2009}, (b) \citet{Orosz-etal2014}, (c) \citet{Jia-etal2023}, (d) \citet{Bhuvana-etal2021}, (e) \citet{Reynolds-Miller2010}
	\end{tablenotes}

\end{center}
\end{table*}

\begin{table*}[h]
\begin{center}
    \footnotesize
	\caption{Reflection modelling of simultaneous SXT and LAXPC data for the source 4U 1543$-$47 using model \texttt{tbabs$\times$(diskbb+relxilllp)$\times$gabs}. Error represents at $90\%$ confidence interval.}
	\label{tab:4u-reflec-param}
	\setlength{\tabcolsep}{1.8pt} 
	\renewcommand{\arraystretch}{1.8} 
	\begin{tabular}{ccccccccccccccc}  
		\hline 
 	\multirow{2}{*}{Epoch} & \multirow{2}{*}{$n_{\rm H}$} &\multicolumn{2}{c}{\texttt{diskbb}}  & \multicolumn{7}{c}{\texttt{relxilllp}} &\multicolumn{3}{c}{\texttt{gabs}} & \multirow{2}{*}{$\chi^2_{\rm red}$}\\
	\cline{3-14} 
    &($\times 10^{22}$ &$kT_{\rm in}$ & \textit{norm} & \textit{h} & $i^{f}$  & $\Gamma$ & $\log$ $\xi $ & $A_{\rm Fe}$ & $R_{\rm f}$ &  \textit{norm} & \textit{line E} & $\sigma$ & \textit{strength} & \\
    & $\rm atoms ~cm^{-2}$) & (keV) & ($ \times 10^3$) & $(GM/c^2)$ & (deg) & & (erg cm s$^{-1}$) & ($A_{\rm Fe,\odot}$) & &  ($\times 10^{-3}$)&(keV) & (keV)& (keV) & \\ 
		\hline

AS1 & $0.46^{+0.02}_{-0.01}$ & $1.01\pm0.01$ & $11.70^{+0.76}_{-0.44}$ & $30.0^{f}$ & $40$ & $3.00^{f}$ & $4.70^{b}$ & $9.4^{b}_{-0.4}$ & $5.59^{+1.85}_{-3.48}$ & $0.32\pm0.08$ & $9.37^{+0.27}_{-0.13}$ & $0.75^{+0.37}_{-0.64}$ & $0.48^{+0.03}_{-0.19}$ &  $1.4$ \\
 
AS2 & $0.49^{+0.01}_{-0.02}$ & $0.99^{+0.07}_{-0.01}$ & $9.23^{+1.46}_{-0.19}$ & $100.0^{f}$ & $40$ & $3.00^{f}$ & $4.30_{-0.21}^{+0.21}$ & $10.0^{b}$ & $7.18^{a}$ & $7.11^{+1.95}_{-4.1}$ & $10.00^{+0.13}_{-0.14}$ & $1.30\pm0.13$ & $1.39^{+0.28}_{-0.11}$  & $1.3$ \\

AS3 & $0.53^{+0.03}_{-0.01}$ & $0.99\pm0.003$ & $6.94^{+0.16}_{-0.27}$ & $8.3^{a}$ & $40$ & $2.46^{+0.36}_{-0.25}$ & $3.70^{+0.75}_{-0.68}$ & $10.0^{b}$ & $9.99^{a}$ & $5.75^{+0.55}_{-0.69}$ & $10.50^{+0.16}_{-0.14}$ & $1.45^{+0.20}_{-0.17}$ & $1.65^{+0.30}_{-0.25}$ & $1.4$ \\

AS4 & $0.44^{+0.03}_{-0.01}$ & $0.98\pm0.004$ & $4.95^{+0.10}_{-0.90}$ &  $70.0^{f}$& $40$ & $2.87^{+0.33}_{-0.24}$ & $4.70^{b}$ & $10.0^{b}$ & $8.10^{a}$ & $1.33^{+0.28}_{-0.12}$ & $10.20^{+0.13}_{-0.12}$ & $1.78^{+0.16}_{-0.19}$ & $2.76^{+0.36}_{-0.38}$  & $1.2$ \\

AS5 & $0.47\pm0.01$ & $0.94^{+0.002}_{-0.003}$ & $5.90^{+0.08}_{-0.09}$ & $44.3^{a}$ & $40$ & $2.25^{f}$ & $3.64^{+0.09}_{-0.18}$ & $3.6^{+0.6}_{-0.5}$ & $8.10^{a}$ & $1.11^{+0.07}_{-0.16}$ & $9.22^{+0.21}_{-0.18}$ & $0.84^{+0.20}_{-0.21}$ & $0.74^{+0.10}_{-0.13}$ & $1.1$ \\

AS6 & $0.48\pm0.01$ & $0.91\pm0.002$ & $7.95^{+0.21}_{-0.12}$ & $6.7^{a}$ & $40$ & $2.61^{+0.08}_{-0.19}$ & $3.96^{+0.07}_{-0.26}$ & $10.0^{b}$ & $10.60^{a}$ & $0.14\pm0.01$ & $10.10_{-0.13}^{+0.13}$ & $1.03^{+0.12}_{-0.13}$ & $1.25\pm0.14$ & $1.2$ \\

		\hline
	\end{tabular}
	\begin{tablenotes}
	\item[] $^a$Parameter uncertainty can't be estimated. $^b$Parameter pegged to the hard limit. $^{f}$ Frozen parameters.\\
	\end{tablenotes}  
\end{center}
\end{table*}

\begin{figure*}[h]
    \centering
    \includegraphics[width=0.45\textwidth]{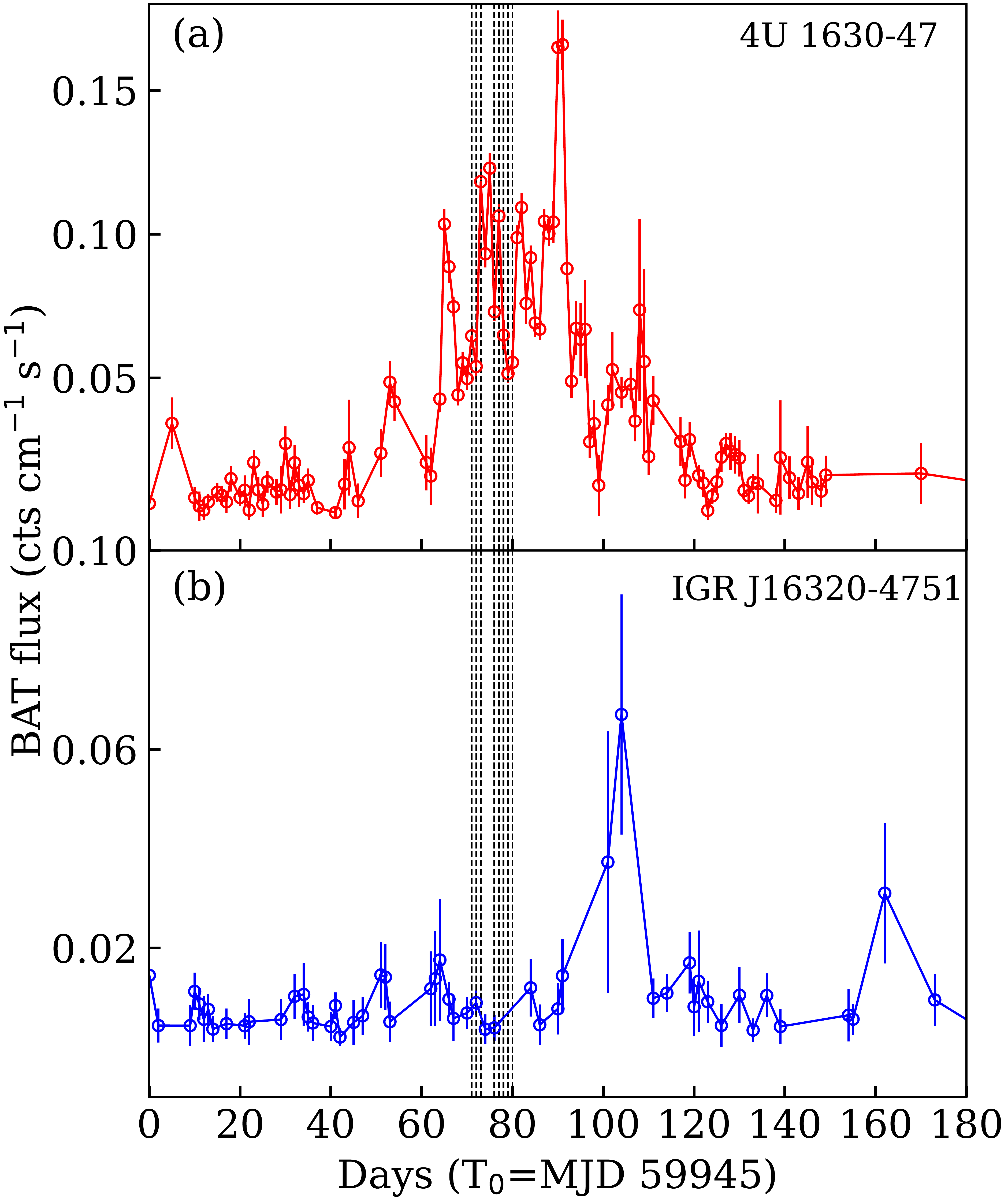}
    \includegraphics[width=0.45\textwidth]{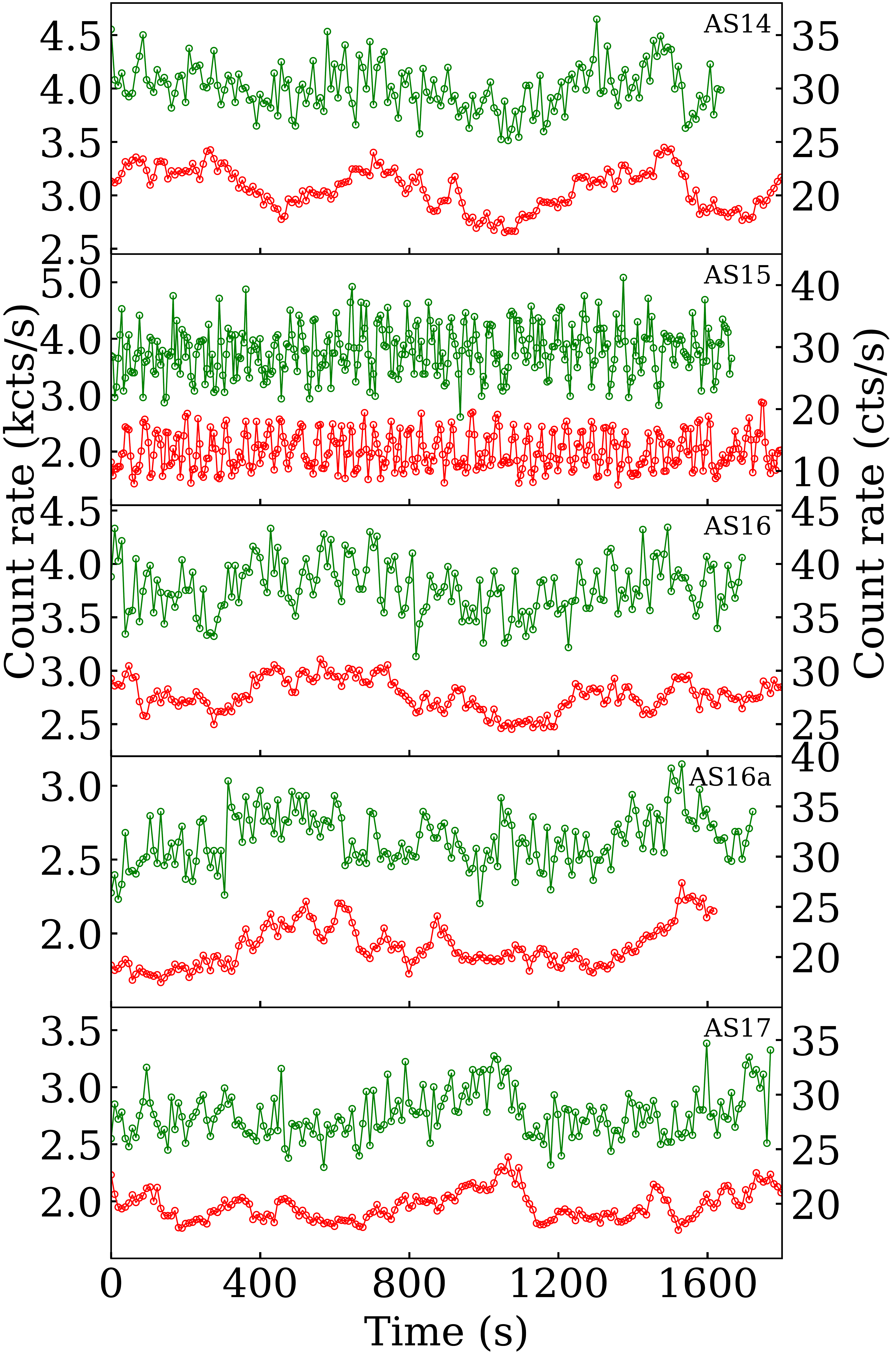}
    \caption{{\it Left:} Evolution of {\it Swift}/BAT flux in the 15$-$50 keV energy range, showing (a) the outburst profile of 4U 1630$-$47 during its 2023 outburst and (b) the simultaneous observation of IGR J16320$-$4751. The dashed lines indicate the epochs of {\it AstroSat} observations. {\it Right:} Lightcurve corresponding to AS14, AS15, AS16, AS16a (17th March 2023, not included in the tables of the manuscript) and AS17 of 4U 1630$-$47, where different variability patterns are observed during 2023 outburst. Lightcurves in green and red correspond to the SXT (right $y$-axis) and LAXPC (left $y$-axis) observations. See text for details.
    }
    \label{fig:fig39}
\end{figure*}

\begin{table*}[h]
	\centering
	\caption{Parameters of MAXI J1820$+$070 obtained from spectral fitting using \texttt{relxill} group of models. Error represents at $90\%$ confidence interval.}
	\label{tab:maxi-simult-refl}
        \setlength{\tabcolsep}{3.7pt} 
	\renewcommand{\arraystretch}{1.5} 
	\begin{tabular}{c c c c c c c c c c c c}  
		\hline
	\multirow{2}{*}{Epoch} & \multicolumn{2}{c}{\texttt{diskbb}}  & \multicolumn{8}{c}{\texttt{relxilllp}/\texttt{relxilllpCp}} & \multirow{2}{*}{$\chi^2_{\rm red}$}\\
	\cline{2-11} 
    & $kT_{\rm in}$ & \textit{norm} & \textit{h} & $i^\dagger$  & $\Gamma$ & log $\xi $ & $A_{\rm Fe}$ & $R_{\rm f}$ & $kT_{\rm e} $ & \textit{norm} & \\
    & (keV) & ($ \times 10^4$) & $(GM/c^2)$ & (deg) & & (erg cm s$^{-1}$) & ($A_{\rm Fe,\odot}$) & & (keV) & ($\times 10^{-2}$)& \\ 
		\hline
		AS1 & $0.33 \pm 0.02$ & $2.30_{-0.50}^{+0.60}$ & $100.0^{\dag}$ & $63$ & $1.54\pm0.01$ & 
$3.70_{-0.04}^{+0.07}$ & $4.4_{-0.8}^{+2.5}$ & $0.70_{-0.30}^{+0.20}$ & $100.00^{\dag}$ & $16.50_{-1.70}^{+2.80}$ & $1.5$ \\

		AS2$^*$ & $0.71 \pm 0.002$ & $1.91_{-0.03}^{+0.04}$ & $3.0 \pm 0.5$ & $63$ & $2.32_{-0.03}^{+0.02} $ & $3.09_{-0.03}^{+0.04}$ & $10.0^{\ddag}$ & $3.60_{-0.20}^{+0.30} $ & --- & $14.80_{-4.40}^{+8.80} $ & $1.6$  \\
        
		AS3$^*$ & $0.69 \pm 0.002$ & $0.55\pm0.01$ & $2.7 \pm 0.5$ & $63$ &  $3.12\pm0.01$  & $4.30\pm0.06$ & $10.0^{\ddag}$ & $3.59\pm0.37$ & --- & $15.80\pm1.00$ & $1.4$
 \\\hline
	\end{tabular}
	\begin{tablenotes}		
		\item[a] $^*$\texttt{relxilllp} model is used instead of \texttt{relxilllpCp} in HSS. 			
		\item[b] $^\dag$Frozen parameter. $^\ddag$Parameter pegged at hard limit.
		\end{tablenotes}  
\end{table*}

\begin{figure*}[h]
    \centering
    \includegraphics[width=0.47\textwidth]{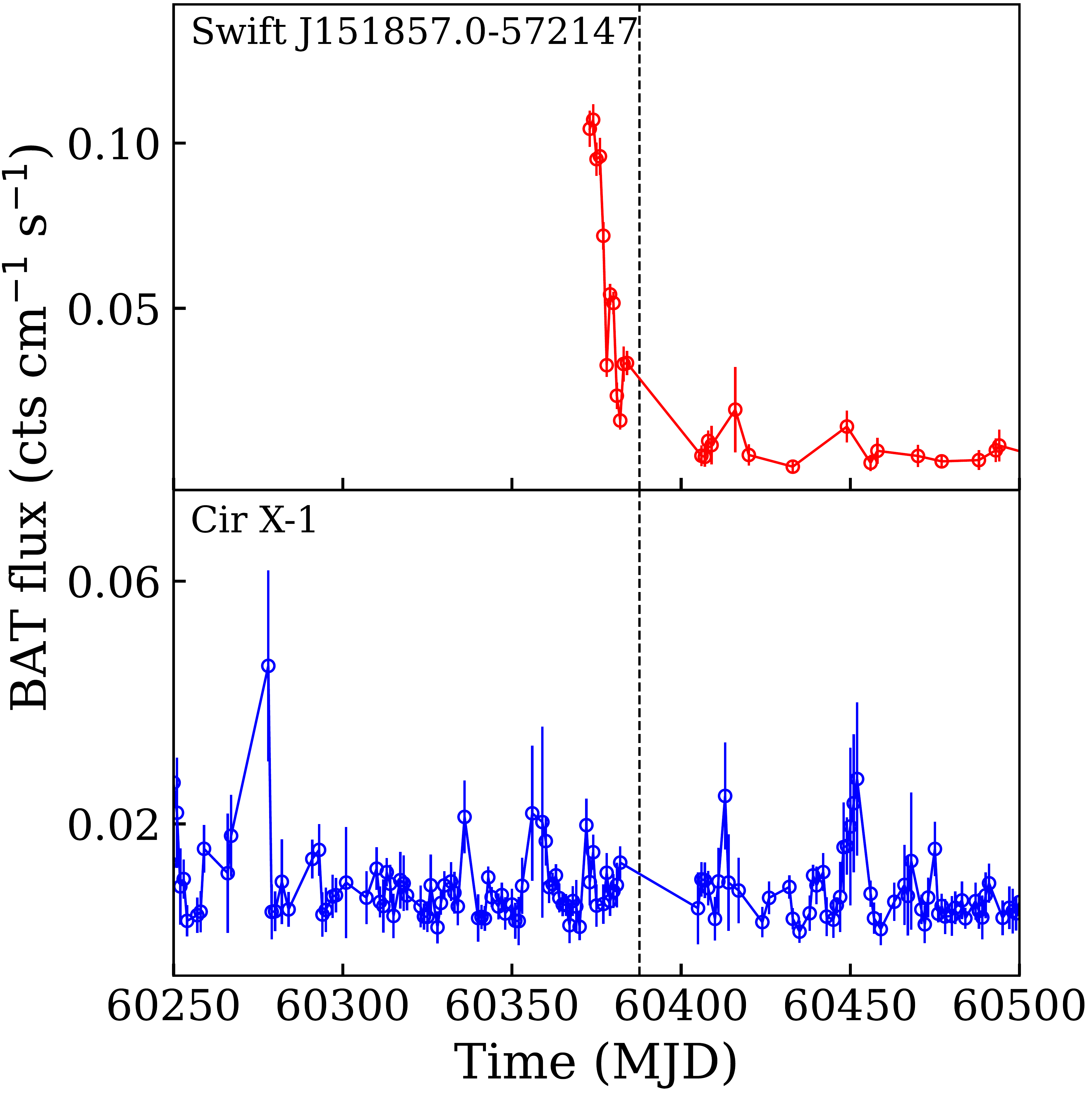}
    \includegraphics[width=0.47\textwidth]{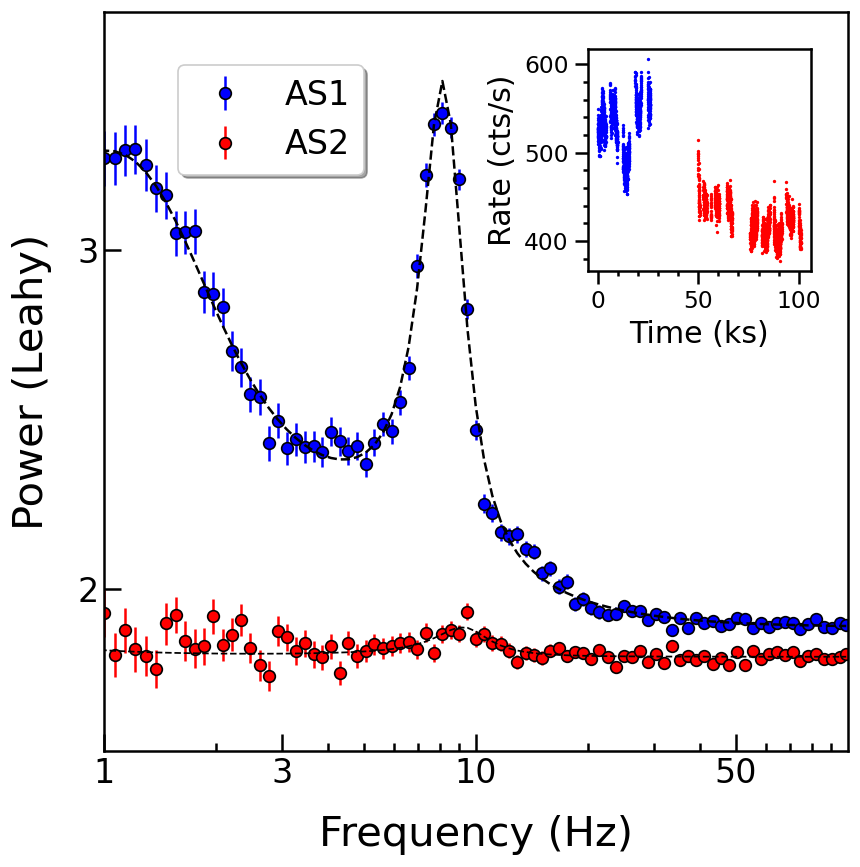}
    \caption{\red \textit{Left:} Hard X-ray ($15-50$ keV) lightcurves from the \textit{Swift}/BAT  monitoring of the outburst of Swift J151857.0$-$572147 (upper panel) and Cir X$-$1 (lower panel). The vertical dashed line marks the epoch of the \textit{AstroSat} observation of Swift J151857.0$-$572147. \textit{Right:} PDS of Swift J151857.0$-$572147 in the $3-25$ keV energy band. The inset shows the corresponding lightcurve with a time bin of $10$ s. Blue and red symbols denote data from AS1 and AS2, respectively. See text for details.
    }
    \label{fig:fig40}
\end{figure*}









\end{document}